\documentclass[amsart,11pt,oneside,english]{article}
\pdfoutput=1
\usepackage{amssymb,amsmath,amsfonts,mathtools,amsthm,euscript,mathrsfs,MnSymbol,verbatim,enumerate,multirow,bbding,slashed,multicol,color,array,esint,babel,tikz,
tikz-cd,tikz-3dplot,
pgfplots,ytableau,graphicx,float,rotating,hyperref,geometry,mathdots,savesym,cite,wasysym,amscd,graphicx,pifont,float,setspace,wrapfig,picture,subfigure,tabularx,datetime,cleveref,makecell,adjustbox,youngtab,parskip,cases,dynkin-diagrams,chronology,longtable}
\usepackage[normalem]{ulem} 
\usepackage[utf8]{inputenc}
\usepackage{prettyref}
\usepackage[T1]{fontenc}
\usepackage[comma,numbers,sort&compress]{natbib}
\numberwithin{equation}{section}
\usetikzlibrary{arrows,positioning,decorations.pathmorphing, decorations.markings, matrix, patterns, shapes}
\hypersetup{colorlinks=true}
\hypersetup{linkcolor=black}
\hypersetup{citecolor=black}
\hypersetup{urlcolor=black}
\theoremstyle{plain}
\newtheorem*{thm*}{Theorem}
\theoremstyle{plain}% default
\newtheorem{thm}{Theorem}[section]

\theoremstyle{definition}
\newtheorem{defn}[thm]{Definition}

\usepackage[tableposition=below]{caption}
\makeatletter
\newbox\LT@firstfoot
\def\endfirstfoot{\LT@end@hd@ft\LT@firstfoot}
\newdimen\LT@footdiff
\def\LT@start{%
  \let\LT@start\endgraf
  \endgraf\penalty\z@
  \vskip\LTpre\endgraf
  \LT@footdiff-\ht\LT@foot
  \advance\LT@footdiff\ht\LT@firstfoot
  \dimen@\pagetotal
  \advance\dimen@ \ht\ifvoid\LT@firsthead\LT@head\else\LT@firsthead\fi
  \advance\dimen@ \dp\ifvoid\LT@firsthead\LT@head\else\LT@firsthead\fi
  \advance\dimen@ \ht\ifvoid\LT@firstfoot\LT@foot\else\LT@firstfoot\fi
  \dimen@ii\vfuzz
  \vfuzz\maxdimen
  \setbox\tw@\copy\z@
  \setbox\tw@\vsplit\tw@ to \ht\@arstrutbox
  \setbox\tw@\vbox{\unvbox\tw@}%
  \vfuzz\dimen@ii
  \advance\dimen@ \ht
      \ifdim\ht\@arstrutbox>\ht\tw@\@arstrutbox\else\tw@\fi
  \advance\dimen@\dp
      \ifdim\dp\@arstrutbox>\dp\tw@\@arstrutbox\else\tw@\fi
  \advance\dimen@ -\pagegoal
  \ifdim \dimen@>\z@\vfil\break\fi
  \global\@colroom\@colht
  \ifvoid\LT@firstfoot
    \ifvoid\LT@foot
    \else
      \advance\vsize-\ht\LT@foot
      \global\advance\@colroom-\ht\LT@foot
      \dimen@\pagegoal\advance\dimen@-\ht\LT@foot\pagegoal\dimen@
      \maxdepth\z@
    \fi
  \else
    \advance\vsize-\ht\LT@firstfoot
    \global\advance\@colroom-\ht\LT@firstfoot
    \dimen@\pagegoal\advance\dimen@-\ht\LT@firstfoot\pagegoal\dimen@
    \maxdepth\z@
  \fi
  \ifvoid\LT@firsthead\copy\LT@head\else\box\LT@firsthead\fi\nobreak
  \output{\LT@output}%
}
\def\LT@output{%
  \ifnum\outputpenalty <-\@Mi
    \ifnum\outputpenalty > -\LT@end@pen
      \LT@err{floats and marginpars not allowed in a longtable}\@ehc
    \else
      \setbox\z@\vbox{\unvbox\@cclv}%
      \ifdim \ht\LT@lastfoot>\ht\LT@foot
        \dimen@\pagegoal
        \advance\dimen@-\ht\LT@lastfoot
        \ifdim\dimen@<\ht\z@
          \setbox\@cclv\vbox{\unvbox\z@\copy\LT@foot\vss}%
          \@makecol
          \@outputpage
          \setbox\z@\vbox{\box\LT@head}%
        \fi
      \fi  
      \global\@colroom\@colht
      \global\vsize\@colht   
      \vbox
        {\unvbox\z@\box\ifvoid\LT@lastfoot\LT@foot\else\LT@lastfoot\fi}%
    \fi
  \else
    \ifvoid\LT@firstfoot
      \setbox\@cclv\vbox{\unvbox\@cclv\copy\LT@foot\vss}%
      \@makecol
      \@outputpage
      \global\vsize\@colroom
    \else
      \setbox\@cclv\vbox{\unvbox\@cclv\box\LT@firstfoot\vss}%
      \@makecol
      \@outputpage
      \global\advance\@colroom\LT@footdiff
      \global\vsize\@colroom
    \fi
    \copy\LT@head\nobreak
  \fi
}
\makeatother
\newcommand{\sug}{$\text{SU}(2)\!\times\!\text{G}_2$}
 \newtheorem{Proposal}{Proposal}[section]

\numberwithin{equation}{section}

\tikzset{
  big arrow/.style={
    decoration={markings,mark=at position 1 with {\arrow[scale=1.5,#1]{>}}},
    postaction={decorate},
    shorten >=0.4pt},
  big arrow/.default=black}

\allowdisplaybreaks
\begin{document}

\begin{titlepage}
\vspace*{-2.2cm} 
\begin{flushright}
{\tt CALT-TH-2020-057}\\
\end{flushright}
\begin{center}
\vspace{1.1cm}
{\Huge\bfseries Matter representations from geometry:\\ 
under the spell of Dynkin 
\\  } 
\vspace{1cm}
{\Large
Mboyo Esole$^{\diamond}$ and Monica Jinwoo Kang$^{\clubsuit \spadesuit}$\\}
\vspace{.6cm}
{\large $^{\diamond}$ Department of Mathematics, Northeastern University}\par
{\large   Boston, MA 02115, U.S.A.}\par
\vspace{.2cm}
{\large $^\clubsuit$ 
Walter Burke Institute for Theoretical Physics, California Institute of Technology}\par
{  Pasadena, CA 91125, U.S.A.}\par
\vspace{.2cm}
{\large $^\spadesuit$ 
Department of Physics, Korea Advanced Institute of Science and Technology}\par
{  Daejeon 34141, Republic of  Korea}\par
\vspace{1.6cm}

{ \bf{Abstract}}\\
\end{center}

In the traditional Katz--Vafa method, matter representations are determined by decomposing the adjoint representation of a parent simple Lie algebra $\mathfrak{m}$ as the direct sum of irreducible representations of a semisimple subalgebra $\mathfrak{g}$. 
The Katz--Vafa method becomes ambiguous as soon as $\mathfrak{m}$ contains several subalgebras isomorphic to $\mathfrak{g}$ but giving different decompositions of the adjoint representation. 
We propose a selection rule that characterizes the matter representations observed in generic constructions in F-theory and M-theory: 
the matter representations in generic F-theory compactifications correspond to linear equivalence classes of subalgebras $\mathfrak{g}\subset \mathfrak{m}$ with Dynkin index one along each simple components of $\mathfrak{g}$. 
 This simple yet elegant selection rule allows us to apply the Katz--Vafa method to a much large class of models. 
We illustrate on numerous examples how this proposal streamlines the derivation of matter representations in F-theory and resolves previously ambiguous cases.

\vfill 
%{ \today\   at \ \currenttime\par}

\end{titlepage}

\tableofcontents
\newpage
\section{Introduction} \label{sec:KatzVafa}

All the known fundamental forces in theoretical physics are modeled by gauge theories. Some of the basic data of a gauge theory are its Lie group $G$ (called the gauge group) and its Lie algebra $\mathfrak{g}$. A gauge theory describes local forces propagating between particles, where the force is carried by a fundamental particle that interacts with itself when the gauge group is non-Abelian. In a given gauge theory, fundamental particles transform under irreducible representations of the gauge group that we call {\em matter representations}.  Determining  matter representations of a given model in theoretical physics  is a fundamental aspect of its description.  Characterizing matter representations occurring in M-theory and F-theory compactifications will be the focus of this paper.

In  string compactifications, matter representations are not arbitrary but are constrained by the the geometry of the compactifying space, in particular, its intersection ring, its singularities, and their degenerations  \cite{Sen:1997kz,Witten:1995ex,Sadov:1996zm}.
In F-theory \cite{Vafa:1996xn,Morrison:1996na,Weigand:2018rez}, the compactifying space is an elliptic fibration \cite{Esole:2017csj}.  In fact, certain weights of the matter representations are  observed geometrically as intersection numbers between fibral divisors and rational curves constituting the irreducible components of singular fibers over particular loci of the discriminant locus of the elliptic fibration \cite{Aspinwall:1996nk,Aspinwall:2000kf, Sen:1997kz,IMS,Hayashi:2014kca, Diaconescu:1998cn, Morrison:2011mb, Esole:2020alo,F4,G2,Marsano:2011hv,Grimm:2011tb}. However, when the gauge algebra is not simple or simply-laced, the subsets of  geometrically-derived weights do not always uniquely determine a representation of the Lie algebra, as pointed out in \cite{SU2G2,SU2SU3,EKY2}. In \cite{Katz:1996xe}, Katz and Vafa propose to deduce the matter representation by considering the decomposition of the adjoint representation in an embedding $\mathfrak{g}\to \mathfrak{m}$ as
 $$
\text{adj}\ \mathfrak{m}=\text{adj}\ \mathfrak{g}\oplus\chi,
 $$
 and  following Dynkin \cite{Dynkin.SubA}, we call $\chi$ the \textit{characteristic representation of $\mathfrak{g}$ in $\mathfrak{m}$}.
The Katz--Vafa method is beloved for its universality and computational ease, but becomes ambiguous  when the Lie algebra $\mathfrak{g}$ have distinct embeddings in $\mathfrak{m}$ leading to distinct representations. Moreover, to explain the observed matter representation, the choice of the isomorphic class of $\mathbf{m}$ is not always consistent with the geometry of the elliptic fibration. This begs for a clear  characterization of the matter representations appearing in M-theory or F-theory compactifications without imposing  any ad hoc rule and in perfect consistency with the restriction derived from the geometry. 

The cases studied in the original Katz--Vafa paper are the  following embeddings between  ADE Lie algebras\footnote{As we will explain later, in the first four cases, there are subtleties for  certain values of $n$ and $k$ that were never scrutinized before.}:
\begin{align}
\begin{cases}
& A_{n-k} \oplus A_{k-1}\   \   \to\    A_n, \qquad
 D_{n-k}\oplus A_{k-1}\   \  \to\     D_n, 
 \qquad\quad\   D_{n -1}\  \   \to\     D_n, \\
& \qquad\quad\   A_{n-1}\  \  \to\     D_n, \qquad
 \qquad\qquad\   D_5\   \   \to\     E_6, 
 \qquad\qquad\  A_5\   \  \to\     E_6, \\
& \qquad\qquad\  E_6\   \  \to\     E_7, \qquad
 \qquad\qquad\  D_6\   \  \to\     E_7, 
 \qquad\qquad\  A_6\     \    \to\     E_7, \\
& \qquad\qquad\  E_7\   \    \to\     E_8.
\end{cases}
\end{align}
These original cases analyzed by Katz--Vafa have the following properties:
\begin{enumerate}
\item
{\em Rank-one enhancement between ADE Lie algebra }:
 The Lie algebra $\mathfrak{m}$ and $\mathfrak{g}\subset\mathfrak{m}$ are both Lie algebras of type ADE with $\mathfrak{m}$ a simple Lie algebra of one rank higher than $\mathfrak{g}$.
\item 
{\em Levi subalgebras}
: the Dynkin diagram of $\mathfrak{g}$ is obtained from that of $\mathfrak{m}$ by removing a single node. 
However, $\mathfrak{g}$ is not necessarily a maximal subalgebra of $\mathfrak{m}$. 
\item  {\em  Minuscule representations}: The non-trivial  irreducible components of the characteristic representation $\chi$ are all minuscule representations. 
\item   {\em regular embedding}: The embedding $\mathfrak{g}\to \mathfrak{m}$ is always a regular embedding. 
\item {\em Dynkin index one}: The Dynkin index of any irreducible component of $\mathfrak{g}$ in $\mathfrak{m}$ is always one. 
\end{enumerate}

There are many models outside of those considered in the original Katz--Vafa method that play a central role in F-theory. The geometry of these models have been studied carefully in the past few years \cite{G2,SU2SU3,F4,Euler,CharMW,Char,SU2G2,SO4,EKY2,EKY,ESY1,ESY2,EY, Esole:2020alo,Esole:2019rzq, USP4,SO356, ES,Chernchar,Esole:2018vnm,Kuramochi:2020jzz,Gu:2020fem, Bhardwaj:2018yhy,Grimm:2011tb,Marsano:2011hv,Kan:2020lbe,Anderson:2017zfm} and they provide valuable and explicit data that will guide us to extend and improve the Katz--Vafa method on determining matter representations. The previous analysis of these models enlightened us that there is no reason to restrict ourselves to rank one enhancements, Levi subalgebras, (quasi)-minuscule representations, or regular embeddings. 
For example, higher rank enhancements occur routinely in well-known F-theory constructions such as those with the following enhancements:
$\text{A}_1\to \text{D}_4$, $\text{A}_1\oplus \text{G}_2\to \text{E}_7$, and  $\text{A}_1\oplus \text{C}_2\to \text{A}_5$ that are individually observed respectively in the SU($2$)-model \cite{ESY1}, the SU($2$)$\times$G$_2$-model \cite{SU2G2}, and the SU($2$)$\times$USp($4$)-model \cite{EKY2}. 

In this paper, we propose a simple (mathematical) selection rule to identify the correct matter representations in generic compactifications of M-theory and F-theory. 
Our approach is rooted in three fundamental ideas introduced by Dynkin in the middle of the 20th century: the notions of linear equivalence, characteristic representations, and the Dynkin embedding index of a subalgebra \cite{Dynkin.SubA,deGraaf,Minchenko,LorenteGruber}. 
We argue that the overwhelming majority of matter representations in F-theory can be identified as the ones of subalgebras  $\mathfrak{g}\to \mathfrak{m}$ of Dynkin embedding index one (see definitions in Section \ref{sec:prelim}). The physical origin of this restriction will be presented in a companion paper \cite{Chernchar}.

The organization of the rest of the paper is as follows. 
In Section \ref{sec:Intro3}, we explain the notions of singleton subalgebras, Dynkin index of embeddings and announce our selection rule for identifying the correct matter representations of M-theory and F-theory. We also give an overview of our methodology.
In Section \ref{sec:Need}, we illustrate  through explicit examples the power of this approach.
In Section  \ref{sec:prelim}, we review basic mathematical definitions and conventions. 
In Section \ref{sec:original}, we describe the relationship between the Katz--Vafa method and Dynkin's characteristic representation and then revisit the original Katz--Vafa cases with our  approach. We separate them into singleton and non-singleton cases and show how all the choices coincide with our selection rule. 
In the rest of the paper, we carefully analyze several cases that are well understood via crepant resolutions of singularities of Weierstrass models.

\subsection{Singleton subalgebras, Dynkin index of embeddings, and the proposal}\label{sec:Intro3}

 We take the following point of view inspired by the work of Dynkin \cite{Dynkin.MSG,Dynkin.SubA}, Minchenko \cite{Minchenko}, de Graaf \cite{deGraaf}, and Lorente--Gruber \cite{LorenteGruber}.
A semisimple Lie algebra $\mathfrak{g}$ can be embedded in many different ways as a subalgebra of a simple Lie algebra $\mathfrak{m}$. Each embedding can lead to a distinct characteristic representation. Thus, it is important to identify an appropriate notion of equivalence of subalgebras.
In his classification of semisimple subalgebras of simple Lie algebras, Dynkin has introduced two notions of equivalence of subalgebras: the equivalence and the linear equivalence. 
Two subalgebras are said to be conjugate or {\it equivalent} if they are related by an inner automorphism of the parent Lie algebra. 
Two subalgebras $\mathfrak{g}_1$ and $\mathfrak{g}_2$ of $\mathfrak{m}$ are said to be {\em linearly equivalent} if every linear representation induces linear conjugate representations of $\mathfrak{g}_1$ and $\mathfrak{g}_2$. Two equivalent subalgebras are always linearly equivalent but the reverse is not always true.  Two linearly equivalent embeddings have the same characteristic representation and the same Dynkin index. 
Since our focus is on branching rules, the coarser notion -- the linear equivalence -- is the appropriate notion for our purpose. We revisit this in detail in Section \ref{sec:equivalence}.

Our proposal can be put succinctly as following.\smallskip

\begin{Proposal}[Matter Representation Selection Rule] \label{SR}  
 In generic configurations in F-theory, up to multiplicities and trivial representations, the  matter representations with respect to a Lie algebra $\mathfrak{g}$  are given by the characteristic representations of embeddings $\mathfrak{g}\to\mathfrak{m}$ having Dynkin index one along each simple components of $\mathfrak{g}$. 
\end{Proposal}

The Dynkin embedding index of a simple subalgebra $\mathfrak{g}$ of a simple Lie algebra $\mathfrak{m}$ was introduced by Dynkin in his classification of semisimple subalgebras of  Lie algebras \cite{Dynkin.SubA}. 
The Dynkin index naturally extends to the case of semisimple subalgebras of a simple Lie algebra by giving the Dynkin index of each of its simple components. 
 We say that a semisimple Lie algebra $\mathfrak{g}=\bigoplus_i \mathfrak{g}_i$ has {\em Dynkin embedding index one} if each of its simple components $\mathfrak{g}_i$ have Dynkin embedding index one in $\mathfrak{m}$.

By ``generic configuration,'' we mean those appearing from Tate's algorithm with gauge groups supported on generic divisors. In particular, these divisors are assumed to be smooth.  
In algebraic geometry, a generic divisor is always smooth. Smooth divisors occur naturally in physics such as in non-Higgsable clusters. In particular, non-singular divisors can occur and produce exotic configurations that do not have to respect the proposed selection rule. We will list specific examples.

This selection rule applies to a broad collection of theories: all generic Tate's models \cite{Bershadsky:1996nh,GM1}, Miranda models \cite{Miranda.smooth,Bershadsky:1996nu}, and more general collisions of singularities without further assumptions \cite{SO4,EKY2,SU2G2,SU2SU3}, even for non-simply-laced Lie algebras. We work out various examples explicitly and meticulously. This selection rule has a nontrivial predictive power, since the subalgebras of Dynkin index one is typically unique in a given isomorphism class of subalgebras $\mathfrak{g}$.\footnote{There are very rare cases when it is not unique; we analyze all the various.} It provides a simple way to identify the correct matter representations in many situations that previously were only done case by case using ad hoc rules.

Our strategy can be summarized as follows.

\begin{itemize}

\item Subalgebras that are unique up to linear equivalence within their semisimple type are particularly nice, as there is no possible ambiguity for the choice of a matter representation since there is a unique possible reduced matter representation. 
We call them {\em singleton subalgebras}. 

\item The next best situation is when a subalgebra is not a singleton subalgebra, but all the non-linearly equivalent subalgebras of the same isomorphism type give the same characteristic representation for the adjoint representation. In such a case, the matter representation is still not ambiguous. We call them {\em adjoint singleton subalgebras}. 
Any singleton subalgebra is also an adjoint singleton subalgebra, but the reverse is not true. 

\item When we are outside of the class of adjoint singleton subalgebras, by definition, 
we have several possible representations to choose from with different branching rules for the adjoint representation. 
In those cases, we need a criterion to identify which linear equivalent class, within a given semisimple type of subalgebra, gives the characteristic representation we expect in F-theory. We argue that in the generic case, F-theory chooses the branching rule that corresponds to  a subalgebra of Dynkin index one along each of its simple factor. 
\end{itemize}

With very rare exceptions listed in \cite{GEO}, adjoint singleton subalgebras are of Dynkin index one. 
A naturally  question is then if there exists any situation in which adjoint singleton subalgebras of higher Dynkin index appear naturally in F-theory or string theory. It would also be interesting to know if  the subalgebras of index one that are not yet used in F-theory could give rise to a special class of matter representations that might be relevant for models of which we are unaware yet. In that case, it would be important to classify all of them systematically.

We note that some subalgebras that are not of Dynkin index one do appear in exotic configurations in F-theory when the divisor supporting the gauge group has singularities \cite{Klevers:2017aku}. In fact, they also appear in some aspects of string theory. For example, we also notice that the bosonic string compactified on the self-dual circle has a gauge enhancement with Lie algebra $\text{A}_1\oplus\text{A}_1$ and matter in the representation $(\mathbf{3},\mathbf{3})$ corresponding to the branching rule $\text{A}^2_1\oplus\text{A}^2_1\to \text{A}_3$, which we explain in Section \ref{sec.bosonic}. 
The characteristic representation of the subalgebra  $\text{A}^1_1\oplus\text{A}^2_1$ of $\text{D}_4$ matches (up to multiplicity) the matter content of the $\text{A}_1\oplus\text{A}_1$ theory discussed  in   \cite[Appendix A]{Ohmori:2015pia}.
 Moreover, in a footnote of \cite[{\S 3.3}]{Intriligator:1995id},  Intriligator and Seiberg point out  that in the study of spontaneous symmetry breaking in supersymmetric gauge theories, a higher Dynkin embedding index introduces subtleties in dealing with instantons. 
 We will come back to this important point  in some detail   in a companion paper \cite{Chernchar}.

\subsection{Application to  the birational geometry of elliptic fibrations}

Matter representations also have a natural application in purely geometric constructions \cite{IMS,Hayashi:2014kca,ESY1,ESY2,EJJN1,EJJN2}. 
A given Weierstrass model can have several distinct crepant resolutions connected by flops. 
Using hints from M-theory compactifications to 5d theories \cite{IMS,Hayashi:2014kca}, the network of these flops can be modeled by a hyperplane arrangement I$(\mathfrak{g},\mathbf{\chi})$ \cite{ESY1,ESY2,EJJN1,EJJN2}, whose open chambers are defined by the connected components of the  fundamental Weyl chamber of $\mathfrak{g}$ after removing the hyperplanes defined by the weights of the characteristic representation $\mathbf{\chi}$. 
The coincidence graph of the chambers of the hyperplane arrangement I$(\mathfrak{g},\mathbf{\chi})$ is expected to be isomorphic to the graph of crepant resolutions connected by flops. When the elliptic fibrations has several distinct matter representations in different locations, we take the direct sum of all the corresponding characteristic representations. 
The weights observed from the geometry by intersection numbers of curves and divisors over codimension-two points in the base are those responsible for the wall between distinct chambers. 
We refer to \cite{IMS, Diaconescu:1998cn, Hayashi:2014kca, G2,SU2SU3,F4, Euler,ES, EFY,CharMW,Char,SU2G2,SO4,EKY2,EKY,ESY1,ESY2,EY,EJJN1,EJJN2,USP4,SO356,Esole:2020alo,Esole:2019rzq,Esole:2018vnm,Chernchar} for explicit examples and applications.

This brings up an interesting fundamental question: given a simple Lie algebra $\mathfrak{g}$, which are the subalgebras $\mathfrak{m}$ up to linear equivalence, is it possible to determine the associated hyperplane arrangement I$(\mathfrak{g},\mathbf{\chi})$ matching the flop diagram of an elliptic fibration with dual fibers corresponding to the simple components of $\mathfrak{g}$? For example, as we will review in the next subsection, in the study of the G$_2$-model \cite{G2}, one would expect an embedding $\text{G}_2\to \text{E}_6$. 
There are  two subalgebras isomorphic to G$_2$ in E$_6$, one would give $\chi=\mathbf{7}$ and the other $\chi=\mathbf{64}$. 
We know that the G$_2$-model does not flop, which is consistent with the fact that I(G$_2$, $\mathbf{7})$ has only one chamber. In contrast, 
I(G$_2$, $64$) has two distinct  chambers and is therefore not consistent with the geometry of the G$_2$-model even though all the weights observed geometrically are both weights of $\mathbf{7}$ and $\mathbf{64}$.

\subsection{Evidence for the proposed selection rule}\label{sec:Need}

In this section, we  illustrate that the evidences for the selection rule presented in Proposal \ref{SR}  are overwhelming. Interestingly, this selection rule is useful even in cases that might sound at first non-ambiguous such as the matter content obtained when A$_2$ is reduced to A$_1$. 
 Our aim here is not to exhaust all possible cases but to present the main ideas through key examples. We will give a more exhaustive approach in \cite{GEO}. 

\begin{enumerate}
\item The ambiguity in choosing the correct branching rule is already apparent in the original Katz--Vafa cases when we focus on boundary cases that give non-singleton subalgebras. Namely,
\begin{align}
\text{A}_1 & \to \text{A}_2, \quad
\text{A}_1\oplus \text{A}_1 \to   \text{A}_3,\quad
A_3\oplus A_{1} \to \text{D}_5,\quad
A_3\oplus A_{1}\oplus A_1 \to \text{D}_6,\quad 
A_3\oplus A_{3} \to \text{D}_7.
\end{align}
In each case, there are multiple possible embeddings giving distinct matter representations and the one chosen in F-theory always corresponds to the one of Dynkin index one. 
The elliptic fibrations for the SU($2$)-model and the SU($2$)$\times$SU($2$)-model are respectively studied in \cite{ESY1, ES}  and \cite{SO4}. 
The characteristic representations for these cases are treated in detail in Section \ref{sec:boundaryKV}.

In the original Katz--Vafa's list, we have the following case 
\begin{equation}
\text{A}_{n-1}\to \text{D}_n.
\end{equation}
This embedding is  ambiguous for  $n=3$ as already mentioned. We should also distinguish the cases when $n=4$, $n=2k+1$ ($k\geq 2$) and the case $n=2k$ with $k\geq 3$. 
Putting them together, we have to consider four distinct cases
\begin{align}
& \text{A}_2\to \text{A}_3, \quad \text{A}_3 \to \text{D}_4, \quad
\text{A}_{2k-1} \to \text{D}_{2k}, \quad
 \text{A}_{2k}\to \text{D}_{2k+1}.\nonumber 
\end{align}
\begin{enumerate}
\item A$_2$ is a singleton subalgebra of A$_3$ discussed in Section \ref{A3D4}. 
\item A$_3$ has  three distinct non-maximal embeddings in D$_4$ connected by triality, all three  have Dynkin index one. 
\item A$_{2k-1}$ has two distinct maximal embeddings in D$_{2k}$, both have index one and give the same characteristic representation, the same branching rule for the vector representation,  but different branching for the half-spin representations. 
They are related by the outer automorphism of D$_{2k}$ that permutes the two half-spin representations. 
Thus, these subalgebras are different but give the same characteristic representation. They are adjoint singleton subalgebras.
\item  A$_{2k}$ a singleton subalgebra of  D$_{2k+1}$ and has Dynkin  index one. It gives a unique possible characteristic representation. 
\end{enumerate}

\item \begin{enumerate}

\item In the original Katz--Vafa cases, the embedding $A_7\to E_8$ is absent in the considered embeddings, while $A_6\to E_7$ and $A_5\to E_6$ are explicitly discussed. From the  perspective and methodology in this paper, this can be understood as follows. While A$_6$ and A$_5$ are singleton subalgebra respectively of $E_7$ and $E_6$,  there are two distinct subalgebras of type  A$_7$ in E$_8$ up to linear equivalence. Moreover, they both have Dynkin index one and yield distinct matter representations. 
It follows that identifying the branching rule for the adjoint of E$_8$ along A$_7$ is ambiguous without further requirement. 

\item The existence of multiple isomorphic subalgebras that are not linearly equivalent but all of embedding index one is expected when one thinks of the possibility that they can be Lie algebras of distinct groups sharing the same simply connected cover. In that case, they differ by taking different discrete quotients of the same underlying connected group. Thus, they might also have distinct characteristic representations. 
We illustrate this point with the two  A$_7$   subalgebras in E$_8$. 
Their characteristic representations are:  
\begin{align}
&  \text{A}_7^{1} \to \text{E}_8  :  \quad  \chi=\mathbf{8}\oplus\mathbf{\overline{8}}\oplus\mathbf{28}\oplus \mathbf{\overline{28}}\oplus\mathbf{56} \oplus\mathbf{\overline{56}}\oplus\mathbf{1},\\
&  \text{A}_7^{1} \to \text{E}_8  :  \quad  \chi=\mathbf{28}^{\oplus 2} \oplus  \mathbf{\overline{28}}^{\oplus 2}\oplus\mathbf{70} \oplus\mathbf{1}^{\oplus 3},
\end{align}
where $\mathbf{8}=V_{\text{A}_7}$ is the first fundamental representation, $\mathbf{28}=\Lambda^2_{\text{A}_7}$, $\mathbf{56}=\Lambda^3_{\text{A}_7}$, and $\mathbf{70}=\Lambda^4_{\text{A}_7}$. 
 The first one factors via A$_7^1$ and the second one factors via E$^1_7$. Both are subalgebras of the maximal subalgebra D$_8$ of E$_8$. 
 The existence of these two options can be understood by the existence of distinct compact group with Lie algebra A$_7$ in E$_8$. 
 Using Borel--Siebenthal techniques, we can show that the first A$^1_7$ is the Lie algebra of a simply-connected group SU($8$)  while the second is the Lie algebra of the quotient  group SU($8$)/$(\pm 1)$. 
 The maximal A$_8$ subalgebra of E$_8$ corresponds to the group SU($9$) mod by its $\mathbb{Z}_3$ subgroup while 
 $\text{A}_1\oplus E_7$ corresponds to the subgroup $(\text{SU}(2)\otimes E_7)/ \pm (1,1)$.  The presence of the non-trivial center suggests that in F-theory, we will be dealing with an elliptic fibration with a non-trivial torsion sector in the Mordell--Weil group. 
\end{enumerate}

\item A poster child for our selection rule appears in the specialization of the $\text{SU}(2)\times\text{SU}(3)$-model where there is an enhancement to an (incomplete) fiber of type III$^*$ or II$^*$. 
SU($2$)$\times$SU($3$) models in which all the matter content are on the same points are characterized by an enhancement $\text{A}_1^{1}\oplus\text{A}_2^{1} \to \text{E}_7$ or 
$\text{A}_1^{1}\oplus\text{A}_2^{1} \to \text{E}_8$. They are given by collisions of the type III+IV$^{\text{s}}$ and their crepant resolutions are studied in \cite{SU2SU3}. 

\begin{enumerate}
\item There are 25 distinct $\text{A}_1\oplus \text{A}_2$ subalgebra inside E$_7$, but only one of them gives the matter content we expect, which corresponds to the one having Dynkin index $(1,1)$: 
\begin{align}
&    \text{A}_1^{1}\oplus\text{A}_2^{1} \to \text{E}_7  :  \quad  \chi=(\mathbf{2},\mathbf{3})^{\oplus 4}\oplus(\mathbf{2},\mathbf{\overline{3}})^{\oplus 4}\oplus (\mathbf{1},\mathbf{3})^{\oplus 7}\oplus(\mathbf{1},\mathbf{\overline{3}})^{\oplus 7}\oplus (\mathbf{2},\mathbf{1})^{\oplus 8}\oplus (\mathbf{1},\mathbf{1})^{\oplus 
16}.
\end{align}
 Thus, the proposed selection rule successfully identifies the correct matter representations without referring to any additional analysis, such as D-brane techniques or anomaly cancellation conditions. 

\item Up to linear equivalence, there are 55 distinct subalgebras of type $\text{A}_1\oplus \text{A}_2$ in E$_8$. However, only one of them have Dynkin embedding index $(1,1)$ and it gives the following characteristic representation: 
\begin{equation}
\text{A}_1^{1}\oplus\text{A}_2^{1} \to \text{E}_8  :  \quad
\chi=(\mathbf{2},\mathbf{3})^{\oplus 6}\oplus(\mathbf{2},\mathbf{\overline{3}})^{\oplus 6}\oplus (\mathbf{1},\mathbf{3})^{\oplus 15}\oplus(\mathbf{1},\mathbf{\overline{3}})^{\oplus 15}\oplus (\mathbf{2},\mathbf{1})^{\oplus 20}\oplus (\mathbf{1},\mathbf{1})^{\oplus 
35}.
\end{equation}

\item As discussed in Section \ref{A4Excep},  the Lie algebra A$_4$ is a singleton subalgebra of index one for both E$_6$ and E$_7$. In the study of the SU($5$)-model, one may also consider specializations in which all the matter appear at points of E$_8$ enhancements. There are exactly two subalgebras of type A$_4$ in E$_8$ up to linear equivalence. One has Dynkin index 1 and the other one has Dynkin index 2. Only the one with Dynkin index one gives the correct matter expected in F-theory and in SU($5$) Grand Unified theories: 
\begin{align}
& \text{A}_4^1\to \text{E}_8:\quad   \chi=
\mathbf{5}^{\oplus 10}\oplus
 \mathbf{\overline{5}}^{\oplus 10}\oplus
 \mathbf{10}^{\oplus 5}\oplus
\mathbf{\overline{10}}^{\oplus 5}\oplus
\mathbf{1}^{\oplus 24}, \\ 
&  \text{A}_4^2\to \text{E}_8:\quad  \chi=
\mathbf{24}
\oplus
\mathbf{5}\oplus\mathbf{\overline{5}}\oplus
\mathbf{10}\oplus\mathbf{\overline{10}}\oplus
\mathbf{40}\oplus\mathbf{\overline{40}}\oplus
\mathbf{45}\oplus
\mathbf{\overline{45}}.
\end{align}
The subalgebra A$^2_4$ factors through the maximal semisimple algebra $\text{A}_4^1\oplus \text{A}_4^1$. 
\end{enumerate}

\item The study of crepant resolutions of Tate's models \cite{Bershadsky:1996nu,Miranda.Smooth,GM1,Bershadsky:1996nh}  provides a rich playground to test our selection rule. 
We add the following interesting list of subalgebras appearing in F-theory: 
\begin{equation}
\begin{cases}
& \text{A}_1\to \text{G}_2,\quad   \text{A}_1\to\text{B}_3, \quad  \text{A}_1\to\text{D}_4, \\
& \text{A}_2\to \text{G}_2,\quad  \text{A}_2\to\text{B}_3, \quad  \text{A}_2\to\text{D}_4, \\ 
& \text{G}_2\to \text{B}_3, \quad
\text{G}_2\to \text{D}_4 ,\quad
\text{G}_2\to \text{D}_5,\\
& \text{G}_2\to \text{F}_4,\quad  \text{G}_2\to \text{B}_4, \quad  \text{G}_2\to \text{E}_6,\\
& \text{C}_2\to \text{A}_3 ,\quad
\text{C}_2\to \text{B}_3 ,\quad
\text{C}_2\to \text{C}_3 ,\quad
\text{C}_2\to \text{A}_4 ,\quad
\text{C}_2\to \text{D}_4, \quad 
\text{C}_2\to \text{D}_5, \\
& \text{B}_3\to \text{D}_4 ,\quad
\text{B}_3\to \text{B}_4 ,\quad
\text{B}_3\to \text{D}_5 ,\quad
\text{B}_3\to \text{F}_4 ,\quad
\text{B}_3\to \text{E}_6, \\
& \text{D}_4\to \text{F}_4 ,\quad
\text{D}_4\to \text{E}_6, \\
& \text{F}_4\to \text{E}_6 ,\quad
\text{F}_4\to \text{E}_7,\quad \text{F}_4\to \text{E}_8, \\
 & \text{E}_6\to \text{E}_7 ,\quad
\text{E}_6\to \text{E}_8, \\
& \text{E}_7\to \text{E}_8, \\
& \text{A}_1\oplus \text{A}_1\to \text{C}_2,\quad   \text{A}_1\oplus \text{A}_1\to \text{G}_2,\quad    \text{A}_1\oplus \text{A}_1\to \text{B}_3,\\
 & \text{A}_1\oplus \text{A}_1\to \text{D}_4,\quad    \text{A}_1\oplus \text{A}_1\to \text{B}_4, \quad  \text{A}_1\oplus \text{A}_1\to \text{D}_5.
\end{cases}
\end{equation}

\item In particular, one of the key examples that motivated this paper is the treatment of non-simply-laced cases. We effectively answer a problem that was raised in the papers \cite{G2,F4}. 
The crepant resolutions of the F$_4$ and G$_2$-model suggest the following embeddings: 
\begin{align}
& \text{F}_4\to  \text{E}_6, \text{E}_7, \text{E}_8,\\
& \text{G}_2\to\text{B}_3,  \text{D}_4, \text{D}_5, \text{D}_6,\\
& \text{B}_3\to \text{D}_4,  \text{F}_4, \text{E}_6, \\
& \text{C}_2 \to \text{A}_3, \text{A}_4, \text{B}_3, \text{C}_3, \text{D}_4, \text{D}_5 .
\end{align}
\begin{enumerate}

\item Since F$_4$ is a singleton subalgebra of E$_6$, E$_7$, and E$_8$, there is no ambiguity in determining the matter content of an F$_4$-model. In each case, we end up with the matter in the representation $\mathbf{26}$. 

\item For the G$_2$-model, the enhancement $\text{G}_2\to \text{E}_6$ can be realized in two distinct ways leading to two distinct characteristic representations, namely, the representation $\mathbf{7}$ and the representation $\mathbf{64}$. 
The one with Dynkin index one gives the representation $\mathbf{7}$ as expected. 

\item In the case of the $\text{B}_3$-model \cite{G2}, there are two possible enhancements to fibers of type $\text{F}_4$ and $\text{E}_6$. B$_3$ is a singleton subalgebra of Dynkin embedding $1$ in  both F$_4$ and E$_6$. In both cases, the reduced characteristic representation is $\mathbf{7}\oplus\mathbf{8}$.  There are three distinct subalgebras of type  B$_3$ in D$_4$, they all have Dynkin embedding index one and  give the same reduced characteristic representation, namely the representation $\mathbf{7}$. Thus B$_3$ in D$_4$ is an example of an adjoint singleton subalgebra as there are several isomorphic but nonlinearly equivalent subalgebras giving the same reduced characteristic representation.

\item The $\text{C}_2$-model is particularly intriguing because it has various types of fiber enhancements allowed and there  are two possible group structures, SO($5$) and USp($4$) \cite{USP4}. In most of the cases, there is no need to refer to the selection rule, as C$_2$ is a singleton subalgebra of A$_3$, B$_3$, C$_3$, and D$_4$. However, in the case $\text{C}_2\to \text{A}_4$, there are two possible isomorphic subalgebras C$_3$: the one with Dynkin index one gives the correct matter representation (namely, the representation $\mathbf{4}\oplus\mathbf{5}$), while the other one gives the representation $\mathbf{14}$.  
In the case $\text{C}_2\to \text{D}_5$, there are two distinct C$_4$ subalgebras up to linear equivalence and they both have Dynkin index one. Interestingly, one of them gives the matter content of an SO($5$) theory, while the other one corresponds to a $\text{Spin}(5)\cong\text{USp}(4)$ theory. 
\end{enumerate}

\item Non-Higgsable models are particularly important models in 5d and 6d theories \cite{Morrison:2012np,Heckman:2018jxk}.
They correspond to the simple Lie algebras of the Deligne exceptional series:
\begin{equation}
\text{A}_1\to \text{A}_2\to \text{G}_2\to \text{F}_4\to \text{D}_4\to \text{E}_6\to \text{E}_7\to \text{E}_8,
\end{equation}
together with the semisimple algebras  $
\text{A}_1\oplus \text{B}_3$ and $\text{A}_1\oplus \text{G}_2$. describing local collisions between two rational curves supporting non-trivial matter representations. At the collision of $\text{A}_1\oplus\text{G}_2$, we find E$_7$ or E$_8$ models for which we get the branching rules of Dynkin index one to give rise to the expected matter representations from F-theory models $\mathbf{R}=(\mathbf{2},\mathbf{7})\oplus(\mathbf{2},\mathbf{1})\oplus(\mathbf{1},\mathbf{7})$, where the matter representations $(\mathbf{2},\mathbf{1})$ and $(\mathbf{1},\mathbf{7})$ are also produced away from the collision \cite{SU2G2}. We further explore in detail all the relevant  branching rules:
\begin{enumerate}
\item $\text{E}_7\to \text{E}_8$ from the original Katz--Vafa as a singleton model in Section \ref{sec:singletonKV},
\item $\text{A}_1\oplus\text{G}_2 \to \text{D}_5, \text{B}_5, \text{D}_6, \text{E}_7,\text{E}_8$ in Section \ref{sec:A1G2},
\item $\text{A}_1\oplus\text{B}_3 \to \text{E}_7, \text{E}_8$ in Section \ref{sec:A1B3},
\item $\text{A}_1\oplus\text{D}_4 \to \text{E}_7, \text{E}_8$ in Section \ref{sec:A1D4}.
\end{enumerate}

\item The proposed selection rule \ref{SR}  can also be used to identify those models that do not have an F-theory origin.
A simple example is given by the gauge theory with Lie algebra $\mathfrak{g}=\text{A}_1\oplus\text{D}_4$ and reduced matter in the representation 
\begin{equation}
\mathbf{R}=(\mathbf{2},\mathbf{8}_{\text{v}})\oplus (\mathbf{1},\mathbf{8}_{\text{s}}),(\mathbf{1},\mathbf{8}_{\text{c}}) 
\end{equation}
discussed in \cite[{\S 3.3}]{Heckman:2018jxk}. 
A quick scan of possibilities show that there is no simple Lie algebra $\mathfrak{m}$ in E$_8$ containing  $\mathfrak{g}$ and giving the expected matter content up to multiplicities of the individual irreducible representations. In F-theory, this Lie algebra will be produced by a collision of type III+I$_0^*$ which is a Miranda model giving a fiber of type III$^*$ or II$^*$. 
That means, the corresponding subalgebra would be  of the type $\text{A}_1\oplus \text{D}_4$ in $\text{E}_7$. The embedding could also be in $\text{E}_8$ for a specialization of th elliptic fibration.  There are three subalgebras of type 
 $\text{A}_1\oplus \text{D}_4$ up to linear equivalence in E$_7$.  Their characteristic representations and Dynkin index are
\begin{align}
\text{A}^1_1\oplus \text{D}^1_4\to \text{E}_7:&
\quad
\chi=
           (\mathbf{2},\mathbf{8}_{\text{v}})^{\otimes 2}\oplus
       (\mathbf{{2}},\mathbf{8}_{\text{s}})^{\otimes 2}\oplus
                      (\mathbf{1},\mathbf{8}_{\text{c}})^{\otimes 4}\oplus
                                    (\mathbf{1},\mathbf{1})^{\otimes 6},
\\
\text{A}^2_1\oplus \text{D}^1_4\to \text{E}_7:&
\quad
\chi= 
 (\mathbf{3},\mathbf{1})\oplus 
 (\mathbf{1},\mathbf{8}_{\text{v}})
 \oplus (\mathbf{3},\mathbf{8}_{\text{v}})
\oplus 
(\mathbf{2},\mathbf{8}_{\text{s}})^{\oplus 2}\oplus (\mathbf{2},\mathbf{8}_{\text{c}})^{\oplus 2}\oplus
 (\mathbf{1},\mathbf{1})^{\oplus 3},
\\
\text{A}^3_1\oplus \text{D}^1_4\to \text{E}_7:&
\quad
\chi=
(\mathbf{3},\mathbf{8}_{\text{v}})  \oplus 
     (\mathbf{3},\mathbf{8}_{\text{s}})\oplus 
 (\mathbf{3},\mathbf{8}_{\text{c}})\oplus 
   (\mathbf{1},\mathbf{8}_{\text{v}})
  \oplus
     (\mathbf{1},\mathbf{8}_{\text{s}})\oplus
      (\mathbf{1},\mathbf{8}_{\text{c}})\oplus
        (\mathbf{3},\mathbf{1})^{\oplus 2}.  
 \end{align}
The first has Dynkin index one and is therefore the one expected in F-theory. In particular, we see that the representation ${\mathbf R}=(\mathbf{2},\mathbf{8}_{\text{v}})\oplus (\mathbf{1},\mathbf{8}_{\text{s}}),(\mathbf{1},\mathbf{8}_{\text{c}})$ is never realized by the Katz--Vafa decomposition of E$_7$. 

There are  eight linearly nonequivalent subalgebras of type $\text{A}_1\oplus\text{D}_4$ in E$_8$. 
These eight embeddings can be distinguished by their Dynkin index.  None of them matches the representation $\mathbf{R}$. However, we note that the unique one of  Dynkin index $(1,1)$ contains $\mathbf{R}$ in a triality-invariant way: 
\begin{align}
\text{A}^1_1 \oplus \text{D}^1_4\to\text{E}_8:\quad &\chi=
(\mathbf{2},\mathbf{8}_{\text{v}})^{\oplus 2}\oplus (\mathbf{2},\mathbf{8}_{\text{s}})^{\oplus 2}\oplus (\mathbf{2},\mathbf{8}_{\text{c}})^{\oplus 2}\oplus (\mathbf{1},\mathbf{8}_{\text{v}})^{\oplus 4}\oplus (\mathbf{1},\mathbf{8}_{\text{s}})^{\oplus 4}\nonumber\\
&\qquad\quad\oplus (\mathbf{1},\mathbf{8}_{\text{c}})^{\oplus 4}\oplus (\mathbf{2},\mathbf{1})^{\oplus 8}\oplus (\mathbf{1},\mathbf{1})^{\oplus 9}.
\end{align}
We also note that the matter discussed in  \cite[Section 3.3]{Heckman:2018jxk} is not compatible with any branching of the adjoint of E$_8$ along  $A_1 \oplus D_4$, as any other embedding of $A_1\oplus D_4\to E_8$  involves representations of A$_1$ of higher spin or the irreducible representations of D$_4$ of dimension $35$. 

\end{enumerate}

\section{Preliminaries}
\label{sec:prelim}

Classifying semisimple subalgebras of a complex semisimple Lie algebra is a fundamental problem arising in many different contexts in mathematics and physics \cite{
Dynkin.MSG,Dynkin.SubA, Minchenko, LorenteGruber, deGraaf, Borel}. The solution to this classical problem was beautifully addressed by Dynkin's seminal papers, which also introduced several key ideas such as the projective and affine Dynkin's diagrams, the equivalence and linear equivalence of subalgebras and embeddings,  the notion of R-subalgebras and S-subalgebras, the Dynkin index of a representation, and the Dynkin index of an embedding of Lie algebras. 

In this section, we review basic notions that  will play a central role in our approach. Our description follows Dynkin \cite{Dynkin.SubA}, Minchenko \cite{Minchenko}, de Graaf \cite{deGraaf}, McKay--Patera \cite{McKayPatera}, and Lorente--Gruber \cite{LorenteGruber}. We use the conventions of Bourbaki \cite{Bourbaki.GLA79}.

In 1950s, Dynkin classified the maximal semisimple subalgebras of exceptional Lie algebras up to equivalence and simple subalgebras only up to linear equivalence   \cite{Dynkin.SubA}. Dynkin also introduced an algorithm to classify regular subalgebras of semisimple classical Lie algebras up to equivalence and provided an explicit classification (up to linear equivalence) of S-subalgebras and simple subalgebras of exceptional Lie algebras. 
Dynkin focused especially on exceptional Lie algebras, since the classical cases can be described by the work of Cartan, Weyl, and Malcev.\footnote{The case of A$_n$ was studied by Cartan and Weyl, while the cases of $B_n$, C$_n$, and D$_n$ were analyzed by Malcev who also investigated representations of G$_2$ and F$_4$. Borel and de Siebenthal studied the maximal semisimple subalgebras of simple Lie algebras \cite{Borel}. }
Dynkin's result has some inaccuracies that are corrected by Minchenko, who further provided a classification of embeddings of semisimple subalgebras of simple Lie algebras of exceptional types up to equivalence rather than just linear equivalence, thereby completing the description of embeddings of semisimple subalgebras of simple Lie algebras \cite{Minchenko}. 
Using Dynkin's algorithm, Lorente and Gruber obtained the detailed classification of subalgebras of simple Lie algebras of classical types up to rank six \cite{LorenteGruber}.
Motivated from problems in physics, de Graaf describes algorithms that give a classification of the semisimple subalgebras of all simple Lie algebra or rank eight or lower together with the inclusion relations between all these semisimple Lie algebras \cite{deGraaf}. de Graaf also correct few inaccuracies in the classification of Lorente and Gruber (in D$_4$, D$_6$, and D$_8$).

\subsection{Embeddings, equivalence, linear equivalence, and branching rules} \label{sec:equivalence}

\begin{defn}[Embeddings]
An embedding $f:\mathfrak{g}\to \mathfrak{m}$ between two  Lie algebras $\mathfrak{g}$ and $\mathfrak{m}$ is an injective homomorphism from $\mathfrak{g}$ to $\mathfrak{m}$. 
\end{defn}

\begin{defn}[Induced representations and branching rules]
Given an embedding $f:\mathfrak{g}\to \mathfrak{m}$, any irreducible linear representation $\varphi$ of $\mathfrak{m}$ induces a linear representation $\varphi\circ f$ of $\mathfrak{g}$. The representation $\varphi\circ f$  is in general reducible and its decomposition into irreducible representations of $\mathfrak{g}$ is called {\em the branching rule of} $\varphi$. 
\end{defn}

The classification of subalgebras starts with a choice of a notion of equivalence. 
There are two different notions of equivalence that are typically considered, where booth are introduced by Dynkin. One is simply called equivalence, which is more natural, and the other is called linear equivalence, which is coarser but more computationally friendly. We give definitions for these two notions as follows. 

\begin{defn}[Equivalent of Lie subalgebras]
Two subalgebras $\mathfrak{g}_1$ and $\mathfrak{g}_2$ of a Lie algebra $\mathfrak{m}$ are said to be {\em equivalent} if they are conjugate by an inner automorphism of the parent Lie algebra $\mathfrak{m}$. 
\end{defn}

\begin{defn}[Linear equivalent of Lie subalgebras]Two subalgebras are {\em linearly equivalent} if for any linear representation  $f:\mathfrak{m}\to \mathfrak{gl}(n)$ of the parent Lie algebra induces representations of the two subalgebras that are conjugated by a linear inner automorphism of $\mathfrak{gl}(n)$.
\end{defn}

When discussing branching rules, linear equivalence is the appropriate notion of equivalence to consider, as two linearly equivalent subalgebras give the same branching rules.  
Two equivalent subalgebras are always linearly equivalent. The reverse is often (but not necessarily)  true. There are known examples of embeddings that are linearly equivalent but not equivalent. 
Linear equivalence and equivalence coincide for A$_n$, B$_n$, C$_n$, G$_2$, and F$_4$.  
Subtleties appears for E$_6$, E$_7$, E$_8$, and D$_n$  ($n\geq 4$) Lie algebras. For example, E$_8$ has $1183$ equivalent subalgebras organized into $155$ linear equivalence classes \cite{deGraaf}.

\subsection{R- and S-subalgebras}

In his classification of subalgebras of simple Lie algebras, Dynkin introduced the notions of a regular subalgebras, R-subalgebras,  and S-subalgebras. We present them as follows.

\begin{defn}[Regular, R-, and S-algebras]
A subalgebra $\mathfrak{g}$ is said to be a {\em regular subalgebra} of a parent Lie algebra $\mathfrak{m}$ when the Dynkin diagram of $\mathfrak{g}$ can be obtained by removing nodes of the extended Dynkin diagram of  $\mathfrak{m}$.
An {\em R-subalgebra} is a subalgebra that is contained in a proper regular subalgebra, whereas an {\em S-subalgebra} is a proper subalgebra that is not an R-subalgebra. 
\end{defn}
Tables listing maximal R- and S-subalgebras of simple Lie algebras can be found in \cite{Slansky:1981yr,McKayPatera,Yamatsu:2015npn}.

\subsection{Normalized Killing form}
Consider a simple Lie algebra $\mathfrak{g}$ of rank $n$ with a choice of simple roots. We denote the set of positive roots as $\Phi^+$.
The Killing form
 \begin{equation}
 \kappa(x,y):= \text{Tr}(\text{ad}\ x, \text{ad}\ y)
 \end{equation}
  defines an invariant bilinear symmetric form on $\mathfrak{g}$.
Up to a multiplicative factor, there exists a unique non-degenerate invariant symmetry bilinear form on $\mathfrak{g}$.
We denote by $(\cdot, \cdot)$ the invariant scalar product in the root space to be normalized such that a root of the longest length has its length squared equals to two.
 In particular, if we denote by $\theta$ the highest root (which is always a long root), we have
\begin{equation}
(\theta,\theta)=2.
\end{equation}
The Lie algebra is semisimple if and only if  the Killing form is negative-definite. 
Given any invariant bilinear symmetric form $B( \cdot, \cdot )$ on $\mathfrak{g}$, the normalized Killing form is given by 
\begin{equation}
(x,y):= \frac{2B(x,y)}{B(\theta,\theta)}.
\end{equation}
 
\subsection{Dynkin's index of a representation}

The notion of an index of a subalgebra was introduced by Dynkin in his classification of semisimple subalgebras of exceptional simple Lie algebra and requires first introducing the notion of the Dynkin index of a representation. 
The index of a simple subalgebra in a simple Lie algebra is always a non-negative integer.  
Two linearly equivalent subalgebras have the same index. Thus, two subalgebras with a different index are necessarily non-equivalent. However, the index is not a complete characterization of a subalgebra as two subalgebras with the same index are not necessarily linearly equivalent, even if they are isomorphic as abstract algebras.

Given a   linear representation   of $\mathfrak{g}$ in a vector space $V$
\begin{equation}
f:\mathfrak{g}\to \mathfrak{gl}(V),
\end{equation}
we define the invariant symmetric bilinear form  in $\mathfrak{g}$ induced by $f$ as follows
\begin{equation}
(x,y)_f:=\mathrm{Tr}(f(x) f(y)),
\end{equation}
where for a square $n\times n$ matrix $M$,  $\mathrm{Tr}(M)=\sum_{i=1}^n M_{ii}$ is the trace of  $M$. 
Since the invariant bilinear symmetric form is unique up to a  multiplicative constant, $(x,y)_f$ should be proportional to the normalized Killing form $(x,y)$ that takes value $2$ on the longest root. The multiplicative constant is independent of $x$ and $y$ and characterizes the representation $f$. Then, it is by definition the Dynkin index of the representation $f$. 

\begin{defn} Given a representation $f:\mathfrak{g}\to \mathfrak{gl}(V)$,  
the {\em Dynkin index of the representation} $f$ is the multiplicative constant  $\ell_f$ defined by the identity: 
\begin{equation}
(x,y)_f=\ell_f  \cdot  (x,y).
\end{equation}
\end{defn}

 If we take $x=y=\theta$, we can express the Dynkin index of a representation as half the length squared of the highest root (or any long root) with respect to the invariant bilinear form induced by the representation $f$:
\begin{equation}
\ell_f=\frac{1}{2}(\theta,\theta)_f.
\end{equation} 
In particular, the index of the  adjoint representation is the {\em dual Coxeter number} of the Lie algebra.

Cartan's theorem states that a weight is the highest weight of an irreducible representation of a simple Lie algebra if and only if its Dynkin indices are all non-negative integers. 
The following formula allows to compute the index of a representation $f$ in terms of its highest weight $\Lambda_f$: 
\begin{equation}\label{Index0}
\ell_f=(\Lambda_f, \Lambda_f+2\rho) \frac{\dim f}{\dim g} ,
\end{equation}
where $\Lambda_d$ is the highest weight of the representation $f$, $\dim f$ is the dimension of the representation $f$ (that is, of the vector space $V$),  $\dim \mathfrak{g}$ is the dimension of the simple Lie algebra $\mathfrak{g}$, and the Weyl vector $\rho$  is half the sum of all positive roots $\Phi^+$ of $\mathfrak{g}$.
\begin{equation}
\rho=\frac{1}{2}\sum_{\alpha\in\Phi^+} \alpha.
\end{equation} 
The number $ (\Lambda, \Lambda+2\rho)$ is the eigenvalue of the second-order Casimir operator. 
The formula \eqref{Index0} is reminiscent of Weyl's formula for the dimension $d_{\Lambda}$ of a representation with the highest weight $\Lambda$:
\begin{equation}
d_{\Lambda}=\sum_{\alpha\in \Phi^+} \frac{(\alpha,\Lambda+\rho)}{(\alpha,\alpha)}.
\end{equation}
The dual Coxeter number of a simple Lie algebra  is the Dynkin's index of its adjoint representation:
\begin{equation}
h^\vee=\ell_{\mathrm{ad}}=(\theta, \theta+2\rho)=\mathrm{Tr} (\mathrm{ad} \theta, \mathrm{ad} \theta).
\end{equation}

xThe Dynkin index of a representation satisfies the following two identities 
\begin{equation}\label{Index1}
\ell_{f_1\oplus f_m}=\ell_{f_1}+\ell_{f_m}.
\end{equation}
\begin{equation}\label{Index2}
\ell_{f_1\otimes f_2}=(\dim f_2) \ell_{f_1}+(\dim f_1)\ell_{f_2}.
\end{equation}

\subsection{Dynkin index of an embedding} 
An embedding  between two simple Lie algebras $\mathfrak{g}$ and $\mathfrak{m}$ is an isomorphism 
\begin{equation}
\varphi: \mathfrak{g}\to \mathfrak{m}
\end{equation}
 that maps $\mathfrak{g}$ onto its image in $\mathfrak{m}$.  
Then, we can think of the embedding $\varphi$ as  a representation of $\mathfrak{g}$ in $\mathfrak{m}$. Analogous to the case of a linear representation, it is natural that we define the Dynkin index of an embedding $\varphi$. 
We assume that both Lie algebras $\mathfrak{g}$ and $\mathfrak{m}$  have their bilinear form to be normalized such that the norm squared of each of their root of highest length is two. 
We can define an invariant bilinear form in $\mathfrak{m}$ via the embeddings $\varphi$. Because such an invariant bilinear form is unique upto a multiplicative constant, we have  a number  $j_\varphi$ such that
 \begin{equation}
(\varphi x, \varphi  y)_{\mathfrak{g}} = j_\varphi \cdot (x,y)_{\mathfrak{m}}.
\end{equation}
The number $j_\varphi$ is called the {\em Dynkin index of the embedding} $\varphi$. The index of a subalgebra is invariant with respect to automorphisms of $\mathfrak{m}$.

The Dynkin index of an embedding can be computed as the ratio between the Dynkin index of an induced representation and its parent representation. 
For any representation $f: \mathfrak{m}\to \mathfrak{gl}(V)$, we have $f\circ \varphi:\mathfrak{m}\to \mathfrak{gl}(V)$ defining an induced representation of the subalgebra $\mathfrak{g}$. We can therefore compute the index of the embedding $\varphi$ in terms of the index of a representation $f$ of $\mathfrak{m}$ and the induced representation on $\mathfrak{g}$ as
\begin{equation}\label{Index3}
j_\varphi= \frac{\ell_{f\circ \varphi}}{\ell_f}.
\end{equation}

The Dynkin index of an embedding is multiplicative in the following sense. 
If $\mathfrak{g}_1$, $\mathfrak{g}_2$, and $\mathfrak{g}_3$ are three simple algebras and $\varphi_i$ is an embedding of  $\mathfrak{g}_i$ into $\mathfrak{g}_{i+1}$ $i=1,2$, then 
\begin{equation}\label{Index4}
j_{\varphi_2 \circ \varphi_1} =j_{\varphi_2}\cdot j_{\varphi_1}.
\end{equation}

If under the embedding $\varphi: \mathfrak{g}\to \mathfrak{m}$ a representation $\mathbf{R}$ of $\mathfrak{m}$  has a decomposition 
\begin{equation}
R= R_1^{\oplus m_1}\oplus \cdots\oplus R_n^{\oplus m_n},
\end{equation}
where $R_i$ is an irreducible representations of $\mathfrak{g}$ with multiplicity $m_i$. Then, the Dynkin index of the embedding $\varphi$ is simply given by the formula 
\begin{equation}
j_\varphi= \frac{1}{\ell_R}(m_1\ell_{R_1}+\cdots  +m_n\ell_{R_n}).
\end{equation}
The following is a key theorem on the Dynkin index. 
\begin{thm}
The Dynkin index of an injective homomorphism between simple Lie algebras  is a positive integer. 
\end{thm}
When  the Dynkin index of a representation $f$ is one, the representation $f$ transforms the longest root of $\mathfrak{g}$ into the longest root of $\mathfrak{m}$, and transforms the corresponding root vectors of $\mathfrak{g}$ into root vectors of $\mathfrak{m}$.
In the language of embedded subalgebras, we can characterize subalgebras of index one  by the following theorem. 

\begin{thm}[{ \cite[Theorem 2.4]{Dynkin.SubA}}]
Let $\mathfrak{g}$ be a simple subalgebra of the simple algebra $\mathfrak{m}$. If the index of $\mathfrak{g}$ in $\mathfrak{m}$ is equal to $1$, then the root of maximal length and the corresponding root vectors in $\mathfrak{g}$ are respectively the roots and the root vectors in the algebra  $\mathfrak{m}$.
\end{thm}

A fundamental representation of a Lie algebra is one for which the highest weight is a simple root. 
The defining (or the standard) representation of a simple Lie algebra is the fundamental representation of the smallest dimension. In physics, it is often called {\em the fundamental representation} of the simple Lie algebra. The Dynkin indices of all fundamental representations of all simple Lie algebras are listed in  Table \ref{DynkinT5}.

We denote a simple subalgebra with its index as a superscript. When the indices  are the same for not linearly equivalent subalgebras, we use (one, two, or three) primes additionally  following the notation of Dynkin in \cite[Table 25]{Dynkin.SubA}.  

\subsection{Dynkin index of a simple Lie algebra}

The Dynkin index of a simple Lie algebra $\mathfrak{g}$, denoted $\ell_{\mathfrak{g}}$ is by definition the greatest common divisor of the Dynkin indices $\ell_{\mathbf{R}}$ where $\mathbf{R}$ runs over all linear representations of $\mathfrak{g}$.
 It is enough to take the greatest common divisor among all fundamental representations of $\mathfrak{g}$ since the Dynkin index of any other representation is a linear combination of them with non-negative integral coefficients.  
 We  list  the Dynkin index of fundamental representations of all simple Lie groups on Table \ref{DynkinT5} while the Dynkin index of simple Lie algebras are listed on Table \ref{tb:normalization}.

\begin{table}[H]
\begin{center}
\begin{tabular}{|c|c|c|c|c|c|c|c|c|c|}
\hline
 $\mathfrak{g}$ & A$_n$ ($n\geq 1$) & B$_n$  ($n\geq 3$) & C$_n$ ($n\geq 2$) & D$_n$  ($n\geq 4$)& G$_2$ & F$_4$ & E$_6$&  E$_7$ & E$_8$ \\
 \hline
  $h$ & $n+1$ & $2n$  & $2n$ & $2n-2$ & $6$ & $12$ & $12$ & $18$ & $30$ \\
 \hline
  $h^\vee$ & $n+1$ & $2n-1$  & $n+1$ & $2n-2$ & $4$ & $9$ & $12$ & $18$ & $30$ \\
 \hline  
 $\lambda_{\mathfrak{g}}$ & $1$ & $2$  & $1$ & $2$ & $2$ & $6$ & $6$ & $12$ & $60$ \\
 \hline  
$\varpi$  & $\varpi_1$ , $\varpi_n$ & $\varpi_1$  & $\varpi_1$ & $\varpi_1$ & $\varpi_1$ & $\varpi_1$ & $\varpi_1$, $\varpi_6$ & $\varpi_7$ & $\varpi_8$ \\
 \hline  
\end{tabular}
\caption{The Dynkin index of  simple Lie algebras. 
The {\em Dynkin index $\lambda_{\mathfrak{g}}$ of a  simple Lie algebra} $\mathfrak{g}$ is the greatest common divisor  $\lambda_{\mathfrak{g}}$ of the Dynkin indexes of its fundamental representations. The numbers $h$ and $h^\vee$ are the  Coxeter and the dual Coxeter number. 
On the last row, we list the fundamental weights (in the conventions of Bourbaki)  whose corresponding representation has Dynkin index $\lambda_{\mathfrak{g}}$.
See also  \cite[Proposition 2.6]{LS97}. 
\label{tb:normalization}}
\end{center}
\end{table}

\subsection{Conventions and basic data on simple Lie algebras and their representations}
\label{sec:Conv}

\begin{table}[H]
\begin{center}
\begin{tabular}{c}
{\scalebox{1.1}{\begin{tikzpicture}
\node  at (-1,0){A$_n$};
\node[draw,circle,thick,scale=1.1, label=right:{$\binom{n-1}{0}$}] (1) at (0,0){};
\node[draw,circle,thick,scale=1.1, label=right:{$\binom{n-1}{1}$}] (2) at (0,-1){};
\node[draw,circle,thick,scale=1.1, label=right:{$\binom{n-1}{k-1}$}] (3) at (0,-2.4){};
\node[draw,circle,thick,scale=1.1, label=right:{$\binom{n-1}{n-2}$}] (4) at (0,-3.8){};
\node[draw,circle,thick,scale=1.1, label=right:{$\binom{n-1}{n-1}$}] (5) at (0,-4.8){};
\draw[thick] (1)--(2);
\draw[thick,dashed] (2)--(3)--(4);
\draw[thick] (4)--(5);
\end{tikzpicture}}
}

{\scalebox{1.1}{\begin{tikzpicture}
\node  at (-1,0){B$_n$};
\node[draw,circle,thick,scale=1.1, label=right:{$2\binom{2n-1}{0}$}] (1) at (0,0){};
\node[draw,circle,thick,scale=1.1, label=right:{$2\binom{2n-1}{1}$}] (2) at (0,-1){};
\node[draw,circle,thick,scale=1.1, label=right:{$2\binom{2n-1}{k-1}$}] (3) at (0,-2.4){};
\node[draw,circle,thick,scale=1.1, label=right:{$2\binom{2n-1}{n-2}$}] (4) at (0,-3.8){};
\node[draw,circle,thick,scale=1.1, label=right:{$2^{n-2}$}] (5) at (0,-4.8){};
\draw[thick] (1)--(2);
\draw[thick,dashed] (2)--(3)--(4);
\draw[thick, double, double distance =3pt] (4)--(5);
		\draw[thick] plot [smooth,tension=1] coordinates { (0,-4.5) (.13,-4.2) (.27,-4.1)};
			\draw[thick] plot [smooth,tension=1] coordinates { (0,-4.5) (-.13,-4.2) (-.27,-4.1)};				

\end{tikzpicture}}
}

{\scalebox{1.1}{\begin{tikzpicture}
\node  at (-1,0){C$_n$};
\node[draw,circle,thick,scale=1, label=right:{$\binom{2n-1}{0}$}] (1) at (0,0){};
\node[draw,circle,thick,scale=1, label=right:{$\binom{2n-1}{1}-\binom{2n-1}{0}$}] (2) at (0,-1){};
\node[draw,circle,thick,scale=1, label=right:{$\binom{2n-1}{k-1}-\binom{2n-1}{k-2}$}] (3) at (0,-2.4){};
\node[draw,circle,thick,scale=1, label=right:{$\binom{2n-1}{n-2}-\binom{2n-1}{n-3}$}] (4) at (0,-3.8){};
\node[draw,circle,thick,scale=1, label=right:{$\binom{2n-1}{n-1}-\binom{2n-1}{n-2}$}] (5) at (0,-4.8){};
\draw[thick] (1)--(2);
\draw[thick,dashed] (2)--(3)--(4);
\draw[thick, double, double distance=3pt] (4)--(5);

\draw[thick] plot [smooth,tension=1] coordinates { (0,-4.1) (.13,-4.4) (.27,-4.5)};
			\draw[thick] plot [smooth,tension=1] coordinates { (0,-4.1) (-.13,-4.4) (-.27,-4.5)};	

\end{tikzpicture}}
}

{\scalebox{1.1}{\begin{tikzpicture}
\node  at (-1,0){D$_n$};
\node[draw,circle,thick,scale=1, label=right:{$2\binom{2n-2}{0}$}] (1) at (0,0){};
\node[draw,circle,thick,scale=1, label=right:{$2\binom{2n-2}{1}$}] (2) at (0,-1){};
\node[draw,circle,thick,scale=1, label=right:{$2\binom{2n-2}{k-1}$}] (3) at (0,-2.4){};
\node[draw,circle,thick,scale=1, label=right:{$2\binom{2n-2}{n-3}$}] (4) at (0,-3.8){};
\node[draw,circle,thick,scale=1, label= left:{$2^{n-3}$}] (5) at (-.5,-4.8){};
\node[draw,circle,thick,scale=1, label= right:{$2^{n-3}$}] (6) at (.5,-4.8){};
\draw[thick] (1)--(2);
\draw[thick,dashed] (2)--(3)--(4);
\draw[thick] (4)--(5);
\draw[thick] (4)--(6);
\end{tikzpicture}}
}
\\
\\
{\scalebox{1.1}{\begin{tikzpicture}
\node  at (-1,-5){G$_2$};
\node[draw,circle,thick,scale=1, label= right:{$2$}] (5) at (0,-5){};
\node[draw,circle,thick,scale=1, label= right:{$8$}] (6) at (0,-6){};
\draw[thick, double, double distance =5 pt] (5)--(6);
\draw[thick] (5)--(6);
			\draw[thick] plot [smooth,tension=1] coordinates { (0,-5.3) (.13,-5.6) (.27,-5.7)};
			\draw[thick] plot [smooth,tension=1] coordinates { (0,-5.3) (-.13,-5.6) (-.27,-5.7)};	
\end{tikzpicture}}}
{\scalebox{1}{\begin{tikzpicture}
\node  at (-1,0){F$_4$};
\node[draw,circle,thick,scale=1, label= right:{$6$}] (1) at (0,0){};
\node[draw,circle,thick,scale=1, label= right:{$126$}] (2) at (0,-1){};
\node[draw,circle,thick,scale=1, label= right:{$882$}] (3) at (0,-2){};
\node[draw,circle,thick,scale=1, label= right:{$18$}] (4) at (0,-3){};
\draw[thick] (1)--(2);
\draw[thick] (3)--(4);
\draw[thick, double, double distance=3pt] (2)--(3);
\draw[thick] plot [smooth,tension=1] coordinates { (0,-1.3) (.13,-1.6) (.27,-1.7)};
			\draw[thick] plot [smooth,tension=1] coordinates { (0,-1.3) (-.13,-1.6) (-.27,-1.7)};						
\end{tikzpicture}}}
{\scalebox{1}{\begin{tikzpicture}
\node  at (-1,0){E$_6$};
\node[draw,circle,thick,scale=1, label= left:{$24$}] (0) at (-1,-2){};
\node[draw,circle,thick,scale=1, label= right:{$6$}] (1) at (0,0){};
\node[draw,circle,thick,scale=1, label= right:{$150$}] (2) at (0,-1){};
\node[draw,circle,thick,scale=1, label= right:{$1800$}] (3) at (0,-2){};
\node[draw,circle,thick,scale=1, label= right:{$150$}] (4) at (0,-3){};
\node[draw,circle,thick,scale=1, label= right:{$6$}] (5) at (0,-4){};
\draw[thick] (1)--(2)--(3)--(4)--(5);
\draw[thick] (0)--(3);
\end{tikzpicture} }}
{\scalebox{1}{\begin{tikzpicture}
\node  at (-1,0){E$_7$};
\node[draw,circle,thick,scale=1, label= left:{$360$}] (0) at (-1,-3){};
\node[draw,circle,thick,scale=1, label= right:{$36$}] (1) at (0,-5){};
\node[draw,circle,thick,scale=1, label= right:{$4680$}] (2) at (0,-4){};
\node[draw,circle,thick,scale=1, label= right:{$297000$}] (3) at (0,-3){};
\node[draw,circle,thick,scale=1, label= right:{$17160$}] (4) at (0,-2){};
\node[draw,circle,thick,scale=1, label= right:{$648$}] (5) at (0,-1){};
\node[draw,circle,thick,scale=1, label= right:{$12$}] (6) at (0,0){};

\draw[thick] (1)--(2)--(3)--(4)--(5)--(6);
\draw[thick] (0)--(3);
\end{tikzpicture}}}

{\scalebox{1}{\begin{tikzpicture}
\node  at (-1,0){E$_8$};
\node[draw,circle,thick,scale=1, label= left:{$85500$}] (0) at (-1,-4){};
\node[draw,circle,thick,scale=1, label= right:{$60$}] (1) at (0,0){};
\node[draw,circle,thick,scale=1, label= right:{$14700$}] (2) at (0,-1){};
\node[draw,circle,thick,scale=1, label= right:{$177840$}] (3) at (0,-2){};
\node[draw,circle,thick,scale=1, label= right:{$141605100$}] (4) at (0,-3){};
\node[draw,circle,thick,scale=1, label= right:{$8345660400$}] (5) at (0,-4){};
\node[draw,circle,thick,scale=1, label= right:{$5292000$}] (6) at (0,-5){};
\node[draw,circle,thick,scale=1, label= right:{$1500$}] (7) at (0,-6){};
\draw[thick] (1)--(2)--(3)--(4)--(5)--(6)--(7);
\draw[thick] (0)--(5);
\end{tikzpicture}}}
\end{tabular}
\end{center}
\caption{Dynkin indices of the fundamental representations of simple Lie algebras. 
This table is obtained from  \cite[Table 5]{Dynkin.SubA} after fixing some  typos and inaccuracies  (index of $\varpi_{n-1}$ for B$_n$ and C$_n$; and the index of $\varpi_{i}$ of E$_8$ for $i=3,4,5,6$).  
\label{DynkinT5}}
\end{table}

\clearpage
\begin{table}[htb]
\begin{center}
\scalebox{.9}{
$
\arraycolsep=3pt\def\arraystretch{1.7}
\begin{array}{|l|c|c|c|c|c|}
\hline
\quad \text{                Algebra} (\mathfrak{g}) & \text{Cartan} & \text{Dynkin} & \text{Condition} & \text{Representation} \\
\hline
\hline
\begin{array}{c} A_{n}\\ n=0,2,3 (\text{mod}\ 4) \end{array} & \multirowcell{5}{\begin{pmatrix} 2 & -1 & 0 & \cdots & 0 & 0 & 0\\ -1 & 2 & -1 & \cdots & 0 & 0 & 0\\ \vdots & \vdots &\vdots &\ddots &\vdots &\vdots &\vdots\\ 0 & 0 & 0 & \cdots & 2 & -1 & 0\\ 0 & 0 & 0 & \cdots & -1 & 2 & -1\\ 0 & 0 & 0 & \cdots & 0 & -1 & 2 \end{pmatrix}}  & \multirowcell{5}{\scalebox{0.7}{\begin{tikzpicture}
\node[draw,circle,thick,scale=1] (X) at (1.1,-2){};
\node[draw,circle,thick,scale=1,label=left: \scalebox{1.2}{$a_{1}$}] (1) at (0,0){};
\node[draw,circle,thick,scale=1,label=left: \scalebox{1.2}{$a_{2}$}] (2) at (0,-1){};
\node[draw,circle,thick,scale=1,label=left: \scalebox{1.2}{$a_{n-1}$}] (4) at (0,-3){};
\node[draw,circle,thick,scale=1,label=left: \scalebox{1.2}{$a_{n}$}] (5) at (0,-4){};
\draw[thick] (X)--(1)--(2)--(0,-1.6);
\draw[thick,dashed] (0,-1.6)--(0,-2.4);
\draw[thick] (0,-2.4)--(4)--(5)--(X);
\end{tikzpicture}}} & \begin{array}{c} (a_1,\cdots ,a_n)\neq (a_n,\cdots ,a_1)\\  (a_1,\cdots ,a_n)=(a_n,\cdots ,a_1) \end{array} & \begin{array}{c}  \text{Complex}\\ \text{Real} \end{array} \\
\cline{1-1} \cline{4-5}
\begin{array}{c} A_{n}\\ n=1\ (\mod\ 4) \end{array}  &  &  & \begin{array}{c} (a_1,\cdots ,a_n)\neq (a_n,\cdots ,a_1)\\ \left\{ \begin{aligned} (a_1,\cdots ,a_n)=(a_n,\cdots ,a_1)\\ a_{(n+1)/2}=0 (\mod\ 2) \end{aligned}\right\} \\  \left\{ \begin{aligned} (a_1\cdots a_n)=(a_n\cdots a_1)\\ a_{(n+1)/2}=1 (\mod\ 2) \end{aligned}\right\} \end{array} & \begin{array}{c}  \text{Complex}\\ \Gape[2ex]{\text{Real}}\\ \text{Pseudo-real} \end{array} \\
\hline
\hline
\begin{array}{c} B_n\\ n=0,3\ (\mod\ 4) \end{array} & \multirowcell{4}{\scalebox{.9}{$\begin{pmatrix} 2 & -1 & 0 & \cdots & 0 & 0 & 0\\ -1 & 2 & -1 & \cdots & 0 & 0 & 0\\ \vdots & \vdots &\vdots &\ddots &\vdots &\vdots &\vdots\\ 0 & 0 & 0 & \cdots & 2 & -1 & 0\\ 0 & 0 & 0 & \cdots & -1 & 2 & -2\\ 0 & 0 & 0 & \cdots & 0 & -1 & 2 \end{pmatrix}$}}  & \raisebox{1.7ex}{\multirowcell{5}{\scalebox{0.7}{\begin{tikzpicture}
\node[draw,circle,thick,scale=1,label=left: \scalebox{1.2}{$a_{1}$}] (1a) at (-0.5,0){};
\node[draw,circle,thick,scale=1] (1b) at (0.5,0){};
\node[draw,circle,thick,scale=1,label=left: \scalebox{1.2}{$a_{2}$}] (2) at (0,-1){};
\node[draw,circle,thick,scale=1,label=left: \scalebox{1.2}{$a_{n-1}$}] (4) at (0,-3.1){};
\node[draw,circle,thick,scale=1,label=left: \scalebox{1.2}{$a_{n}$}] (5) at (0,-4.2){};
\draw[thick] (1a)--(2);
\draw[thick] (1b)--(2)--(0,-1.6);
\draw[thick,dashed] (0,-1.6)--(0,-2.4);
\draw[thick] (0,-2.4)--(4);
\draw[double,thick,double distance=3pt] (4)--(5);
\begin{scope}[shift={(0,1.8)}]
\draw[thick] plot [smooth,tension=1] coordinates { (0,-5.3) (.13,-5.6) (.27,-5.7)};
\draw[thick] plot [smooth,tension=1] coordinates { (0,-5.3) (-.13,-5.6) (-.27,-5.7)};
\end{scope}
\end{tikzpicture}}}} & \text{None} & \Gape[5ex]{\text{Real}} \\
\cline{1-1} \cline{4-5}
\begin{array}{c} B_n \\ n=1,2\ (\mod\ 4) \end{array}  &  &  & \begin{array}{c} a_n=0\ (\mod\ 2)\\ a_n=1(\mod\ 2) \end{array} & \Gape[3ex]{\begin{array}{c} \text{Real}\\ \text{Pseudo-real} \end{array}} \\
\hline
\hline
\multirowcell{4}{C_n} & \multirowcell{1}{\begin{pmatrix} 2 & -1 & 0 & \cdots & 0 & 0 & 0\\ -1 & 2 & -1 & \cdots & 0 & 0 & 0\\ \vdots & \vdots &\vdots &\ddots &\vdots &\vdots &\vdots\\ 0 & 0 & 0 & \cdots & 2 & -1 & 0\\ 0 & 0 & 0 & \cdots & -1 & 2 & -1\\ 0 & 0 & 0 & \cdots & 0 & -2 & 2 \end{pmatrix}}  & \raisebox{3.5ex}{\multirowcell{2}{\scalebox{0.7}{\begin{tikzpicture}
\node[draw,circle,thick,scale=1] (X) at (0,1.2){};
\node[draw,circle,thick,scale=1,label=left: \scalebox{1.2}{$a_1$}] (1) at (0,0.1){};
\node[draw,circle,thick,scale=1,label=left: \scalebox{1.2}{$a_2$}] (2) at (0,-1){};
\node[draw,circle,thick,scale=1,label=left: \scalebox{1.2}{$a_{n-1}$}] (4) at (0,-3.1){};
\node[draw,circle,thick,scale=1,label=left: \scalebox{1.2}{$a_{n}$}] (5) at (0,-4.2){};
\draw[double,thick,double distance=3pt] (X)--(1);
\draw[thick] (1)--(2)--(0,-1.6);
\draw[thick,dashed] (0,-1.6)--(0,-2.4);
\draw[thick] (0,-2.4)--(4);
\draw[double,thick,double distance=3pt] (4)--(5);
 \begin{scope}[shift={(0,1.5)}]
\draw[thick] plot [smooth,tension=1] coordinates { (0,-5.3) (.13,-5) (.27,-4.9)};
\draw[thick] plot [smooth,tension=1] coordinates { (0,-5.3) (-.13,-5) (-.27,-4.9)};
\end{scope}
\end{tikzpicture}}}} & \begin{array}{c} a_1+\cdots +a_{n}=0(\mod\ 2) \end{array} & \Gape[6.5ex]{\text{Real}} \\
\cline{4-5}
 & & & \begin{array}{c} a_1+\cdots +a_{n}=1(\mod\ 2) \end{array} &  \Gape[6.5ex]{\text{Pseudo-real}} \\
\hline
\hline

\begin{array}{c} D_n \\ n=1\ (\mod\ 2) \end{array} & \multirowcell{5}{\scalebox{.9}{$\begin{pmatrix} 2 & -1 & 0 & \cdots & 0 & 0 & 0\\ -1 & 2 & -1 & \cdots & 0 & 0 & 0\\ \vdots & \vdots &\vdots &\ddots &\vdots &\vdots &\vdots\\ 0 & 0 & 0 & \cdots & 2 & -1 & -1\\ 0 & 0 & 0 & \cdots & -1 & 2 & 0\\ 0 & 0 & 0 & \cdots & -1 & 0 & 2 \end{pmatrix}$}}  & \multirowcell{5}{\scalebox{0.7}{\begin{tikzpicture}
\node[draw,circle,thick,scale=1] (1a) at (0.5,0){};
\node[draw,circle,thick,scale=1,label=left:\scalebox{1.2}{$a_1$}] (1b) at (-0.5,0){};
\node[draw,circle,thick,scale=1,label=left:\scalebox{1.2}{$a_2$}] (2) at (0,-1){};
\node[draw,circle,thick,scale=1,label=left:\scalebox{1.2}{$a_{n-2}$}] (3) at (0,-3){};
\node[draw,circle,thick,scale=1,label=below:\scalebox{1.2}{$a_{n-1}$}] (4) at (0.5,-4){};
\node[draw,circle,thick,scale=1,label=below:\scalebox{1.2}{$a_n$}] (5) at (-0.5,-4){};
\draw[thick] (1a)--(2);
\draw[thick] (1b)--(2)--(0,-1.6);
\draw[thick,dashed] (0,-1.6)--(0,-2.4);
\draw[thick] (0,-2.4)--(3)--(4);
\draw[thick] (3)--(5);
\end{tikzpicture}}} &\begin{array}{c} a_{n-1}\neq a_n \\ a_{n-1}=a_n \end{array}  & \Gape[0ex]{\begin{array}{c}  \text{Complex}\\ \text{Real} \end{array}} \\
\cline{1-1} \cline{4-5}
\begin{array}{c} D_n\\ n=0\ (\mod\ 4) \end{array}  &  &  & \text{None} & \Gape[0ex]{\text{Real}} \\
\cline{1-1} \cline{4-5}
\begin{array}{c} D_n\\ n=2\ (\mod\ 4) \end{array}  &  &  & \begin{array}{c} a_{n-1}+a_n=0\ (\mod\ 2)\\ a_{n-1}+a_n=1\  (\mod\ 2) \end{array} & \Gape[0ex]{\begin{array}{c} \text{Real}\\ \text{Pseudo-real} \end{array}} \\
\hline
\end{array}
$}
\end{center}
\label{tb:representation}
\caption{Cartan matrices, Dynkin diagrams, and reality conditions of representations of  classical simple Lie algebras. }
\end{table}
\clearpage \newpage

\begin{table}[htb]
\begin{center}
\scalebox{.9}{
$
\arraycolsep=3pt\def\arraystretch{1.4}
\begin{array}{|c|c|c|c|c|c|}
\hline
\mathfrak{g} & \text{Cartan} & \text{Dynkin} & \text{Condition} & \text{Representation} \\
\hline
\hline
E_6 & \raisebox{0ex}{\scalebox{.9}{$\begin{pmatrix} 2 & -1 & 0 & 0 & 0 & 0 \\ -1 & 2 & -1 & 0 & 0 & 0 \\ 0 & -1 & 2 & -1 & 0 & -1 \\ 0 & 0 & -1 & 2 & -1 & 0 \\ 0 & 0 & 0 & -1 & 2 & 0 \\ 0 & 0 & -1 & 0 & 0 & 2 \end{pmatrix}$}} & \raisebox{-8ex}{{\scalebox{0.7}{\begin{tikzpicture}
\node[draw,circle,thick,scale=1] (X) at (2,-2){};
\node[draw,circle,thick,scale=1, label=left:\scalebox{1.2}{$a_1$}] (1) at (0,0){};
\node[draw,circle,thick,scale=1, label=left:\scalebox{1.2}{$a_2$}] (2) at (0,-1){};
\node[draw,circle,thick,scale=1, label=left:\scalebox{1.2}{$a_3$}] (3) at (0,-2){};
\node[draw,circle,thick,scale=1, label=left:\scalebox{1.2}{$a_4$}] (4) at (0,-3){};
\node[draw,circle,thick,scale=1, label=left:\scalebox{1.2}{$a_5$}] (5) at (0,-4){};
\node[draw,circle,thick,scale=1, label=above:\scalebox{1.2}{$a_6$}] (6) at (1,-2){};
\draw[thick] (1)--(2)--(3)--(4)--(5);
\draw[thick] (3)--(6)--(X);
\end{tikzpicture}}}}

& \begin{array}{c} {\begin{array}{c} (a_1,a_2,a_3,a_4,a_5,a_6)\\ \neq (a_5,a_4,a_3,a_2,a_1,a_6)\end{array}} \\ \\
 {\begin{array}{c} (a_1,a_2,a_3,a_4,a_5,a_6)\\ =(a_5,a_4,a_3,a_2,a_1,a_6)\end{array}} \end{array} & \Gape[1ex]{\begin{array}{c} \text{Complex}\\ \\ \\ \text{Real}\\ \end{array}} \\

\hline
E_7 & \raisebox{0ex}{{$\begin{pmatrix} 2 & -1 & 0 & 0 & 0 & 0 & 0 \\ -1 & 2 & -1 & 0 & 0 & 0 & 0 \\ 0 & -1 & 2 & -1 & 0 & 0 & -1\\ 0 & 0 & -1 & 2 & -1 & 0 & 0 \\ 0 & 0 & 0 & -1 & 2 & -1 & 0 \\ 0 & 0 & 0 & 0 & -1 & 2 & 0\\ 0 & 0 & -1 & 0 & 0 & 0 & 2 \end{pmatrix}$}} & \raisebox{-12ex}{{\scalebox{0.7}{\begin{tikzpicture}
\node at (0,1.5){};  \node at (0,-5.5){};
\node[draw,circle,thick,scale=1] (X) at (0,1){};

\node[draw,circle,thick,scale=1, label=left:\scalebox{1.2}{$a_1$}] (1) at (0,0){};
\node[draw,circle,thick,scale=1, label=left:\scalebox{1.2}{$a_2$}] (2) at (0,-1){};
\node[draw,circle,thick,scale=1, label=left:\scalebox{1.2}{$a_3$}] (3) at (0,-2){};
\node[draw,circle,thick,scale=1, label=left:\scalebox{1.2}{$a_4$}] (4) at (0,-3){};
\node[draw,circle,thick,scale=1, label=left:\scalebox{1.2}{$a_5$}] (5) at (0,-4){};
\node[draw,circle,thick,scale=1, label=left:\scalebox{1.2}{$a_6$}] (6) at (0,-5){};
\node[draw,circle,thick,scale=1, label=right:\scalebox{1.2}{$a_7$}] (7) at (1,-2){};
\draw[thick] (X)--(1)--(2)--(3)--(4)--(5)--(6);
\draw[thick] (3)--(7);
\end{tikzpicture}}}} &

 {\begin{array}{c} a_4+a_6+a_7=0\  (\mod\ 2)\\ \\  \\  
 a_4+a_6+a_7=1\ (\mod\ 2)
 \end{array}}  & \Gape[1ex]{\begin{array}{c} \text{Real }\\ \\ \\ \text{Pseudo-real}\\ \end{array}} \\

\hline
E_8 & \raisebox{12ex}{\multirowcell{1}{\scalebox{1}{$\begin{pmatrix} 2 & -1 & 0 & 0 & 0 & 0 & 0 & 0\\ -1 & 2 & -1 & 0 & 0 & 0 & 0 & 0\\ 0 & -1 & 2 & -1 & 0 & 0 & 0 & -1\\ 0 & 0 & -1 & 2 & -1 & 0 & 0 & 0\\ 0 & 0 & 0 & -1 & 2 & -1 & 0 & 0\\ 0 & 0 & 0 & 0 & -1 & 2 & -1 & 0\\ 0 & 0 & 0 & 0 & 0 & -1 & 2 & 0\\ 0 & 0 & -1 & 0 & 0 & 0 & 0 & 2 \end{pmatrix}$}}} & \raisebox{14ex}{\multirowcell{2}{\scalebox{0.7}{\begin{tikzpicture}
\node[draw,circle,thick,scale=1] (X) at (0,-7){};
\node[draw,circle,thick,scale=1, label=left:\scalebox{1.2}{$a_1$}] (1) at (0,0){};
\node[draw,circle,thick,scale=1, label=left:\scalebox{1.2}{$a_2$}] (2) at (0,-1){};
\node[draw,circle,thick,scale=1, label=left:\scalebox{1.2}{$a_3$}] (3) at (0,-2){};
\node[draw,circle,thick,scale=1, label=left:\scalebox{1.2}{$a_4$}] (4) at (0,-3){};
\node[draw,circle,thick,scale=1, label=left:\scalebox{1.2}{$a_5$}] (5) at (0,-4){};
\node[draw,circle,thick,scale=1, label=left:\scalebox{1.2}{$a_6$}] (6) at (0,-5){};
\node[draw,circle,thick,scale=1, label=left:\scalebox{1.2}{$a_7$}] (7) at (0,-6){};
\node[draw,circle,thick,scale=1, label=right:\scalebox{1.2}{$a_8$}] (8) at (1,-2){};
\draw[thick] (1)--(2)--(3)--(4)--(5)--(6)--(7)--(X);
\draw[thick] (3)--(8);
\end{tikzpicture}}}} & \text{None} & \Gape[17ex]{\text{Real}} \\
\hline
\arraycolsep=2pt\def\arraystretch{2}
F_4 & \raisebox{3ex}{\scalebox{1}{$\begin{pmatrix} 2 & -1 & 0 & 0\\ -1 & 2 & -2 & 0\\ 0 & -1 & 2 & -1\\ 0 & 0 & -1 & 2 \end{pmatrix}$}} & \raisebox{-6.5ex}{{\scalebox{0.7}{\begin{tikzpicture}
\node  at (0,1.5){};
\node  at (0,-3.5){};
\node[draw,circle,thick,scale=1] (X) at (0,1){};
\node[draw,circle,thick,scale=1, label=right:{$a_1$}] (1) at (0,0){};
\node[draw,circle,thick,scale=1, label= right:{$a_2$}] (2) at (0,-1){};
\node[draw,circle,thick,scale=1, label= right:{$a_3$}] (3) at (0,-2.2){};
\node[draw,circle,thick,scale=1, label=right:{$a_4$}] (4) at (0,-3.2){};
\draw[thick] (X)--(1)--(2);
\draw[thick] (3)--(4);
\draw[thick, double, double distance=3pt] (2)--(3);
\draw[double,thick,double distance=3pt] (2)--(3);
\draw[thick] plot [smooth,tension=1] coordinates { (0,-1.5) (.13,-1.8) (.27,-1.9)};
			\draw[thick] plot [smooth,tension=1] coordinates { (0,-1.5) (-.13,-1.8) (-.27,-1.9)};	

\end{tikzpicture}}}} & \text{None} & \Gape[0ex]{\text{Real}} \\
\hline
G_2 & \raisebox{1ex}{\multirowcell{1}{$\begin{pmatrix} 2 & -3 \\ -1 & 2 \end{pmatrix}$}} & \raisebox{-6ex}{{\scalebox{0.7}{\begin{tikzpicture}
\node at (0,-3.5){};
\node at (0,-6.7){};
\node[draw,circle,thick,scale=1] (X) at (0,-4){};
\node[draw,circle,thick,scale=1, label= right:\scalebox{1.2}{$a_1$}] (5) at (0,-5){};
\node[draw,circle,thick,scale=1, label= right:\scalebox{1.2}{$a_2$}] (6) at (0,-6.2){};
\draw[thick, double, double distance =5 pt] (5)--(6);
\draw[thick] (X)--(5)--(6);
			\draw[thick] plot [smooth,tension=1] coordinates { (0,-5.3) (.13,-5.6) (.27,-5.7)};
			\draw[thick] plot [smooth,tension=1] coordinates { (0,-5.3) (-.13,-5.6) (-.27,-5.7)};	
\end{tikzpicture}}}} & \text{None} & \Gape[3ex]{\text{Real}} \\
\hline
\end{array}
$}
\end{center}
\label{tb:representation2}
\caption{Cartan matrices, Dynkin diagrams, and reality conditions of representations of  exceptional Lie algebras.}
\end{table}
\clearpage

\begin{table}[htb]
\begin{center}
\scalebox{.93}{
$
\arraycolsep=3pt\def\arraystretch{1.3}
\begin{array}{|c|c|c|c|}
\hline
\text{Algebra} & \text{Highest weight } & \text{Name} & \text{Comment} \\
\hline
\hline
\text{A}_1 & 2    &  \mathbf{3} &   \text{adjoint} \\
& 1 &  \mathbf{2} & \text{defining (fundamental) rep., pseudo-real} \\
\hline 
\text{A}_{n\geq 2} & [1,0,\ldots, 0,1]    &  \mathbf{n(n+2)} &   \text{adjoint} \\
& [1,0\ldots,0] &  \mathbf{n+1} & \multirowcell{2}{\text{defining (fundamental) rep., complex}} \\
& [0,\ldots,0,1] &\mathbf{\overline{n+1}}  &   \\
\hline
\text{B}_{n\geq 2} & [0,1,\ldots, 0]  & \mathbf{n(2n + 1)} &  \text{adjoint} \\
 & [1,0,\ldots, 0]  & \mathbf{2n+1} &  \text{defining (vector) representation, real}  \\
  & [0,\ldots, 0, 1]  & \mathbf{2^n} & \text{spin representation} \\
\hline
\text{C}_{n\geq 2} &   [2,0,\ldots, 0]  &\mathbf{n(2n + 1)} &   \text{adjoint}  \\
&   [1,0,\ldots, 0]   & \mathbf{2n}  &    \text{defining (fundamental) rep., pseudo-real }    \\
&   [0,1,0\ldots, 0]  & \mathbf{(n-1) (2 n+1)}   &   \text{traceless antisymmetric  rep., pseudo-real }     \\
 \hline
\text{D}_{n\geq 4} & [0, 1, 0, \ldots, 0] 
&  \mathbf{ n(2n -1)}&  \text{adjoint}    \\
 & [1,0,0,\ldots,0]  
 & \mathbf{2n} & \text{defining (vector) rep., real} \\
&[0,\cdots, 0, 1,0] 
&  \mathbf{2^{n-1}} & \multirowcell{2}{\text{half-spin}} \\
&[0,\cdots, 0,0, 1] 
&   \mathbf{\overline{2^{n-1}}}&  \\
\hline
\text{G}_2 & [0,1]  \dynkin[labels={0,1}]{G}{2} &  \mathbf{14} &  \text{adjoint}  \\
& [1,0]    \dynkin[labels={1,0}]{G}{2}&  \mathbf{7} &  \text{defining (vector) representation, real}   \\
\hline
\text{F}_4 & [ 1,0, 0, 0]  \dynkin[labels={1,0,0,0}]{F}{4} &  \mathbf{52} & \text{adjoint}  \\
 & [0,  0, 0, 1]   
 \dynkin[labels={0,0,0,1}]{F}{4}
 &   \mathbf{26} &   \text{defining representation, real}   \\
\hline
\text{E}_6 & [0, 1, 0, 0, 0, 0] 
\dynkin[labels={0,1,0,0,0,0}]{E}{6}
 & \mathbf{78} & \text{adjoint}   \\
 &[1,0,0,0,0,0]
 \dynkin[labels={1,0,0,0,0,0}]{E}{6}
  & \mathbf{27} &   \multirowcell{2}{\text{defining representation, complex.}}  \\
  & [0,0,0,0,0,1] 
  \dynkin[labels={0,0,0,0,0, 1}]{E}{6}
  &  \mathbf{\overline{27}}  & \\
\hline
\text{E}_7 & [1,0,0,0,0,0,0,0] 
\dynkin[labels={1,0,0,0,0, 0, 0}]{E}{7}
 &  \mathbf{133}  &  \text{adjoint}   \\
 & [0,0,0,0,0,0,0,1]    
 \dynkin[labels={0,0,0,0,0, 0,1}]{E}{7}
 &   \mathbf{56}  &   \text{defining representation, pseudo-real.}  \\
\hline
\text{E}_8 & [0, 0, 0, 0, 0, 0, 0, 1] \dynkin[labels={0,0,0,0,0, 0, 0,1}]{E}{8}
& \mathbf{248} &  \text{adjoint}  \\
\hline
\end{array}
$}
\end{center}
\caption{Highest weights and names of the most common representations of simple Lie algebras. 
The highest weights are given in the basis of fundamental weights and the labeling follows the conventions of Bourbaki \cite{Bourbaki.GLA79}. 
The defining representation is always the shortest. 
For D$_4$, there are three representations that are of equal dimension: $[1,0,0,0]=\mathbf{8}_{\text{v}}$, 
$[0,0,0,1]=\mathbf{8}_{\text{s}}$, and $[0,0,1,0]=\mathbf{8}_{\text{s}}$.
 For E$_8$, the shortest representation is the adjoint representation. 
 For ADE algebras, the adjoint representation is quasi-minuscule. 
 For B$_n$, G$_2$, and F$_4$ algebras, the vector representation is quasi-minuscule. For C$_n$, the traceless antisymmetric representation is quasi-minuscule. 
 The half-spin representations of D$_n$ are complex conjugate of each other if $n=\pm 1 \mod \ 4$, they are independent of each other and self-dual otherwise. More precisely, they are real when $n=0 \mod\ 4$ and pseudo-real if $n=2 \mod 4$. }
\label{tb:BasicReps}
\end{table}
%\clearpage
\clearpage

\begin{table}[htb]
\begin{center}
$
\arraycolsep=4pt\def\arraystretch{1.7}
\begin{array}{|c|c|c|}
\hline
\text{Algebras} & \text{Minuscule representations} & \text{Quasi-minuscule representations} \\
\hline
\hline
\text{A}_n & \begin{pmatrix} \mathbf{n+1}\\ \mathbf{k} \end{pmatrix}\ ( 0\leq k\leq n ) & \mathbf{n^2+2n}\   (\text{adjoint}) \\
\hline
\text{B}_n\    (n\geq 2)& \mathbf{1}\, (\text{trivial}),\ \mathbf{2^n}\ (\text{spin}) & \mathbf{2n+1}\    (\text{vector}) \\
\hline
\text{C}_n\    (n\geq 3) & \mathbf{1}\, (\text{trivial}),\ \mathbf{2n}\ (\text{vector}) &{ \mathbf{(n-1) (2 n+1)}} \  (\text{traceless antisymmetric}) \\
 \hline
 & \mathbf{1}\, (\text{trivial}),\ \mathbf{2n}\ (\text{vector})& \\
\text{D}_n\  (n\geq 4) & \mathbf{2^{n-1}}\    (\text{half-spin})& \mathbf{2n^2-n}\     (\text{adjoint})  \\
  &  \mathbf{\overline{2^{n-1}}}\    (\text{half-spin})&  \\
\hline
\text{G}_2 & \mathbf{1}\, (\text{trivial}) & \mathbf{7} \\
\hline
\text{F}_4 & \mathbf{1}\, (\text{trivial}) & \mathbf{26} \\
\hline
\text{E}_6 & \mathbf{1}\, (\text{trivial}),\quad \mathbf{27},\quad \overline{\mathbf{27}} & \mathbf{78} (\text{adjoint}) \\
\hline
\text{E}_7 & \mathbf{1}\, (\text{trivial}),\quad \mathbf{56} & \mathbf{133} (\text{adjoint}) \\
\hline
\text{E}_8 & \mathbf{1}\, & \mathbf{248} (\text{adjoint}) \\
\hline
\end{array}
$
\end{center}
\caption{Minuscule and quasi-minuscule representations of the simple Lie algebras. 
A minuscule representation is one for which all the weights are in the same orbit of the Weyl group. 
A quasi-minuscule representation is one for which all the non-zero weights are in the same orbit of the Weyl group.}
\label{tb:Minuscule}
\end{table}

\begin{table}[H]
\arraycolsep=4pt\def\arraystretch{1.7}
\begin{center}
\begin{tabular}{| c | c | }
\hline 
Exceptional Lie algebras & Maximal S-subalgebras\\
\hline
$\text{G}_2$ & $\text{A}^{28}_1$ \\
\hline 
$\text{F}_4$ & $\text{A}^{156}_1$, $\text{A}^8_1\oplus \text{G}^1_2$ \\
\hline 
$\text{E}_6$ & $\text{A}^9_1$, $\text{G}^3_2$, $\text{C}^1_4$, $\text{A}_2^{2''}\oplus\text{G}^1_2$, $\text{F}^1_4$ \\
\hline 
$\text{E}_7$ & $\text{A}^{231}_1$, 
$\text{A}^{399}_1$,
$\text{A}^{21}_2$,
$\text{A}^{15}_1 \oplus \text{A}^{24}_1$,
$\text{A}^7_1\oplus\text{G}^2_2$,
$\text{G}^1_2 \oplus \text{C}^{1''}_3$, 
$\text{A}^{3''}_1\oplus\text{F}^1_4$
\\
\hline 
$\text{E}_8$  &
$\text{A}^{1240}_1$,  
$\text{A}^{760}_1$, 
$\text{A}^{520}_1$, 
$\text{A}^{6'}_2\oplus\text{ A}^{16}_1$, 
 $\text{B}^{12}_2$, $\text{G}^1_2 \oplus\text{F}^1_4$
\\
\hline 
\end{tabular}
\end{center}
\caption{The maximal S-subalgebras of the simple exceptional Lie algebras up to linear equivalence.
The superscript gives the Dynkin index of the subalgebra.  See \cite[Theorem 4.2]{Dynkin.MSG}.}
\label{tb:maxS}
\end{table}

\begin{table}[htb]
\begin{center}
$
\begin{array}{| c | c | c | c |}
\hline
\text{Rank} & \text{Algebra }\mathfrak{m} & \text{Maximal R-subalgebras }\mathfrak{g} & \text{Maximal S-subalgebras }\mathfrak{g} \\
\hline
\hline
1 & \text{A}_1 & \mathfrak{u}_1 & \\
\hline
2 &\text{A}_2 & \text{A}_1\oplus\mathfrak{u}_1 & \text{A}_1 \\
&\text{C}_2 & \text{A}_1\oplus\text{A}_1,\,\text{A}_1\oplus\mathfrak{u}_1 & \text{A}_1 \\
& \text{G}_2 &\text{A}_2,\,\text{A}_1\oplus\text{A}_1 & \text{A}_1 \\
\hline 
3 &\text{A}_3 &\text{A}_2\oplus\mathfrak{u}_1,\,\text{A}_1\oplus \text{A}_1\oplus\mathfrak{u}_1, &\text{C}_2,\,\text{A}_1\oplus\text{A}_1 \\
& \text{B}_3 &\text{A}_3,\,\text{A}_1\oplus\text{A}_1\oplus\text{A}_1,\,\text{C}_2\oplus\mathfrak{u}_1 & \text{G}_2 \\
& \text{C}_3 &\text{A}_2\oplus\mathfrak{u}_1,\,\text{A}_1\oplus\text{C}_2 & \text{A}_1,\,\text{A}_1\oplus\text{A}_1 \\
\hline
4 &\text{A}_4 &\text{A}_3\oplus\mathfrak{u}_1, \text{A}_2\oplus\text{A}_1\oplus\mathfrak{u}_1 &\text{C}_2 \\
& \text{D}_4 &\text{A}_3\oplus\mathfrak{u}_1,\,\text{A}_1\oplus\text{A}_1\oplus\text{A}_1\oplus\text{A}_1 &\text{A}_2,\,\text{B}_3,\,\text{A}_1\oplus\text{C}_2 \\
& \text{B}_4 & \text{D}_4,\,\text{A}_1\oplus\text{A}_1\oplus\text{C}_2,\,\text{A}_3\oplus\text{A}_1,\,\text{B}_3\oplus\mathfrak{u}_1 & \text{A}_1,\,\text{A}_1\oplus\text{A}_1 \\
& \text{C}_4 &\text{A}_3\oplus\mathfrak{u}_1,\,\text{A}_1\oplus \text{C}_3,\,\text{C}_2\oplus\text{C}_2 & \text{A}_1,\,\text{A}_1\oplus\text{A}_1\oplus\text{A}_1 \\
& \text{F}_4 & \text{B}_4,\,\text{A}_2\oplus\text{A}_2,\,\text{A}_1\oplus\text{C}_3 & \text{A}_1,\,\text{A}_1\oplus\text{G}_2 \\
\hline
5 &\text{A}_5 &\text{A}_4\oplus\mathfrak{u}_1,\,\text{A}_3\oplus\text{A}_1\oplus\mathfrak{u}_1,\,\text{A}_2\oplus\text{A}_2\oplus\mathfrak{u}_1 &\text{A}_2,\,\text{A}_3,\,\text{C}_3,\,\text{A}_2\oplus\text{A}_1 \\
& \text{D}_5 &\text{A}_4\oplus\mathfrak{u}_1,\,\text{A}_1\oplus\text{A}_1\oplus\text{A}_3,\,\text{D}_4\oplus\mathfrak{u}_1 &\text{C}_2,\,\text{B}_4,\,\text{A}_1\oplus\text{B}_3,\,\text{C}_2\oplus\text{C}_2 \\
& \text{B}_5 & \text{D}_5,\,\text{D}_4\oplus\text{A}_1,\,\text{A}_3\oplus\text{C}_2,\, & \text{A}_1 \\
& & \text{A}_1\oplus\text{A}_1\oplus\text{B}_3,\,\text{B}_4\oplus\mathfrak{u}_1 & \\
& \text{C}_5 &\text{A}_4\oplus\mathfrak{u}_1,\,\text{A}_1\oplus \text{C}_4,\,\text{C}_2\oplus\text{C}_3 & \text{A}_1,\,\text{A}_1\oplus\text{C}_2 \\
\hline
6 &\text{A}_6 &\text{A}_5\oplus\mathfrak{u}_1,\,\text{A}_4\oplus\text{A}_1\oplus\mathfrak{u}_1,\,\text{A}_3\oplus\text{A}_2\oplus\mathfrak{u}_1 & \text{B}_3 \\
& \text{D}_6 &\text{A}_5\oplus\mathfrak{u}_1,\,\text{A}_1\oplus\text{A}_1\oplus\text{D}_4,\,\text{D}_5\oplus\mathfrak{u}_1 &\text{C}_2,\,\text{B}_5,\,\text{A}_1\oplus\text{B}_4,\,\text{C}_2\oplus\text{B}_3, \\
& &\text{A}_3\oplus\text{A}_3 & \text{A}_1\oplus\text{C}_3 \\
& \text{B}_6 & \text{D}_6,\,\text{D}_5\oplus\text{A}_1,\,\text{D}_4\oplus\text{C}_2,\, & \text{A}_1 \\
& &\text{A}_3\oplus\text{B}_3,\,\text{A}_1\oplus\text{A}_1\oplus\text{B}_4,\,\text{B}_5\oplus\mathfrak{u}_1 & \\
& \text{C}_6 &\text{A}_5\oplus\mathfrak{u}_1,\,\text{A}_1\oplus \text{C}_5,\,\text{C}_2\oplus\text{C}_4,\,& \text{A}_1,\,\text{A}_1\oplus\text{A}_3,\,\text{A}_1\oplus\text{C}_2 \\
& & \text{C}_3\oplus\text{C}_3  & \\
& \text{E}_6 & \text{D}_5\oplus\mathfrak{u}_1,\,\text{A}_5\oplus\text{A}_1,\,\text{A}_2\oplus\text{A}_2\oplus\text{A}_2 & \text{C}_4,\,\text{F}_4,\,\text{G}_2,\,\text{A}_2,\,\text{A}_2\oplus\text{G}_2\\
\hline
7 &\text{A}_7 &\text{A}_6\oplus\mathfrak{u}_1,\,\text{A}_5\oplus\text{A}_1\oplus\mathfrak{u}_1,\,\text{A}_4\oplus\text{A}_2\oplus\mathfrak{u}_1,\, & \text{D}_4,\,\text{C}_4,\,\text{A}_3\oplus\text{A}_1 \\
& &\text{A}_3\oplus\text{A}_3\oplus\mathfrak{u}_1 & \\
& \text{D}_7 &\text{A}_6\oplus\mathfrak{u}_1,\,\text{A}_1\oplus\text{A}_1\oplus\text{D}_5,\,\text{D}_6\oplus\mathfrak{u}_1 & \text{G}_2,\,\text{C}_2,\,\text{C}_3,\,\text{B}_6,\,\text{C}_2\oplus\text{B}_4, \\
& &\text{A}_3\oplus\text{D}_4 & \text{A}_1\oplus\text{B}_5,\,\text{B}_3\oplus\text{B}_3\\
& \text{B}_7 & \text{D}_7,\,\text{B}_6\oplus\mathfrak{u}_1,\,\text{D}_6\oplus\text{A}_1,\,\text{D}_5\oplus\text{C}_2,\, & \text{A}_1,\,\text{A}_3,\,\text{A}_1\oplus\text{C}_2 \\
& & \text{D}_4\oplus\text{B}_3,\,\text{A}_3\oplus\text{B}_4,\,\text{A}_1\oplus\text{A}_1\oplus\text{B}_5 & \\
& \text{C}_7 &\text{A}_6\oplus\mathfrak{u}_1,\,\text{A}_1\oplus \text{C}_6,\,\text{C}_2\oplus\text{C}_5,\, & \text{A}_1,\,\text{A}_1\oplus\text{B}_3, \text{C}_3 \\
& & \text{C}_3\oplus\text{C}_4 & \\
& \text{E}_7 &\text{A}_7,\,\text{E}_6\oplus\mathfrak{u}_1,\,\text{A}_5\oplus\text{A}_2,\,\text{D}_6\oplus\text{A}_1 &\text{A}_2,\,\text{A}_1,\,\text{A}_1\oplus \text{F}_4,\,\text{G}_2\oplus\text{C}_3,\,\\
& & & \text{A}_1\oplus\text{G}_2,\,\text{A}_1\oplus\text{A}_1 \\
\hline
8 &\text{A}_8 &\text{A}_7\oplus\mathfrak{u}_1,\,\text{A}_6\oplus\text{A}_1\oplus\mathfrak{u}_1,\,\text{A}_5\oplus\text{A}_2\oplus\mathfrak{u}_1,\, & \text{B}_4,\,\text{A}_2\oplus\text{A}_2 \\
& &\text{A}_4\oplus\text{A}_3\oplus\mathfrak{u}_1 & \\
& \text{D}_8 &\text{A}_7\oplus\mathfrak{u}_1,\,\text{A}_1\oplus\text{A}_1\oplus\text{D}_6,\,\text{D}_7\oplus\mathfrak{u}_1 & \text{B}_4,\,\text{B}_7,\,\text{C}_2\oplus\text{C}_2,\,\text{A}_1\oplus\text{C}_4 \\
& &\text{A}_3\oplus\text{D}_5,\,\text{D}_4\oplus\text{D}_4 &\text{C}_2\oplus\text{B}_5,\, \text{A}_1\oplus\text{B}_6,\,\text{B}_3\oplus\text{B}_4 \\
& \text{B}_8 & \text{D}_8,\,\text{B}_7\oplus\mathfrak{u}_1,\,\text{D}_7\oplus\text{A}_1,\,\text{D}_6\oplus\text{C}_2,\, & \text{A}_1 \\
& & \text{D}_5\oplus\text{B}_3,\,\text{A}_3\oplus\text{B}_5,\,\text{D}_4\oplus\text{B}_4, & \\
& & \text{A}_1\oplus\text{A}_1\oplus\text{B}_6 & \\
& \text{C}_8 &\text{A}_7\oplus\mathfrak{u}_1,\,\text{A}_1\oplus \text{C}_7,\,\text{C}_2\oplus\text{C}_6 & \text{A}_1,\,\text{C}_2,\,\text{A}_1\oplus\text{D}_4 \\
& & \text{C}_3\oplus\text{C}_5,\,\text{C}_4\oplus\text{C}_4 & \\
& \text{E}_8 &\text{A}_8,\,\text{D}_8,\,\text{E}_7\oplus\text{A}_1,\,\text{E}_6\oplus\text{A}_2,\,\text{A}_4\oplus\text{A}_4 & \text{A}_1,\,\text{C}_2,\,\text{A}_1\oplus\text{A}_2,\,\text{G}_2\oplus\text{F}_4 \\
\hline
\end{array}
$
\end{center}
\caption{Maximal subalgebras of the simple Lie algebras upto rank 8.\cite{Yamatsu:2015npn}}
\label{tb:maxRSsub}
\end{table}
\clearpage

\section{Curating the original Katz--Vafa cases with singleton subalgebras} 
\label{sec:original}

During the second string revolution, Katz and Vafa proposed a new method for determining matter representations without relying on a duality argument between F-theory and the Heterotic string theory. 
Historically, the Katz--Vafa method was defined for rank-one enhancements between Lie algebras of type ADE \cite{Katz:1996xe}. 
The Katz--Vafa method is intuitive and user-friendly to most physicists, as it is rooted in the concept of Higgsing and utilizes the method of branching rules familiar from the techniques of Grand Unified Theories since the 1960s. 
One of the appeals of the Katz--Vafa method is its simplicity and versatility: it relies on relatively  simple data and can be applied to M-theory, Type IIA theory, and F-theory compactifications.

The Katz--Vafa method determines the matter representation via the branching rule for the decomposition of the adjoint representation of a parent Lie algebra $\mathfrak{m}$ along one of its proper subalgebras $\mathfrak{g}$.  Given an embedding $\mathfrak{g}\to \mathfrak{m}$ between two Lie algebras $\mathfrak{m}$ and $\mathfrak{g}$, we define the {\em characteristic representation} $\chi$ as
\begin{equation}
\text{adj}\ \mathfrak{m}=\text{adj}\ \mathfrak{g}\oplus \chi ,
\end{equation}  
and the characteristic representation can 
be written as a sum of irreducible linear representations $\mathbf{r}_i$ with multiplicities $m_i$ and $\mathbf{r}_0$ is the trivial representation:
\begin{align}
\chi=\sum_{i\geq 0} \mathbf{r}_i^{\oplus m_i} .
\end{align}
Since an adjoint representation is always real, the characteristic representation $\chi$ is also a real representation and necessarily takes the form 
\begin{equation}
\chi=\chi_r\oplus\chi_{q}^{\oplus 2}\oplus\chi_c\oplus \overline{\chi}_c,
\end{equation}
where $\chi_r$ is a real representation, $\chi_q$ is a pseudo-real representation, $\chi_c$ is a complex representation, and $\overline{\chi}_c$  is the complex conjugate of $\chi_c$.

The characteristic representation $\chi$ associated to an embedding  was introduced by Dynkin in his classification of semisimple subalgebras of simple Lie algebras \cite[Chap 1 \S1 No 6]{Dynkin.SubA}. 
  If we pass to Lie groups, the characteristic representation $\chi$ corresponds to the isotropic group of the homogeneous space $M/G$ such that $\mathfrak{g}=\text{Lie}(G)$ and $\mathfrak{m}=\text{Lie}(M)$. 
  We introduce the notion of the {\em reduced characteristic representation} as the representation obtained from $\chi$ by removing the trivial representations and all the multiplicities.

This characteristic representation is meaningful both in physical and in mathematical sense.
Physically, the characteristic representation  $\chi$ corresponds to the matter representation $\mathbf{R}$  up to trivial representations and multiplicities. 
Mathematically, $\chi$ is more informative than $\mathbf{R}$: for example, when $\mathfrak{m}$ is a simple Lie algebra, we can compute the Dynkin index of $\mathfrak{g}$ in $\mathfrak{m}$ from the representation $\chi$.
The multiplicities of each irreducible representation is not physical, so we introduce the notion of the {\em reduced characteristic  representation} 
\begin{equation}
\chi_{\text{red}}=\sum_{i>0} \mathbf{r}_i,
\end{equation} 
which is the sum of the non-trivial component of $\chi$ without any multiplicity. 
The Lie algebra $\mathfrak{g}$ is the Lie algebra determined by the  gauge algebra, while  $\mathfrak{m}$ is an ``enhancement''  characterizing certain codimension-two loci in the base of the elliptic fibration and responsible for the matter. Then it is natural to think that two distinct branching rules that give different characteristic representations might give the same reduced characteristic representation.

 The original Katz--Vafa method considers rank one enhancements between simply-laced Lie algebras, listed in Table \ref{tb:MirandaMatter}.\footnote{In geometric configurations, both Lie algebras $\mathfrak{g}$ and $\mathfrak{m}$ are associated with some singularities and their degenerations.} It is important to note that, in the cases analyzed in the original Katz--Vafa paper, the branching rules were not explicitly computed but inferred using intersecting branes techniques and the results were checked against explicit Heterotic string theory constructions. We note that the irreducible components of $\chi$ are (quasi)-minuscule representations. 

\begin{table}[H]
\begin{center}
\scalebox{.95}{
$
\arraycolsep=3pt\def\arraystretch{1.5}
\begin{array}{|l |l|}
\hline
\text{$\mathfrak{g}\longrightarrow\mathfrak{m}$} & \text{Matter representation $\mathbf{R}$} \\
\hline
\hline
A_{n-k} \oplus A_{k-1} \longrightarrow A_n  & \mathbf{R}=\left(\bf{n-k+1},\bf{\overline{k}}\right)\oplus\left(\bf{\overline{n-k+1}},\bf{k}\right) \\
 \hline
 D_{n-2}\oplus A_{1} \longrightarrow D_n & \mathbf{R}=\left(\bf{2(n-2),2}\right)\\
 \hline
D_{n-4}\oplus A_{3} \longrightarrow D_n&\mathbf{R}= \left(\bf{2(n-4),4}\right)\oplus  \left(\bf{2(n-4),\overline{4}}\right)\oplus\left(\bf{1,6}\right)\\
 \hline
D_{n-k}\oplus A_{k-1} \longrightarrow D_n  & \mathbf{R}=\left(\bf{2(n-k),k}\right)\oplus\left(\bf{2(n-k),\overline{k}}\right)\oplus \left(\bf{1,\frac{1}{2} k(k-1)}\right)\oplus \left(\bf{1,\overline{\frac{1}{2} k(k-1)}}\right)\\
 \hline
D_{n-1} \longrightarrow D_{n} & \mathbf{R}=\bf{2(n-1)}\\
 \hline
A_{n-1} \longrightarrow D_n& \mathbf{R}=\bf{\frac{1}{2} n(n-1)}\oplus\bf{\overline{\frac{1}{2} n(n-1)}} \\
 \hline
D_5 \longrightarrow E_6& \mathbf{R}=\bf{16}\oplus\bf{\overline{16}} \\
 \hline
A_5 \longrightarrow E_6&\mathbf{R}= \bf{20}\\
 \hline
E_6 \longrightarrow E_7 & \mathbf{R}=\bf{27}\oplus\bf{\overline{27}} \\
\hline
D_6 \longrightarrow E_7 &\mathbf{R}= \bf{32} \\
\hline
A_6 \longrightarrow E_7 &\mathbf{R}= \mathbf{35}\oplus \mathbf{\overline{35}}\oplus\mathbf{7}\oplus \mathbf{\overline{7}}  \\	
\hline
E_7\longrightarrow  E_8 & \mathbf{R}=\bf{56}  \\
 \hline	
\end{array}
$}
\end{center}
\caption{Matter representations from gauge enhancements of type ADE considered in the original Katz--Vafa's paper \cite{Katz:1996xe}. \label{tb:MirandaMatter}}
\end{table}

When $\mathfrak{g}$ is the only subalgebra in its isomorphic class in $\mathfrak{m}$ up to linear equivalence, there is only one possible branching rule and the Katz--Vafa method is perfectly well defined.  In such cases, we say that $\mathfrak{g}$ is a {\em singleton subalgebra} of $\mathfrak{m}$. However, singleton subalgebras are a minority of subalgebras in a given simple Lie algebra $\mathfrak{m}$. 

In particular, we emphasize that we find non-singleton subalgebras even at the boundary of the cases initially studied in the Katz--Vafa's paper. For example, the Lie subalgebra $A_1\oplus A_1$ has two distinct embeddings in A$_3$, leading to two distinct branching rules for the adjoint representation of A$_3$. These two embeddings also have different Dynkin indices, and the one that produces the representation expected in F-theory has Dynkin embedding index ($1,1$). This is a manifestation of what we think is a key data on the Katz--Vafa method: even when multiple choices are possible, it always selects the branching rule corresponding to an embedding $\mathfrak{g}\to \mathfrak{m}$ with Dynkin index one along all the simple components of $\mathfrak{g}$. With a slight abuse of nomenclature,  we will simply call them {\em subalgebras of Dynkin (embedding) index one}.

This section aims to curate the original Katz--Vafa cases to shed light on how the Dynkin index clearly shows a preference for index one subalgebras whenever there is an ambiguity. We will separate the Katz--Vafa cases  into two sectors: singleton and non-singleton subalgebras. The singleton sector has no ambiguities for determining the matter representations and consists of subalgebras of Dynkin embedding index one. For non-singleton subalgebras, we will list all the possible matter content that could be derived from isomorphic but non-linearly equivalent subalgebras. We then observe that the choice consistent with F-theory and M-theory construction is always the ones corresponding to subalgebras of Dynkin's embedding index one.

\subsection{Singleton sectors of the original Katz--Vafa cases} \label{sec:singletonKV}

 In this section, we give a more detailed description of the cases that are originally studied by Katz and Vafa and listed them in Table \ref{Table:KV} after screening to allow parameters such that the branching is consistent with a singleton subalgebra. 
For each embedding $\mathfrak{g}\to \mathfrak{m}$, we  give explicitly the  Dynkin indices of the irreducible components of the subalgebra $\mathfrak{g}$
and the characteristic representation $\chi$ such that
\begin{align}
[\mathbf{adj}\   \mathfrak{m}]=  [\mathbf{adj}\  \mathfrak{g}]\oplus \chi.
\end{align}

In Table \ref{Table:KV}, we used the traditional physics notations  to identify representations. We explain these notations below following the order of the entries in Table \ref{Table:KV}.
\begin{enumerate}
\item The defining representation of A$_{n-1}$ is denoted as $\mathbf{n}$ and its complex conjugate is denoted as $\mathbf{\overline{n}}$. Both are minuscule representations. The minuscule and quasi-minuscule representations of the simple Lie algebras are summarized in Table \ref{tb:Minuscule}.
\item The second antisymmetric representation of A$_n$ is denoted as $\mathbf{\frac{1}{2}n (n-1)}$ and its complex conjugate is denoted as $\mathbf{\overline{\frac{1}{2}n (n-1)}}$; both are minuscule representations.  
In the case of A$_3$, the second antisymmetric representation of A$_3$ is real and denoted as $\mathbf{6}$. Thus, in that case, $\mathbf{\overline{6}}=\mathbf{6}$. 
\item The defining representation of D$_n$ is minuscule, real, and denoted as $\mathbf{2n}$. 
\item The spinor representation of D$_{5}$  of the highest weight $[0,0,0,1,0]$ is  complex and denoted as $\mathbf{16}$,  its complex conjugate has the highest weight $[0,0,0,0,1]$ and is denoted as $\mathbf{\overline{16}}$. 
Both representations $\mathbf{16}$ and $\mathbf{\overline{16}}$ are  minuscule and half-spinor representations of D$_5$. 
\item  The triple antisymmetric representation of A$_5$ (the highest weight $[0,0,1,0,0]$) is minuscule,  pseudo-real, and denoted as $\mathbf{20}$.
\item The defining representation of E$_6$ is complex and denoted as $\mathbf{27}$ and its complex conjugate is $\mathbf{\overline{27}}$. 
Both representations $\mathbf{27}$ and  $\mathbf{\overline{27}}$ of E$_6$ are minuscule representations. 
\item The spinor representation of D$_6$  of the highest weight $[0,0,0,0,1,0]$ (resp. $[0,0,0,0,0,1]$) is pseudo-real and denoted as $\mathbf{32}$ (resp. $\mathbf{\overline{32}}$); both are minuscule representations. 
\item The triple antisymmetric representation of A$_6$ is the complex representation $\mathbf{35}$ of the highest weight $[0,0,1,0,0,0]$. Its complex conjugate is the representation $\mathbf{\overline{35}}$ of the highest weight $[0,0,0,1,0,0]$. 
The representations $\mathbf{35}$ and  $\mathbf{\overline{35}}$ of A$_6$ are both minuscule. 
\item  The defining representation of E$_7$ is pseudoreal,  quasi-minuscule, and  denoted as $\mathbf{56}$. 
\end{enumerate}

\begin{table}[H]
\begin{center}
\scalebox{.95}{
$
\arraycolsep=3pt\def\arraystretch{1.5}
\begin{array}{|l|l|c|}
\hline
\quad\mathfrak{g}\to \mathfrak{m} & \text{Characteristic representation $\chi$ } & \text{Index} \\
\hline
\hline
{\begin{array}{l}
A_{n-k} \oplus A_{k-1}\to A_n\\
(n> k> 1, n\geq 4)
\end{array} }& 
\chi=
(\mathbf{n-k+1},\mathbf{\overline{k}}) \oplus (\mathbf{\overline{n-k+1}},\mathbf{k})\oplus (\mathbf{1},\mathbf{1})
%\end{array} 
& (1,1) \\
\hline
\begin{array}{l}
 D_{n-2}\oplus A_1  \to D_n \\
(n\geq 6)
\end{array}
 & 
\chi= (\mathbf{2n-4},\mathbf{2})^{\oplus 2} \oplus  \mathbf{(1,1)}^{\oplus 3}
 & (1,1) \\
\hline
\begin{array}{c}  D_{n-4}\oplus A_3 \to     D_n \\
(n\geq 8)
\end{array}
  & 
 \chi=(\mathbf{2n-8}, \mathbf{1})^{\oplus 2}\oplus (\mathbf{1},\mathbf{6})^{\oplus 2} \oplus (\mathbf{2n-8},\mathbf{6})\oplus \mathbf{(1,1)}
 & (1,1) \\
 
 \hline\hline
\begin{array}{l}
 D_{n-k}\oplus A_{k-1} \to D_n   \\
(k\geq 3, n-k\geq 4)
\end{array} & 
\chi= (\mathbf{2n-2k},\mathbf{k})
\oplus (\mathbf{2n-2k},\mathbf{\overline{k}})\oplus (\mathbf{1},  \mathbf{\frac{1}{2} k(k-1)}) \oplus (\mathbf{1}, \mathbf{\overline{\frac{1}{2} k(k-1)}}) \oplus \mathbf{(1,1)}
& (1,1) \\
\hline
\hline
%:
\ A_{n-1} \to A_{n}\ { (n\geq  4)}  &  
\chi=\mathbf{n} \oplus\mathbf{\overline{n}}\oplus\mathbf{1} & 1 \\
\hline
\ A_{n-1} \to D_n\  (n\geq 5) &
\chi=
\mathbf{\frac{1}{2} n(n-1)}
\oplus \mathbf{\overline{\frac{1}{2} n(n-1)}}\oplus \mathbf{1} & 1 \\
\hline
\ D_{n-1} \to  D_{n}\  (n\geq 5) &
\chi= \mathbf{(2n-2)}^{\oplus 2}\oplus \mathbf{1}  & 1 \\
\hline
\  D_5\to E_6 &
\chi=\mathbf{16}\oplus \mathbf{\overline{16}}\oplus \mathbf{1}  & 1 \\
 \hline
\ A_5\to E_6 &
\chi=\mathbf{20}^{\oplus 2}\oplus \mathbf{1}^{\oplus 3}  & 1 \\
 \hline
\ E_6 \to E_7& 
\chi= \mathbf{27}\oplus \mathbf{\overline{27}}\oplus \mathbf{1}  & 1 \\
\hline
\ D_6 \to  E_7&
\chi=\mathbf{32}^{\oplus 2}\oplus \mathbf{1}^{\oplus 3}  & 1 \\
\hline
\ A_6\to E_7 & 
\chi= \mathbf{35}\oplus \mathbf{\overline{35}}\oplus\mathbf{7}\oplus \mathbf{\overline{7}} & 1 \\	
\hline
\ E_7\to E_8 &
\chi=
  \mathbf{56}^{\oplus 2} \oplus  \mathbf{1}^{\oplus 3} & 1 \\
 \hline	
\end{array}
$}
\end{center}
\label{tb:branching}
\caption{Complete branching rules of the embeddings $\mathfrak{g}\rightarrow  \mathfrak{m}$ described in the Katz--Vafa's paper. 
 By definition   $  [\mathbf{adj}\   \mathfrak{m}]=  [\mathbf{adj}\  \mathfrak{g}]\oplus \chi$. 
In the last column, we indicate the Dynkin index of the embedding $\mathfrak{g}\to \mathfrak{m}$.  We restrict the values of $k$ and $n$ to ensure that the Lie subalgebra $\mathfrak{g}$ is unique in its isomorphic class up to linear equivalence in $\mathfrak{m}$.
\label{Table:KV}
}
\end{table}

\subsection{Non-singleton subalgebras among the orignal Katz--Vafa's cases} \label{sec:boundaryKV}

 In this section, we scrutinize  Katz--Vafa cases  in which the subalgebra $\mathfrak{g}$ is not unique up to linear equivalence in its isomorphic class inside the parent subalgebra $\mathfrak{m}$. 
These are the {\em non-singleton subalgebras}. They are all boundary cases of the generic singleton cases of the Katz--Vafa cases. 
First we show that such subalgebras do exist.  Two of the Katz--Vafa's cases, namely 
\begin{align}
& \quad \text{D}^1_{n-4}\oplus \text{A}^1_{3} \to \text{D}_n\quad\text{and} \quad  \text{D}_{n-k}\oplus \text{A}_{k-1} \to \text{D}_n ,
\end{align}
share an intersection, which has not been studied in the literature. When $k=4$ (or $k=n-3$), they both give 
an embedding  $D^1_{n-4}\oplus A^1_3\to D_n$. However, they give distinct characteristic representations, namely: 
\begin{align}
& [\mathbf{adj}_{\mathfrak{so}_{2n}}] =[ (\mathbf{adj}_{\mathfrak{so}_{2n-8}},\mathbf{1})\oplus (\mathbf{1},\mathbf{15})]  \oplus (\mathbf{1},\mathbf{6})^{\oplus 2}  \oplus \mathbf{(1,1)} \oplus (\mathbf{2n-8},\mathbf{6})  \oplus (\mathbf{2n-8}, \mathbf{1})^{\oplus 2}, \label{eq:DAtoDch1} \\
& [\mathbf{adj}_{\mathfrak{so}_{2n}}] =[ (\mathbf{adj}_{\mathfrak{so}_{2n-8}},\mathbf{1})\oplus (\mathbf{1},\mathbf{15})]   \oplus  (\mathbf{1},  \mathbf{6})^{\oplus 2}
  \oplus \mathbf{(1,1)}  \oplus (\mathbf{2n-8},\mathbf{\overline{4}})  \oplus (\mathbf{2n-8},\mathbf{4}). \label{eq:DAtoDch2}
\end{align}
The first one uses the representation $\mathbf{6}$ of A$_3$ and does not use the spin representation of A$_3$, while the second one explicitly uses the spin representation $\mathbf{4}$ of A$_3$. 
Using the  small rank isomorphism $\text{Spin}(6)\cong\text{SU}(4)$, we determine that the first branching rule given in the equation \eqref{eq:DAtoDch1} is compatible with a gauge group $\text{SO}(2n-8)\times\text{SO}(6)$, while the second one given by the equation \eqref{eq:DAtoDch2} is compatible with a gauge group $\text{SO}(2n-8)\times\text{SU}(4)$.

All the other cases are related to one of the following  rank isomorphisms involving D$_n$: 
\begin{equation}
D_1 \cong A_1, \quad 
 D_2 \cong A_1\oplus A_1, \quad
D_3\cong A_3.
\end{equation}
Utlizing these isomorphisms, we get the following boundary cases: 
\begin{align}
A_1 & \to \text{A}_2 \\
 A_1\oplus A_{1} & \to \text{A}_3,\\
 A_3  & \to \text{D}_4, \\
A_3\oplus A_{1} & \to \text{D}_5,\\
A_3\oplus A_{1}\oplus A_1 & \to \text{D}_6,\\
A_3\oplus A_{3} &\to \text{D}_7.
\end{align}
In each case, the subalgebra is not unique up to linear equivalence. We will scrutinize each case in the rest of this section.

\subsection{$A_1\to A_2$} \label{sec:A1toA2}

Despite the simple appearance, the case of $A_1\to A_2$ gives rise to two distinct embeddings:
\begin{align}
& \text{A}_1^1\to \text{A}_2 : \quad [\mathbf{8}]=[\mathbf{3}]\oplus\chi ,\quad \chi=\mathbf{2^{\oplus 2}}\oplus\mathbf{1}, \\
& \text{A}_1^2\to \text{A}_2 : \quad [\mathbf{8}]=[\mathbf{3}]\oplus\chi ,\quad \chi=\mathbf{5}.
\end{align}
The first one give rise to an embedding of Dynkin index one, however, the other one corresponds to an embedding with a Dynkin index two. The matter representation from the geometric perspective corresponds to the first one, which demonstrates the effectiveness of our proposal as a selection rule. Having two distinct matter contents are not surprising since these corresponds to the gauge groups SU($2$) and SO($3$) respectively that are both subgroups of SU($3$).

\subsection{The  $A_1\oplus A_1\to A_3$ case} \label{sec.bosonic}
Up to linear equivalence, there are two classes of  embeddings of type  $A_1\oplus A_1\to A_3$:
\begin{align}
& \text{A}_1^1\oplus\text{A}^1_1\to \text{A}_3 : \quad [\mathbf{15}]=[\mathbf{(3,1)}\oplus\mathbf{(1,3)}]\oplus (\mathbf{2,2})^{\oplus 2} \oplus\mathbf{(1,1)},
\label{D2toD3index11} \\
& \text{A}_1^2\oplus\text{A}^2_1\to \text{A}_3 : \quad [\mathbf{15}]=[\mathbf{(3,1)}\oplus \mathbf{(1,3)}]\oplus \mathbf{(3,3)}.
\label{D2toD3index22}
\end{align}
They have distinct Dynkin indices and characteristic representations.
The embedding of index one, namely  $\text{A}_1^1\oplus\text{A}^1_1\to \text{A}_3$, gives the matter content that we expect at the collision of two branes carrying each an  SU($2$) gauge group with the bifundamental matter  $(\mathbf{2,2})$.  
In F-theory, this representation is observed for the  Spin($4$) and SO($4$)-model studied in \cite{SO4}.

We note that this is exactly the case considered in the original Katz--Vafa for the A-series where the Levi rule is used to ruled out  equation \eqref{D2toD3index22} and remain the equation \eqref{D2toD3index11} solely as the correct matter representation listed in Table \ref{Table:KV}.

However, despite that the embedding of equation \eqref{D2toD3index22} does not produce the bifundamental representation one would expect from D-brane construction, it does correspond to a representation familiar in string theory compactification. In fact, the matter transforming in the representation $(\mathbf{3},\mathbf{3})$ appears  naturally when the bosonic string is compactified on a  circle of  self-dual  radius as discussed in \cite[Section 8.3]{Polchinski:V1}. 
 
\subsection{$\text{A}_3\to \text{D}_4$}\label{A3D4}

There are three nonequivalent subalgebras of type $\text{A}_3$ in D$_4$. 
They are connected by triality as each copy of $\text{A}_3$ corresponds to removing one of the legs of D$_4$. 
They all have Dynkin index one and give the same branching rule for the adjoint representation of D$_4$:
\begin{align}
&  \text{A}_3^1\to \text{D}_4:\quad [\mathbf{28}]=[\mathbf{15}]\oplus \chi, \quad \chi=
\mathbf{6}\oplus \mathbf{1}. 
\end{align}
Unlike the adjoint representation, they have different branchings for the vector and spinor representations of D$_4$:
\begin{align}
&  \text{A}_3^1\to \text{D}_4:
\quad  [\mathbf{28}]=[\mathbf{15}]\oplus
\mathbf{6}\oplus \mathbf{1}, 
\quad 
   \    
\mathbf{8}_{\text{v}}=
\mathbf{4}\oplus \mathbf{\overline{4}},
\quad \mathbf{8}_{\text{s}}=
\mathbf{4}\oplus \mathbf{\overline{4}}, 
\quad 
\mathbf{8}_{\text{c}}=
\mathbf{6}\oplus\mathbf{1}^{\otimes 2}, \\
&  
\text{A}_3^1\to \text{D}_4:
\quad 
 [\mathbf{28}]=[\mathbf{15}]\oplus
\mathbf{6}\oplus \mathbf{1}, \quad
\mathbf{8}_{\text{v}}=
\mathbf{4}\oplus \mathbf{\overline{4}},
\quad 
   \      
\mathbf{8}_{\text{s}}=
\mathbf{6}\oplus\mathbf{1}^{\otimes 2}, 
\quad \mathbf{8}_{\text{c}}=
\mathbf{4}\oplus \mathbf{\overline{4}}, 
\\
&  \text{A}_3^1\to \text{D}_4:\quad
  [\mathbf{28}]=[\mathbf{15}]\oplus
\mathbf{6}\oplus \mathbf{1}, \quad
 \mathbf{8}_{\text{v}}=\mathbf{6}\oplus\mathbf{1}^{\otimes 2}
\quad \mathbf{8}_{\text{s}}=
\mathbf{4}\oplus \mathbf{\overline{4}}, 
\quad 
\quad \mathbf{8}_{\text{c}}=
\mathbf{4}\oplus \mathbf{\overline{4}}.
\end{align}

These three enhancements are familiar from the antisymmetric representation of the SU($4$)-model  \cite{ESY1} and the full matter representation of the SO($6$)-model \cite{SO356}. 
 
\subsection{$A_1\oplus A_3\to D_5$} 

There are two linearly inequivalent subalgebras isomorphic to $A_1\oplus A_3$ in D$_5$ and they differ by their Dynkin indices and their characteristic representations:
\begin{align} 
& \text{A}_1^1\oplus\text{A}^1_3\to \text{D}_5 : 
\quad [\mathbf{45}]=[(\mathbf{3},\mathbf{1})\oplus(\mathbf{1},\mathbf{15})]\oplus (\mathbf{2},\mathbf{6})^{\oplus 2}\oplus(\mathbf{1},\mathbf{1})^{\oplus 3}, 
\label{eq:A1A1toD5case1}\\ 
& \text{A}_1^2\oplus\text{A}^1_3\to \text{D}_5 : 
\quad [\mathbf{45}]=[(\mathbf{3},\mathbf{1})\oplus(\mathbf{1},\mathbf{15})]\oplus (\mathbf{3},\mathbf{6})\oplus(\mathbf{3},\mathbf{1})\oplus(\mathbf{1},\mathbf{6}).
\label{eq:A1A1toD5case2}
\end{align}

The branching rule in equation \eqref{eq:A1A1toD5case1} has Dynkin index one and has bifundamental representation as expected for a generic model $\text{SU}(2)\times\text{SU}(4)/\mathbb{Z}_2$, which is derived geometrically in \cite{EKY2}.  We do not know any model satisfying the  branching rule given in equation \eqref{eq:A1A1toD5case2}, however, it is compatible with the matter content of an   $\text{SO}(3)\times\text{SO}(6)$-model since there is no appearance of the spin representation $\mathbf{2}$ of SU($2$).

\subsection{$A_3 \oplus A_1\oplus A_1\to D_6$}

There are three distinct embeddings of the Lie algebra $A_3 \oplus A_1\oplus A_1$ as a subalgebra of D$_6$. 
They have the following Dynkin indices and branching rule for the adjoint representation: 
 \begin{align} 
& \begin{aligned}
\text{A}_3^1\oplus \text{A}_1^1\oplus\text{A}^1_1\to \text{D}_6 : 
\quad & [\mathbf{66}]=
[(\mathbf{15},\mathbf{1},\mathbf{1})\oplus  (\mathbf{3},\mathbf{1},\mathbf{1})\oplus(\mathbf{1},\mathbf{3}, \mathbf{1})]
\oplus \chi ,\\
& \chi = (\mathbf{4},\mathbf{2},\mathbf{2})\oplus(\mathbf{\overline{4}},\mathbf{2},\mathbf{2})\oplus (\mathbf{6},\mathbf{1},\mathbf{1})^{\oplus 2}\oplus (\mathbf{1},\mathbf{1},\mathbf{1}) ,
\end{aligned}
\label{eq:A3A1A2toD6ch1} \\
& \begin{aligned} 
\text{A}_3^1\oplus \text{A}_1^1\oplus\text{A}^1_1\to \text{D}_6 : 
\quad & [\mathbf{66}]=
[(\mathbf{15},\mathbf{1},\mathbf{1})\oplus  (\mathbf{3},\mathbf{1},\mathbf{1})\oplus(\mathbf{1},\mathbf{3}, \mathbf{1})]
\oplus \chi ,\\
& \chi = (\mathbf{1},\mathbf{2},\mathbf{2})^{\oplus 2}
\oplus (\mathbf{6},\mathbf{2},\mathbf{2})
\oplus (\mathbf{6},\mathbf{1},\mathbf{1})^{\oplus 2}\oplus (\mathbf{1},\mathbf{1},\mathbf{1}) ,
\end{aligned}
\label{eq:A3A1A2toD6ch2} \\
& \begin{aligned} 
\text{A}_3^1\oplus \text{A}_1^2\oplus\text{A}^2_1\to \text{D}_6 : 
\quad & [\mathbf{66}]=
[(\mathbf{15},\mathbf{1},\mathbf{1})\oplus  (\mathbf{3},\mathbf{1},\mathbf{1})\oplus(\mathbf{1},\mathbf{3}, \mathbf{1})]
\oplus \chi ,\\
& \chi = (\mathbf{1},\mathbf{3}, \mathbf{3})
\oplus (\mathbf{6},\mathbf{3},\mathbf{1})
\oplus (\mathbf{6},\mathbf{1},\mathbf{3}) .
\end{aligned}
\label{eq:A3A1A2toD6ch3}
\end{align}
In the equations \eqref{eq:A3A1A2toD6ch1} and \eqref{eq:A3A1A2toD6ch2}, we have two subalgebras $A_3 \oplus A_1\oplus A_1$ of index one in D$_6$ but with different characteristic representations. 
Such situations are to be expected as A$_3$ is the Lie algebra of both SU($4$) and SO($6$) and the resulting representations in a gauge theory will be different. This is easy to be witnessed as the equation \eqref{eq:A3A1A2toD6ch1} yields the representations $\mathbf{4}$ and $\mathbf{\overline{4}}$ corresponding to the gauge algebra SU($4$) whereas the equation \eqref{eq:A3A1A2toD6ch2} yields the representation $\mathbf{6}$ corresponding to the gauge algebra SO($6$).

\subsection{$A_3\oplus A_3\to D_7$}

There are two linearly nonequivalent subalgebras $A_3\oplus A_3$ in D$_7$ and they both have Dynkin index one for each of their simple components. But they have distinct characteristic representations:  
 \begin{align} 
&\text{A}_3^1\oplus\text{A}^1_3\to \text{D}_7 : 
\quad [\mathbf{91}]=[(\mathbf{15},\mathbf{1})\oplus(\mathbf{1},\mathbf{15})]
\oplus (\mathbf{4},\mathbf{6})\oplus(\mathbf{\overline{4}},\mathbf{6})
\oplus  (\mathbf{6},\mathbf{1})^{\oplus 2}  
\oplus (\mathbf{1},\mathbf{1}), \label{eq:A3A3toD7ch1} \\
 &\text{A}_3^1\oplus\text{A}^1_3\to \text{D}_7 : 
\quad [\mathbf{91}]=[(\mathbf{15},\mathbf{1})\oplus(\mathbf{1},\mathbf{15})]
\oplus(\mathbf{6},\mathbf{6})
\oplus  (\mathbf{1},\mathbf{6})^{\oplus 2}
\oplus  (\mathbf{6},\mathbf{1})^{\oplus 2}  
\oplus (\mathbf{1},\mathbf{1}). \label{eq:A3A3toD7ch2}
\end{align}
This shows that in a given isomorphism class, the subalgebra of embedding index one is not necessarily unique. 
Such cases are rare but do exist. When the parent Lie algebra has a nontrivial center, the global structure of the Lie group can help select one embedding over another. For example, in the current case, the first branching rule given in equation \eqref{eq:A3A3toD7ch1} involves the spin representations $\mathbf{4}$ and $\overline{\mathbf{4}}$ of D$_3$ while the other one given in equation \eqref{eq:A3A3toD7ch2} does not.

\section{Beyond  the Katz--Vafa cases: singleton subalgebras}
\label{sec:singleton}
In this section, we explore branching rules from the embeddings $\mathfrak{g}\to \mathfrak{m}$ of type ADE that are not in the original cases of Katz--Vafa.
To ensure that the Katz--Vafa method is not ambiguous, we consider subalgebras $\mathfrak{g}\subset\mathfrak{m}$ that admit a unique embedding $\mathfrak{g}\to \mathfrak{m}$. More precisely, we consider singleton embeddings, which is defined as following.
\begin{defn} 
\label{def:singleton}
An embedding $\mathfrak{g}\subset\mathfrak{m}$ is said to be a {\em singleton embedding} if there is no other embedding $\mathfrak{s}\to \mathfrak{m}$  such that $\mathfrak{g}$ is isomorphic to $\mathfrak{s}$, while  $\mathfrak{g}$ and $\mathfrak{s}$ are not linearly equivalent.  
\end{defn}
All the singleton subalgebras of simple Lie algebras up to rank eight are classified in \cite{GEO}. 
In the context of this section, we will only consider those that are ADE semisimple Lie subalgebras inside E$_6$, E$_7$, and E$_8$. They are listed in Table \ref{Table:SingletonADE}.

We focus on exploring the branching rules for the embeddings $\mathfrak{g}\to\mathfrak{m}$  given by Miranda's models. 
Miranda's models have seven types of collisions of Kodaira fibers that we consider here as  ADE cases \cite{Miranda.Smooth, Bershadsky:1996nu}. 
These embeddings are a natural extension of the cases considered by Katz--Vafa and are inspired by the geometry of elliptic threefolds defined by Weierstrass models:
\begin{align}
\begin{cases}
\begin{array}{rl}
\text{I$_n$+I$_m$} &\to\   \    ``\text{I}_{n+m}",\quad (n, m> 0),\\
\text{I$_n$+I$^*_m$} &\to\  \   `` \text{I}^*_{n+m}", \quad (n>0, m>0) \\
\text{II+IV} &\to\   \   ``\text{I}_0^*" \\
\text{II+I$_0^*$} &\to\   \   ``\text{IV}^* ",\\
\text{III+I$^*_0$} &\to\   \  ``\text{III}^*", \\
\text{II+IV$^*$}& \to\  \   ``\text{II}^*",\\
\text{IV+I$^*_0$}& \to\   \    ``\text{II}^*".
\end{array}
\end{cases}
 \end{align}

These collisions of singularities often appear in F-theory models \cite{Bershadsky:1996nu, SU2G2}. In contrast to those of Katz--Vafa, they are  not always rank-one enhancements. 
The dual graphs of these fibers suggest the following embeddings when all Kodaira fibers are understood to be of split types: 
\begin{align}
\begin{cases}
\begin{array}{l}
\text{A}_{n-1}\oplus \text{A}_{m-1}\to \text{A}_{n+m-1}, \quad (n, m> 0),\\
\text{A}_{n-1}\oplus \text{D}_{4+m}\to \text{D}_{n+m+4}, \quad (n, m> 0),\\
\text{A}_2\to \text{D}_4, \\
 \text{D}_4\to \text{E}_6,\\
 \text{A}_1\oplus \text{D}_4\to \text{E}_7, \\
 \text{E}_6\to \text{E}_8,\\
\text{A}_2\oplus \text{D}_4\to \text{E}_8,\\
\text{A}_1\oplus \text{D}_4\to \text{E}_8.
\end{array}
\end{cases}
 \end{align}
We have already studied the three embeddings in Section \ref{sec:original} as a part of the original  Katz--Vafa's cases. We consider the rest of the new embeddings that corresponds to the following five cases: 
\begin{align}
\text{D}_4\   \to\    \text{E}_6,\quad
\text{A}_1\oplus \text{D}_4\   \to\    \text{E}_7,\quad
\text{E}_6\   \to\    \text{E}_8, 
\quad  \text{A}_2\oplus \text{D}_4\  \to\    \text{E}_8, 
\quad  \text{A}_1\oplus \text{D}_4\  \to\    \text{E}_8.
 \end{align}
They are rank-two  enhancements.
We can use our understanding of the geography of semisimple Lie subalgebras of  exceptional Lie algebras to streamline the analysis \cite{GEO}. 
There is no possible ambiguity for applying the Katz--Vafa method to these five cases as they are all  singleton embeddings (the singleton embeddings are embeddings of the type $\mathfrak{g}\to \mathfrak{m}$ where $\mathfrak{g}$ is unique in $\mathfrak{m}$ up to linear equivalence, see Definition \ref{def:singleton}). Moreover, each case has  Dynkin index one. 

\begin{table}[H]
\arraycolsep=3pt\def\arraystretch{1.5}
\begin{center}
\begin{tabular}{|c|p{14cm}|}
\hline 
$\mathfrak{g}$ & \qquad\qquad\qquad ADE semisimple singleton subalgebras $\mathfrak{g}\subset\mathfrak{m}$\\
\hline 
A$_2$ & none \\
\hline 
A$_3$ & A$_2$  \\
\hline 
A$_4$ & A$_2$, A$_3$, $\text{A}_1\oplus\text{A}_2$ \\
\hline 
D$_4$ & $\text{A}_1^{\oplus 4}$ \\
\hline
A$_5$ & $\text{A}_1\oplus\text{A}_3$, $\text{A}_2^{\oplus 2}$, A$_4$\\
\hline 
D$_5$& 
$\text{A}_1^{\oplus 2}\oplus \text{A}_2$, $\text{A}_1^{\oplus 2} \oplus \text{A}_3$, A$_4$, D$_4$\\
\hline 
E$_6$ &
$\text{A}_1\oplus \text{A}_1\oplus\text{A}_3$, $\text{A}_1\oplus\text{A}_4$, $\text{A}_1\oplus\text{A}_5$, $\text{A}^{\oplus 3}_2$, $\text{A}_4$, $\text{A}_5$, $\text{D}_4$, $\text{D}_5$
 \\
\hline
E$_7$ &
A$_1^{\oplus 7}$,  $\text{A}_1^{\oplus 3} \oplus \text{D}_4$, $\text{A}_1\oplus \text{A}_2\oplus\text{A}_3$, $\text{A}_1 \oplus \text{A}_3^{\oplus 2}$,  $\text{A}_1\oplus\text{D}_5$, $\text{A}_1\oplus\text{D}_6$,
$\text{A}_2^{\oplus 3}$, $\text{A}_2\oplus\text{A}_4$, $\text{A}_2\oplus \text{A}_5$, $\text{A}_3^{\oplus 2}$, $\text{A}_4$, A$_6$, A$_7$, D$_5$, D$_6$, E$_6$
\\   
\hline   
E$_8$ & 
A$_1^{\oplus 8}$, 
$\text{A}_1^{\oplus 4}\oplus \text{D}_4$, 
$\text{A}_1^{\oplus 2} \oplus \text{A}_2 \oplus \text{A}_3$,
$\text{A}_1^{\oplus 2}\oplus \text{A}_3^{\oplus 2}$, 
$\text{A}_1^{\oplus 2}\oplus \text{D}_6$, $\text{A}_1\oplus \text{A}_2\oplus \text{A}_4$, 
$\text{A}_1\oplus\text{A}_2\oplus \text{A}_5$,  
$\text{A}_1\oplus\text{A}_6$, $\text{A}_1\oplus\text{A}_7$,
$\text{A}_1\oplus\text{E}_7$,
$\text{A}_2^{\oplus 4}$, $\text{A}_2\oplus\text{A}_4$, $\text{A}_2\oplus\text{A}_5$, 
$\text{A}_2\oplus \text{D}_5$, $\text{A}_2 \oplus \text{E}_6$,
$\text{A}_3\oplus \text{A}_4$,
$\text{A}_3\oplus \text{D}_4$, 
$\text{A}_3\oplus\text{D}_5$,
$\text{A}_4^{\oplus 2}$, 
$\text{D}_4\oplus\text{D}_4$, 
$\text{A}_5$, $\text{A}_6$, $\text{A}_8$,  
D$_5$, D$_6$, D$_7$, D$_8$, E$_6$, E$_7$
\\
\hline
\end{tabular}
\end{center}
\caption{ Singleton ADE semisimple subalgebras of A$_5$, D$_5$, E$_6$, E$_7$, and E$_8$. Each subalgebra is the only subalgebra (up to linear equivalence) in its  isomorphism class. See  \cite{GEO} for the full list of all singleton semisimple subalgebras of simple Lie algebras of rank up to eight. }
\label{Table:SingletonADE}
\end{table}

 \subsection{$ \text{D}_4\to \text{E}_6$} 
 
 The algebra E$_6$ has 118 proper Lie subalgebras up to linear equivalence, among whom only one subalgebra is isomorphic to D$_4$. It follows that D$_4$ is a singleton subalgebra of E$_6$. Moreover, it has Dynkin index one and gives the following branching rule for the adjoint representation:  
 \begin{align}
\text{D}^1_4\to \text{E}_6:&\quad [\mathbf{78}]=[\mathbf{28}]\oplus\chi, \quad \chi= \mathbf{8}_{\text{v}}^{\oplus 2}\oplus \mathbf{8}_{\text{s}}^{\oplus 2}\oplus \mathbf{8}_{\text{c}}^{\oplus 2}\oplus\mathbf{1}^{\oplus 2}.
 \end{align}
 
 The Spin($8$)-model is studied in \cite{G2}. 
  \subsection{$ \text{A}_1\oplus\text{D}_4\to \text{E}_7$}

\begin{align}
\begin{split}
\text{A}_1^1\oplus\text{D}^1_4\to \text{E}_7:&\quad [\mathbf{133}]=[(\mathbf{3},\mathbf{1})\oplus (\mathbf{1},\mathbf{28})]\oplus \chi, \quad \\
&\quad \chi=(\mathbf{1}, \mathbf{8}_{\text{v}})^{\oplus 4}\oplus (\mathbf{2}, \mathbf{8}_{\text{s}})^{\oplus 2}\oplus (\mathbf{2}, \mathbf{8}_{\text{c}})^{\oplus 2}\oplus (\mathbf{1}, \mathbf{1})^{\oplus 6} ,\\
\end{split} \\
\begin{split}
\text{A}_1^1\oplus\text{D}^1_4\to \text{E}_7:&\quad [\mathbf{133}]=[(\mathbf{3},\mathbf{1})\oplus (\mathbf{1},\mathbf{28})]\oplus \chi, \quad \\
&\quad \chi=(\mathbf{2}, \mathbf{8}_{\text{v}})^{\oplus 2}\oplus (\mathbf{1}, \mathbf{8}_{\text{s}})^{\oplus 4}\oplus (\mathbf{2}, \mathbf{8}_{\text{c}})^{\oplus 2}\oplus (\mathbf{1}, \mathbf{1})^{\oplus 6} ,\\
\end{split} \\
\begin{split}
\text{A}_1^1\oplus\text{D}^1_4\to \text{E}_7:&\quad [\mathbf{133}]=[(\mathbf{3},\mathbf{1})\oplus (\mathbf{1},\mathbf{28})]\oplus \chi, \quad \\
&\quad \chi=(\mathbf{2}, \mathbf{8}_{\text{v}})^{\oplus 2}\oplus (\mathbf{2}, \mathbf{8}_{\text{s}})^{\oplus 2}\oplus (\mathbf{1}, \mathbf{8}_{\text{c}})^{\oplus 4}\oplus (\mathbf{1}, \mathbf{1})^{\oplus 6} .\\
\end{split}
\end{align}

All the branching rules do not respect the triality with three representations $\mathbf{8}_{\text{v}}$, $\mathbf{8}_{\text{c}}$, and $\mathbf{8}_{\text{s}}$. In fact, the natural branching rule to respect the triality to have $\chi=(\mathbf{2}, \mathbf{8}_{\text{v}})^{\oplus 2}\oplus (\mathbf{2}, \mathbf{8}_{\text{s}})^{\oplus 2}\oplus (\mathbf{2}, \mathbf{8}_{\text{c}})^{\oplus 2}\oplus (\mathbf{1}, \mathbf{1})^{\oplus 6}$ does not exist as a branching rule; this is consistent with that this will yield a fractional Dynkin index.

 \subsection{$ \text{E}_6\to \text{E}_8$}

The subalgebra E$_6$ is a singleton subalgebra of E$_8$ of Dynkin index one, and its characteristic representation  $\chi$ is 
 \begin{align}
\text{E}^1_6\to \text{E}_8:&\quad 
[\mathbf{248}]=[\mathbf{78}]\oplus\chi, \quad
\chi=\mathbf{27}^{\oplus 3}
\oplus\mathbf{\overline{27}}^{\oplus 3}
\oplus\mathbf{1}^{\oplus 8}.
 \end{align}

 \subsection{$\text{A}_3\oplus \text{D}_4\to \text{E}_8$} 
 
The subalgebra  $\text{A}_3\oplus \text{D}_4$   is a singleton subalgebra of E$_8$. Its characteristic representation  $\chi$ is:
\begin{align}
\begin{split}
\text{A}^1_3\oplus \text{D}^1_4\to \text{E}_8:&\quad [\mathbf{248}]=[(\mathbf{15},\mathbf{1})\oplus(\mathbf{1},\mathbf{28})]\oplus\chi, \quad \\
&\quad \chi=(\mathbf{6},\mathbf{8}_{\text{v}})
\oplus(\mathbf{1},\mathbf{8}_{\text{v}})^{\oplus 2}
\oplus(\mathbf{6},\mathbf{1})^{\oplus 2}
\oplus (\mathbf{\overline{4}},\mathbf{8}_{\text{c}})
\oplus(\overline{\mathbf{4}},\mathbf{8}_{\text{s}})
\oplus(\mathbf{4},\mathbf{8}_{\text{s}})
\oplus(\mathbf{4},\mathbf{8}_{\text{c}})
\oplus (\mathbf{1},\mathbf{1}).
\end{split}
\end{align}
The branching rules do not respect the triality with three representations $\mathbf{8}_{\text{v}}$, $\mathbf{8}_{\text{c}}$, and $\mathbf{8}_{\text{s}}$. We note that this is compatible with an $\text{SU}(4)\times\text{Spin}(8)$-model since we have spinors representations $(\mathbf{4},\mathbf{8}_{\text{s}})$ and $(\mathbf{4},\mathbf{8}_{\text{c}})$.

\subsection{$\text{A}_1\oplus \text{A}_1\to \text{D}_4$}

There is a unique subalgebra  of the type $\text{A}_1\oplus \text{A}_1$ in $\text{D}_4$  \cite{GEO} and it has  Dynkin index $(1,1$) and characteristic representation $\chi$ given as follows: 
 \begin{align}
\text{A}^1_1\oplus\text{A}^1_1\to \text{D}_4:&\quad 
[\mathbf{28}]=(\mathbf{3},\mathbf{1})\oplus(\mathbf{1},\mathbf{3})\oplus\chi, \quad 
\chi=
(\mathbf{2},\mathbf{2})^{\oplus 4}\oplus (\mathbf{1},\mathbf{1})^{\oplus 6}.
\end{align}
These representations occur in the  SO($4$)-models and Spin($4$)-models \cite{SO4}. 

\section{The SU($5$)-model}
\label{sec:nonsingleton}
 There are models from F-theory that give rise to non-singleton subalgebras. We illustrate this point in this section with  geometrically well-understood models such as SU(5)-models  \cite{EY,ESY1,ESY2} and SO(6)-models \cite{SO356}.

\begin{figure}[H]
\centering
\scalebox{.8}{
\begin{tikzpicture}
\node at (-1.5,0) { $\mathfrak{su}_5$ };
\node at (11,0) { $\mathfrak{su}_6$ };
\node[draw,circle,thick,scale=1.25,fill=black,xshift=1cm] (B1) at (90:1.2) {};
\node[draw,circle,thick,scale=1.25,xshift=1cm] (B2) at (162:1.2) {};
\node[draw,circle,thick,scale=1.25,xshift=1cm] (B3) at (234:1.2) {};
\node[draw,circle,thick,scale=1.25,xshift=1cm] (B4) at (306:1.2) {};
\node[draw,circle,thick,scale=1.25,xshift=1cm] (B5) at (378:1.2) {};
\draw[thick] (B1)--(B2)--(B3)--(B4)--(B5)--(B1);
\draw[->,>=stealth',thick=4mm]  (3.5,0)--(5.5,0);
\node[draw,circle,thick,scale=1.25,fill=black,xshift=6cm] (B1) at (180:1.2) {};
\node[draw,circle,thick,scale=1.25,xshift=6cm] (B2) at (240:1.2) {};
\node[draw,circle,thick,scale=1.25,xshift=6cm] (B3) at (300:1.2) {};
\node[draw,circle,thick,scale=1.25,xshift=6cm] (B4) at (360:1.2) {};
\node[draw,circle,thick,scale=1.25,xshift=6cm] (B5) at (60:1.2) {};
\node[draw,circle,thick,scale=1.25,xshift=6cm] (B6) at (120:1.2) {};
\draw[thick] (B1)--(B2)--(B3)--(B4)--(B5)--(B6)--(B1);
\end{tikzpicture}}
\vspace{3mm}

\scalebox{.8}{
\begin{tikzpicture}
\node at (-1.5,0) { $\mathfrak{su}_5$ };
\node at (11,0) { $\mathfrak{so}_{10}$ };
\node[draw,circle,thick,scale=1.25,fill=black,xshift=1cm] (B1) at (90:1.2) {};
\node[draw,circle,thick,scale=1.25,xshift=1cm] (B2) at (162:1.2) {};
\node[draw,circle,thick,scale=1.25,xshift=1cm] (B3) at (234:1.2) {};
\node[draw,circle,thick,scale=1.25,xshift=1cm] (B4) at (306:1.2) {};
\node[draw,circle,thick,scale=1.25,xshift=1cm] (B5) at (378:1.2) {};
\draw[thick] (B1)--(B2)--(B3)--(B4)--(B5)--(B1);
\draw[->,>=stealth',thick=4mm]  (3.5,0)--(5.5,0);
\node[draw,circle,thick,scale=1.25,fill=black,label=left:{1}] (0b) at (6.5,.8){};
\node[draw,circle,thick,scale=1.25,label=left:{1}] (0a) at (6.5,-.8){};
\node[draw,circle,thick,scale=1.25,label=below:{2}] (1a) at (6.5+.8,0){};
\node[draw,circle,thick,scale=1.25,label=below:{2}] (2a) at (6.5+.8*2.2,0){};
\node[draw,circle,thick,scale=1.25,label=right:{1}] (3a) at (6.5+.8*3.2,-.8){};
\node[draw,circle,thick,scale=1.25,label=right:{1}] (3b) at (6.5+.8*3.2,.8){};
\draw[thick] (0a)--(1a)--(2a)--(3a);
\draw[thick] (0b)--(1a);
\draw[thick] (2a)--(3b);
\end{tikzpicture}}
\vspace{3mm}

\scalebox{.8}{
\begin{tikzpicture}
\node at (-1.5,0) { $\mathfrak{su}_5$ };
\node at (11,0) { $\mathfrak{e}_6$ };
\node[draw,circle,thick,scale=1.25,fill=black,xshift=1cm] (B1) at (90:1.2) {};
\node[draw,circle,thick,scale=1.25,xshift=1cm] (B2) at (162:1.2) {};
\node[draw,circle,thick,scale=1.25,xshift=1cm] (B3) at (234:1.2) {};
\node[draw,circle,thick,scale=1.25,xshift=1cm] (B4) at (306:1.2) {};
\node[draw,circle,thick,scale=1.25,xshift=1cm] (B5) at (378:1.2) {};
\draw[thick] (B1)--(B2)--(B3)--(B4)--(B5)--(B1);
\draw[->,>=stealth',thick=4mm]  (3.5,0)--(5.5,0);
\node[draw,circle,thick,scale=1.25,fill=black,label=below:{1}] (0a) at (6.5,-.8){};
\node[draw,circle,thick,scale=1.25,label=below:{2}] (1a) at (6.5+.8,-.8){};
\node[draw,circle,thick,scale=1.25,label=below:{3}] (2a) at (6.5+.8*2,-.8){};
\node[draw,circle,thick,scale=1.25,label=below:{2}] (3a) at (6.5+.8*3,-.8){};
\node[draw,circle,thick,scale=1.25,label=below:{1}] (4a) at (6.5+.8*4,-.8){};
\node[draw,circle,thick,scale=1.25,label=left:{2}] (3b) at (6.5+.8*2,0){};
\node[draw,circle,thick,scale=1.25,label=left:{1}] (4b) at (6.5+.8*2,.8){};
\draw[thick] (0a)--(1a)--(2a)--(3a)--(4a);
\draw[thick] (2a)--(3b)--(4b);
\end{tikzpicture}}
\vspace{3mm}

\scalebox{.8}{
\begin{tikzpicture}
\node at (-1.5,0) { $\mathfrak{su}_5$ };
\node at (11.3,0.8) { $\mathfrak{e}_7$ };
\node[draw,circle,thick,scale=1.25,fill=black,xshift=1cm] (B1) at (90:1.2) {};
\node[draw,circle,thick,scale=1.25,xshift=1cm] (B2) at (162:1.2) {};
\node[draw,circle,thick,scale=1.25,xshift=1cm] (B3) at (234:1.2) {};
\node[draw,circle,thick,scale=1.25,xshift=1cm] (B4) at (306:1.2) {};
\node[draw,circle,thick,scale=1.25,xshift=1cm] (B5) at (378:1.2) {};
\draw[thick] (B1)--(B2)--(B3)--(B4)--(B5)--(B1);
\draw[->,>=stealth',thick=4mm]  (3.5,0)--(5.5,0);
\node[draw,circle,thick,scale=1.25,fill=black,label=below:{1}] (0a) at (6.5,0){};
\node[draw,circle,thick,scale=1.25,label=below:{2}] (1a) at (6.5+.8,0){};
\node[draw,circle,thick,scale=1.25,label=below:{3}] (2a) at (6.5+.8*2,0){};
\node[draw,circle,thick,scale=1.25,label=below:{4}] (3a) at (6.5+.8*3,0){};
\node[draw,circle,thick,scale=1.25,label=left:{2}] (4b) at (6.5+.8*3,.8){};
\node[draw,circle,thick,scale=1.25,label=below:{3}] (4a) at (6.5+.8*4,0){};
\node[draw,circle,thick,scale=1.25,label=below:{2}] (5a) at (6.5+.8*5,0){};
\node[draw,circle,thick,scale=1.25,label=below:{1}] (6a) at (6.5+.8*6,0){};
\draw[thick] (0a)--(1a)--(2a)--(3a)--(4a)--(5a)--(6a);
\draw[thick] (3a)--(4b);
\end{tikzpicture}}
\vspace{3mm}

\quad\quad\quad\scalebox{.8}{
\begin{tikzpicture}
\node at (-1.5,0) { $\mathfrak{su}_5$ };
\node at (12.2,0.8) { $\mathfrak{e}_8$ };
\node[draw,circle,thick,scale=1.25,fill=black,xshift=1cm] (B1) at (90:1.2) {};
\node[draw,circle,thick,scale=1.25,xshift=1cm] (B2) at (162:1.2) {};
\node[draw,circle,thick,scale=1.25,xshift=1cm] (B3) at (234:1.2) {};
\node[draw,circle,thick,scale=1.25,xshift=1cm] (B4) at (306:1.2) {};
\node[draw,circle,thick,scale=1.25,xshift=1cm] (B5) at (378:1.2) {};
\draw[thick] (B1)--(B2)--(B3)--(B4)--(B5)--(B1);
\draw[->,>=stealth',thick=4mm]  (3.5,0)--(5.5,0);
\node[draw,circle,thick,scale=1.25,fill=black,label=below:{1}] (0a) at (6.5,0){};
\node[draw,circle,thick,scale=1.25,label=below:{2}] (1a) at (6.5+.8,0){};
\node[draw,circle,thick,scale=1.25,label=below:{3}] (2a) at (6.5+.8*2,0){};
\node[draw,circle,thick,scale=1.25,label=below:{4}] (3a) at (6.5+.8*3,0){};
\node[draw,circle,thick,scale=1.25,label=below:{5}] (4a) at (6.5+.8*4,0){};
\node[draw,circle,thick,scale=1.25,label=left:{3}] (4b) at (6.5+.8*5,.8){};
\node[draw,circle,thick,scale=1.25,label=below:{6}] (5a) at (6.5+.8*5,0){};
\node[draw,circle,thick,scale=1.25,label=below:{4}] (6a) at (6.5+.8*6,0){};
\node[draw,circle,thick,scale=1.25,label=below:{2}] (7a) at (6.5+.8*7,0){};
\draw[thick] (0a)--(1a)--(2a)--(3a)--(4a)--(5a)--(6a)--(7a);
\draw[thick] (5a)--(4b);
\end{tikzpicture}}
\vspace{3mm}
\caption{The expected gauge group enhancement of SU(5).}
\label{fig:su5enhancements}
\end{figure}
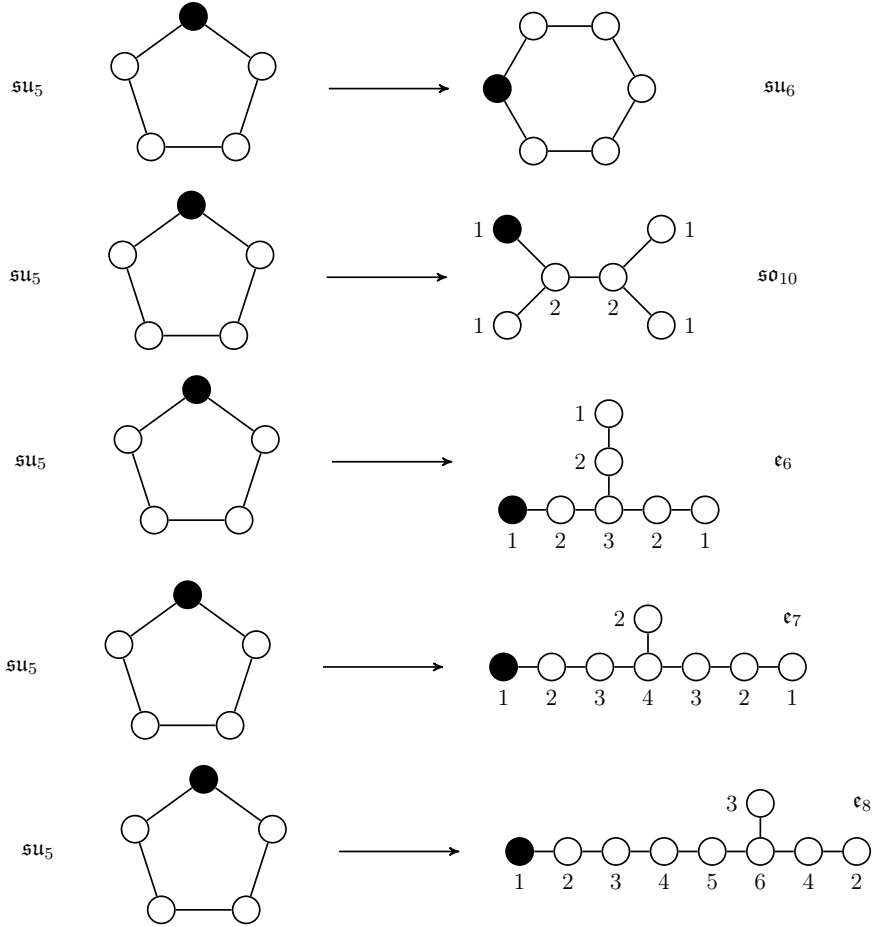

\subsection{ $\text{A}_4\to \text{A}_5, \text{D}_5$}
In this section, we study the branching rule of the Lie algebra A$_4$ in  A$_5$, E$_6$, E$_7$, and E$_8$, as represented in Figure \ref{fig:su5enhancements}. 
In this way, we will reproduce the known result of the SU($5$)-model with respect to its enhancement to E$_6$ and E$_7$. We will also illustrate how the Dynkin index helps identify the ``generic" configuration expected in F-theory when SU($5$) enhances to E$_8$. 

The geometry of the SU($5$)-model proposes the following enhancement over codimension two loci, as represented as the first and second rows of Figure \ref{fig:su5enhancements}: 
\begin{equation}
\begin{cases}
\text{I}^{\text{s}}_5\to \ \text{I}^{\text{s}}_6, \\
 \text{I}^{\text{s}}_5\to \ ``\text{I}_1^{*\text{s}}".
   \end{cases}
\end{equation}
These fiber enhancements indicate that the matter representations will be induced by the following embeddings. 
\begin{align}
& \text{A}_4\to \text{A}_5, \\
 &\text{A}_4\to \text{D}_5.
\end{align}
As we can see from Table \ref{Table:SingletonADE}, A$_4$ is a singleton subalgebra of both A$_5$ and D$_5$.  Thus,  the isomorphism class of the subalgebra is enough to completely determine the embedding and the characteristic representation.
They are already treated in Table \ref{Table:KV} to yield the characteristic representations as
\begin{align}
& \text{A}_4\to \text{A}_5:\quad \chi= \mathbf{5}\oplus\mathbf{\overline{5}}\oplus\mathbf{1}, \\
 &\text{A}_4\to \text{D}_5:\quad 
 \chi=
\mathbf{10}
\oplus \mathbf{\overline{10}}\oplus \mathbf{1}.
\end{align}

\subsection{ $\text{A}_4\to \text{E}_6, \text{E}_7,\text{E}_8$ }\label{A4Excep}
It is also possible to have both representations $\bf{5}$ and $\bf{10}$ and their complex conjugates by requiring an enhancement of the type 
 \begin{equation}
\text{I}^{\text{s}}_5\to \text{IV}^{*\text{s}},\quad \text{III}^*,\quad \text{or}\quad \text{II}^*.
\end{equation}
In terms of corresponding algebras, this represents an enhancement from A$_4$ to E$_6$, E$_7$, or E$_8$, which are represented as the third to fifth rows of Figure \ref{fig:su5enhancements}. 
There is a unique subalgebra of type A$_4$ inside E$_6$ and the same is true inside E$_7$. In both cases, the A$_4$ subalgebra has Dynkin index one. 
For the case of the enhancement to E$_8$, the situation is a bit more complicated. There are two nonlinearly equivalent subalgebras of type A$_4$ inside E$_8$: one subalgebra has Dynkin index one and the other subalgebra has Dynkin index two.
  
The characteristic representations for the subalgebras of index one are
\begin{align}
&  \text{A}^1_4\to \text{E}_6:\quad \   \  
[\mathbf{78}]=[\mathbf{24}]\oplus \chi, \quad
 \chi=
\mathbf{5}\oplus
\mathbf{\overline{5}}\oplus
\mathbf{10}^{\oplus 2}\oplus
\mathbf{\overline{10}}^{\oplus 2}\oplus
\mathbf{1}^{\oplus 4},\\
&  \text{A}_4^1\to \text{E}_7:\quad 
[\mathbf{133}]=[\mathbf{24}]\oplus \chi, \quad
 \chi=\mathbf{5}^{\oplus 4}
\oplus
\mathbf{\overline{5}}^{\oplus 4}
\oplus \mathbf{10}^{\oplus 3}\oplus
\mathbf{\overline{10}}^{\oplus 3}\oplus
\mathbf{1}^{\oplus 9},\\
 & 
 \text{A}_4^1\to \text{E}_8:\quad  [\mathbf{248}]=[\mathbf{24}]\oplus \chi, \quad  \chi=
\mathbf{5}^{\oplus 10}\oplus
 \mathbf{\overline{5}}^{\oplus 10}\oplus
 \mathbf{10}^{\oplus 5}\oplus
\mathbf{\overline{10}}^{\oplus 5}\oplus
\mathbf{1}^{\oplus 24}. 
\end{align}
They all give rise to the same matter representations $\mathbf{5}$, $\mathbf{10}$, and $\mathbf{24}$.

The subalgebra A$^2_4$ in E$_8$ is unique up to linear equivalence and has the following branching rule and characteristic representation: 
\begin{align}
&  \text{A}_4^2\to \text{E}_8:\quad [\mathbf{248}]=[\mathbf{24}]\oplus \chi, \quad \chi=
\mathbf{24}
\oplus
\mathbf{5}\oplus\mathbf{\overline{5}}\oplus
\mathbf{10}\oplus\mathbf{\overline{10}}\oplus
\mathbf{40}\oplus\mathbf{\overline{40}}\oplus
\mathbf{45}\oplus
\mathbf{\overline{45}}.
\end{align}
We can also distinguish A$^1_4$ and A$^2_4$ in E$_8$ by the fact that  A$^2_4$ is a  semisimple subalgebra of  $\text{A}_4^1\oplus \text{A}^1_4$, while $\text{A}^1_4$ is a subalgebra of E$_7$: 
 \begin{align}
&  \text{A}_4^2\to \text{A}_4^1\oplus \text{A}_4^1 \to E_8, \\
& \text{A}_4^1\to \text{A}_5^1\to 
\text{D}_5^1\to \text{E}_6^1 \to \text{E}_7^1\to E_8. 
 \end{align}

\section{G$_2$ and F$_4$-models}\label{sec:G2F4}

In \cite{GM1}, the following branching rules are listed to explain the appearance of the fundamental matters for G$_2$ and F$_4$-models:
\begin{equation}
 \begin{array}{lll}
\text{G}_2\to \text{D}_4 \quad: \quad \mathbf{28}=\mathbf{14}\oplus \mathbf{7}^{\oplus 2},\qquad\quad\quad
 \text{F}_4\to \text{E}_6 \quad: \quad \mathbf{78}=\mathbf{52}\oplus \mathbf{26}.
\end{array}
\end{equation}
 While these branching rules  give the correct matter content observed in F-theory (namely, the representation $\mathbf{7}$ for G$_2$ \cite{G2} and the representation $\mathbf{26}$ for F$_4$ \cite{F4}), they are incompatible with  the geometry of the localized enhancements of the fibers of these geometries. 
For the F$_4$-model, the fiber IV$^{* \text{ns}}$ has localized enhancements not to an incomplete E$_6$ but rather to an incomplete E$_7$ (whose fiber is III$^*$) or an incomplete E$_8$ (whose fiber is II$^*$) depending on the valuations of the Weierstrass coefficients. 

This choice is also not minimal as we can get the representation $\mathbf{7}$ from the unique embedding of G$_2$ in B$_3$ where G$_2$ is a maximal subalgebra. 

The crepant resolution of the Weierstrass models of G$_2$ and F$_4$-models are studied in detail respectively in \cite{G2} and \cite{F4}. 
The geometry of these fibrations suggests that the matter content derived by the Katz--Vafa algorithm should be based on the branching rules for the following embeddings:
\begin{align}
\text{G}_2 \to \text{D}_5,\quad 
\text{G}_2 \to \text{E}_6,\quad
\text{F}_4 \to \text{E}_7,\quad 
\text{F}_4 \to \text{E}_8.
\end{align}
Our discussion will rely on the  geography of the subalgebras of D$_5$, E$_6$, E$_7$, and E$_8$ \cite{GEO}. 
It would be desirable to identify branching rules that explain the observed matter representation using the Katz--Vafa picture while being consistent with the geometry. 

\begin{itemize}
\item The Lie algebra G$_2$ is a singleton subalgebra of B$_3$, D$_4$,  and  D$_5$, and has  Dynkin index one in all of them. 
Thus, the embedding of G$_2$ in these Lie groups give a unique characteristic representation in each case. 

\item The Lie algebra G$_2$ has two non-linearly equivalent embeddings in E$_6$: One is an R-subalgebra of Dynkin index one and the other one is a maximal S-subalgebra of E$_6$ with Dynkin index three. 

\item 
The Lie subalgebra F$_4$ is a singleton subalgebra of Dynkin index one of E$_6$, E$_7$, and E$_8$. 
It follows that the characteristic representation for F$_4$ in all the exceptional subalgebras is unique.  
\end{itemize}

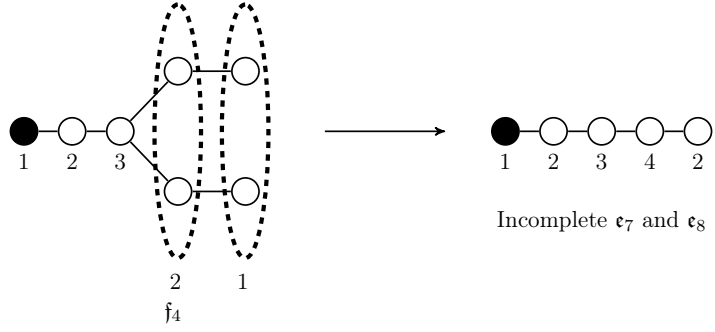
\begin{figure}[H]
\centering
\begin{center}

\scalebox{.8}{
\begin{tikzpicture}
\node[draw,circle,thick,scale=1.25,fill=black,label=below:{1}] (0) at (0,0){};
\node[draw,circle,thick,scale=1.25,label=below:{2}] (1) at (.8,0){};
\node[draw,circle,thick,scale=1.25,label=below:{3}] (2) at (.8*2,0){};
\node[draw,circle,thick,scale=1.25] (3) at (.8*3.2,-1){};
\node[draw,circle,thick,scale=1.25] (4) at (.8*4.6,-1){};
\node[draw,circle,thick,scale=1.25] (5) at (.8*3.2,1){};
\node[draw,circle,thick,scale=1.25] (6) at (.8*4.6,1){};
\node at (.8*3.15,0)[draw,dashed, line width=2pt, ellipse, minimum width=120pt, minimum height=22pt,rotate=90,yshift=-1pt]{};
\node at (.8*4.55,0)[draw,dashed, line width=2pt, ellipse, minimum width=120pt, minimum height=22pt,rotate=90,yshift=-1pt]{};
\node at (.8*3.15,-2.5) {2};
\node at (.8*4.55,-2.5) {1};
\node at (.8*3.1,-3) { $\mathfrak{f}_4$ };
\draw[thick] (0)--(1)--(2)--(3)--(4);
\draw[thick] (2)--(5)--(6);
\draw[->,>=stealth',thick=4mm]  (5,0) -- (7,0);
\node[draw,circle,thick,scale=1.25,fill=black,label=below:{1}] (0a) at (8,0){};
\node[draw,circle,thick,scale=1.25,label=below:{2}] (1a) at (8+.8,0){};
\node[draw,circle,thick,scale=1.25,label=below:{3}] (2a) at (8+.8*2,0){};
\node[draw,circle,thick,scale=1.25,label=below:{4}] (3a) at (8+.8*3,0){};
\node[draw,circle,thick,scale=1.25,label=below:{2}] (4a) at (8+.8*4,0){};
\draw[thick] (0a)--(1a)--(2a)--(3a)--(4a);
\node at (8+.8*2,-1.5) { Incomplete $\mathfrak{e}_7$ and $\mathfrak{e}_8$ };
\end{tikzpicture}}
\end{center}
\caption{ Degeneration of the F$_4$ fiber at the  non-transverse collision  $\text{IV}^{*\text{ns}}+\text{I}_1$.
 }
\label{fig:fiberenhancements}
\end{figure}

\begin{figure}[H]
\centering
\begin{center}
\scalebox{.8}{
\begin{tikzpicture}
\node[draw,circle,thick,scale=1.25,fill=black,label=below:{1}] (1) at (.8,0){};
\node[draw,circle,thick,scale=1.25,label=below:{2}] (2) at (.8*2,0){};
\node[draw,circle,thick,scale=1.25] (3) at (.8*3.2,-1){};
\node[draw,circle,thick,scale=1.25] (4) at (.8*3.2,0){};
\node[draw,circle,thick,scale=1.25] (5) at (.8*3.2,1){};
\node at (.8*3.15,0)[draw,dashed, line width=2pt, ellipse, minimum width=120pt, minimum height=22pt,rotate=90,yshift=-1pt]{};
\node at (.8*3.15,-2.5) {1};
\node at (.8*2.1,-2.8) { $\mathfrak{g}_2$ };
\draw[thick] (1)--(2)--(3);
\draw[thick] (2)--(4);
\draw[thick] (2)--(5);
\draw[->,>=stealth',thick=4mm]  (4.5,0) -- (6.5,0);
\node[draw,circle,thick,scale=1.25,fill=black,label=left:{1}] (0b) at (8+.8,.8){};
\node[draw,circle,thick,scale=1.25,label=left:{1}] (0a) at (8+.8,-.8){};
\node[draw,circle,thick,scale=1.25,label=below:{2}] (1a) at (8+.8*2,0){};
\node[draw,circle,thick,scale=1.25,label=below:{2}] (2a) at (8+.8*3.2,0){};
\draw[thick] (0a)--(1a)--(2a);
\draw[thick] (0b)--(1a);
\node at (8+.8*1.6,-2) { Incomplete $\mathfrak{so}_{10}$ };
\end{tikzpicture}}
\vspace{.7cm}

\scalebox{.8}{
\begin{tikzpicture}
\node[draw,circle,thick,scale=1.25,fill=black,label=below:{1}] (1) at (.8,0){};
\node[draw,circle,thick,scale=1.25,label=below:{2}] (2) at (.8*2,0){};
\node[draw,circle,thick,scale=1.25] (3) at (.8*3.2,-1){};
\node[draw,circle,thick,scale=1.25] (4) at (.8*3.2,0){};
\node[draw,circle,thick,scale=1.25] (5) at (.8*3.2,1){};
\node at (.8*3.15,0)[draw,dashed, line width=2pt, ellipse, minimum width=120pt, minimum height=22pt,rotate=90,yshift=-1pt]{};
\node at (.8*3.15,-2.5) {1};
\node at (.8*2.1,-2.8) { $\mathfrak{g}_2$ };
\draw[thick] (1)--(2)--(3);
\draw[thick] (2)--(4);
\draw[thick] (2)--(5);
\draw[->,>=stealth',thick=4mm]  (4.5,0) -- (6.5,0);
\node[draw,circle,thick,scale=1.25,fill=black,label=below:{1}] (0a) at (8+.8,0){};
\node[draw,circle,thick,scale=1.25,label=below:{2}] (1a) at (8+.8*2,0){};
\node[draw,circle,thick,scale=1.25,label=below:{3}] (2a) at (8+.8*3,0){};
\draw[thick] (0a)--(1a)--(2a);
\node at (8+.8*2,-1.5) { Incomplete $\mathfrak{e}_6$ };
\end{tikzpicture}}
\vspace{.7cm}

\end{center}
\caption{Degeneration of the G$_2$ fiber.
 }
\label{fig:fiberenhancements2}
\end{figure}
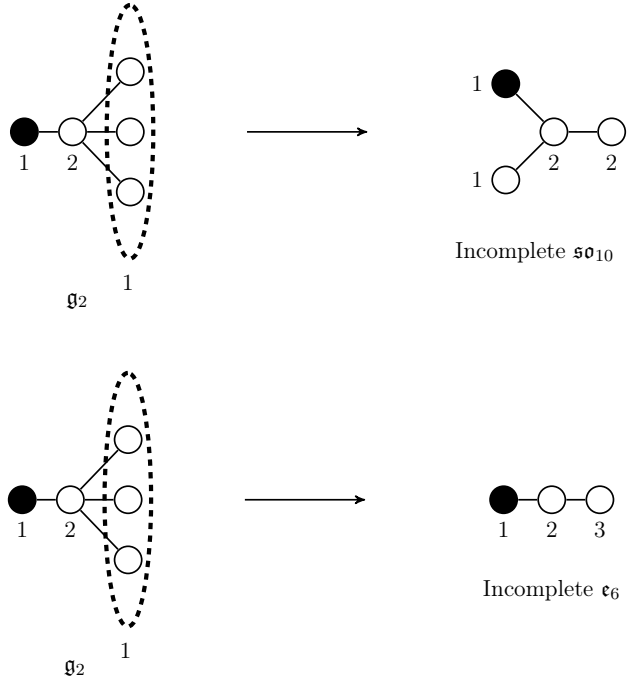

\subsection{G$_2\to \text{B}_3, \text{D}_4, \text{D}_5, \text{E}_6$}
The list of all possible branching rules are
\begin{align}
& \text{G}^1_2\to \text{B}_3: \quad[ \mathbf{21}]=[\mathbf{14}]\oplus\chi, \quad \chi= \mathbf{7},\\
& \text{G}^1_2\to \text{D}_4 : \quad[ \mathbf{28}]=[\mathbf{14}]\oplus\chi, \quad \chi= \mathbf{7}^{\oplus 2},\\
& \text{G}^1_2 \to \text{D}_5:\quad [\mathbf{45}]=[\mathbf{14}]\oplus \chi, \quad \chi=\mathbf{7}^{\oplus 4}\oplus\mathbf{1}^3,\\
& \text{G}^1_2 \to \text{E}_6:\quad
[\mathbf{78}]=[\mathbf{14}]\oplus \chi, \quad \chi= \mathbf{7}^{8}  \oplus\mathbf{1}^{8},\\
&\text{G}^3_2 \to \text{E}_6:\quad [\mathbf{78}]=[\mathbf{14}]\oplus \chi, \quad \chi= \mathbf{64}.
\end{align}
All the embeddings of G$_2$ in B$_3$, D$_4$, and D$_5$ have  Dynkin index one and the same reduced characteristic representation, namely the representation $\mathbf{7}$.
However, there are two linearly inequivalent representations of G$_2$ in E$_6$ with Dynkin index one and two.    
The one with Dynkin index one gives again the representation $\mathbf{7}$. However, the one of Dynkin index two gives the isotropic characteristic representation $\mathbf{64}$.

\subsection{F$_4\to \text{E}_6, \text{E}_7, \text{E}_8$}
\begin{align}
 &\text{F}^1_4\to \text{E}_6 : \quad [\mathbf{78}]=[\mathbf{52}]\oplus\chi, \quad \chi= \mathbf{26},\\
&\text{F}^1_4 \to \text{E}_7:\quad   [\mathbf{133}]=[\mathbf{52}]\oplus\chi, \quad \chi= \mathbf{26}^{\oplus 3}\oplus\mathbf{1}^{\oplus 3}, \\ 
&\text{F}^1_4\to \text{E}_8:\quad  [\mathbf{248}]=[\mathbf{52}]\oplus \chi, \quad \chi= \mathbf{26}^{\oplus 7}\oplus\mathbf{1}^{\oplus 14}.
\end{align}
In contrast to the case of G$_2$, for F$_4$-models, there is no room for ambiguity: all the embeddings of F$_4$ in E$_n$ ($n=6,7,8$) have  Dynkin index one and give the same reduced characteristic representations, namely the representation $\mathbf{26}$.

\section{The  Spin($7$)-model}\label{Sub:Spin7}

The geometry of the Spin($7$)-model is studied in \cite{G2}. The possible degenerations of the generic fiber I$_0^{*\text{ss}}$ suggest the following embeddings:
\begin{align}
& \text{B}_3 \to \text{D}_4, \quad 
 \text{B}_3 \to \text{B}_4, \quad 
  \text{B}_3 \to \text{D}_5, \quad 
    \text{B}_3 \to \text{F}_4, \quad 
      \text{B}_3 \to \text{E}_6.
\end{align}
We will rely on the geography of semisimple subalgebras of D$_4$, B$_4$, D$_5$, F$_4$, and E$_6$ to quickly explain these embeddings \cite{GEO}. 
We will number the embeddings according to how they appear in de Graaf's database. 
There is a particular implication to these cases: the triality and its breaking by an uplift. 
We will also see that F$_4$ and E$_6$ contains Spin($7$) as a subgroup as they clearly require the presence of the spin representation of Spin($7$). 

\subsection{$\text{B}_3^1\to \text{D}_4$}
The subalgebra B$_3$ is embedded in D$_4$ in three distinct ways corresponding to three maximal S-subalgebra of Dynkin index one connected by triality. 
The embedding of B$_3$ in D$_4$ is an example of an adjoint singleton subalgebra: the embeddings are not linearly equivalent but they give the same  characteristic representation, namely the representation $\mathbf{7}$ of B$_3$: 
\begin{align}
\text{B}_3^1\to \text{D}_4: \quad [\mathbf{28}]=[\mathbf{21}]\oplus \chi, \quad \chi=\mathbf{7}. 
\end{align}
We note that we get an isotropic characteristic representation ($\chi$ is irreducible). 

\subsection{$\text{B}_3^1\to \text{B}_4$ or B$_5$}
When D$_4$ is embedded in B$_4$, we see the breaking of triality in two different ways, which gives two different groups compatible whose algebra is B$_3$. 
Two of the three non-linearly equivalent subalgebra B$_3$ of D$_4$ becomes linearly equivalent in B$_4$. 
It follows that B$_3$ can be embedded in two different ways in B$_4$, both have Dynkin index one and are contained in D$_4$:  
\begin{align}
 & \text{B}^1_ 3 \to \text{B}_4: \quad [\mathbf{36}] =[\mathbf{21}]\oplus\chi, \quad \chi= \mathbf{7}\oplus \mathbf{8}, \\
 & \text{B}^1_3\to \text{B}_4: \quad [\mathbf{36}]=[\mathbf{21}]\oplus\chi, \quad \chi=\mathbf{7}^{\oplus 2}\oplus\mathbf{1}. 
\end{align}
The distinction between the two cases is clear when we consider the possible global structure of a compact group with Lie algebra B$_3$. 
The first embedding is compatible with the group Spin($7$) as it involves the spin representation $\mathbf{8}$ of B$_3$, while the second embedding is only compatible with SO($7$) since it involves only the vector and trivial representation of B$_3$.
 
There are two distinct embeddings of B$_3$ in D$_5$, whose Dynkin index is one but can be differentiated from each other by the presence of the spin representation $\mathbf{8}$: 
\begin{align}
& \text{B}^1_ 3 \to \text{D}_5: \quad [\mathbf{45}] =[\mathbf{21}]\oplus\chi, \quad \chi= \mathbf{7}\oplus \mathbf{8}^{\oplus 2}\oplus \mathbf{1}, \\
 & \text{B}^1_3\to \text{D}_5: \quad [\mathbf{45}]=[\mathbf{21}]\oplus\chi, \quad \chi=\mathbf{7}^{\oplus 3}\oplus\mathbf{1}^{\oplus 3}. 
\end{align}

\subsection{$\text{B}_3^1\to \text{F}_4$ or E$_6$ }
The Lie algebra B$_3$ is a singleton subalgebra of both F$_4$ and E$_6$ and it has index one in both of them.  In both cases, we get both vector, spin, and trivial representations of B$_3$: 
\begin{align}
& \text{B}^1_ 3 \to \text{F}_4: \quad [\mathbf{52}] =[\mathbf{21}]\oplus\chi, \quad \chi= \mathbf{7}^{\oplus 2}\oplus \mathbf{8}^{\oplus 2}\oplus \mathbf{1}, \\
 & \text{B}^1_ 3 \to \text{E}_6: \quad [\mathbf{78}] =[\mathbf{21}]\oplus\chi, \quad \chi= \mathbf{7}^{\oplus 3}\oplus \mathbf{8}^{\oplus 4}\oplus \mathbf{1}^{\oplus 4}.
\end{align}

\section{The USp($4$)-model}\label{Sec:USp4}

The USp($4$)-model is interesting in many ways \cite{USP4}. 
We recall that USp($4$) is isomorphic to Spin($5$) and is the universal over of SO($5$). 
This is related to the isomorphism between C$_2$ and B$_2$. 
The analysis of the fibers of USp($4$)-model shows that the localized matter contains representations $\mathbf{5}$ and $\mathbf{4}$ of C$_2$. 
The SO($5$)-model contains matter only in the representation $\mathbf{5}$. 
The fiber appearing geometrically are of type I$_5$, I$^*_0$, and I$^*_1$. The generic fiber is itself I$_4^{\text{ns}}$ with an I$_4$ as a geometric fiber. 
It is therefore natural to investigate the embedding of C$_2$ in  A$_3$, C$_3$,   B$_3$,  A$_4$, D$_4$, D$_5$. 

The geographies of all these simple Lie algebras are studied in \cite{GEO}. The key points are the following: 

\begin{itemize}
\item  C$_2$ is a singleton subalgebra of A$_3$, B$_3$, and C$_3$. 
\item There are two subalgebras  of A$_4$ of type C$_2$ with distinct Dynkin $1$ and $2$. 
\item There are three subalgebras  of D$_4$ of type C$_2$ and they are related by triality and are adjoint singletons. 
\item There are four subalgebras of D$_5$ of type C$_2$, two have index one, one has index 2, and the last has index 3.
\end{itemize}

\subsection{$ \text{C}_2\to \text{A}_3$}
There is  a unique embedding of C$_2$ in A$_3$ up to linear equivalence. Moreover, this embedding has Dynkin embedding index one and its characteristic representation is isotropic: 
\begin{align}
& \text{C}_2^1\to \text{A}_3: \quad   [\mathbf{15}]=[\mathbf{10}]\oplus \chi, \quad \chi=\mathbf{5}, 
\end{align}

\subsection{$ \text{C}_2\to \text{B}_3$}

There is  a unique embedding of C$_2$ in B$_3$ up to linear equivalence and its  Dynkin embedding index is one: 
\begin{align}
& \text{C}_2^1\to \text{B}_3: \quad  [\mathbf{21}]=[\mathbf{10}]\oplus \chi, \quad \chi=\mathbf{5}^{\oplus 2}\oplus\mathbf{1},
\end{align}

\subsection{$ \text{C}_2\to \text{C}_3$}

There is  a unique embedding of C$_2$ in C$_3$ up to linear equivalence and its  Dynkin embedding index is one: 
\begin{align}
& \text{C}_2^1\to \text{C}_3: \quad [\mathbf{21}]=[\mathbf{10}]\oplus \chi, \quad \chi=\mathbf{4}^{\oplus 2}\oplus\mathbf{1}^{\oplus 3}, 
\end{align}

\subsection{$ \text{C}_2\to \text{A}_4$}
There are two embeddings of C$_2$ in B$_3$ up to linear equivalence, including one with   Dynkin embedding index one: 
\begin{align}
& \text{C}_2^1\to \text{A}_4: \quad  [\mathbf{24}]=[\mathbf{10}]\oplus\chi, \quad \chi=\mathbf{5}\oplus\mathbf{4}^{\oplus 2}\oplus\mathbf{1},\\
& \text{C}_2^2\to \text{A}_4: \quad  [\mathbf{24}]=[\mathbf{10}]\oplus\chi, \quad \chi=\mathbf{14},
\end{align}

\subsection{$ \text{C}_2\to \text{D}_4$}
There is  a unique embedding of C$_2$ in D$_4$ up to linear equivalence and its  Dynkin embedding index is one: 
\begin{align}
& \text{C}_2^1\to \text{D}_4: \quad [\mathbf{28}]=[\mathbf{10}]\oplus \chi, \quad \chi=\mathbf{5}^{\oplus 3}  \oplus \mathbf{1}^{\oplus 3},
\end{align}

\subsection{$ \text{C}_2\to \text{D}_5$}
There are four embeddings of C$_2$ in D$_5$ up to linear equivalence, one of them has Dynkin index one: 
\begin{align}
& \text{C}_2^1\to \text{D}_5: \quad [\mathbf{45}]=[\mathbf{10}]\oplus \chi, \quad \chi=\mathbf{5}^{\oplus 3} \oplus \mathbf{4}^{\oplus 4}  \oplus \mathbf{1}^{\oplus 4},\\
& \text{C}_2^1\to \text{D}_5: \quad [\mathbf{45}]=[\mathbf{10}]\oplus \chi, \quad \chi= \mathbf{5}^{\oplus 5}  \oplus \mathbf{1}^{\oplus 10},\\
& \text{C}_2^2\to \text{D}_5: \quad [\mathbf{45}]=[\mathbf{10}]\oplus \chi, \quad \chi=\mathbf{14}\oplus \mathbf{10}^{\oplus 2} \oplus \mathbf{1}, \\
& \text{C}_2^3\to \text{D}_5: \quad [\mathbf{45}]=[\mathbf{10}]\oplus \chi, \quad \chi= \mathbf{35}.
\end{align}
We see that the two $\text{C}_2^1\to \text{D}_5$ have different characteristic representations. They also correspond to different sets of compatible groups. 
The first one need the group to be USp($4$) while the second could have an SO($5$) as well since the $\mathbf{4}$ of C$_2$ does not appear.

\section{$\text{A}_1\oplus\text{A}_1$: the SU($2$)$\times$SU($2$)-models} \label{sec:collisionA1A1}
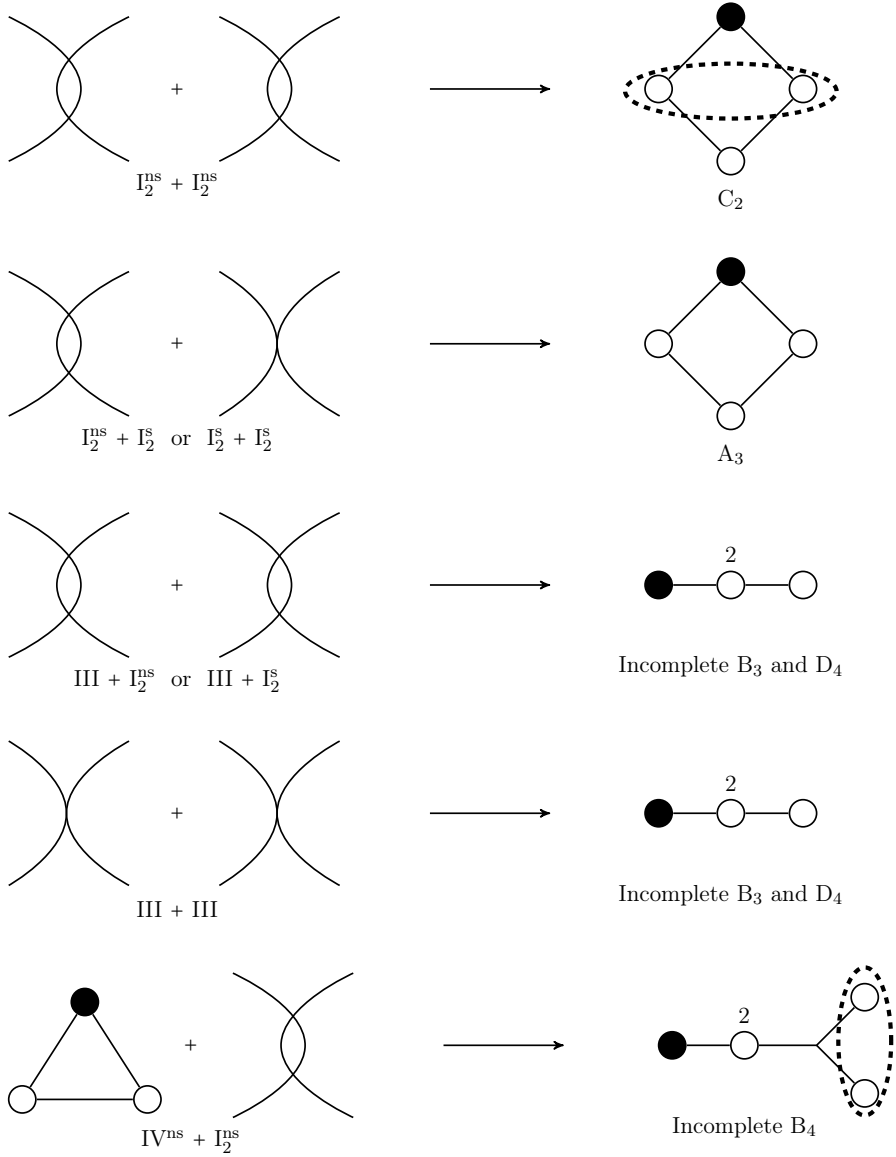
\begin{figure}[H]
\qquad\qquad\scalebox{.8}{
\begin{tikzpicture}
\node at (.8*3.5,-.8*2) { I$_2^{\text{ns}}$ $+$ I$_2^{\text{ns}}$ };
\node at (10+.8*2.5,-.8*2.3) { $\text{C}_2$ };
\draw[thick=4mm] plot [smooth, tension=1] coordinates {(0,.8*1.5) (.8*1.5,0) (0,-.8*1.5)};
\draw[thick=4mm] plot [smooth, tension=1] coordinates {(.8*2.5,.8*1.5) (.8*1,0) (.8*2.5,-.8*1.5)};
\node[draw=none] at (.8*3.5,0){$+$};
\draw[thick=4mm] plot [smooth, tension=1,xshift=3.5cm] coordinates {(0,.8*1.5) (.8*1.5,0) (0,-.8*1.5)};
\draw[thick=4mm] plot [smooth, tension=1,xshift=3.5cm] coordinates {(.8*2.5,.8*1.5) (.8*1,0) (.8*2.5,-.8*1.5)};
\draw[->,>=stealth',thick=4mm]  (7,0)--(9,0);
\node[draw,circle,thick,scale=1.25,fill=black] (A1) at (10+.8*2.5,.8*1.5) {};
\node[draw,circle,thick,scale=1.25] (A2) at (10+.8,0) {};
\node[draw,circle,thick,scale=1.25] (A3) at (10+.8*4,0) {};
\node[draw,circle,thick,scale=1.25] (A4) at (10+.8*2.5,-.8*1.5) {};
\draw[thick] (A1)--(A2)--(A4)--(A3)--(A1);
\node at (10+.8*2.5,0)[draw,dashed,line width=2pt,ellipse,minimum width=100pt,minimum height=26pt, yshift=-1pt]{};
\end{tikzpicture}}
\vspace{5mm}

\qquad\qquad\scalebox{.8}{
\begin{tikzpicture}
\node at (.8*3.5,-.8*2) { I$_2^{\text{ns}}$ $+$ I$_2^{\text{s}}\ $  or  $\ $I$_2^{\text{s}}$ $+$ I$_2^{\text{s}}$ };
\node at (10+.8*2.5,-.8*2.3) { $\text{A}_3$ };
\draw[thick=4mm] plot [smooth, tension=1] coordinates {(0,.8*1.5) (.8*1.5,0) (0,-.8*1.5)};
\draw[thick=4mm] plot [smooth, tension=1] coordinates {(.8*2.5,.8*1.5) (.8*1,0) (.8*2.5,-.8*1.5)};
\node[draw=none] at (.8*3.5,0){$+$};
\draw[thick=4mm] plot [smooth, tension=1,xshift=3.5cm] coordinates {(0,.8*1.5) (.8*1.2,0) (0,-.8*1.5)};
\draw[thick=4mm] plot [smooth, tension=1,xshift=3.5cm] coordinates {(.8*2.5,.8*1.5) (.8*1.2,0) (.8*2.5,-.8*1.5)};
\draw[->,>=stealth',thick=4mm]  (7,0)--(9,0);
\node[draw,circle,thick,scale=1.25,fill=black] (A1) at (10+.8*2.5,.8*1.5) {};
\node[draw,circle,thick,scale=1.25] (A2) at (10+.8,0) {};
\node[draw,circle,thick,scale=1.25] (A3) at (10+.8*4,0) {};
\node[draw,circle,thick,scale=1.25] (A4) at (10+.8*2.5,-.8*1.5) {};
\draw[thick] (A1)--(A2)--(A4)--(A3)--(A1);
\end{tikzpicture}}
\vspace{5mm}

\qquad\qquad\scalebox{.8}{
\begin{tikzpicture}
\node at (.8*3.5,-.8*2) { III $+$ I$_2^{\text{ns}}\ $  or  $\ $III $+$ I$_2^{\text{s}}$ };
\node at (10+.8*2.5,-.8*1.7) { Incomplete $\text{B}_3$ and $\text{D}_4$ };
\draw[thick=4mm] plot [smooth, tension=1] coordinates {(0,.8*1.5) (.8*1.5,0) (0,-.8*1.5)};
\draw[thick=4mm] plot [smooth, tension=1] coordinates {(.8*2.5,.8*1.5) (.8*1,0) (.8*2.5,-.8*1.5)};
\node[draw=none] at (.8*3.5,0){$+$};
\draw[thick=4mm] plot [smooth, tension=1,xshift=3.5cm] coordinates {(0,.8*1.5) (.8*1.5,0) (0,-.8*1.5)};
\draw[thick=4mm] plot [smooth, tension=1,xshift=3.5cm] coordinates {(.8*2.5,.8*1.5) (.8*1,0) (.8*2.5,-.8*1.5)};
\draw[->,>=stealth',thick=4mm]  (7,0)--(9,0);
\node[draw,circle,thick,scale=1.25,fill=black] (B1) at (10+.8,0) {};
\node[draw,circle,thick,scale=1.25,label=above:2] (B2) at (10+.8*2.5,0) {};
\node[draw,circle,thick,scale=1.25] (B3) at (10+.8*4,0) {};
\draw[thick] (B1)--(B2)--(B3);
\end{tikzpicture}}
\vspace{5mm}

\qquad\qquad\scalebox{.8}{
\begin{tikzpicture}
\node at (.8*3.5,-.8*2) { III $+$ III };
\node at (10+.8*2.5,-.8*1.7) { Incomplete $\text{B}_3$ and $\text{D}_4$ };
\draw[thick=4mm] plot [smooth, tension=1] coordinates {(0,.8*1.5) (.8*1.2,0) (0,-.8*1.5)};
\draw[thick=4mm] plot [smooth, tension=1] coordinates {(.8*2.5,.8*1.5) (.8*1.2,0) (.8*2.5,-.8*1.5)};
\node[draw=none] at (.8*3.5,0){$+$};
\draw[thick=4mm] plot [smooth, tension=1,xshift=3.5cm] coordinates {(0,.8*1.5) (.8*1.2,0) (0,-.8*1.5)};
\draw[thick=4mm] plot [smooth, tension=1,xshift=3.5cm] coordinates {(.8*2.5,.8*1.5) (.8*1.2,0) (.8*2.5,-.8*1.5)};
\draw[->,>=stealth',thick=4mm]  (7,0)--(9,0);
\node[draw,circle,thick,scale=1.25,fill=black] (B1) at (10+.8,0) {};
\node[draw,circle,thick,scale=1.25,label=above:2] (B2) at (10+.8*2.5,0) {};
\node[draw,circle,thick,scale=1.25] (B3) at (10+.8*4,0) {};
\draw[thick] (B1)--(B2)--(B3);
\end{tikzpicture}}
\vspace{5mm}

\qquad\qquad\scalebox{.8}{
\begin{tikzpicture}
\node at (.8*3.5,-.8*2) { IV$^{\text{ns}}$ $+$ I$_2^{\text{ns}}$ };
\node at (10+.8*2.5,-.8*1.7) { Incomplete $\text{B}_4$ };
\node[draw,circle,thick,scale=1.25,fill=black] (A1) at (.8*1.3,.8*.9) {};
\node[draw,circle,thick,scale=1.25] (A2) at (0,-.9) {};
\node[draw,circle,thick,scale=1.25] (A3) at (.8*2.6,-.9) {};
\draw[thick] (A1)--(A2)--(A3)--(A1);
\node[draw=none] at (.8*3.5,0){$+$};
\draw[thick=4mm] plot [smooth, tension=1,xshift=3.5cm] coordinates {(0,.8*1.5) (.8*1.5,0) (0,-.8*1.5)};
\draw[thick=4mm] plot [smooth, tension=1,xshift=3.5cm] coordinates {(.8*2.5,.8*1.5) (.8*1,0) (.8*2.5,-.8*1.5)};
\draw[->,>=stealth',thick=4mm]  (7,0)--(9,0);
\node[draw,circle,thick,scale=1.25,fill=black] (B1) at (10+.8,0) {};
\node[draw,circle,thick,scale=1.25,label=above:2] (B2) at (10+.8*2.5,0) {};
\node[draw,circle,thick,scale=1.25] (B3) at (10+.8*5,.8) {};
\node[draw,circle,thick,scale=1.25] (B4) at (10+.8*5,-.8) {};
\draw[thick] (B1)--(B2)--(10+.8*4,0)--(B3);
\draw[thick] (10+.8*4,0)--(B4);
\node at (10+.8*5,0.1)[draw,dashed,line width=2pt,ellipse,minimum width=25pt,minimum height=70pt, yshift=-1pt]{};
\end{tikzpicture}}
\caption{Degeneration of the SO(4) fibers, which is established via SU(2)$\times$SU(2) fibers with a Mordell-Weil group $\mathbb{Z}_2$.}
\label{fig:su2su2enhancementG}
\end{figure}

In this section, we consider the collisions 
\begin{align}
& \text{I}_2+\text{I}_2\to \text{I}_4,\\
& \text{I}_2+\text{I}_2\to \text{I}^*_0.
\end{align}
They correspond to SO(4)- or Spin(4)-models where their geometry is studied in \cite{SO4}, where the degenerations of the fibers are presented in Figure \ref{fig:su2su2enhancementG}.
Taking into account all the possible split, non-split, and semi-split options for the Kodaira fibers of type I$_4$ and I$^*_0$, we get the following types of embeddings: 
\begin{align}
\text{A}_1\oplus\text{A}_1\to \text{A}_3, \\
\text{A}_1\oplus\text{A}_1\to \text{C}_2, \\
\text{A}_1\oplus\text{A}_1\to \text{G}_2, \\
\text{A}_1\oplus\text{A}_1\to \text{B}_3, \\
\text{A}_1\oplus\text{A}_1\to \text{D}_4.
\end{align}

We will analyze each case using the geography of semisimple subalgebras of A$_3$, C$_2$, G$_2$, B$_3$, and D$_4$. They will be considered in Section \ref{sec:A1A1} for its links of the channels of maximal subalgebras.

\subsection{$\text{A}_1\oplus\text{A}_1\to \text{A}_3$} \label{sec:collisionA1A1toA3}
The A$_3$ has two distinct embeddings of type $\text{A}_1\oplus\text{A}_1$: one is an S-subalgebra of Dynkin index $(1,1)$ and contained in the S-maximal subalgebra B$_2^1$  of A$_3$, where as the other  one is a maximal S-subalgebra of Dynkin index  $(2,2)$.   
\begin{align}
&7: \quad \text{A}_1^1\oplus\text{A}_1^1\to \text{A}_3 : \quad [\mathbf{15}]=[(\mathbf{3},\mathbf{1})\oplus (\mathbf{1},\mathbf{3})]\oplus\chi, \quad \chi=(\mathbf{2},\mathbf{2})^{\oplus 2}\oplus (\mathbf{1},\mathbf{1}),\\
&8: \quad \text{A}_1^2\oplus\text{A}_1^2\to \text{A}_3 : \quad [\mathbf{15}]=[(\mathbf{3},\mathbf{1})\oplus (\mathbf{1},\mathbf{3})]\oplus\chi, \quad \chi=(\mathbf{3},\mathbf{3}). \end{align}

\subsection{$\text{A}_1\oplus\text{A}_1\to \text{C}_2$}

The $\text{A}_1\oplus\text{A}_1$ is a singleton subalgebra of C$_2$ of Dynkin index (1,1): 
\begin{align}
&4: \quad \text{A}_1^1\oplus\text{A}_1^1\to \text{C}_2 : \quad [\mathbf{10}]=[(\mathbf{3},\mathbf{1})\oplus (\mathbf{1},\mathbf{3})]\oplus\chi, \quad \chi=(\mathbf{2},\mathbf{2}).
\end{align}
\subsection{$\text{A}_1\oplus\text{A}_1\to \text{G}_2$}
$\text{A}_1\oplus\text{A}_1$ is a singleton subalgebra of G$_2$ of Dynkin index (3,1): 
\begin{align}
&6: \quad \text{A}_1^3\oplus\text{A}_1^1\to \text{G}_2 : \quad [\mathbf{10}]=[(\mathbf{3},\mathbf{1})\oplus (\mathbf{1},\mathbf{3})]\oplus\chi, \quad \chi=(\mathbf{4},\mathbf{2}).
\end{align}
This exotic matter shows that the enhancement to G$_2$ is not generic for a $\text{A}_1\oplus\text{A}_1$ model.

\subsection{$\text{A}_1\oplus\text{A}_1\to \text{B}_3$}

$\text{A}_1\oplus\text{A}_1$ can be embedded in three different ways inside B$_3$, each case is distinguished by a different Dynkin index: (1,1), (1,2), or (1,3):
\begin{align}
& 10: \quad \text{A}_1^1\oplus\text{A}_1^1\to \text{B}_3 : \quad [\mathbf{21}]=[(\mathbf{3},\mathbf{1})\oplus (\mathbf{1},\mathbf{3})]\oplus\chi, \quad \chi=(\mathbf{2},\mathbf{2})^{\oplus 3}
\oplus (\mathbf{1},\mathbf{1})^{\oplus 3}, \\
& 11: \quad \text{A}_1^1\oplus\text{A}_1^2\to \text{B}_3 : \quad [\mathbf{21}]=[(\mathbf{3},\mathbf{1})\oplus (\mathbf{1},\mathbf{3})]\oplus\chi, \quad \chi=(\mathbf{2},\mathbf{3})^{\oplus 2}\oplus (\mathbf{1},\mathbf{1})^{\oplus 3},\\
& 12: \quad \text{A}_1^1\oplus\text{A}_1^2\to \text{B}_3 : \quad [\mathbf{21}]=[(\mathbf{3},\mathbf{1})\oplus (\mathbf{1},\mathbf{3})]\oplus\chi, \quad \chi=(\mathbf{2},\mathbf{4})\oplus (\mathbf{1},\mathbf{3})\oplus 
(\mathbf{2},\mathbf{2}). 
\end{align}
We see that the subalgebra of Dynkin index one gives the matter representations we expect from considering the technique of intersecting branes (the bifundamental representation).

\subsection{$\text{A}_1\oplus\text{A}_1\to \text{D}_4$}

The $\text{A}_1\oplus\text{A}_1$ can be embedded in 13 different ways inside D$_4$ some of which are related by triality. 
 The embeddings of index one are related by triality and gives the same characteristic representation. 
 The three  embeddings of index (1,2) (resp. (2,2), and (2,10)) are also related by triality. 
 The embedding of index (1,3) is a singleton embedding. 
\begin{align}
&\text{A}_1^1\oplus\text{A}_1^1\to \text{D}_4: \  \quad [\mathbf{28}]=[(\mathbf{3},\mathbf{1})\oplus (\mathbf{1},\mathbf{3})]\oplus\chi, \quad \chi=(\mathbf{2},\mathbf{2})^{\oplus 4}\oplus 
(\mathbf{1},\mathbf{1})^{\oplus 6}, \\
& \text{A}_1^1\oplus\text{A}_1^2\to \text{D}_4:\   \quad [\mathbf{28}]=[(\mathbf{3},\mathbf{1})\oplus (\mathbf{1},\mathbf{3})]\oplus\chi, \quad \chi=(\mathbf{2},\mathbf{3})^{\oplus 2}\oplus 
(\mathbf{1},\mathbf{3})\oplus
(\mathbf{2},\mathbf{1})^{\oplus 2}\oplus
(\mathbf{1},\mathbf{1})^{\oplus 3},\\
& \text{A}_1^2\oplus\text{A}_1^2\to \text{D}_4: \quad\   [\mathbf{28}]=[(\mathbf{3},\mathbf{1})\oplus (\mathbf{1},\mathbf{3})]\oplus\chi, \quad 
\chi=(\mathbf{3},\mathbf{3})\oplus (\mathbf{3},\mathbf{1})^{\oplus 2}\oplus(\mathbf{1},\mathbf{3})^{\oplus 2}\oplus(\mathbf{1},\mathbf{1}),\\
& \text{A}_1^{2}\oplus\text{A}_1^{10}\to \text{D}_4: \quad [\mathbf{28}]=[(\mathbf{3},\mathbf{1})\oplus (\mathbf{1},\mathbf{3})]\oplus\chi, \quad 
\chi=(\mathbf{1},\mathbf{7})\oplus (\mathbf{3},\mathbf{5}),\\
& \text{A}_1^{1}\oplus\text{A}_1^{3}\to \text{D}_4: \quad\    [\mathbf{28}]=[(\mathbf{3},\mathbf{1})\oplus (\mathbf{1},\mathbf{3})]\oplus\chi, \quad 
\chi=(\mathbf{2},\mathbf{4})\oplus (\mathbf{1},\mathbf{3})^{\oplus 2}\oplus (\mathbf{2},\mathbf{2})^{\oplus 2}. 
\end{align}

\section{$\text{A}_1\oplus\text{A}_2$: the SU($2$)$\times$SU($3$)-model}
The geometry of the $ \text{A}_1\oplus\text{A}_2$-models realized by crepant resolutions of a Weierstrass model  indicates that the 
following embedding are at play:
\begin{align}
& \text{A}_1\oplus\text{A}_2\to \text{A}_4,\\
& \text{A}_1\oplus\text{A}_2\to \text{A}_5,\\
& \text{A}_1\oplus\text{A}_2\to \text{D}_5,\\
& \text{A}_1\oplus\text{A}_2\to \text{E}_6,\\
& \text{A}_1\oplus\text{A}_2\to \text{E}_7.
\end{align}
There is a unique embedding of $ \text{A}_1\oplus\text{A}_2$ in A$_4$ already covered as one of the original Katz--Vafa  cases. 
There are three distinct embeddings of 
$ \text{A}_1\oplus\text{A}_2$ in A$_5$, two embeddings in D$_5$, twelve in E$_6$, and twenty-six embedding in E$_7$.

\subsection{$ \text{A}_1\oplus\text{A}_2\to \text{A}_5$ }

\begin{align}
 &  \text{A}^1_1\oplus \text{A}^1_2 \to \text{A}_5: \quad 
\chi=  (\mathbf{2},\mathbf{3})\oplus  (\mathbf{1},\mathbf{3})\oplus 
 (\mathbf{2},\mathbf{1})^{\oplus 2}\oplus  (\mathbf{2},\mathbf{\overline{3}})
 \oplus  (\mathbf{1},\mathbf{\overline{3}}) \oplus  (\mathbf{1},\mathbf{1})^{\oplus 2},
\\
 &  \text{A}^4_1\oplus \text{A}^1_2 \to \text{A}_5: \quad 
\chi=  (\mathbf{3},\mathbf{3})\oplus  (\mathbf{5},\mathbf{1})\oplus  (\mathbf{3},\mathbf{\overline{3}})\oplus  (\mathbf{1},\mathbf{1}),\\
 &   \text{A}^3_1\oplus \text{A}^2_2 \to \text{A}_5: \quad    
 \chi= (\mathbf{3},\mathbf{8}). 
\end{align}
\subsection{$ \text{A}_1\oplus\text{A}_2\to \text{D}_5$ }

\begin{align}
 &  \text{A}^1_1\oplus \text{A}^1_2 \to \text{D}_5: \quad 
\chi=  (\mathbf{2},\mathbf{3})^{\oplus 2}
\oplus  (\mathbf{2},\mathbf{\overline{3}})^{\oplus 2}
\oplus  (\mathbf{1},\mathbf{3})
\oplus  (\mathbf{1},\mathbf{\overline{3}})
\oplus  (\mathbf{1},\mathbf{1})^{\oplus 4}
 ,
\\
 &  \text{A}^2_1\oplus \text{A}^1_2 \to \text{D}_5: \quad 
 \chi=  (\mathbf{3},\mathbf{3})
\oplus  (\mathbf{3},\mathbf{\overline{3}})
\oplus  (\mathbf{1},\mathbf{\overline{3}})^{\oplus 2}
\oplus  (\mathbf{3},\mathbf{1})
\oplus  (\mathbf{1},\mathbf{3})^{\oplus 2}
\oplus  (\mathbf{1},\mathbf{1}).
\end{align}

\subsection{$ \text{A}_1\oplus\text{A}_2\to \text{E}_6$ }
\begin{align}
&    \text{A}_1^{1}\oplus\text{A}_2^{1}\to\text{E}_6  :  \quad  \chi=(\mathbf{2},\mathbf{3})^{\oplus 3}\oplus(\mathbf{1},\mathbf{\overline{3}})^{\oplus 3}\oplus(\mathbf{2},\mathbf{\overline{3}})^{\oplus 3}\oplus  (\mathbf{1},\mathbf{1})^{\oplus 9}\oplus(\mathbf{
2},\mathbf{1})^{\oplus 2}\oplus(\mathbf{1},\mathbf{3})^{\oplus 3},\\
&    \text{A}_1^{2}\oplus\text{A}_2^{1} \to\text{E}_6 :  \quad  \chi=(\mathbf{3},\mathbf{3})\oplus(\mathbf{2},\mathbf{\overline{3}})^{\oplus 2}\oplus(\mathbf{3},\mathbf{\overline{3}})\oplus(\mathbf{3},\mathbf{1})\oplus(\mathbf{2},\mathbf{1})^{\oplus 
4}\oplus(\mathbf{1},\mathbf{\overline{3}})^{\oplus 2}
\oplus(\mathbf{2},\mathbf{3})^{\oplus 2}
\\  \nonumber 
& \hspace{3.5cm} 
\oplus(\mathbf{1},\mathbf{3})^{\oplus 2}
\oplus (\mathbf{1},\mathbf{1})^{\oplus 2},\\
&    \text{A}_1^{4}\oplus\text{A}_2^{1}\to\text{E}_6  :  \quad  \chi=(\mathbf{3},\mathbf{3})^{\oplus 3}\oplus(\mathbf{3},\mathbf{\overline{3}})^{\oplus 3}\oplus(\mathbf{5},\mathbf{1})\oplus  (\mathbf{1},\mathbf{1})^{\oplus 8},\\
&    \text{A}_1^{5}\oplus\text{A}_2^{1} \to\text{E}_6 :  \quad  \chi=(\mathbf{4},\mathbf{3})\oplus(\mathbf{3},\mathbf{\overline{3}})\oplus(\mathbf{4},\mathbf{\overline{3}})\oplus(\mathbf{5},\mathbf{1})\oplus(\mathbf{3},\mathbf{1})\oplus(\mathbf{
2},\mathbf{1})^{\oplus 2}\oplus(\mathbf{2},\mathbf{\overline{3}})\oplus(\mathbf{2},\mathbf{3})
\\  \nonumber 
& \hspace{3.5cm} 
\oplus(\mathbf{3},\mathbf{3})\oplus (\mathbf{1},\mathbf{1}),\\
&    \text{A}_1^{8}\oplus\text{A}_2^{1} \to\text{E}_6 :  \quad  \chi=(\mathbf{5},\mathbf{3})\oplus(\mathbf{5},\mathbf{\overline{3}})\oplus(\mathbf{3},\mathbf{\overline{3}})\oplus(\mathbf{5},\mathbf{1})^{\oplus 2}\oplus(\mathbf{3},\mathbf{1})\oplus(\mathbf{
1},\mathbf{\overline{3}})\oplus(\mathbf{3},\mathbf{3})\oplus(\mathbf{1},\mathbf{3}),\\
&    \text{A}_1^{1}\oplus\text{A}_2^{2}  \to\text{E}_6:  \quad  \chi=(\mathbf{2},\mathbf{8})^{\oplus 2}\oplus(\mathbf{1},\mathbf{8})^{3}\oplus (\mathbf{1},\mathbf{1})^{\oplus 3}\oplus(\mathbf{2},\mathbf{1})^{\oplus 4},\\
&    \text{A}_1^{3}\oplus\text{A}_2^{2} \to\text{E}_6 :  \quad  \chi=(\mathbf{3},\mathbf{8})\oplus(\mathbf{2},\mathbf{8})^{\oplus 2}\oplus (\mathbf{4},\mathbf{1})^{\oplus 2}\oplus (\mathbf{1},\mathbf{1})^{\oplus 3},\\
&    \text{A}_1^{4}\oplus\text{A}_2^{2} \to\text{E}_6 :  \quad  \chi=(\mathbf{3},\mathbf{8})^{\oplus 2}\oplus(\mathbf{1},\mathbf{8})\oplus(\mathbf{5},\mathbf{1})\oplus(\mathbf{3},\mathbf{1})^{2},\\
&    \text{A}_1^{28}\oplus\text{A}_2^{2} \to\text{E}_6 :  \quad  \chi=(\mathbf{7},\mathbf{8})\oplus (\mathbf{11},\mathbf{1}),\\
&    \text{A}_1^{1}\oplus\text{A}_2^{2} \to\text{E}_6 :  \quad  \chi=(\mathbf{2},\mathbf{6})\oplus(\mathbf{1},\mathbf{\overline{6}})\oplus(\mathbf{1},\mathbf{3})\oplus(\mathbf{2},\mathbf{\overline{6}})\oplus(\mathbf{2},\mathbf{3})\oplus (\mathbf{
2},\mathbf{1})^{\oplus 2}\oplus(\mathbf{1},\mathbf{8})\oplus(\mathbf{2},\mathbf{\overline{3}})
\\  \nonumber   & 
\hspace{3.5cm}
\oplus(\mathbf{1},\mathbf{6}) 
\oplus(\mathbf{1},\mathbf{\overline{3}})\oplus(\mathbf{1},\mathbf{1}).
\end{align}

\subsection{$ \text{A}_1\oplus\text{A}_2\to \text{E}_7$ }

\begin{align}
&    \text{A}_1^{1}\oplus\text{A}_2^{1} \to \text{E}_7  :  \quad  \chi=(\mathbf{2},\mathbf{3})^{\oplus 4}\oplus(\mathbf{1},\mathbf{\overline{3}})^{\oplus 7}\oplus (\mathbf{2},\mathbf{1})^{\oplus 8}\oplus (\mathbf{1},\mathbf{1})^{\oplus 
16}\oplus(\mathbf{2},\mathbf{\overline{3}})^{\oplus 4}\oplus (\mathbf{1},\mathbf{3})^{\oplus 7},\\
&    \text{A}_1^{2}\oplus\text{A}_2^{1} \to \text{E}_7  :  \quad  \chi=(\mathbf{3},\mathbf{3})\oplus(\mathbf{3},\mathbf{\overline{3}})\oplus(\mathbf{1},\mathbf{\overline{3}})^{\oplus 4}\oplus(\mathbf{3},\mathbf{1})^{\oplus 3}\oplus(\mathbf{2},\mathbf{1})^{\oplus 8}\oplus (\mathbf{1},\mathbf{1})^{\oplus 7}\\
\nonumber &  \hspace{3.5cm}  \oplus(\mathbf{2},\mathbf{\overline{3}})^{\oplus 4}\oplus(\mathbf{2},\mathbf{3})^{\oplus 4}\oplus(\mathbf{1},\mathbf{3})^{\oplus 4},\\
&    \text{A}_1^{3}\oplus\text{A}_2^{1} \to \text{E}_7  :  \quad  \chi=(\mathbf{3},\mathbf{3})^{\oplus 3}\oplus(\mathbf{3},\mathbf{\overline{3}})^{\oplus 3}\oplus(\mathbf{1},\mathbf{\overline{3}})^{\oplus 6}\oplus(\mathbf{3},\mathbf{1})^{\oplus 8}\oplus (\mathbf{1},\mathbf{1})^{\oplus 8}\oplus(\mathbf{1},\mathbf{3})^{\oplus 6},\\
&    \text{A}_1^{4}\oplus\text{A}_2^{1} \to \text{E}_7  :  \quad  \chi=(\mathbf{3},\mathbf{3})^{\oplus 4}\oplus(\mathbf{3},\mathbf{\overline{3}})^{\oplus 4}\oplus(\mathbf{1},\mathbf{\overline{3}})^{\oplus 3}\oplus(\mathbf{5},\mathbf{1})\oplus(\mathbf{3},\mathbf{1})^{\oplus 6}\oplus (\mathbf{1},\mathbf{1})^{\oplus 9}\oplus (\mathbf{1},\mathbf{3})^{\oplus 3},\\
&    \text{A}_1^{5}\oplus\text{A}_2^{1} \to \text{E}_7  :  \quad  \chi=(\mathbf{3},\mathbf{3})^{\oplus 2}\oplus(\mathbf{3},\mathbf{\overline{3}})^{\oplus 2}\oplus(\mathbf{1},\mathbf{\overline{3}})\oplus(\mathbf{5},\mathbf{1})\oplus(\mathbf{
4},\mathbf{1})^{\oplus 2}\oplus(\mathbf{3},\mathbf{1})^{\oplus 3} \\ 
\nonumber &  \hspace{3.5cm}  \oplus(\mathbf{1},\mathbf{1})^{\oplus 2}\oplus(\mathbf{2},\mathbf{1})^{\oplus 4}\oplus(\mathbf{4},\mathbf{\overline{3}})\oplus(\mathbf{2},\mathbf{\overline{3}})^{\oplus 2}\oplus(\mathbf{4},\mathbf{3})\oplus(\mathbf{2},\mathbf{3})^{\oplus 2}\oplus(\mathbf{1},\mathbf{3}),\\
&    \text{A}_1^{10}\oplus\text{A}_2^{1} \to \text{E}_7  :  \quad  \chi=(\mathbf{4},\mathbf{3})^{\oplus 2}\oplus(\mathbf{5},\mathbf{\overline{3}})\oplus(\mathbf{1},\mathbf{\overline{3}})^{\oplus 2}\oplus(\mathbf{7},\mathbf{1})\oplus(\mathbf{
5},\mathbf{1})\oplus(\mathbf{4},\mathbf{1})^{\oplus 4} \\
\nonumber &  \hspace{3.5cm}   \oplus(\mathbf{1},\mathbf{1})^{\oplus 4}\oplus(\mathbf{4},\mathbf{\overline{3}})^{\oplus 2}\oplus(\mathbf{5},\mathbf{3})\oplus(\mathbf{1},\mathbf{3})^{\oplus 2},\\
&    \text{A}_1^{8}\oplus\text{A}_2^{1} \to \text{E}_7  :  \quad  \chi=(\mathbf{5},\mathbf{3})\oplus(\mathbf{5},\mathbf{\overline{3}})\oplus(\mathbf{3},\mathbf{\overline{3}})^{\oplus 3}\oplus(\mathbf{1},\mathbf{\overline{3}})\oplus(\mathbf{5},\mathbf{1})^{\oplus 4}\oplus(\mathbf{3},\mathbf{1})^{\oplus 3}\\
  \nonumber &  \hspace{3.5cm}  \oplus(\mathbf{1},\mathbf{1})^{\oplus 3}\oplus(\mathbf{3},\mathbf{3})^{\oplus 3}\oplus(\mathbf{1},\mathbf{3}),\\
&    \text{A}_1^{11}\oplus\text{A}_2^{1} \to \text{E}_7  :  \quad  \chi=(\mathbf{5},\mathbf{3})^{\oplus 2}\oplus(\mathbf{5},\mathbf{\overline{3}})^{\oplus 2}\oplus(\mathbf{3},\mathbf{\overline{3}})\oplus(\mathbf{1},\mathbf{\overline{3}})^{\oplus 
2}\oplus(\mathbf{7},\mathbf{1})\oplus(\mathbf{5},\mathbf{1})^{\oplus 3}\oplus(\mathbf{3},\mathbf{1})^{\oplus 3} \\
\nonumber &  \hspace{3.5cm}  \oplus (\mathbf{1},\mathbf{1})\oplus(\mathbf{3},\mathbf{3})\oplus(\mathbf{1},\mathbf{3})^{\oplus 2},\\
&    \text{A}_1^{20}\oplus\text{A}_2^{1} \to \text{E}_7  :  \quad  \chi=(\mathbf{7},\mathbf{3})\oplus(\mathbf{5},\mathbf{3})\oplus(\mathbf{7},\mathbf{\overline{3}})\oplus(\mathbf{5},\mathbf{\overline{3}})\oplus(\mathbf{
3},\mathbf{\overline{3}})\oplus(\mathbf{9},\mathbf{1})\oplus(\mathbf{7},\mathbf{1}) \\ 
\nonumber &  \hspace{3.5cm}  \oplus(\mathbf{5},\mathbf{1})^{\oplus 3}\oplus  (\mathbf{1},\mathbf{1})\oplus(\mathbf{3},\mathbf{3}),\\
&    \text{A}_1^{35}\oplus\text{A}_2^{1} \to \text{E}_7  :  \quad  \chi=(\mathbf{9},\mathbf{3})\oplus(\mathbf{9},\mathbf{\overline{3}})\oplus(\mathbf{5},\mathbf{\overline{3}})\oplus(\mathbf{1},\mathbf{\overline{3}})\oplus(\mathbf{
11},\mathbf{1})\oplus(\mathbf{9},\mathbf{1})\oplus(\mathbf{7},\mathbf{1}) \\   
\nonumber &  \hspace{3.5cm}  \oplus(\mathbf{5},\mathbf{1})\oplus (\mathbf{5},\mathbf{3})\oplus(\mathbf{1},\mathbf{3}),\\
&    \text{A}_1^{1}\oplus\text{A}_2^{2} \to \text{E}_7  :  \quad  \chi=(\mathbf{1},\mathbf{6})\oplus(\mathbf{1},\mathbf{\overline{6}})\oplus(\mathbf{2},\mathbf{8})^{\oplus 2}\oplus (\mathbf{1},\mathbf{8})^{3}\oplus(\mathbf{2},\mathbf{1})^{\oplus 
4}\oplus(\mathbf{1},\mathbf{1})^{\oplus 4} \\  
\nonumber &  \hspace{3.5cm} \oplus (\mathbf{1},\mathbf{3})^{\oplus 3}\oplus(\mathbf{2},\mathbf{\overline{3}})^{\oplus 2}\oplus(\mathbf{1},\mathbf{\overline{3}})^{\oplus 3}\oplus(\mathbf{2},\mathbf{3})^{\oplus 
2},\\
&    \text{A}_1^{3}\oplus\text{A}_2^{2} \to \text{E}_7  :  \quad  \chi=(\mathbf{1},\mathbf{6})\oplus(\mathbf{1},\mathbf{\overline{6}})\oplus(\mathbf{2},\mathbf{8})^{\oplus 2}\oplus(\mathbf{4},\mathbf{1})^{\oplus 2}\oplus(\mathbf{
3},\mathbf{8}) ,\\
\nonumber &  \hspace{3.5cm}  \oplus(\mathbf{1},\mathbf{1})^{\oplus 4}\oplus(\mathbf{3},\mathbf{\overline{3}})\oplus(\mathbf{2},\mathbf{\overline{3}})^{\oplus 2}\oplus(\mathbf{3},\mathbf{3})\oplus(\mathbf{2},\mathbf{3})^{\oplus 2},\\
&    \text{A}_1^{4}\oplus\text{A}_2^{2} \to \text{E}_7  :  \quad  \chi=(\mathbf{1},\mathbf{6})\oplus(\mathbf{1},\mathbf{\overline{6}})\oplus(\mathbf{3},\mathbf{8})^{\oplus 2}\oplus(\mathbf{5},\mathbf{1})\oplus(\mathbf{1},\mathbf{8})\oplus(\mathbf{3},\mathbf{1})^{\oplus 2}\oplus(\mathbf{1},\mathbf{1})\\
\nonumber &  \hspace{3.5cm}  \oplus(\mathbf{3},\mathbf{3})^{\oplus 2}\oplus(\mathbf{1},\mathbf{3})\oplus(\mathbf{3},\mathbf{\overline{3}})^{\oplus 2}\oplus(\mathbf{1},\mathbf{\overline{3}}),\\
&    \text{A}_1^{28}\oplus\text{A}_2^{2} \to \text{E}_7  :  \quad  \chi=(\mathbf{1},\mathbf{6})\oplus(\mathbf{1},\mathbf{\overline{6}})\oplus(\mathbf{7},\mathbf{8})\oplus(\mathbf{11},\mathbf{1})\oplus  (\mathbf{
7},\mathbf{\overline{3}})\oplus (\mathbf{7},\mathbf{3})\oplus(\mathbf{1},\mathbf{1}),\\
&    \text{A}_1^{1}\oplus\text{A}_2^{2} \to \text{E}_7  :  \quad  \chi=(\mathbf{1},\mathbf{6})\oplus(\mathbf{1},\mathbf{8})^{\oplus 3}\oplus(\mathbf{2},\mathbf{\overline{6}})\oplus(\mathbf{2},\mathbf{3})^{\oplus 3}\oplus (\mathbf{1},\mathbf{1})^{\oplus 4}\oplus(\mathbf{1},\mathbf{\overline{6}})\\
\nonumber &  \hspace{3.5cm}  \oplus(\mathbf{2},\mathbf{6})\oplus(\mathbf{2},\mathbf{\overline{3}})^{\oplus 3}\oplus(\mathbf{1},\mathbf{3})^{\oplus 3}\oplus(\mathbf{2},\mathbf{1})^{\oplus 2}\oplus(\mathbf{
1},\mathbf{\overline{3}})^{\oplus 3},\\
&    \text{A}_1^{3}\oplus\text{A}_2^{2} \to \text{E}_7  :  \quad  \chi=(\mathbf{1},\mathbf{6})^{\oplus 3}\oplus(\mathbf{3},\mathbf{8})\oplus(\mathbf{3},\mathbf{3})^{\oplus 3}\oplus(\mathbf{1},\mathbf{\overline{6}})^{\oplus 3}\oplus  (\mathbf{
3},\mathbf{\overline{3}})^{\oplus 3} \oplus(\mathbf{1},\mathbf{1})^{\oplus 8},\\
&    \text{A}_1^{4}\oplus\text{A}_2^{2} \to \text{E}_7  :  \quad  \chi=(\mathbf{3},\mathbf{6})\oplus(\mathbf{1},\mathbf{8})^{\oplus 3}\oplus(\mathbf{3},\mathbf{\overline{6}})\oplus(\mathbf{3},\mathbf{3})^{\oplus 3}\oplus(\mathbf{5},\mathbf{1})\oplus (\mathbf{1},\mathbf{1})^{\oplus 3}\oplus (\mathbf{3},\mathbf{\overline{3}})^{\oplus 3},\\
&    \text{A}_1^{4}\oplus\text{A}_2^{2} \to \text{E}_7  :  \quad  \chi=(\mathbf{2},\mathbf{6})\oplus(\mathbf{3},\mathbf{8})\oplus(\mathbf{4},\mathbf{3})\oplus(\mathbf{2},\mathbf{\overline{6}})\oplus(\mathbf{2},\mathbf{3})\oplus(\mathbf{
3},\mathbf{3})\oplus(\mathbf{1},\mathbf{\overline{6}})\\  
\nonumber &  \hspace{3.5cm}   \oplus(\mathbf{3},\mathbf{1})\oplus(\mathbf{4},\mathbf{\overline{3}})\oplus(\mathbf{2},\mathbf{\overline{3}})\oplus(\mathbf{3},\mathbf{\overline{3}})\oplus(\mathbf{
1},\mathbf{6})\oplus(\mathbf{2},\mathbf{1})^{\oplus 2}\oplus(\mathbf{1},\mathbf{1}),\\
&    \text{A}_1^{7}\oplus\text{A}_2^{2} \to \text{E}_7  :  \quad  \chi=(\mathbf{3},\mathbf{6})\oplus(\mathbf{3},\mathbf{8})\oplus(\mathbf{3},\mathbf{\overline{6}})\oplus(\mathbf{5},\mathbf{3})\oplus(\mathbf{3},\mathbf{3})\oplus(\mathbf{
1},\mathbf{3}) \\ 
\nonumber &  \hspace{3.5cm} \oplus (\mathbf{5},\mathbf{1})\oplus(\mathbf{3},\mathbf{1})\oplus(\mathbf{5},\mathbf{\overline{3}})\oplus(\mathbf{3},\mathbf{\overline{3}})\oplus(\mathbf{1},\mathbf{\overline{3}}),
\\
&    \text{A}_1^{3}\oplus\text{A}_2^{3} \to \text{E}_7  :  \quad  \chi=(\mathbf{3},\mathbf{\overline{6}})\oplus(\mathbf{3},\mathbf{3})\oplus(\mathbf{1},\mathbf{15})\oplus(\mathbf{1},\mathbf{3})\oplus(\mathbf{3},\mathbf{6})\oplus(\mathbf{
3},\mathbf{\overline{3}})\oplus(\mathbf{1},\mathbf{\overline{15}}) \\ 
\nonumber &  \hspace{3.5cm} \oplus (\mathbf{1},\mathbf{\overline{3}})\oplus(\mathbf{3},\mathbf{8})\oplus (\mathbf{1},\mathbf{8}),\\
&    \text{A}_1^{1}\oplus\text{A}_2^{5} \to \text{E}_7  :  \quad  \chi=(\mathbf{1},\mathbf{27})\oplus(\mathbf{2},\mathbf{\overline{15}})\oplus (\mathbf{1},\mathbf{\overline{15}})\oplus(\mathbf{2},\mathbf{1})^{\oplus 2}\oplus (\mathbf{
1},\mathbf{15})\oplus(\mathbf{2},\mathbf{15})\oplus(\mathbf{1},\mathbf{1}),\\
&    \text{A}_1^{4}\oplus\text{A}_2^{5} \to \text{E}_7  :  \quad  \chi=(\mathbf{1},\mathbf{27})\oplus(\mathbf{3},\mathbf{\overline{15}})\oplus(\mathbf{5},\mathbf{1})\oplus  (\mathbf{3},\mathbf{15}),\\
&    \text{A}_1^{1}\oplus\text{A}_2^{5} \to \text{E}_7  :  \quad  \chi=(\mathbf{2},\mathbf{10})\oplus(\mathbf{2},\mathbf{\overline{6}})\oplus(\mathbf{1},\mathbf{15})\oplus(\mathbf{1},\mathbf{27})\oplus (\mathbf{1},\mathbf{\overline{15}})\oplus(\mathbf{2},\mathbf{\overline{10}})    \\
\nonumber & \hspace{3.5cm} 
\oplus 
 (\mathbf{2},\mathbf{6})\oplus(\mathbf{1},\mathbf{1}),\\
&    \text{A}_1^{1}\oplus\text{A}_2^{3} \to \text{E}_7  :  \quad  \chi=(\mathbf{1},\mathbf{10})\oplus(\mathbf{2},\mathbf{8})^{\oplus 4}\oplus(\mathbf{1},\mathbf{\overline{10}})\oplus(\mathbf{1},\mathbf{8})^{\oplus 4}\oplus (\mathbf{1},\mathbf{1})^{\oplus 6},\\
&    \text{A}_1^{2}\oplus\text{A}_2^{3} \to \text{E}_7  :  \quad  \chi=(\mathbf{1},\mathbf{10})\oplus(\mathbf{3},\mathbf{8})\oplus(\mathbf{2},\mathbf{8})^{\oplus 4}\oplus(\mathbf{
1},\mathbf{\overline{10}})\oplus(\mathbf{1},\mathbf{8})\oplus(\mathbf{3},\mathbf{1})\oplus(\mathbf{1},\mathbf{1})^{\oplus 3},\\
&    \text{A}_1^{3}\oplus\text{A}_2^{3} \to \text{E}_7  :  \quad  \chi=(\mathbf{1},\mathbf{10})\oplus(\mathbf{3},\mathbf{8})^{\oplus 3}\oplus(\mathbf{1},\mathbf{\overline{10}})\oplus(\mathbf{1},\mathbf{8})^{\oplus 3}\oplus(\mathbf{3},\mathbf{1})^{\oplus 2}.
\end{align}

\begin{figure}[H]
\vspace{5mm}
\qquad\qquad\scalebox{.9}{
\begin{tikzpicture}
\node at (.8*3.5,-.8*2) { $\mathfrak{su}_2\oplus\mathfrak{su}_3$ };
\node at (10+.8*3.2,-.8*2) { $\mathfrak{su}_5$ };
\draw[thick=4mm] plot [smooth, tension=1] coordinates {(0,.8*1.5) (.8*1.5,0) (0,-.8*1.5)};
\draw[thick=4mm] plot [smooth, tension=1] coordinates {(.8*2.5,.8*1.5) (.8*1,0) (.8*2.5,-.8*1.5)};
\node[draw=none] at (.8*3.5,0){$+$};
\node[draw,circle,thick,scale=1.25,fill=black] (A1) at (.8*6.5,.9) {};
\node[draw,circle,thick,scale=1.25] (A2) at (.8*5,-.8) {};
\node[draw,circle,thick,scale=1.25] (A3) at (.8*8,-.8) {};
\draw[thick] (A1)--(A2)--(A3)--(A1);
\draw[->,>=stealth',thick=4mm]  (8,0)--(10,0);
\node[draw,circle,thick,scale=1.25,fill=black,xshift=10cm] (B1) at (90:1.2) {};
\node[draw,circle,thick,scale=1.25,xshift=10cm] (B2) at (162:1.2) {};
\node[draw,circle,thick,scale=1.25,xshift=10cm] (B3) at (234:1.2) {};
\node[draw,circle,thick,scale=1.25,xshift=10cm] (B4) at (306:1.2) {};
\node[draw,circle,thick,scale=1.25,xshift=10cm] (B5) at (378:1.2) {};
\draw[thick] (B1)--(B2)--(B3)--(B4)--(B5)--(B1);
\end{tikzpicture}}
\vspace{8mm}

\qquad\qquad\scalebox{.9}{
\begin{tikzpicture}
\node at (.8*3.5,-.8*2) { $\mathfrak{su}_2\oplus\mathfrak{su}_3$ };
\node at (10+.8*3.2,-.8*2) { $\mathfrak{su}_6$ };
\draw[thick=4mm] plot [smooth, tension=1] coordinates {(0,.8*1.5) (.8*1.5,0) (0,-.8*1.5)};
\draw[thick=4mm] plot [smooth, tension=1] coordinates {(.8*2.5,.8*1.5) (.8*1,0) (.8*2.5,-.8*1.5)};
\node[draw=none] at (.8*3.5,0){$+$};
\node[draw,circle,thick,scale=1.25,fill=black] (A1) at (.8*6.5,.9) {};
\node[draw,circle,thick,scale=1.25] (A2) at (.8*5,-.8) {};
\node[draw,circle,thick,scale=1.25] (A3) at (.8*8,-.8) {};
\draw[thick] (A1)--(A2)--(A3)--(A1);
\draw[->,>=stealth',thick=4mm]  (8,0)--(10,0);
\node[draw,circle,thick,scale=1.25,fill=black,xshift=10cm] (B1) at (180:1.2) {};
\node[draw,circle,thick,scale=1.25,xshift=10cm] (B2) at (240:1.2) {};
\node[draw,circle,thick,scale=1.25,xshift=10cm] (B3) at (300:1.2) {};
\node[draw,circle,thick,scale=1.25,xshift=10cm] (B4) at (360:1.2) {};
\node[draw,circle,thick,scale=1.25,xshift=10cm] (B5) at (60:1.2) {};
\node[draw,circle,thick,scale=1.25,xshift=10cm] (B6) at (120:1.2) {};
\draw[thick] (B1)--(B2)--(B3)--(B4)--(B5)--(B6)--(B1);
\end{tikzpicture}}
\vspace{8mm}

\qquad\qquad\scalebox{.9}{
\begin{tikzpicture}
\node at (.8*3.5,-.8*2) { $\mathfrak{su}_2\oplus\mathfrak{su}_3$ };
\draw[thick=4mm] plot [smooth, tension=1] coordinates {(0,.8*1.5) (.8*1.5,0) (0,-.8*1.5)};
\draw[thick=4mm] plot [smooth, tension=1] coordinates {(.8*2.5,.8*1.5) (.8*1,0) (.8*2.5,-.8*1.5)};
\node[draw=none] at (.8*3.5,0){$+$};
\node[draw,circle,thick,scale=1.25,fill=black] (A1) at (.8*6.5,.9) {};
\node[draw,circle,thick,scale=1.25] (A2) at (.8*5,-.8) {};
\node[draw,circle,thick,scale=1.25] (A3) at (.8*8,-.8) {};
\draw[thick] (A1)--(A2)--(A3)--(A1);
\draw[->,>=stealth',thick=4mm]  (8,0)--(10,0);
\node[draw,circle,thick,scale=1.25,fill=black,label=left:{1},xshift=2.6cm] (0b) at (8,.8){};
\node[draw,circle,thick,scale=1.25,label=left:{1},xshift=2.6cm] (0a) at (8,-.8){};
\node[draw,circle,thick,scale=1.25,label=below:{2},xshift=2.6cm] (1a) at (8+.8,0){};
\node[draw,circle,thick,scale=1.25,label=below:{2},xshift=2.6cm] (2a) at (8+.8*2.2,0){};
\node[draw,circle,thick,scale=1.25,label=right:{1},xshift=2.6cm] (3a) at (8+.8*3.2,-.8){};
\node[draw,circle,thick,scale=1.25,label=right:{1},xshift=2.6cm] (3b) at (8+.8*3.2,.8){};
\draw[thick] (0a)--(1a)--(2a)--(3a);
\draw[thick] (0b)--(1a);
\draw[thick] (2a)--(3b);
\node at (10+.8*3.2,-.8*1.7) { $\mathfrak{so}_{10}$ };
\end{tikzpicture}}
\vspace{8mm}

\qquad\qquad\scalebox{.9}{
\begin{tikzpicture}
\node at (.8*3.5,-.8*2) { $\mathfrak{su}_2\oplus\mathfrak{su}_3$ };
\draw[thick=4mm] plot [smooth, tension=1] coordinates {(0,.8*1.5) (.8*1.5,0) (0,-.8*1.5)};
\draw[thick=4mm] plot [smooth, tension=1] coordinates {(.8*2.5,.8*1.5) (.8*1,0) (.8*2.5,-.8*1.5)};
\node[draw=none] at (.8*3.5,0){$+$};
\node[draw,circle,thick,scale=1.25,fill=black] (A1) at (.8*6.5,.9) {};
\node[draw,circle,thick,scale=1.25] (A2) at (.8*5,-.8) {};
\node[draw,circle,thick,scale=1.25] (A3) at (.8*8,-.8) {};
\draw[thick] (A1)--(A2)--(A3)--(A1);
\draw[->,>=stealth',thick=4mm]  (8,0)--(10,0);
\node[draw,circle,thick,scale=1.25,fill=black,label=below:{1},xshift=2.3cm] (0a) at (8,-.8){};
\node[draw,circle,thick,scale=1.25,label=below:{2},xshift=2.3cm] (1a) at (8+.8,-.8){};
\node[draw,circle,thick,scale=1.25,label=below:{3},xshift=2.3cm] (2a) at (8+.8*2,-.8){};
\node[draw,circle,thick,scale=1.25,label=below:{2},xshift=2.3cm] (3a) at (8+.8*3,-.8){};
\node[draw,circle,thick,scale=1.25,label=below:{1},xshift=2.3cm] (4a) at (8+.8*4,-.8){};
\node[draw,circle,thick,scale=1.25,label=left:{2},xshift=2.3cm] (3b) at (8+.8*2,0){};
\node[draw,circle,thick,scale=1.25,label=left:{1},xshift=2.3cm] (4b) at (8+.8*2,.8){};
\draw[thick] (0a)--(1a)--(2a)--(3a)--(4a);
\draw[thick] (2a)--(3b)--(4b);
\node at (10+.8*3.2,-.8*2.2) { $\mathfrak{e}_6$ };
\end{tikzpicture}}
\vspace{11mm}

\qquad\qquad\scalebox{.9}{
\begin{tikzpicture}
\node at (.8*3.5,-.8*2) { $\mathfrak{su}_2\oplus\mathfrak{su}_3$ };
\draw[thick=4mm] plot [smooth, tension=1] coordinates {(0,.8*1.5) (.8*1.5,0) (0,-.8*1.5)};
\draw[thick=4mm] plot [smooth, tension=1] coordinates {(.8*2.5,.8*1.5) (.8*1,0) (.8*2.5,-.8*1.5)};
\node[draw=none] at (.8*3.5,0){$+$};
\node[draw,circle,thick,scale=1.25,fill=black] (A1) at (.8*6.5,.9) {};
\node[draw,circle,thick,scale=1.25] (A2) at (.8*5,-.8) {};
\node[draw,circle,thick,scale=1.25] (A3) at (.8*8,-.8) {};
\draw[thick] (A1)--(A2)--(A3)--(A1);
\draw[->,>=stealth',thick=4mm]  (7.5,0)--(9.5,0);
\node[draw,circle,thick,scale=1.25,fill=black,label=below:{1},xshift=2cm] (0a) at (8,0){};
\node[draw,circle,thick,scale=1.25,label=below:{2},xshift=2cm] (1a) at (8+.8,0){};
\node[draw,circle,thick,scale=1.25,label=below:{3},xshift=2cm] (2a) at (8+.8*2,0){};
\node[draw,circle,thick,scale=1.25,label=below:{4},xshift=2cm] (3a) at (8+.8*3,0){};
\node[draw,circle,thick,scale=1.25,label=left:{2},xshift=2cm] (4b) at (8+.8*3,.8){};
\node[draw,circle,thick,scale=1.25,label=below:{3},xshift=2cm] (4a) at (8+.8*4,0){};
\node[draw,circle,thick,scale=1.25,label=below:{2},xshift=2cm] (5a) at (8+.8*5,0){};
\node[draw,circle,thick,scale=1.25,label=below:{1},xshift=2cm] (6a) at (8+.8*6,0){};
\draw[thick] (0a)--(1a)--(2a)--(3a)--(4a)--(5a)--(6a);
\draw[thick] (3a)--(4b);
\node at (10+.8*3.6,-.8*1.6) { $\mathfrak{e}_7$ };
\end{tikzpicture}}
\vspace{5mm}
\caption{The expected gauge group enhancement of SU(2)$\times$SU(3).}
\label{fig:su2su3enhancements}
\end{figure}
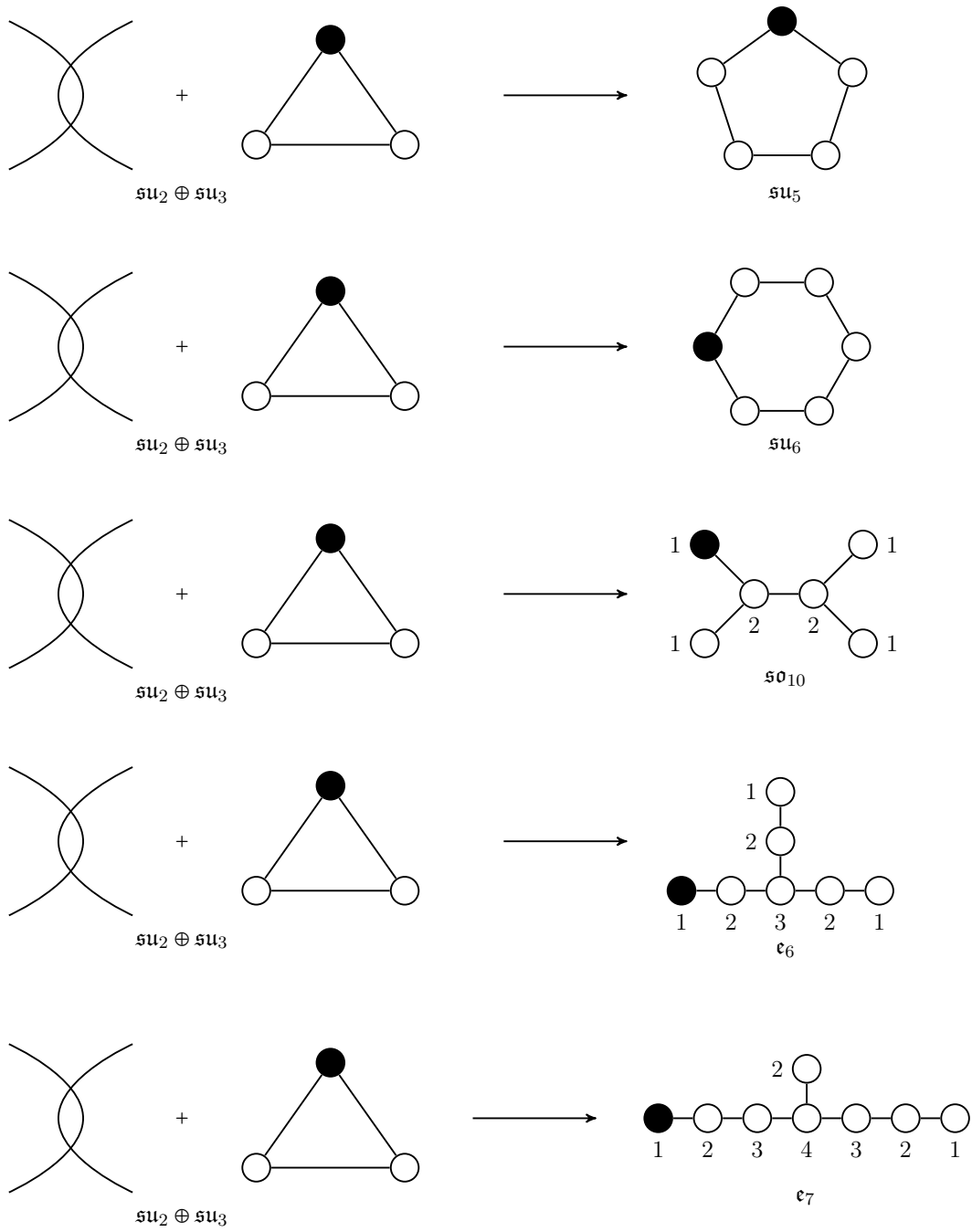
%\clearpage\newpage

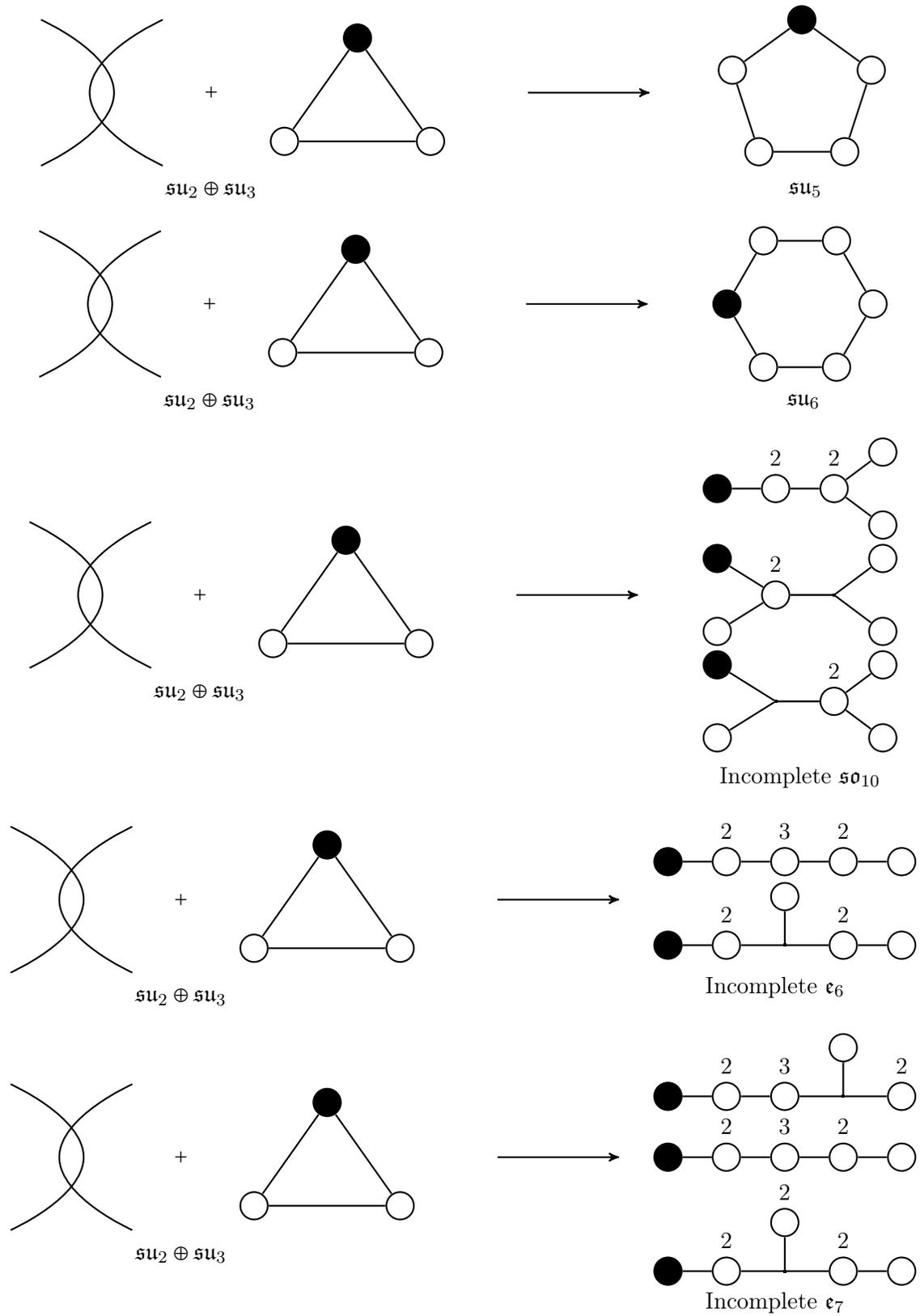
\begin{figure}[H]
\begin{center}
\begin{tikzpicture}
\node at (.8*3.5,-.8*2) { $\mathfrak{su}_2\oplus\mathfrak{su}_3$ };
\node at (10+.8*3.2,-.8*2) { $\mathfrak{su}_5$ };
\draw[thick=4mm] plot [smooth, tension=1] coordinates {(0,.8*1.5) (.8*1.5,0) (0,-.8*1.5)};
\draw[thick=4mm] plot [smooth, tension=1] coordinates {(.8*2.5,.8*1.5) (.8*1,0) (.8*2.5,-.8*1.5)};
\node[draw=none] at (.8*3.5,0){$+$};
\node[draw,circle,thick,scale=1.25,fill=black] (A1) at (.8*6.5,.9) {};
\node[draw,circle,thick,scale=1.25] (A2) at (.8*5,-.8) {};
\node[draw,circle,thick,scale=1.25] (A3) at (.8*8,-.8) {};
\draw[thick] (A1)--(A2)--(A3)--(A1);
\draw[->,>=stealth',thick=4mm]  (8,0)--(10,0);
\node[draw,circle,thick,scale=1.25,fill=black,xshift=10cm] (B1) at (90:1.2) {};
\node[draw,circle,thick,scale=1.25,xshift=10cm] (B2) at (162:1.2) {};
\node[draw,circle,thick,scale=1.25,xshift=10cm] (B3) at (234:1.2) {};
\node[draw,circle,thick,scale=1.25,xshift=10cm] (B4) at (306:1.2) {};
\node[draw,circle,thick,scale=1.25,xshift=10cm] (B5) at (378:1.2) {};
\draw[thick] (B1)--(B2)--(B3)--(B4)--(B5)--(B1);
\end{tikzpicture}
\vspace{3mm}

\begin{tikzpicture}
\node at (.8*3.5,-.8*2) { $\mathfrak{su}_2\oplus\mathfrak{su}_3$ };
\node at (10+.8*3.2,-.8*2) { $\mathfrak{su}_6$ };
\draw[thick=4mm] plot [smooth, tension=1] coordinates {(0,.8*1.5) (.8*1.5,0) (0,-.8*1.5)};
\draw[thick=4mm] plot [smooth, tension=1] coordinates {(.8*2.5,.8*1.5) (.8*1,0) (.8*2.5,-.8*1.5)};
\node[draw=none] at (.8*3.5,0){$+$};
\node[draw,circle,thick,scale=1.25,fill=black] (A1) at (.8*6.5,.9) {};
\node[draw,circle,thick,scale=1.25] (A2) at (.8*5,-.8) {};
\node[draw,circle,thick,scale=1.25] (A3) at (.8*8,-.8) {};
\draw[thick] (A1)--(A2)--(A3)--(A1);
\draw[->,>=stealth',thick=4mm]  (8,0)--(10,0);
\node[draw,circle,thick,scale=1.25,fill=black,xshift=10cm] (B1) at (180:1.2) {};
\node[draw,circle,thick,scale=1.25,xshift=10cm] (B2) at (240:1.2) {};
\node[draw,circle,thick,scale=1.25,xshift=10cm] (B3) at (300:1.2) {};
\node[draw,circle,thick,scale=1.25,xshift=10cm] (B4) at (360:1.2) {};
\node[draw,circle,thick,scale=1.25,xshift=10cm] (B5) at (60:1.2) {};
\node[draw,circle,thick,scale=1.25,xshift=10cm] (B6) at (120:1.2) {};
\draw[thick] (B1)--(B2)--(B3)--(B4)--(B5)--(B6)--(B1);
\end{tikzpicture}
\vspace{3mm}

\begin{tikzpicture}
\node at (.8*3.5,-.8*2) { $\mathfrak{su}_2\oplus\mathfrak{su}_3$ };
\node at (10.1+.8*3.2,-3) { Incomplete $\mathfrak{so}_{10}$ };
\draw[thick=4mm] plot [smooth, tension=1] coordinates {(0,.8*1.5) (.8*1.5,0) (0,-.8*1.5)};
\draw[thick=4mm] plot [smooth, tension=1] coordinates {(.8*2.5,.8*1.5) (.8*1,0) (.8*2.5,-.8*1.5)};
\node[draw=none] at (.8*3.5,0){$+$};
\node[draw,circle,thick,scale=1.25,fill=black] (A1) at (.8*6.5,.9) {};
\node[draw,circle,thick,scale=1.25] (A2) at (.8*5,-.8) {};
\node[draw,circle,thick,scale=1.25] (A3) at (.8*8,-.8) {};
\draw[thick] (A1)--(A2)--(A3)--(A1);
\draw[->,>=stealth',thick=4mm]  (8,0)--(10,0);
\node[draw,circle,thick,scale=1.25,fill=black,yshift=1.4cm] (A1) at (10.5+.8,0) {};
\node[draw,circle,thick,scale=1.25,label=above:2,yshift=1.4cm] (A2) at (10.5+.8*2.2,0) {};
\node[draw,circle,thick,scale=1.25,label=above:2,yshift=1.4cm] (A3) at (10.5+.8*3.4,0) {};
\node[draw,circle,thick,scale=1.25,yshift=1.4cm] (A5) at (10.5+.8*4.4,.6) {};
\node[draw,circle,thick,scale=1.25,yshift=1.4cm] (A4) at (10.5+.8*4.4,-.6) {};
\draw[thick] (A1)--(A2)--(A3)--(A5);
\draw[thick] (A4)--(A3);
\node[draw,circle,thick,scale=1.25,fill=black] (B1) at (10.5+.8,.6) {};
\node[draw,circle,thick,scale=1.25] (B1b) at (10.5+.8,-.6) {};
\node[draw,circle,thick,scale=1.25,label=above:2] (B2) at (10.5+.8*2.2,0) {};
\node[draw=none,scale=1.25,fill=black,inner sep=0,outer sep=0] (B3) at (10.5+.8*3.4,0) {};
\node[draw,circle,thick,scale=1.25] (B5) at (10.5+.8*4.4,.6) {};
\node[draw,circle,thick,scale=1.25] (B4) at (10.5+.8*4.4,-.6) {};
\draw[thick] (B1)--(B2)--(B3)--(B5);
\draw[thick] (B4)--(B3)--(B2)--(B1b);
\node[draw,circle,thick,scale=1.25,fill=black,yshift=-1.4cm] (C1) at (10.5+.8,.6) {};
\node[draw,circle,thick,scale=1.25,yshift=-1.4cm] (C1b) at (10.5+.8,-.6) {};
\node[draw=none,scale=1.25,fill=black,inner sep=0,outer sep=0,yshift=-1.4cm] (C2) at (10.5+.8*2.2,0) {};
\node[draw,circle,thick,scale=1.25,label=above:2,yshift=-1.4cm] (C3) at (10.5+.8*3.4,0) {};
\node[draw,circle,thick,scale=1.25,yshift=-1.4cm] (C5) at (10.5+.8*4.4,.6) {};
\node[draw,circle,thick,scale=1.25,yshift=-1.4cm] (C4) at (10.5+.8*4.4,-.6) {};
\draw[thick] (C1)--(C2)--(C3)--(C5);
\draw[thick] (C4)--(C3)--(C2)--(C1b);
\end{tikzpicture}
\vspace{3mm}

\begin{tikzpicture}
\node at (.8*3.5,-.8*2) { $\mathfrak{su}_2\oplus\mathfrak{su}_3$ };
\node at (10+.8*3.2,-.8*1.8) {Incomplete $\mathfrak{e}_{6}$ };
\draw[thick=4mm] plot [smooth, tension=1] coordinates {(0,.8*1.5) (.8*1.5,0) (0,-.8*1.5)};
\draw[thick=4mm] plot [smooth, tension=1] coordinates {(.8*2.5,.8*1.5) (.8*1,0) (.8*2.5,-.8*1.5)};
\node[draw=none] at (.8*3.5,0){$+$};
\node[draw,circle,thick,scale=1.25,fill=black] (A1) at (.8*6.5,.9) {};
\node[draw,circle,thick,scale=1.25] (A2) at (.8*5,-.8) {};
\node[draw,circle,thick,scale=1.25] (A3) at (.8*8,-.8) {};
\draw[thick] (A1)--(A2)--(A3)--(A1);
\draw[->,>=stealth',thick=4mm]  (8,0)--(10,0);
\node[draw,circle,thick,scale=1.25,fill=black,yshift=-.6cm] (B1) at (10+.8,0) {};
\node[draw,circle,thick,scale=1.25,label=above:2,yshift=-.6cm] (B2) at (10+.8*2.2,0) {};
\node[draw,circle,thick,scale=1.25,yshift=-.6cm] (B3b) at (10+.8*3.4,.8) {};
\node[draw=none,scale=1.25,yshift=-.6cm,fill=black,inner sep=0,outer sep=0] (B3) at (10+.8*3.4,0) {};
\node[draw,circle,thick,scale=1.25,label=above:2,yshift=-.6cm] (B4) at (10+.8*4.6,0) {};
\node[draw,circle,thick,scale=1.25,yshift=-.6cm] (B5) at (10+.8*5.8,0) {};
\draw[thick] (B1)--(B2)--(B3)--(B4)--(B5);
\draw[thick] (B3)--(B3b);
\node[draw,circle,thick,scale=1.25,fill=black,yshift=.5cm] (C1) at (10+.8,0) {};
\node[draw,circle,thick,scale=1.25,label=above:2,yshift=.5cm] (C2) at (10+.8*2.2,0) {};
\node[draw,circle,thick,scale=1.25,label=above:3,yshift=.5cm] (C3) at (10+.8*3.4,0) {};
\node[draw,circle,thick,scale=1.25,label=above:2,yshift=.5cm] (C4) at (10+.8*4.6,0) {};
\node[draw,circle,thick,scale=1.25,yshift=.5cm] (C5) at (10+.8*5.8,0) {};
\draw[thick] (C1)--(C2)--(C3)--(C4)--(C5);
\end{tikzpicture}
\vspace{3mm}

\begin{tikzpicture}
\node at (.8*3.5,-.8*2) { $\mathfrak{su}_2\oplus\mathfrak{su}_3$ };
\node at (10+.8*3.2,-.8*3) {Incomplete $\mathfrak{e}_{7}$ };
\draw[thick=4mm] plot [smooth, tension=1] coordinates {(0,.8*1.5) (.8*1.5,0) (0,-.8*1.5)};
\draw[thick=4mm] plot [smooth, tension=1] coordinates {(.8*2.5,.8*1.5) (.8*1,0) (.8*2.5,-.8*1.5)};
\node[draw=none] at (.8*3.5,0){$+$};
\node[draw,circle,thick,scale=1.25,fill=black] (A1) at (.8*6.5,.9) {};
\node[draw,circle,thick,scale=1.25] (A2) at (.8*5,-.8) {};
\node[draw,circle,thick,scale=1.25] (A3) at (.8*8,-.8) {};
\draw[thick] (A1)--(A2)--(A3)--(A1);
\draw[->,>=stealth',thick=4mm]  (8,0)--(10,0);
\node[draw,circle,thick,scale=1.25,fill=black,yshift=-1.5cm] (B1) at (10+.8,0) {};
\node[draw,circle,thick,scale=1.25,label=above:2,yshift=-1.5cm] (B2) at (10+.8*2.2,0) {};
\node[draw,circle,thick,scale=1.25,label=above:2,yshift=-1.5cm] (B3b) at (10+.8*3.4,.8) {};
\node[draw=none,scale=1.25,yshift=-1.5cm,fill=black,inner sep=0,outer sep=0] (B3) at (10+.8*3.4,0) {};
\node[draw,circle,thick,scale=1.25,label=above:2,yshift=-1.5cm] (B4) at (10+.8*4.6,0) {};
\node[draw,circle,thick,scale=1.25,yshift=-1.5cm] (B5) at (10+.8*5.8,0) {};
\draw[thick] (B1)--(B2)--(B3)--(B4)--(B5);
\draw[thick] (B3)--(B3b);
\node[draw,circle,thick,scale=1.25,fill=black] (C1) at (10+.8,0) {};
\node[draw,circle,thick,scale=1.25,label=above:2] (C2) at (10+.8*2.2,0) {};
\node[draw,circle,thick,scale=1.25,label=above:3] (C3) at (10+.8*3.4,0) {};
\node[draw,circle,thick,scale=1.25,label=above:2] (C4) at (10+.8*4.6,0) {};
\node[draw,circle,thick,scale=1.25] (C5) at (10+.8*5.8,0) {};
\draw[thick] (C1)--(C2)--(C3)--(C4)--(C5);
\node[draw,circle,thick,scale=1.25,fill=black,yshift=.8cm] (D1) at (10+.8,0) {};
\node[draw,circle,thick,scale=1.25,label=above:2,yshift=.8cm] (D2) at (10+.8*2.2,0) {};
\node[draw,circle,thick,scale=1.25,label=above:3,yshift=.8cm] (D3) at (10+.8*3.4,0) {};
\node[draw=none,scale=1.25,yshift=.8cm,fill=black,inner sep=0,outer sep=0] (D4) at (10+.8*4.6,0) {};
\node[draw,circle,thick,scale=1.25,yshift=.8cm] (D4b) at (10+.8*4.6,.8) {};
\node[draw,circle,thick,scale=1.25,label=above:2,yshift=.8cm] (D5) at (10+.8*5.8,0) {};
\draw[thick] (D1)--(D2)--(D3)--(D4)--(D5);
\draw[thick] (D4)--(D4b);
\end{tikzpicture}
\vspace{-4mm}
\end{center}
\caption{Degeneration of the SU(2)$\times$SU(3) fibers.}
\label{fig:su2su3enhancementG}
\end{figure}

 \section{$\text{A}_1\oplus\text{C}_2$: the SU($2$)$\times$USp($4$)-model}

\subsection{$\text{A}_1\oplus\text{C}_2\to \text{D}_4$}

Up to linear equivalence, there are three embeddings of $\text{A}_1\oplus\text{C}_2$ in D$_4$. They are connected by triality and have Dynkin index $(2,1)$  and the same characteristic representation:
\begin{equation}
\text{A}^2_1\oplus\text{C}^1_2:\quad \chi=(\mathbf{3},\mathbf{5}). 
\end{equation}

\subsection{$\text{A}_1\oplus\text{C}_2\to \text{C}_3$}

Up to linear equivalence, there is a unique  embeddings of $\text{A}_1\oplus\text{C}_2$ in C$_3$. It has Dynkin index $(1,1)$ and the characteristic representation is isotropic: 
\begin{equation}
\text{A}^1_1\oplus\text{C}^1_2:\quad \chi=(\mathbf{2},\mathbf{4}).
\end{equation}

\subsection{$\text{A}_1\oplus\text{C}_2\to \text{C}_4$}

Up to linear equivalence, there are three   embeddings of $\text{A}_1\oplus\text{C}_2$ in C$_4$:
\begin{align}
& \text{A}^1_1\oplus\text{C}^1_2:\quad \chi=(\mathbf{1},\mathbf{4})^{\oplus 2}\oplus(\mathbf{2},\mathbf{1})^{\oplus 2}\oplus(\mathbf{2},\mathbf{4})\oplus(\mathbf{1},\mathbf{1})^{\oplus 3},\\
& \text{A}^2_1\oplus\text{C}^1_2:\quad \chi=(\mathbf{3},\mathbf{1})^{\oplus 2}\oplus(\mathbf{2},\mathbf{4})^{\oplus 2}\oplus(\mathbf{1},\mathbf{1}),\\
& \text{A}^1_1\oplus\text{C}^2_2:\quad \chi=(\mathbf{7},\mathbf{1})\oplus(\mathbf{4},\mathbf{4}).
\end{align}
The embedding of Dynkin index $(1,1)$ gives the matter expected in F-theory.

\subsection{$\text{A}_1\oplus\text{C}_2\to \text{A}_5$}
Up to linear equivalence, there is a unique embedding of $\text{A}_1\oplus\text{C}_2$ into A$_5$. It has Dynkin index $(1,1)$:
\begin{align}
& \text{A}^1_1\oplus\text{C}^1_2\to\text{A}_5:\quad 
\chi=(\mathbf{1},\mathbf{5})\oplus(\mathbf{2},\mathbf{4})^{\oplus 2}\oplus(\mathbf{1},\mathbf{1}).
\end{align}
\subsection{$\text{A}_1\oplus\text{C}_2\to \text{D}_5$}
Up to linear equivalence, there are three   embeddings of $\text{A}_1\oplus\text{C}_2$ in D$_5$ that are distinguishable by their Dynkin index:
\begin{align}
&    \text{A}_1^{2}\oplus\text{C}_2^{1} \to \text{D}_5  :  \quad  \chi= (\mathbf{2},\mathbf{4})^{\oplus 2}\oplus(\mathbf{3},\mathbf{5})\oplus (\mathbf{
1},\mathbf{1}),\\
&    \text{A}_1^{1}\oplus\text{C}_2^{1} \to \text{D}_5  :  \quad  \chi=(\mathbf{2},\mathbf{5})^{\oplus 2}\oplus (\mathbf{2},\mathbf{1})^{\oplus 2}\oplus (\mathbf{
1},\mathbf{5})\oplus(\mathbf{1},\mathbf{1})^{\oplus 3},\\
&    \text{A}_1^{2}\oplus\text{C}_2^{1} \to \text{D}_5  :  \quad  \chi=(\mathbf{3},\mathbf{5})\oplus (\mathbf{1},\mathbf{5})^{\oplus 2}\oplus(\mathbf{3},\mathbf{1})^{\oplus 2}\oplus(\mathbf{1},\mathbf{1}),\\
&    \text{A}_1^{10}\oplus\text{C}_2^{1} \to \text{D}_5  :  \quad  \chi=(\mathbf{5},\mathbf{5})\oplus(\mathbf{7},\mathbf{1}). 
\end{align}

\begin{figure}[H]
\begin{center}
\vspace{16mm}
\begin{tikzpicture}
\node at (.8*3.5,-.8*2.5) { $\mathfrak{su}_2\oplus\mathfrak{usp}_4$ };
\node at (10+.8*3.2,-.8*2.5) { $\mathfrak{usp}_6$ };
\draw[thick=4mm] plot [smooth, tension=1] coordinates {(0,.8*1.5) (.8*1.5,0) (0,-.8*1.5)};
\draw[thick=4mm] plot [smooth, tension=1] coordinates {(.8*2.5,.8*1.5) (.8*1,0) (.8*2.5,-.8*1.5)};
\node[draw=none] at (.8*3.5,0){$+$};
\node[draw,circle,thick,scale=1.25,fill=black] (A1) at (.8*6.5,.8*1.5) {};
\node[draw,circle,thick,scale=1.25] (A2) at (.8*5,0) {};
\node[draw,circle,thick,scale=1.25] (A3) at (.8*8,0) {};
\node[draw,circle,thick,scale=1.25] (A4) at (.8*6.5,-.8*1.5) {};
\draw[thick] (A1)--(A2)--(A4)--(A3)--(A1);
\node at (.8*6.5,0)[draw,dashed,line width=2pt,ellipse,minimum width=100pt,minimum height=26pt, yshift=-1pt]{};
\draw[->,>=stealth',thick=4mm]  (8,0)--(10,0);
\node[draw,circle,thick,scale=1.25,fill=black] (B1) at (10+.8,0) {};
\node[draw,circle,thick,scale=1.25] (B2) at (10+.8*2.5,.8) {};
\node[draw,circle,thick,scale=1.25] (B3) at (10+.8*4,.8) {};
\node[draw,circle,thick,scale=1.25] (B2b) at (10+.8*2.5,-.8) {};
\node[draw,circle,thick,scale=1.25] (B3b) at (10+.8*4,-.8) {};
\node[draw,circle,thick,scale=1.25] (B4) at (10+.8*5.5,0) {};
\node at (10+.8*2.5,0)[draw,dashed,line width=2pt,ellipse,minimum width=70pt,minimum height=26pt, rotate=90,yshift=-1pt]{};
\node at (10+.8*4,0)[draw,dashed,line width=2pt,ellipse,minimum width=70pt,minimum height=26pt, rotate=90,yshift=-1pt]{};
\draw[thick] (B1)--(B2)--(B3)--(B4);
\draw[thick] (B1)--(B2b)--(B3b)--(B4);
\end{tikzpicture}
\vspace{8mm}

\begin{tikzpicture}
\node at (.8*3.5,-.8*2.5) { $\mathfrak{su}_2\oplus\mathfrak{usp}_4$ };
\node at (10+.8*3.2,-.8*2.5) {Incomplete $\mathfrak{so}_{10}$ };
\draw[thick=4mm] plot [smooth, tension=1] coordinates {(0,.8*1.5) (.8*1.5,0) (0,-.8*1.5)};
\draw[thick=4mm] plot [smooth, tension=1] coordinates {(.8*2.5,.8*1.5) (.8*1,0) (.8*2.5,-.8*1.5)};
\node[draw=none] at (.8*3.5,0){$+$};
\node[draw,circle,thick,scale=1.25,fill=black] (A1) at (.8*6.5,.8*1.5) {};
\node[draw,circle,thick,scale=1.25] (A2) at (.8*5,0) {};
\node[draw,circle,thick,scale=1.25] (A3) at (.8*8,0) {};
\node[draw,circle,thick,scale=1.25] (A4) at (.8*6.5,-.8*1.5) {};
\draw[thick] (A1)--(A2)--(A4)--(A3)--(A1);
\node at (.8*6.5,0)[draw,dashed,line width=2pt,ellipse,minimum width=100pt,minimum height=26pt, yshift=-1pt]{};
\draw[->,>=stealth',thick=4mm]  (8,0)--(10,0);
\node[draw,circle,thick,scale=1.25,fill=black] (B1) at (10+.8,0) {};
\node[draw,circle,thick,scale=1.25,label=above:2] (B2) at (10+.8*2.5,0) {};
\node[draw,circle,thick,scale=1.25,label=above:2] (B3) at (10+.8*4,0) {};
\node[draw,circle,thick,scale=1.25] (B4) at (10+.8*5.5,0) {};
\draw[thick] (B1)--(B2)--(B3)--(B4);
\end{tikzpicture}
\vspace{8mm}

\begin{tikzpicture}
\node at (.8*3.5,-.8*2.5) { $\mathfrak{su}_2\oplus\mathfrak{usp}_4$ };
\node at (10+.8*3.2,-.8*2.5) { $\mathfrak{su}_6$ };
\draw[thick=4mm] plot [smooth, tension=1] coordinates {(0,.8*1.5) (.8*1.5,0) (0,-.8*1.5)};
\draw[thick=4mm] plot [smooth, tension=1] coordinates {(.8*2.5,.8*1.5) (.8*1,0) (.8*2.5,-.8*1.5)};
\node[draw=none] at (.8*3.5,0){$+$};
\node[draw,circle,thick,scale=1.25,fill=black] (A1) at (.8*6.5,.8*1.5) {};
\node[draw,circle,thick,scale=1.25] (A2) at (.8*5,0) {};
\node[draw,circle,thick,scale=1.25] (A3) at (.8*8,0) {};
\node[draw,circle,thick,scale=1.25] (A4) at (.8*6.5,-.8*1.5) {};
\draw[thick] (A1)--(A2)--(A4)--(A3)--(A1);
\node at (.8*6.5,0)[draw,dashed,line width=2pt,ellipse,minimum width=100pt,minimum height=26pt, yshift=-1pt]{};
\draw[->,>=stealth',thick=4mm]  (8,0)--(10,0);
\node[draw,circle,thick,scale=1.25,fill=black] (B1) at (10+.8*1.3,0) {};
\node[draw,circle,thick,scale=1.25] (B2) at (10+.8*2.3,.8*1.5) {};
\node[draw,circle,thick,scale=1.25] (B3) at (10+.8*4,.8*1.5) {};
\node[draw,circle,thick,scale=1.25] (B2b) at (10+.8*2.3,-.8*1.5) {};
\node[draw,circle,thick,scale=1.25] (B3b) at (10+.8*4,-.8*1.5) {};
\node[draw,circle,thick,scale=1.25] (B4) at (10+.8*5,0) {};
\draw[thick] (B1)--(B2)--(B3)--(B4);
\draw[thick] (B1)--(B2b)--(B3b)--(B4);
\end{tikzpicture}
\vspace{8mm}

\begin{tikzpicture}
\node at (.8*3.5,-.8*2.5) { $\mathfrak{su}_2\oplus\mathfrak{usp}_4$ };
\node at (10+.8*3.2,-.8*2.5) { Incomplete $\mathfrak{usp}_8$ };
\draw[thick=4mm] plot [smooth, tension=1] coordinates {(0,.8*1.5) (.8*1.5,0) (0,-.8*1.5)};
\draw[thick=4mm] plot [smooth, tension=1] coordinates {(.8*2.5,.8*1.5) (.8*1,0) (.8*2.5,-.8*1.5)};
\node[draw=none] at (.8*3.5,0){$+$};
\node[draw,circle,thick,scale=1.25,fill=black] (A1) at (.8*6.5,.8*1.5) {};
\node[draw,circle,thick,scale=1.25] (A2) at (.8*5,0) {};
\node[draw,circle,thick,scale=1.25] (A3) at (.8*8,0) {};
\node[draw,circle,thick,scale=1.25] (A4) at (.8*6.5,-.8*1.5) {};
\draw[thick] (A1)--(A2)--(A4)--(A3)--(A1);
\node at (.8*6.5,0)[draw,dashed,line width=2pt,ellipse,minimum width=100pt,minimum height=26pt, yshift=-1pt]{};
\draw[->,>=stealth',thick=4mm]  (8,0)--(10,0);
\node[draw,circle,thick,scale=1.25,fill=black] (A1) at (10+.8,0) {};
\node[draw,circle,thick,scale=1.25] (A2) at (10+.8*2,.8) {};
\node[draw,circle,thick,scale=1.25] (A3) at (10+.8*3.5,.8) {};
\node[draw,circle,thick,scale=1.25] (A4) at (10+.8*5,.8) {};
\node[draw,circle,thick,scale=1.25] (A7) at (10+.8*5,-.8) {};
\node[draw,circle,thick,scale=1.25] (A5) at (10+.8*3.5,-.8) {};
\node[draw,circle,thick,scale=1.25] (A6) at (10+.8*2,-.8) {};
\draw[thick] (A1)--(A2)--(A3)--(A4)--(A7)--(A5)--(A6)--(A1);
\node at (10+.8*2,0)[draw,dashed,line width=2pt,ellipse,minimum width=70pt,minimum height=26pt, rotate=90,yshift=-1pt]{};
\node at (10+.8*3.5,0)[draw,dashed,line width=2pt,ellipse,minimum width=70pt,minimum height=26pt, rotate=90,yshift=-1pt]{};
\node at (10+.8*5,0)[draw,dashed,line width=2pt,ellipse,minimum width=70pt,minimum height=26pt, rotate=90,yshift=-1pt]{};
\end{tikzpicture}
\vspace{8mm}

\end{center}
\caption{Degeneration of the SU(2)$\times$USp(4) and SU(2)$\times$USp(4)/$\mathbb{Z}_2$ fibers. 
}
\label{fig:su2usp4enhancement}
\end{figure}
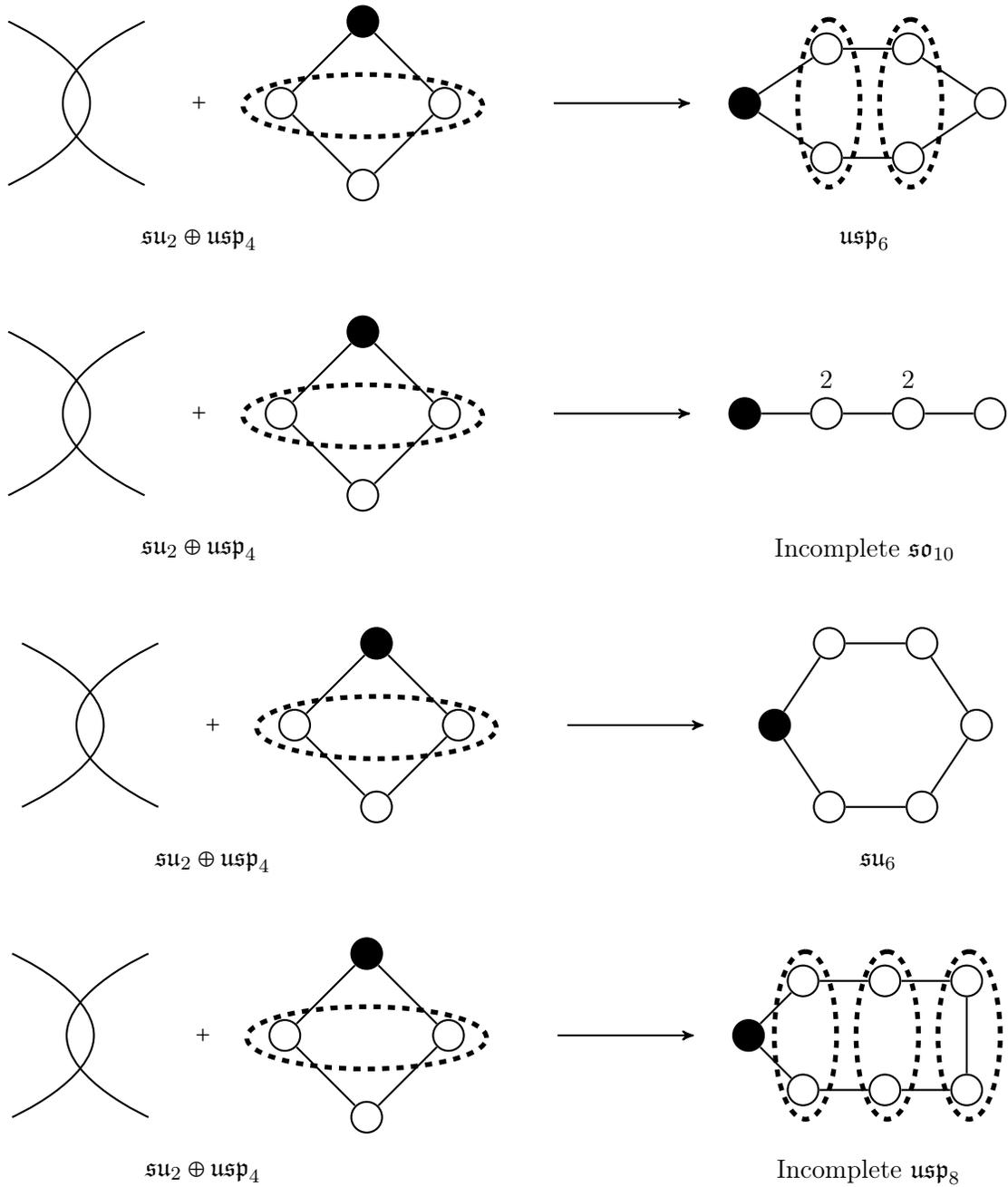

\section{$\text{A}_1\oplus\text{G}_2$: the SU($2$)$\times$G$_2$-model} \label{sec:A1G2}
\subsection{$\text{A}_1\oplus\text{G}_2\to\text{D}_5$}

$\text{A}_1\oplus\text{G}_2$ is a singleton subalgebra of $\text{D}_5$, it has Dynkin index (2,1) and characteristic representation: 
\begin{equation}
\text{A}^2_1\oplus\text{G}^1_2:\quad \chi=(\mathbf{3},\mathbf{7})\oplus (\mathbf{1},\mathbf{7}). 
\end{equation}
 This is a particularly interesting example, since there is no embedding of $\text{A}_1\oplus\text{G}_2\to \text{D}_5$ with  Dynkin index $(1,1)$. 
This is consistent with the fact that in F-theory there is no enhancement of types  $I_2+I^*_0\to I^*_1$, $III +\text{I}^{*}_1\to \text{I}^*_1$, or $IV^{\text{ns}}+I^*_0\to \text{I}^{*}_1$. 
This is because if such an enhancement exist, the discriminant must vanish at least at orders $2$, $6$, $7$, respectively for $\text{A}_1$, G$_2$, and D$_5$, but this is not plausible as $7< 2+6$. 

\subsection{$\text{A}_1\oplus\text{G}_2\to\text{B}_5$}
There are two subalgebras of type $\text{A}_1\oplus\text{G}_2$ in B$_5$: one has Dynkin index (1,1) and the other has Dynkin index (2,1). 
Their characteristic representations are: 
\begin{align}
&\text{A}^1_1\oplus\text{G}^1_2\to \text{B}_5:\quad \chi=(\mathbf{2},\mathbf{7})^{\oplus 2}\oplus (\mathbf{1},\mathbf{7})\oplus (\mathbf{
1},\mathbf{1})^{\oplus 3},\\
& \text{A}^{2}_1\oplus\text{G}^1_2\to \text{B}_5: \quad  \chi= 
 (\mathbf{3},\mathbf{7})\oplus (\mathbf{1},\mathbf{7})^{\oplus 2}
 \oplus  (\mathbf{3},\mathbf{
1}).
\end{align}

\subsection{$\text{A}_1\oplus\text{G}_2\to\text{D}_6$}

The enhancement that happens in F-theory is $\text{A}_1\oplus\text{G}_2\to\text{D}_6$, for example with I$_2$+I$_0^{*\text{ns}}$. 
There are three subalgebras of type $\text{A}_1\oplus\text{G}_2$ and one of them has Dynkin index (1,1): 
\begin{align}
&\text{A}^1_1\ \oplus\text{G}^1_2\to \text{D}_6:\quad \chi=(\mathbf{2},\mathbf{7})^{\oplus 2}\oplus (\mathbf{1},\mathbf{7})^{\oplus 2}\oplus (\mathbf{
2},\mathbf{1})^{\oplus 2}\oplus (\mathbf{1},\mathbf{1})^{\oplus 3},\\
& \text{A}^{2}_1\ \oplus\text{G}^1_2\to \text{D}_6: \quad  \chi= 
 (\mathbf{3},\mathbf{7})\oplus (\mathbf{1},\mathbf{7})^{\oplus 3}
 \oplus  (\mathbf{3},\mathbf{
1})^{\oplus 2} 
 \oplus  (\mathbf{1},\mathbf{
1}),\\
& \text{A}^{10}_1\oplus\text{G}^1_2\to \text{D}_6: \quad  \chi= 
 (\mathbf{5},\mathbf{7})\oplus  (\mathbf{7},\mathbf{
1})\oplus (\mathbf{1},\mathbf{7}).
\end{align}

\subsection{$\text{A}_1\oplus\text{G}_2\to\text{E}_7$} \label{sec:A2G2toE7}

There are eight embeddings of $\text{A}_1\oplus\text{G}_2$ inside E$_7$ up to linear equivalence and they each have a distinct Dynkin index 
\begin{align}
& \text{A}^1_1\ \oplus\text{G}^1_2\to \text{E}_7: \quad \chi=
(\mathbf{2},\mathbf{7})^{\oplus 4}\oplus (\mathbf{1},\mathbf{7})^{\oplus 6}\oplus (\mathbf{
2},\mathbf{1})^{\oplus 4}\oplus (\mathbf{1},\mathbf{1})^{\oplus 10},\\
& \text{A}^2_1\ \oplus\text{G}^1_2\to \text{E}_7: \quad \chi=
 (\mathbf{3},\mathbf{7})\oplus (\mathbf{2},\mathbf{7})^{\oplus 4}\oplus  (\mathbf{1},\mathbf{7})^{\oplus 3}\oplus (\mathbf{3},\mathbf{
1})^{\oplus 2}\oplus (\mathbf{2},\mathbf{1})^{\oplus 4}\oplus (\mathbf{1},\mathbf{1})^{\oplus 4},\\
& \text{A}^3_1\ \oplus\text{G}^1_2\to \text{E}_7: \quad
\chi=(\mathbf{3},\mathbf{7})^{\oplus 3}\oplus (\mathbf{1},\mathbf{7})^{\oplus 5}\oplus (\mathbf{3},\mathbf{1})^{\oplus 5}\oplus (\mathbf{1},\mathbf{
1})^{\oplus 3},\\
& \text{A}^{8}_1\ \oplus\text{G}^1_2\to \text{E}_7: \quad  \chi= 
  (\mathbf{5},\mathbf{7})\oplus (\mathbf{3},\mathbf{7})^{\oplus 3}\oplus 
  (\mathbf{5},\mathbf{1})^{\oplus 3}\oplus (\mathbf{1},\mathbf{1})^{\oplus 3}, \\
& \text{A}^{10}_1\oplus\text{G}^1_2\to \text{E}_7: \quad  \chi= 
(\mathbf{4},\mathbf{7})^{\oplus 2}\oplus (\mathbf{5},\mathbf{7})\oplus (\mathbf{1},\mathbf{7})\oplus (\mathbf{7},\mathbf{
1})\oplus (\mathbf{4},\mathbf{1})^{\oplus 2}\oplus (\mathbf{1},\mathbf{1})^{\oplus 3},\\
& \text{A}^{11}_1\oplus\text{G}^1_2\to \text{E}_7: \quad  \chi= 
 (\mathbf{5},\mathbf{7})^{\oplus 2}\oplus (\mathbf{3},\mathbf{7})\oplus (\mathbf{1},\mathbf{7})\oplus (\mathbf{7},\mathbf{
1})\oplus (\mathbf{5},\mathbf{1})\oplus (\mathbf{3},\mathbf{1})^{\oplus 2},\\
& \text{A}^{35}_1\oplus\text{G}^1_2\to \text{E}_7: \quad  \chi= 
  (\mathbf{9},\mathbf{7})\oplus (\mathbf{5},\mathbf{7})\oplus (\mathbf{11},\mathbf{1})\oplus (\mathbf{7},\mathbf{
1}),\\
& \text{A}^{7}_1\ \oplus\text{G}^2_2\to \text{E}_7: \quad  \chi= 
 (\mathbf{3},\mathbf{27})\oplus (\mathbf{5},\mathbf{7}). 
 \end{align}
 The first one with Dynkin index (1,1) corresponds to the matter content observed at the collision of the two divisors of an SU($2$)$\times$G$_2$-model realize by the collision III+I$_0^{*\text{ns}}$.   The last one with Dynkin index (7,1) is the only one that is a maximal embedding.  

 \subsection{$\text{A}_1\oplus\text{G}_2\to\text{E}_8$} \label{sec:A2G2toE8}
Up to linear equivalence, there are twenty different embeddings of  $\text{A}_1\oplus \text{G}_2$ in E$_8$ and they all have a different  Dynkin embedding index. 
The unique one with Dynkin embedding index (1,1) has the reduced matter representation that we see in F-theory, namely the fundamental representations $(\mathbf{2},\mathbf{1})$, 
$(\mathbf{1},\mathbf{7})$, and the bifundamental representation $(\mathbf{2},\mathbf{7})$. 
The branching of the adjoint of E$_8$ along representations of $\text{A}_1\oplus \text{G}_2$ is 
$$
[\mathbf{248}]= [(\mathbf{3},\mathbf{1})\oplus   (\mathbf{1},\mathbf{14})]\oplus \chi, 
$$
where  $\chi$ is as  follows for each of the twenty possible embeddings:
 \begin{align}
& \text{A}_1^{1}\oplus \text{G}_2^{1}  : & &     \chi=(\mathbf{2},\mathbf{1})^{\oplus 
14}\oplus  (\mathbf{1},\mathbf{7})^{\oplus 14}  \oplus(\mathbf{2},\mathbf{7})^{\oplus 6} \oplus(\mathbf{1},\mathbf{1})^{\oplus 21}\\
&  \text{A}_1^{2}\oplus \text{G}_2^{1}  : & & \chi=(\mathbf{3},\mathbf{7}) \oplus   (\mathbf{1},\mathbf{7})^{\oplus 7}\oplus(\mathbf{2},\mathbf{7})^{\oplus 8}\oplus(\mathbf{2},\mathbf{1})^{\oplus 
8}\oplus(\mathbf{3},\mathbf{1})^{\oplus 6}\oplus(\mathbf{1},\mathbf{1})^{\oplus 15}\\
&  \text{A}_1^{3}\oplus \text{G}_2^{1}  : & & \chi=(\mathbf{3},\mathbf{7})^{\oplus 3}  \oplus(\mathbf{1},\mathbf{7})^{\oplus 5}\oplus(\mathbf{3},\mathbf{1})^{\oplus 5}\oplus(\mathbf{1},\mathbf{1})^{\oplus 
6}\oplus(\mathbf{2},\mathbf{7})^{\oplus 6}\oplus(\mathbf{4},\mathbf{1})^{\oplus 2}\oplus(\mathbf{2},\mathbf{1})^{\oplus 10}\\
&   \text{A}_1^{4}\oplus \text{G}_2^{1}  : & &  \chi=(\mathbf{3},\mathbf{7})^{\oplus 6}\oplus(\mathbf{1},\mathbf{7})^{\oplus 8}\oplus(\mathbf{5},\mathbf{1}) \oplus(\mathbf{3},\mathbf{1})^{\oplus 
12}\oplus(\mathbf{1},\mathbf{1})^{\oplus 8}\\
&  \text{A}_1^{6}\oplus \text{G}_2^{1}  : & & \chi=(\mathbf{4},\mathbf{7})^{\oplus 2}\oplus(\mathbf{2},\mathbf{7})^{\oplus 4}  \oplus(\mathbf{5},\mathbf{1})^{\oplus 3}\oplus(\mathbf{3},\mathbf{7})^{\oplus 
3}\oplus(\mathbf{1},\mathbf{7}) \oplus(\mathbf{3},\mathbf{1})^{\oplus 5}
\nonumber \\
& & & \quad \quad
\oplus(\mathbf{1},\mathbf{1})^{\oplus 3}\oplus(\mathbf{4},\mathbf{1})^{\oplus 2}\oplus(\mathbf{2},\mathbf{1})^{\oplus 4}\\
&  \text{A}_1^{10}\oplus \text{G}_2^{1}  : & & \chi=(\mathbf{5},\mathbf{7}) \oplus   (\mathbf{4},\mathbf{7})^{\oplus 4}\oplus(\mathbf{7},\mathbf{1}) \oplus(\mathbf{5},\mathbf{1})^{\oplus 
4}\oplus(\mathbf{1},\mathbf{7})^{\oplus 5} \oplus   (\mathbf{1},\mathbf{1})^{\oplus 6}\oplus(\mathbf{4},\mathbf{1})^{\oplus 4}\\
&  \text{A}_1^{8}\oplus \text{G}_2^{1}  : & & \chi=(\mathbf{5},\mathbf{7}) \oplus   (\mathbf{3},\mathbf{7})^{\oplus 7}\oplus(\mathbf{5},\mathbf{1})^{\oplus 7} \oplus   (\mathbf{1},\mathbf{1})^{\oplus 14}
\\
& \text{A}_1^{9}\oplus \text{G}_2^{1}  : & & \chi=(\mathbf{4},\mathbf{7})^{\oplus 2}  \oplus(\mathbf{5},\mathbf{7}) \oplus(\mathbf{3},\mathbf{7})^{\oplus 3}\oplus(\mathbf{5},\mathbf{1})^{\oplus 
3}\oplus(\mathbf{6},\mathbf{1})^{\oplus 2}
\nonumber \\
& & & \quad \quad
\oplus(\mathbf{4},\mathbf{1})^{\oplus 2}\oplus(\mathbf{2},\mathbf{7})^{\oplus 2}\oplus(\mathbf{2},\mathbf{1})^{\oplus 4}\oplus(\mathbf{3},\mathbf{1})\oplus(\mathbf{1},\mathbf{1})^{\oplus 3}\\
& \text{A}_1^{11}\oplus \text{G}_2^{1}  : & & \chi=(\mathbf{5},\mathbf{7})^{\oplus 2}\oplus(\mathbf{3},\mathbf{7}) \oplus   (\mathbf{4},\mathbf{7})^{\oplus 2}\oplus(\mathbf{2},\mathbf{7})^{\oplus 
2}\oplus(\mathbf{6},\mathbf{1})^{\oplus 2}\oplus(\mathbf{1},\mathbf{7}) \nonumber \\
& & & \quad \quad
 \oplus(\mathbf{7},\mathbf{1}) \oplus(\mathbf{5},\mathbf{1}) \oplus(\mathbf{3},\mathbf{1})^{\oplus 2}\oplus(\mathbf{1},\mathbf{1})^{\oplus 
3}\oplus(\mathbf{4},\mathbf{1})^{\oplus 4}\\
&  \text{A}_1^{12}\oplus \text{G}_2^{1}  : & & \chi=(\mathbf{5},\mathbf{7})^{\oplus 3} \oplus(\mathbf{3},\mathbf{7})^{\oplus 3}\oplus(\mathbf{1},\mathbf{7})^{\oplus 2}\oplus(\mathbf{7},\mathbf{1})^{\oplus 2}\oplus(\mathbf{5},\mathbf{1})^{\oplus 4}\oplus(\mathbf{3},\mathbf{1})^{\oplus 5}\\
&  \text{A}_1^{28}\oplus \text{G}_2^{1}  : & & \chi=(\mathbf{7},\mathbf{7})^{\oplus 3}\oplus  \oplus(\mathbf{11},\mathbf{1}) \oplus(\mathbf{7},\mathbf{1})^{\oplus 5} \oplus   (\mathbf{1},\mathbf{7})^{\oplus 5}\oplus(\mathbf{1},\mathbf{1})^{\oplus 3}\\
& \text{A}_1^{35}\oplus \text{G}_2^{1}  : & & \chi=(\mathbf{9},\mathbf{7}) \oplus   (\mathbf{5},\mathbf{7}) \oplus(\mathbf{11},\mathbf{1}) 
\nonumber \\
& & & \quad \quad 
\oplus 
(\mathbf{7},\mathbf{1})  \oplus(\mathbf{10},\mathbf{1})^{\oplus 2}\oplus(\mathbf{6},\mathbf{7})^{\oplus 2}\oplus(\mathbf{4},\mathbf{1})^{\oplus 2} \oplus   (\mathbf{1},\mathbf{1})^{\oplus 3}
\\
&  \text{A}_1^{36}\oplus \text{G}_2^{1} : & &  \chi= (\mathbf{9},\mathbf{7}) \oplus(\mathbf{7},\mathbf{7}) \oplus(\mathbf{5},\mathbf{7})^{\oplus 2}\oplus(\mathbf{11},\mathbf{1})^{\oplus 
2}\oplus(\mathbf{9},\mathbf{1}) \oplus(\mathbf{7},\mathbf{1}) \oplus(\mathbf{5},\mathbf{1}) \oplus(\mathbf{3},\mathbf{1})^{\oplus 2}
\\
& 
 \text{A}_1^{60}\oplus \text{G}_2^{1}  : & & \chi=(\mathbf{11},\mathbf{7}) \oplus(\mathbf{9},\mathbf{7}) \oplus(\mathbf{15},\mathbf{1}) \oplus(\mathbf{11},\mathbf{1})^{\oplus 2}\oplus(\mathbf{7},\mathbf{1}) \oplus(\mathbf{5},\mathbf{7}) \oplus(\mathbf{5},\mathbf{1})  \oplus(\mathbf{1},\mathbf{7}) \\
& \text{A}_1^{156}\oplus \text{G}_2^{1}  : & & \chi=(\mathbf{17},\mathbf{7}) \oplus(\mathbf{9},\mathbf{7}) \oplus(\mathbf{23},\mathbf{1}) \oplus(\mathbf{15},\mathbf{1})  \oplus(\mathbf{11},\mathbf{1}) , \\
&   \text{A}_1^{1}\oplus \text{G}_2^{2}  : & &    \chi=(\mathbf{1},\mathbf{27})^{\oplus 3}\oplus(\mathbf{2},\mathbf{14})^{\oplus 2}\oplus(\mathbf{2},\mathbf{7})^{\oplus 4} \oplus(\mathbf{1},\mathbf{7})^{\oplus 
5} \oplus   (\mathbf{1},\mathbf{1})^{\oplus 3}\\
&  \text{A}_1^{7}\oplus \text{G}_2^{2}  : & &  \chi= (\mathbf{3},\mathbf{27}) \oplus(\mathbf{5},\mathbf{7}) \oplus(\mathbf{4},\mathbf{7})^{\oplus 2}\oplus(\mathbf{2},\mathbf{14})^{\oplus 2}   \oplus(\mathbf{1},\mathbf{1})^{\oplus 3}\\
& \text{A}_1^{8}\oplus \text{G}_2^{2}  : & & \chi=(\mathbf{3},\mathbf{27}) \oplus(\mathbf{5},\mathbf{7})^{\oplus 2}\oplus(\mathbf{3},\mathbf{14}) \oplus(\mathbf{3},\mathbf{7}) \oplus(\mathbf{1},\mathbf{14})\oplus(\mathbf{3},\mathbf{1})\\
&  \text{A}_1^{1}\oplus \text{G}_2^{3}  : & &  \chi=(\mathbf{1},\mathbf{64}) \oplus(\mathbf{1},\mathbf{27})^{\oplus 2}\oplus(\mathbf{2},\mathbf{27})^{\oplus 2}\oplus(\mathbf{2},\mathbf{1})^{\oplus 2}\oplus(\mathbf{1},\mathbf{1}) \\\
&  \text{A}_1^{4}\oplus \text{G}_2^{3}  : & &  \chi=(\mathbf{1},\mathbf{64}) \oplus(\mathbf{3},\mathbf{27})^{\oplus 2} \oplus(\mathbf{5},\mathbf{1}).
  \end{align}

\begin{figure}[H]
\scalebox{1}{
\ \begin{tikzpicture}
\node at (.8*3.5,-.8*2) { $\mathfrak{su}_2\oplus\mathfrak{g}_2$ };
\draw[thick=4mm] plot [smooth, tension=1] coordinates {(0,.8*1.5) (.8*1.2,0) (0,-.8*1.5)};
\draw[thick=4mm] plot [smooth, tension=1] coordinates {(.8*2.5,.8*1.5) (.8*1.2,0) (.8*2.5,-.8*1.5)};
\node[draw=none] at (.8*3.5,0){$+$};
\node[draw,circle,thick,scale=1.25,fill=black,label=below:{1}] (1) at (3.2+.8,0){};
\node[draw,circle,thick,scale=1.25,label=below:{2}] (2) at (3.2+.8*2,0){};
\node[draw,circle,thick,scale=1.25] (3) at (3.2+.8*3.2,-1){};
\node[draw,circle,thick,scale=1.25] (4) at (3.2+.8*3.2,0){};
\node[draw,circle,thick,scale=1.25] (5) at (3.2+.8*3.2,1){};
\node at (3.2+.8*3.15,0)[draw,dashed, line width=2pt, ellipse, minimum width=90pt, minimum height=24pt,rotate=90,yshift=-1pt]{};
\node at (3.2+.8*4,.5) {1};
\draw[thick] (1)--(2)--(3);
\draw[thick] (2)--(4);
\draw[thick] (2)--(5);
\draw[->,>=stealth',thick=4mm]  (7.2,0)--(9.2,0);
\node[draw,circle,thick,scale=1.25,fill=black,label=below:{1},xshift=2cm] (0a) at (8,0){};
\node[draw,circle,thick,scale=1.25,label=below:{2},xshift=2cm] (1a) at (8+.8,0){};
\node[draw,circle,thick,scale=1.25,label=below:{3},xshift=2cm] (2a) at (8+.8*2,0){};
\node[draw,circle,thick,scale=1.25,label=below:{4},xshift=2cm] (3a) at (8+.8*3,0){};
\node[draw,circle,thick,scale=1.25,label=left:{2},xshift=2cm] (4b) at (8+.8*3,.8){};
\node[draw,circle,thick,scale=1.25,label=below:{3},xshift=2cm] (4a) at (8+.8*4,0){};
\node[draw,circle,thick,scale=1.25,label=below:{2},xshift=2cm] (5a) at (8+.8*5,0){};
\node[draw,circle,thick,scale=1.25,label=below:{1},xshift=2cm] (6a) at (8+.8*6,0){};
\draw[thick] (0a)--(1a)--(2a)--(3a)--(4a)--(5a)--(6a);
\draw[thick] (3a)--(4b);
\node at (10+.8*3.6,-.8*1.6) { $\mathfrak{e}_7$ };
\end{tikzpicture}}

\scalebox{1}{
\ \begin{tikzpicture}
\node at (.8*3.5,-.8*2) { $\mathfrak{su}_2\oplus\mathfrak{g}_2$ };
\draw[thick=4mm] plot [smooth, tension=1] coordinates {(0,.8*1.5) (.8*1.2,0) (0,-.8*1.5)};
\draw[thick=4mm] plot [smooth, tension=1] coordinates {(.8*2.5,.8*1.5) (.8*1.2,0) (.8*2.5,-.8*1.5)};
\node[draw=none] at (.8*3.5,0){$+$};
\node[draw,circle,thick,scale=1.25,fill=black,label=below:{1}] (1) at (3.2+.8,0){};
\node[draw,circle,thick,scale=1.25,label=below:{2}] (2) at (3.2+.8*2,0){};
\node[draw,circle,thick,scale=1.25] (3) at (3.2+.8*3.2,-1){};
\node[draw,circle,thick,scale=1.25] (4) at (3.2+.8*3.2,0){};
\node[draw,circle,thick,scale=1.25] (5) at (3.2+.8*3.2,1){};
\node at (3.2+.8*3.15,0)[draw,dashed, line width=2pt, ellipse, minimum width=90pt, minimum height=24pt,rotate=90,yshift=-1pt]{};
\node at (3.2+.8*4,.5) {1};
\draw[thick] (1)--(2)--(3);
\draw[thick] (2)--(4);
\draw[thick] (2)--(5);
\draw[->,>=stealth',thick=4mm]  (7.2,0)--(9.2,0);
\node[draw,circle,thick,scale=1.25,fill=black,label=below:{1},xshift=1.8cm] (0a) at (8,0){};
\node[draw,circle,thick,scale=1.25,label=below:{2},xshift=1.8cm] (1a) at (8+.8,0){};
\node[draw,circle,thick,scale=1.25,label=below:{3},xshift=1.8cm] (2a) at (8+.8*2,0){};
\node[draw,circle,thick,scale=1.25,label=below:{4},xshift=1.8cm] (3a) at (8+.8*3,0){};
\node[draw,circle,thick,scale=1.25,label=below:{5},xshift=1.8cm] (4a) at (8+.8*4,0){};
\node[draw,circle,thick,scale=1.25,label=left:{3},xshift=1.8cm] (4b) at (8+.8*5,.8){};
\node[draw,circle,thick,scale=1.25,label=below:{6},xshift=1.8cm] (5a) at (8+.8*5,0){};
\node[draw,circle,thick,scale=1.25,label=below:{4},xshift=1.8cm] (6a) at (8+.8*6,0){};
\node[draw,circle,thick,scale=1.25,label=below:{2},xshift=1.8cm] (7a) at (8+.8*7,0){};
\draw[thick] (0a)--(1a)--(2a)--(3a)--(4a)--(5a)--(6a)--(7a);
\draw[thick] (5a)--(4b);
\node at (10+.8*3.6,-.8*1.6) { $\mathfrak{e}_8$ };
\end{tikzpicture}}
\caption{The expected gauge group enhancement of \sug .}
\label{fig:su2g2enhancements}
\end{figure}
\vspace{-5mm}
\begin{figure}[H]
\begin{center}
\begin{tikzpicture}
\node at (.8*3.5,-.8*2) { $\mathfrak{su}_2\oplus\mathfrak{g}_2$ };
\node at (10+.8*3.2,-.8*3.7) {Incomplete $\mathfrak{e}_{7}$ };
\draw[thick=4mm] plot [smooth, tension=1] coordinates {(0,.8*1.5) (.8*1.2,0) (0,-.8*1.5)};
\draw[thick=4mm] plot [smooth, tension=1] coordinates {(.8*2.5,.8*1.5) (.8*1.2,0) (.8*2.5,-.8*1.5)};
\node[draw=none] at (.8*3.5,0){$+$};
\node[draw,circle,thick,scale=1.25,fill=black,label=below:{1}] (1) at (3.2+.8,0){};
\node[draw,circle,thick,scale=1.25,label=below:{2}] (2) at (3.2+.8*2,0){};
\node[draw,circle,thick,scale=1.25] (3) at (3.2+.8*3.2,-1){};
\node[draw,circle,thick,scale=1.25] (4) at (3.2+.8*3.2,0){};
\node[draw,circle,thick,scale=1.25] (5) at (3.2+.8*3.2,1){};
\node at (3.2+.8*3.15,0)[draw,dashed, line width=2pt, ellipse, minimum width=90pt, minimum height=24pt,rotate=90,yshift=-1pt]{};
\node at (3.2+.8*4,.5) {1};
\draw[thick] (1)--(2)--(3);
\draw[thick] (2)--(4);
\draw[thick] (2)--(5);
\draw[->,>=stealth',thick=4mm]  (7.2,0)--(9.2,0);
\node[draw,circle,thick,scale=1.25,fill=black,yshift=-1.7cm] (B1) at (10+.8,0) {};
\node[draw,circle,thick,scale=1.25,label=above:2,yshift=-1.7cm] (B2) at (10+.8*2.2,0) {};
\node[draw,circle,thick,scale=1.25,label=above:2,yshift=-1.7cm] (B3b) at (10+.8*3.4,.8) {};
\node[draw=none,scale=1.25,yshift=-1.7cm,fill=black,inner sep=0,outer sep=0] (B3) at (10+.8*3.4,0) {};
\node[draw,circle,thick,scale=1.25,label=above:2,yshift=-1.7cm] (B4) at (10+.8*4.6,0) {};
\node[draw,circle,thick,scale=1.25,yshift=-1.7cm] (B5) at (10+.8*5.8,0) {};
\draw[thick] (B1)--(B2)--(B3)--(B4)--(B5);
\draw[thick] (B3)--(B3b);
\node[draw,circle,thick,scale=1.25,fill=black,yshift=-.2cm] (C1) at (10+.8,0) {};
\node[draw,circle,thick,scale=1.25,label=above:2,yshift=-.2cm] (C2) at (10+.8*2.2,0) {};
\node[draw,circle,thick,scale=1.25,label=above:3,yshift=-.2cm] (C3) at (10+.8*3.4,0) {};
\node[draw,circle,thick,scale=1.25,label=above:2,yshift=-.2cm] (C4) at (10+.8*4.6,0) {};
\node[draw,circle,thick,scale=1.25,yshift=-.2cm] (C5) at (10+.8*5.8,0) {};
\draw[thick] (C1)--(C2)--(C3)--(C4)--(C5);
\node[draw,circle,thick,scale=1.25,fill=black,yshift=.6cm] (D1) at (10+.8,0) {};
\node[draw,circle,thick,scale=1.25,label=above:2,yshift=.6cm] (D2) at (10+.8*2.2,0) {};
\node[draw,circle,thick,scale=1.25,label=above:3,yshift=.6cm] (D3) at (10+.8*3.4,0) {};
\node[draw=none,scale=1.25,yshift=.6cm,fill=black,inner sep=0,outer sep=0] (D4) at (10+.8*4.6,0) {};
\node[draw,circle,thick,scale=1.25,yshift=.6cm] (D4b) at (10+.8*4.6,.8) {};
\node[draw,circle,thick,scale=1.25,label=above:2,yshift=.6cm] (D5) at (10+.8*5.8,0) {};
\draw[thick] (D1)--(D2)--(D3)--(D4)--(D5);
\draw[thick] (D4)--(D4b);
\node[draw,circle,thick,scale=1.25,fill=black,yshift=1.7cm] (E1) at (10+.8,0) {};
\node[draw=none,scale=1.25,yshift=1.7cm,fill=black,inner sep=0,outer sep=0] (E2) at (10+.8*2.2,0) {};
\node[draw,circle,thick,scale=1.25,label=above:2,yshift=1.7cm] (E2b) at (10+.8*2.2,.8) {};
\node[draw,circle,thick,scale=1.25,label=above:3,yshift=1.7cm] (E3) at (10+.8*3.4,0) {};
\node[draw,circle,thick,scale=1.25,label=above:2,yshift=1.7cm] (E4) at (10+.8*4.6,0) {};
\node[draw,circle,thick,scale=1.25,yshift=1.7cm] (E5) at (10+.8*5.8,0) {};
\draw[thick] (E1)--(E2)--(E3)--(E4)--(E5);
\draw[thick] (E2)--(E2b);
\end{tikzpicture}
%\vspace{3mm}
\begin{tikzpicture}
\node at (.8*3.5,-.8*2) { $\mathfrak{su}_2\oplus\mathfrak{g}_2$ };
\node at (10+.8*3.2,-.8*3.7) {Incomplete $\mathfrak{e}_{8}$ };
\draw[thick=4mm] plot [smooth, tension=1] coordinates {(0,.8*1.5) (.8*1.2,0) (0,-.8*1.5)};
\draw[thick=4mm] plot [smooth, tension=1] coordinates {(.8*2.5,.8*1.5) (.8*1.2,0) (.8*2.5,-.8*1.5)};
\node[draw=none] at (.8*3.5,0){$+$};
\node[draw,circle,thick,scale=1.25,fill=black,label=below:{1}] (1) at (3.2+.8,0){};
\node[draw,circle,thick,scale=1.25,label=below:{2}] (2) at (3.2+.8*2,0){};
\node[draw,circle,thick,scale=1.25] (3) at (3.2+.8*3.2,-1){};
\node[draw,circle,thick,scale=1.25] (4) at (3.2+.8*3.2,0){};
\node[draw,circle,thick,scale=1.25] (5) at (3.2+.8*3.2,1){};
\node at (3.2+.8*3.15,0)[draw,dashed, line width=2pt, ellipse, minimum width=90pt, minimum height=24pt,rotate=90,yshift=-1pt]{};
\node at (3.2+.8*4,.5) {1};
\draw[thick] (1)--(2)--(3);
\draw[thick] (2)--(4);
\draw[thick] (2)--(5);
\draw[->,>=stealth',thick=4mm]  (7.2,0)--(9.2,0);
\node[draw,circle,thick,scale=1.25,fill=black,yshift=-1.7cm] (B1) at (10+.8,0) {};
\node[draw,circle,thick,scale=1.25,label=above:2,yshift=-1.7cm] (B2) at (10+.8*2.2,0) {};
\node[draw,circle,thick,scale=1.25,label=above:2,yshift=-1.7cm] (B3b) at (10+.8*3.4,.8) {};
\node[draw=none,scale=1.25,yshift=-1.7cm,fill=black,inner sep=0,outer sep=0] (B3) at (10+.8*3.4,0) {};
\node[draw,circle,thick,scale=1.25,label=above:3,yshift=-1.7cm] (B4) at (10+.8*4.6,0) {};
\draw[thick] (B1)--(B2)--(B3)--(B4);
\draw[thick] (B3)--(B3b);
\node[draw,circle,thick,scale=1.25,fill=black,yshift=-.2cm] (C1) at (10+.8,0) {};
\node[draw,circle,thick,scale=1.25,label=above:2,yshift=-.2cm] (C2) at (10+.8*2.2,0) {};
\node[draw,circle,thick,scale=1.25,label=above:3,yshift=-.2cm] (C3) at (10+.8*3.4,0) {};
\node[draw,circle,thick,scale=1.25,label=above:3,yshift=-.2cm] (C4) at (10+.8*4.6,0) {};
\draw[thick] (C1)--(C2)--(C3)--(C4);
\node[draw,circle,thick,scale=1.25,fill=black,yshift=.6cm] (D1) at (10+.8,0) {};
\node[draw,circle,thick,scale=1.25,label=above:2,yshift=.6cm] (D2) at (10+.8*2.2,0) {};
\node[draw,circle,thick,scale=1.25,label=above:4,yshift=.6cm] (D3) at (10+.8*3.4,0) {};
\node[draw,circle,thick,scale=1.25,label=above:2,yshift=.6cm] (D4) at (10+.8*4.6,0) {};
\draw[thick] (D1)--(D2)--(D3)--(D4);
\node[draw,circle,thick,scale=1.25,fill=black,yshift=1.4cm] (E1) at (10+.8,0) {};
\node[draw=none,scale=1.25,yshift=1.4cm,fill=black,inner sep=0,outer sep=0] (E2) at (10+.8*2.2,0) {};
\node[draw,circle,thick,scale=1.25,label=above:3,yshift=1.4cm] (E2b) at (10+.8*2.2,.8) {};
\node[draw,circle,thick,scale=1.25,label=above:4,yshift=1.4cm] (E3) at (10+.8*3.4,0) {};
\node[draw,circle,thick,scale=1.25,label=above:2,yshift=1.4cm] (E4) at (10+.8*4.6,0) {};
\node[draw,circle,thick,scale=1.25,yshift=1.4cm] (E5) at (10+.8*5.8,0) {};
\draw[thick] (E1)--(E2)--(E3)--(E4)--(E5);
\draw[thick] (E2)--(E2b);
\end{tikzpicture}
\vspace{-4mm}
\end{center}
\caption{Degeneration of the \sug\ fibers at the collision III+I$_0^{\text{ns}}$.}
\label{fig:su2g2enhancementG}
\end{figure}
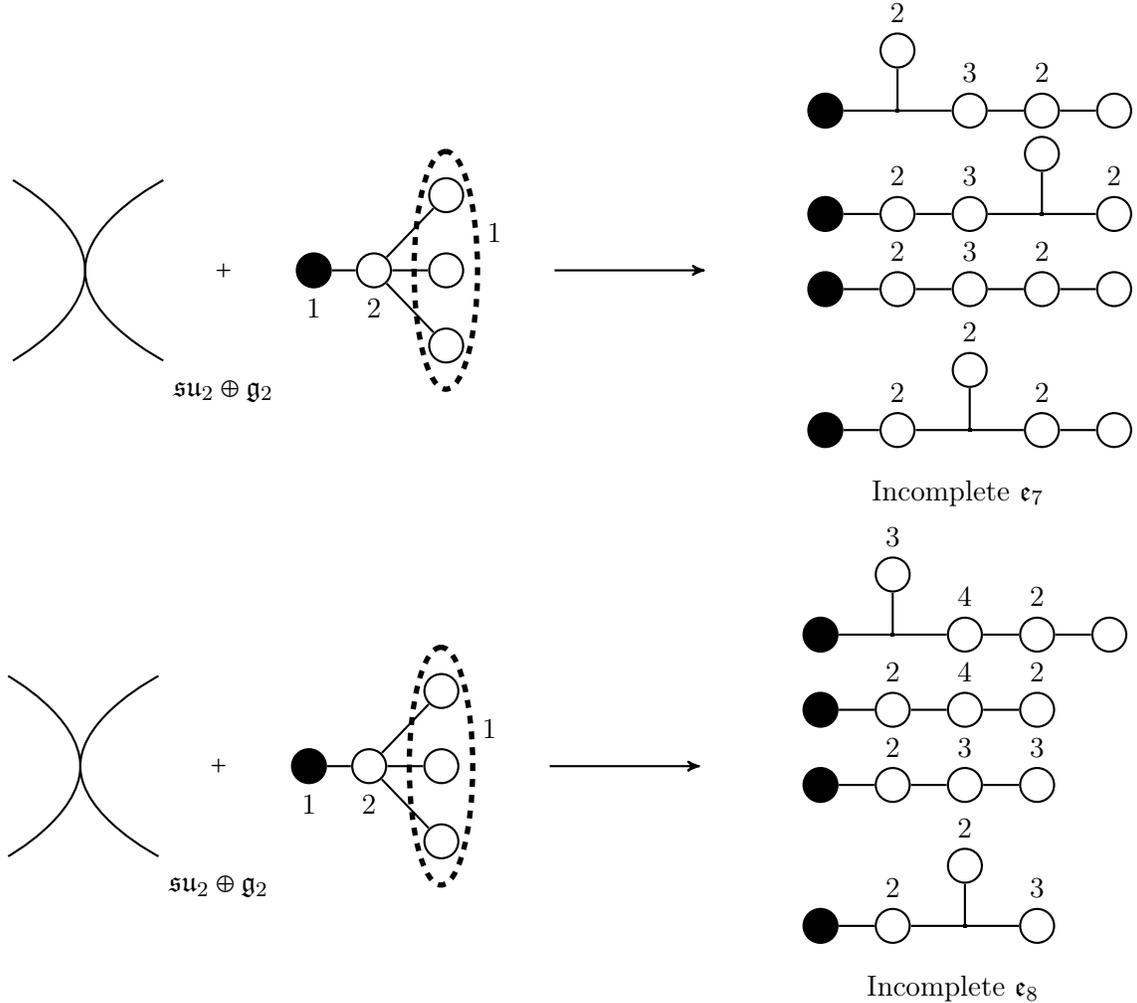

\section{$\text{A}_1\oplus \text{B}_3$: the SU($2$)$\times$Spin($7$)-model}\label{sec:A1B3}

\subsection{$\text{A}_1\oplus \text{B}_3 \to\text{E}_7$}

There are seven subalgebras of type $\text{A}_1\oplus \text{B}_3$ in E$_7$ including two with Dynking embedding index one. 
The characteristic representations for the two subalgebras of Dynkin embedding index one are: 
\begin{align}
&\text{A}_1^1\oplus \text{B}^1_3\to \text{E}_7: \quad 
\chi= (\mathbf{2}, \mathbf{8})^{\oplus 4} \oplus (\mathbf{1}, \mathbf{7})^{\oplus 5} \oplus (\mathbf{1}, \mathbf{1})^{\oplus 10}, \\
&\text{A}_1^1\oplus \text{B}^1_3\to \text{E}_7: \quad 
\chi= (\mathbf{2}, \mathbf{8})^{\oplus 2} \oplus (\mathbf{2}, \mathbf{7})^{\oplus 2}  \oplus  (\mathbf{1}, \mathbf{7}) \oplus (\mathbf{1}, \mathbf{1})^{\oplus 6}
 \oplus (\mathbf{1}, \mathbf{8})^{\oplus 4} \oplus (\mathbf{2}, \mathbf{1})^{\oplus 2}. 
\end{align}
We can understand the difference between these two subalgebras by the fact that they correspond to distinct groups sharing the same Lie algebra.  
The characteristic representations for the subalgebras of higher Dynkin embedding index are: 
\begin{align}
&\text{A}_1^2\ \oplus \text{B}^1_3\to \text{E}_7: \quad
\chi= (\mathbf{3}, \mathbf{8})  \oplus (\mathbf{2}, \mathbf{7})^{\oplus 2} \oplus (\mathbf{2}, \mathbf{8})^{\oplus 2} \oplus (\mathbf{1}, \mathbf{8})\oplus (\mathbf{1}, \mathbf{7}) \oplus (\mathbf{3}, \mathbf{1}) \oplus (\mathbf{2}, \mathbf{1})^{\oplus 2} \oplus (\mathbf{1}, \mathbf{1})^{\oplus 3} 
, \nonumber \\
&\text{A}_1^2\ \oplus \text{B}^1_3\to \text{E}_7: \quad 
\chi= (\mathbf{2}, \mathbf{8})^{\oplus 4}  \oplus (\mathbf{3}, \mathbf{7}) \oplus (\mathbf{1}, \mathbf{7})^{\oplus 2} \oplus (\mathbf{3}, \mathbf{1})^{\oplus 2}\oplus (\mathbf{1}, \mathbf{1})^{\oplus 4},\nonumber \\
&\text{A}_1^3\ \oplus \text{B}^1_3\to \text{E}_7: \quad 
\chi= (\mathbf{3}, \mathbf{8})^{\oplus 2}  \oplus (\mathbf{3}, \mathbf{7}) \oplus (\mathbf{1}, \mathbf{8})^{\oplus 2} \oplus (\mathbf{1}, \mathbf{7})^{\oplus 2}\oplus (\mathbf{3}, \mathbf{1})^{\oplus 3} \oplus (\mathbf{1}, \mathbf{1})
,  \\
&\text{A}_1^{10}\oplus \text{B}^1_3\to \text{E}_7: \quad 
\chi= (\mathbf{4}, \mathbf{8})^{\oplus 2}  \oplus (\mathbf{5}, \mathbf{7}) \oplus (\mathbf{7}, \mathbf{1}) \oplus (\mathbf{1}, \mathbf{1})^{\oplus 3},\nonumber\\
&\text{A}_1^{11}\oplus \text{B}^1_3\to \text{E}_7: \quad 
\chi= (\mathbf{5}, \mathbf{8})  \oplus (\mathbf{3}, \mathbf{8}) \oplus (\mathbf{5}, \mathbf{7}) \oplus (\mathbf{3}, \mathbf{1}).\nonumber
\end{align}

\subsection{$\text{A}_1\oplus \text{B}_3 \to\text{E}_8$}

\begin{align}
& \text{A}_1^1\ \oplus \text{B}_3^1: \quad \chi=
 (\mathbf{2},\mathbf{7})^{\oplus 2}\oplus (\mathbf{1},\mathbf{7})^{\oplus 
5}\oplus (\mathbf{2},\mathbf{1})^{\oplus 10}\oplus (\mathbf{1},\mathbf{1})^{\oplus 13}\oplus (\mathbf{1},\mathbf{8})^{\oplus 8}\oplus (\mathbf{2},\mathbf{
8})^{\oplus 4},\\
&  \text{A}_1^2\ \oplus \text{B}_3^1: \quad\chi= (\mathbf{2},\mathbf{7})^{\oplus 4}\oplus (\mathbf{3},\mathbf{1})^{\oplus 
5}\oplus (\mathbf{1},\mathbf{7})\oplus (\mathbf{1},\mathbf{1})^{\oplus 10}\oplus (\mathbf{2},\mathbf{1})^{\oplus 4}\oplus (\mathbf{2},\mathbf{8})^{\oplus 
4}\oplus (\mathbf{3},\mathbf{8})\oplus (\mathbf{1},\mathbf{8})^{\oplus 5}, \\
& \text{A}_1^2\ \oplus \text{B}_3^1: \quad\chi= (\mathbf{3},\mathbf{7})\oplus (\mathbf{1},\mathbf{7})^{\oplus 
6}\oplus (\mathbf{3},\mathbf{1})^{\oplus 6}\oplus (\mathbf{1},\mathbf{1})^{\oplus 15}\oplus (\mathbf{2},\mathbf{8})^{\oplus 8},\\
& \text{A}_1^3\ \oplus \text{B}_3^1: \quad\chi=\oplus (\mathbf{3},\mathbf{7})   \oplus (\mathbf{4},\mathbf{1})^{\oplus 
2}\oplus (\mathbf{2},\mathbf{7})^{\oplus 2}\oplus (\mathbf{2},\mathbf{1})^{\oplus 6}\oplus (\mathbf{3},\mathbf{1})^{\oplus 3}\oplus (\mathbf{1},\mathbf{7})^{\oplus 
2}\oplus (\mathbf{1},\mathbf{1})^{\oplus 4} \nonumber\\
& \quad\quad \quad\quad\quad\quad\quad \oplus (\mathbf{2},\mathbf{8})^{\oplus 4}\oplus (\mathbf{3},\mathbf{8})^{\oplus 2}\oplus (\mathbf{1},\mathbf{8})^{\oplus 2}, \\
& \text{A}_1^4\ \oplus \text{B}_3^1: \quad\chi= (\mathbf{3},\mathbf{7})^{\oplus 2}\oplus (\mathbf{5},\mathbf{1})\oplus (\mathbf{1},\mathbf{7})^{\oplus 3}\oplus (\mathbf{3},\mathbf{1})^{\oplus 8}\oplus (\mathbf{1},\mathbf{1})^{\oplus 4}\oplus (\mathbf{3},\mathbf{8})^{\oplus 
4}\oplus (\mathbf{1},\mathbf{8})^{\oplus 4}, \\
& \text{A}_1^6\ \oplus \text{B}_3^1: \quad\chi= (\mathbf{3},\mathbf{7})^{\oplus 3}\oplus (\mathbf{5},\mathbf{1})^{\oplus 
3}\oplus (\mathbf{3},\mathbf{1})^{\oplus 5}\oplus (\mathbf{1},\mathbf{1})^{\oplus 3}\oplus (\mathbf{4},\mathbf{8})^{\oplus 2}\oplus (\mathbf{2},\mathbf{8})^{\oplus 4}\oplus \\
& \text{A}_1^{10}\oplus \text{B}_3^1: \quad\chi= (\mathbf{5},\mathbf{7})   \oplus (\mathbf{7},\mathbf{1})\oplus (\mathbf{1},\mathbf{7})^{\oplus 4}\oplus (\mathbf{5},\mathbf{1})^{\oplus 4}\oplus (\mathbf{1},\mathbf{1})^{\oplus 
6}\oplus (\mathbf{4},\mathbf{8})^{\oplus 4}, \\
& \text{A}_1^{10}\oplus \text{B}_3^1: \quad\chi= (\mathbf{4},\mathbf{7})^{\oplus 2}\oplus (\mathbf{7},\mathbf{1})\oplus (\mathbf{5},\mathbf{1})^{\oplus 3}\oplus (\mathbf{1},\mathbf{7})   \oplus (\mathbf{1},\mathbf{1})^{\oplus 
3}\oplus (\mathbf{4},\mathbf{1})^{\oplus 2}\oplus (\mathbf{4},\mathbf{8})^{\oplus 2}\nonumber\\
& \quad\quad \quad\quad\quad\quad\quad \oplus (\mathbf{5},\mathbf{8})   \oplus (\mathbf{1},\mathbf{8})^{\oplus 3}, \\
& \text{A}_1^{11}\oplus \text{B}_3^1: \quad\chi= (\mathbf{5},\mathbf{7})   \oplus (\mathbf{6},\mathbf{1})^{\oplus 
2}\oplus (\mathbf{7},\mathbf{1})   \oplus (\mathbf{3},\mathbf{1})\oplus (\mathbf{4},\mathbf{1})^{\oplus 2}\oplus (\mathbf{2},\mathbf{7})^{\oplus 2}\oplus (\mathbf{1},\mathbf{1})^{\oplus 3}\oplus (\mathbf{4},\mathbf{8})^{\oplus 2}\nonumber\\
& \quad\quad \quad\quad\quad\quad\quad \oplus (\mathbf{5},\mathbf{8})   \oplus (\mathbf{3},\mathbf{8})   , \\
& \text{A}_1^{12}\oplus \text{B}_3^1: \quad\chi= (\mathbf{5},\mathbf{7})   \oplus (\mathbf{7},\mathbf{1})^{\oplus 
2}\oplus (\mathbf{3},\mathbf{7})   \oplus (\mathbf{5},\mathbf{1})^{\oplus 2}\oplus (\mathbf{3},\mathbf{1})^{\oplus 3}\oplus (\mathbf{1},\mathbf{7})\oplus (\mathbf{5},\mathbf{8})^{\oplus 2}\oplus (\mathbf{3},\mathbf{8})^{\oplus 2},\\
& \text{A}_1^{28}\oplus \text{B}_3^1: \quad\chi= (\mathbf{7},\mathbf{7})   \oplus (\mathbf{11},\mathbf{1})\oplus (\mathbf{7},\mathbf{1})^{\oplus 3}\oplus (\mathbf{1},\mathbf{7})^{\oplus 2}\oplus (\mathbf{1},\mathbf{1})\oplus (\mathbf{7},\mathbf{8})^{\oplus 2}\oplus (\mathbf{1},\mathbf{8})^{\oplus 2}, \\
& \text{A}_1^{60}\oplus \text{B}_3^1: \quad\chi= (\mathbf{9},\mathbf{7})   \oplus (\mathbf{15},\mathbf{1})\oplus (\mathbf{11},\mathbf{1})   \oplus (\mathbf{7},\mathbf{1})   \oplus (\mathbf{11},\mathbf{8})\oplus (\mathbf{5},\mathbf{8})   , \\
& \text{A}_1^1\ \oplus \text{B}_3^2: \quad\chi=(\mathbf{1},\mathbf{35})^{\oplus 2}\oplus  (\mathbf{1},\mathbf{27})\oplus  (\mathbf{1},\mathbf{7})^{\oplus 2}\oplus (\mathbf{1},\mathbf{1})   \oplus (\mathbf{2},\mathbf{21})^{\oplus 
2}\oplus (\mathbf{2},\mathbf{7})^{\oplus 2}.
\end{align}

\section{$\text{A}_1\oplus \text{D}_4$: the SU($2$)$\times$Spin($8$)-model}  \label{sec:A1D4}

\subsection{$\text{A}_1\oplus \text{D}_4 \to \text{E}_7$}
Up to linear equivalence, there are three subalgebras of type $\text{A}_1\oplus \text{D}_4$ in E$_7$ with only one having Dynkin index $(1,1)$. 
As expected from our proposal, $ \text{A}_1^1\oplus \text{D}_4^1\to \text{B}_6$ produces the matter we expect to see in F-theory. 
\begin{align}
& \text{A}_1^1\oplus \text{D}_4^1\to \text{B}_6: \quad \chi=
 (\mathbf{2},\mathbf{8}_{\text{c}})^{\oplus 2}\oplus
   (\mathbf{2},\mathbf{8}_{\text{v}})^{\oplus 2}\oplus 
  (\mathbf{1},\mathbf{8}_{\text{s}})^{\oplus 4}\oplus 
  (\mathbf{1},\mathbf{8}_{\text{v}})\oplus
   (\mathbf{1},\mathbf{1})^{\oplus 6},\\
& \text{A}_1^2\oplus \text{D}_4^1\to \text{B}_6: \quad \chi=
 (\mathbf{2},\mathbf{8}_{\text{c}})^{\oplus 2}\oplus
 (\mathbf{2},\mathbf{8}_{\text{s}})^{\oplus 2}\oplus
  (\mathbf{3},\mathbf{1})\oplus
   (\mathbf{3},\mathbf{8}_{\text{v}})\oplus 
 (\mathbf{1},\mathbf{8}_{\text{v}})\oplus
   (\mathbf{1},\mathbf{1})^{\oplus 3},\\ 
& \text{A}_1^3\oplus \text{D}_4^1\to \text{B}_6: \quad \chi= 
(\mathbf{2},\mathbf{1})^{\oplus 2}\oplus
  (\mathbf{3},\mathbf{8}_{\text{v}})\oplus   
 (\mathbf{3},\mathbf{8}_{\text{c}})\oplus 
  (\mathbf{3},\mathbf{8}_{\text{s}})\oplus 
 (\mathbf{1},\mathbf{8}_{\text{c}})\oplus
 (\mathbf{1},\mathbf{8}_{\text{s}}).
\end{align}

\subsection{$\text{A}_1\oplus \text{D}_4 \to \text{E}_8$}

There are 13 distinct subalgebras of type $\text{A}_1\oplus \text{B}_3$ in E$_7$ up to linear equivalence. 
Only one of them has Dynkin index one along all A$_1$ and B$_3$.  
\begin{align}
\text{A}_1^1\ \oplus \text{D}^1_4\to \text{E}_8:& \quad \chi=(\mathbf{2}, \mathbf{8}_{\text{s}})^{\oplus 2} \oplus (\mathbf{1}, \mathbf{8}_{\text{s}})^{\oplus 4} \oplus (\mathbf{2}, \mathbf{1})^{\oplus 8} \oplus (\mathbf{1}, \mathbf{1})^{\oplus 9} \oplus (\mathbf{2}, \mathbf{8}_{\text{v}})^{\oplus 2} \oplus (\mathbf{1}, \mathbf{8}_{\text{v}})^{\oplus 4} \nonumber \\
&\qquad\quad \oplus (\mathbf{2}, \mathbf{8}_{\text{c}})^{\oplus 2} \oplus (\mathbf{1}, \mathbf{8}_{\text{c}})^{\oplus 4}, \\
\text{A}_1^2\ \oplus \text{D}^1_4\to \text{E}_8:& \quad \chi=(\mathbf{2}, \mathbf{8}_{\text{s}})^{\oplus 4} \oplus (\mathbf{3}, \mathbf{1})^{\oplus 5} \oplus (\mathbf{1}, \mathbf{1})^{\oplus 10} \oplus (\mathbf{2}, \mathbf{8}_{\text{v}})^{\oplus 4} \oplus (\mathbf{3}, \mathbf{8}_{\text{c}}) \oplus (\mathbf{1}, \mathbf{8}_{\text{c}})^{\oplus 5}, \\
\text{A}_1^3\ \oplus \text{D}^1_4\to \text{E}_8:& \quad \chi=(\mathbf{3}, \mathbf{8}_{\text{s}}) \oplus (\mathbf{4}, \mathbf{1})^{\oplus 2} \oplus (\mathbf{2}, \mathbf{8}_{\text{s}})^{\oplus 2} \oplus (\mathbf{2}, \mathbf{1})^{\oplus 4} \oplus (\mathbf{3}, \mathbf{1})^{\oplus 2} \oplus (\mathbf{1}, \mathbf{8}_{\text{s}}) \oplus (\mathbf{1}, \mathbf{1})^{\oplus 3} \nonumber\\
&\qquad\quad \oplus (\mathbf{2}, \mathbf{8}_{\text{v}})^{\oplus 2} \oplus (\mathbf{3}, \mathbf{8}_{\text{v}}) \oplus (\mathbf{1}, \mathbf{8}_{\text{v}}) \oplus (\mathbf{2}, \mathbf{8}_{\text{c}})^{\oplus 2} \oplus (\mathbf{3}, \mathbf{8}_{\text{c}}) \oplus (\mathbf{1}, \mathbf{8}_{\text{c}}), \\
\text{A}_1^4\ \oplus \text{D}^1_4\to \text{E}_8:& \quad \chi=(\mathbf{3}, \mathbf{8}_{\text{s}})^{\oplus 2} \oplus (\mathbf{5}, \mathbf{1}) \oplus (\mathbf{1}, \mathbf{8}_{\text{s}})^{\oplus 2} \oplus (\mathbf{3}, \mathbf{1})^{\oplus 6} \oplus (\mathbf{1}, \mathbf{1})^{\oplus 2} \oplus (\mathbf{3}, \mathbf{8}_{\text{v}})^{\oplus 2} \nonumber\\
&\qquad\quad \oplus (\mathbf{1}, \mathbf{8}_{\text{v}})^{\oplus 2} \oplus (\mathbf{3}, \mathbf{8}_{\text{c}})^{\oplus 2} \oplus (\mathbf{1}, \mathbf{8}_{\text{c}})^{\oplus 2}, \\
\text{A}_1^{10}\ \oplus \text{D}^1_4\to \text{E}_8:& \quad \chi=(\mathbf{5}, \mathbf{8}_{\text{s}}) \oplus (\mathbf{7}, \mathbf{1}) \oplus (\mathbf{1}, \mathbf{8}_{\text{s}})^{\oplus 3} \oplus (\mathbf{5}, \mathbf{1})^{\oplus 3} \oplus (\mathbf{1}, \mathbf{1})^{\oplus 3} \oplus (\mathbf{4}, \mathbf{8}_{\text{v}})^{\oplus 2} \oplus (\mathbf{4}, \mathbf{8}_{\text{c}})^{\oplus 2}, \\
\text{A}_1^{12}\ \oplus \text{D}^1_4\to \text{E}_8:& \quad \chi=(\mathbf{5}, \mathbf{8}_{\text{s}}) \oplus (\mathbf{7}, \mathbf{1})^{\oplus 2} \oplus (\mathbf{3}, \mathbf{8}_{\text{s}}) \oplus (\mathbf{5}, \mathbf{1}) \oplus (\mathbf{3}, \mathbf{1})^{\oplus 2} \oplus (\mathbf{5}, \mathbf{8}_{\text{v}}) \oplus (\mathbf{3}, \mathbf{8}_{\text{v}}) \nonumber\\
&\qquad\quad \oplus (\mathbf{5}, \mathbf{8}_{\text{c}}) \oplus (\mathbf{3}, \mathbf{8}_{\text{c}}), \\
\text{A}_1^{28}\ \oplus \text{D}^1_4\to \text{E}_8:& \quad \chi=(\mathbf{7}, \mathbf{8}_{\text{s}}) \oplus (\mathbf{10}, \mathbf{1}) \oplus (\mathbf{7}, \mathbf{1})^{\oplus 2} \oplus (\mathbf{1}, \mathbf{8}_{\text{s}}) \oplus (\mathbf{7}, \mathbf{8}_{\text{v}}) \oplus (\mathbf{1}, \mathbf{8}_{\text{v}}) \oplus (\mathbf{7}, \mathbf{8}_{\text{c}}) \oplus (\mathbf{1}, \mathbf{8}_{\text{c}}), \\
\text{A}_1^2\ \oplus \text{D}^1_4\to \text{E}_8:& \quad \chi=(\mathbf{1}, \mathbf{35}_{\text{v}}) \oplus (\mathbf{1}, \mathbf{35}_{\text{c}}) \oplus (\mathbf{1}, \mathbf{35}_{\text{s}}) \oplus (\mathbf{2}, \mathbf{28})^{\oplus 2}.
\end{align}

\section{The art of channeling}

In this paper, we have used the geography of subalgebras of a given simple Lie algbra  \cite{Dynkin.SubA,GEO,deGraaf}. 
This is a luxury we have only for simple Lie algebras up to rank eight  thanks to the classification of de Graaf \cite{deGraaf}. 
In more general cases, all we can do is compute characteristic representations and then Dynkin indices by considering all the ways a given subalgebra can be embedded into another by a sequence of successive maximal Lie subalgebras. 
We call such a sequence a {\em channel of embeddings}.
A channel is comparable to an itinerary as different channels can lead to the same  subalgebra. 
But unlike the approach of Dynkin and de Graaf \cite{Dynkin.SubA,deGraaf,GEO},  channels do not track linear equivalence.
But channels do not  keep track of linear equivalence. However, if two channels give distinct characteristic representations or Dynkin index, they are clearly not linearly equivalent.

\subsection*{$\text{A}_1\to\text{G}_2$}
\label{sec:A1toG2}

\begin{figure}[H]
\begin{center}
\begin{tikzpicture}
\node at (0,0) { $\text{A}_1$ };
\node at (12,0) { $\text{G}_2$ };

\node[draw,circle,thick,scale=1.25,fill=black] (6) at (.8,0){};
\node[draw,circle,thick,scale=1.25] (7) at (.8*3.2,.8){};
\node[draw,circle,thick,scale=1.25] (8) at (.8*3.2,-.8){};
\draw[thick] (6)--(.8*2,0)--(7); 
\draw[thick] (8)--(.8*2,0); 
\node at (.8*3.15,0)[draw,dashed, line width=2pt, ellipse, minimum width=120pt, minimum height=22pt,rotate=90,yshift=-1pt]{};
%\draw[thick=4mm] (2,.5)--(4,-.5);
%\draw[thick=4mm] (2,-.5)--(4,.5);
%\draw[thick=4mm] (3,1.5)--(3,-1.5);
%\node at (3,0)[draw,dashed, line width=2pt, ellipse, minimum width=100pt, minimum height=46pt,yshift=-1pt]{};
\draw[->,>=stealth',thick=4mm]  (5,0) -- (7,0);
\node[draw,circle,thick,scale=1.25,fill=black,label=below:{1}] (1) at (8+.8,0){};
\node[draw,circle,thick,scale=1.25,label=below:{2}] (2) at (8+.8*2,0){};
\node[draw,circle,thick,scale=1.25] (3) at (8+.8*3.2,-1){};
\node[draw,circle,thick,scale=1.25] (4) at (8+.8*3.2,0){};
\node[draw,circle,thick,scale=1.25] (5) at (8+.8*3.2,1){};
\node at (8+.8*3.15,0)[draw,dashed, line width=2pt, ellipse, minimum width=120pt, minimum height=22pt,rotate=90,yshift=-1pt]{};
%\node at (8+.8*3.15,-2.5) {1};
\draw[thick] (1)--(2)--(3);
\draw[thick] (2)--(4);
\draw[thick] (2)--(5);
\end{tikzpicture}
\end{center}
\caption{The gauge enhancement $\text{A}_1\to\text{G}_2$ from IV$^{\text{ns}}\to \text{I}^{*\text{ns}}_0$.}
\label{fig:g2enhancements}
\end{figure}
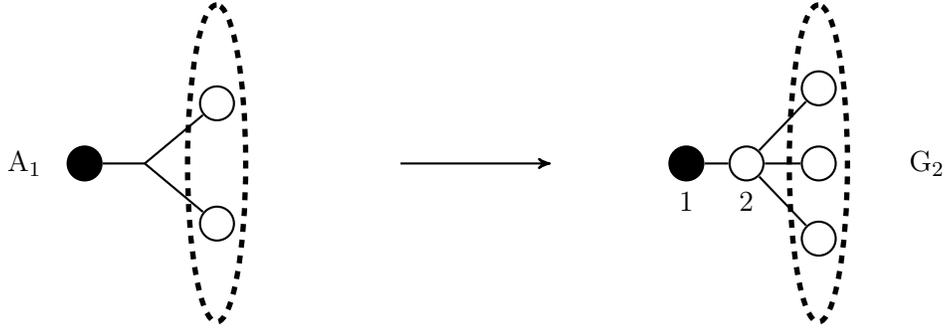

In this section, we consider the enhancement $\text{A}_1\to\text{G}_2$, which is motivated by the geometry of the SU($2$)-model with Kodaira fibers of type IV$^{\text{ns}}$, which is one of the enhancements considered in \cite{SU2SU3} and portrayed in Figure \ref{fig:g2enhancements}.
As expected, there are several non-linearly equivalent subalgebras of type $\text{A}_1$ in $\text{G}_2$. There are six distinct channels to embed $\text{A}_1$ in $\text{G}_2$ by a chain of maximal embeddings and they result in four distinct branching rules as summarized in Table \ref{tb:su2tog2enhancements}. All the embeddings of $\text{A}_1$ factor through one of the three  maximal subalgebras of  $\text{G}_2$ whose branching rules for the adjoint representation are as follows:
\begin{align}
 \label{eq:su2tog2viasu3}
& \begin{tikzcd}[column sep=1.6cm, ampersand replacement=\&] \text{G}_2 \arrow[leftarrow]{r} {\displaystyle (R)} \& \text{A}_2^1 \end{tikzcd}
\quad\quad && \mathbf{14}=\mathbf{3\oplus \overline{3}\oplus 8},\\
 \label{eq:su2tog2viasu2}
& \begin{tikzcd}[column sep=1.6cm, ampersand replacement=\&] \text{G}_2  \arrow[leftarrow]{r} {\displaystyle (R)} \& \text{A}_1^1\oplus\text{A}_1^3 \end{tikzcd}
\quad\quad && \mathbf{14}=\mathbf{(3,1)\oplus (2,4)\oplus (1,3)},\\
 \label{eq:su2tog2maxsu2}
& \begin{tikzcd}[column sep=1.6cm, ampersand replacement=\&] \text{G}_2  \arrow[leftarrow]{r} {\displaystyle (S)} \& \text{A}_1^{28} \end{tikzcd}
\quad\quad && \mathbf{14}=\mathbf{11\oplus 3}.
\end{align}
The maximal embedding  $\text{A}_2\to\text{G}_2$ leads to two distinct channels, each corresponding to an embedding of $\text{A}_1$ with Dynkin index $1$ or $4$ respectively.
\begin{align}
& \begin{tikzcd}[column sep=1cm, ampersand replacement=\&] 
\text{G}_2 \arrow[leftarrow]{r} {(R)}   \&
\text{A}_2^1 \arrow[leftarrow]{r} {(R)} \& 
\text{A}_1^1\oplus \text{U}_1 \arrow[leftarrow]{r} {} \& 
\text{A}_1^1
\end{tikzcd},\label{eq:G2toA1RR}\\
& \begin{tikzcd}[column sep=1cm, ampersand replacement=\&] 
\text{G}_2 \arrow[leftarrow]{r} {(R)}   \&
\text{A}_2^1 \arrow[leftarrow]{r} {(S)} \& 
\text{A}_1^4
\end{tikzcd}.\label{eq:G2toA1RS}
\end{align}
The branching rules corresponding to the first and second channel provide two different decompositions of the adjoint of $\text{G}_2$ via $\text{A}_2$. The decompostion via the first channel (equation \eqref{eq:G2toA1RR}) is given by
\begin{align*}
\mathbf{14} & = \mathbf{3\oplus \overline{3}\oplus 8}\quad  && \text{along} \quad \text{A}_2^1, \\
&= (\mathbf{2})(1)\oplus (\mathbf{1})(-2)\oplus (\mathbf{2})(-1)\oplus (\mathbf{1})(2)\oplus (\mathbf{3})(0)\oplus (\mathbf{2})(3)\oplus (\mathbf{2})(-3)\oplus (\mathbf{1})(0)\quad &&\text{along} \quad  \text{A}_1^1\oplus\text{U}_1 \\
&= \mathbf{3}\oplus \mathbf{2}^{\bigoplus 4}\oplus \mathbf{1}^{\bigoplus 3}\quad &&\text{along} \quad  \text{A}_1^{1},
\end{align*}
where the superindex $\ell=1$ denotes its Dynkin index, whereas the decomposition via the second channel (equation \eqref{eq:G2toA1RS}) is
\begin{align}
\mathbf{14} & = \mathbf{3\oplus \overline{3}\oplus 8}\quad  && \text{along} \quad \text{A}_2^1, \\
&=\mathbf{3}^{\bigoplus 3}\oplus \mathbf{5}\quad &&\text{along} \quad \text{A}_1^4.
\label{eq:su2tog2ch2}
\end{align}
The third, fourth, and fifth channels utilizes the chain of maximal subalgebras via two copies of $\text{A}_1$
\begin{align}
\begin{tikzcd}[column sep=1cm, ampersand replacement=\&] 
\text{G}_2 \arrow[leftarrow]{r} {(R)}   \&
\text{A}_1\oplus\widetilde{\text{A}}_1 \arrow[leftarrow]{r} {(S)} \& 
\text{A}_1,\ \widetilde{\text{A}}_1,\ \text{or}\ (\text{A}_1\oplus\widetilde{\text{A}}_1)_{diag}
\end{tikzcd},
\end{align}
where the three difference choices of $\text{A}_1$ yield different chains of maximal subalgebras.\footnote{For the $\text{A}_1\to\text{G}_2$ embedding, we find them to carry indices as $A_1^1$, $A_1^3$, and $A_1^4$ respectively.} First of all, these two copies of $\text{A}_1$ are non-isomorphic and they yield different decompositions of the adjoint representation of $\text{G}_2$. By choosing the $\text{A}_1$, the decomposition is given by
\begin{align}
\mathbf{14} & = \mathbf{3}\oplus \mathbf{2}^{\bigoplus 4}\oplus \mathbf{1}^{\bigoplus 3}\quad &&\text{along} \quad  \text{A}_1^1,
\end{align}
which is identical to the decomposition observed from the first channel, whereas by picking the $\text{A}_1$ on the right, we get the following decomposition
\begin{align}
\mathbf{14} & = \mathbf{3}\oplus \mathbf{4}^{\bigoplus 2}\oplus \mathbf{1}^{\bigoplus 3}\quad &&\text{along} \quad  \widetilde{\text{A}}_1^3.
\end{align}
Now choose the diagonal comonent residing as $(\text{A}_1\oplus\widetilde{\text{A}}_1)_{diag}$. The decomposition of this channel yields
\begin{align*}
\mathbf{14} & = \mathbf{(3,1)\oplus (2,4)\oplus (1,3)}\quad  && \text{along} \quad \text{A}_1^1\oplus\widetilde{\text{A}}_1^3, \\
&=\mathbf{3}^{\bigoplus 3} \oplus \mathbf{5}\quad  && \text{along} \quad (\text{A}_1\times\widetilde{\text{A}}_1)_{diag}^4,
\end{align*}
which matches with the result from the second channel in equation \eqref{eq:su2tog2ch2}.

All six channels result in four different decompositions of the adjoint of $\text{G}_2$. 
Only a single branching rule provides a Dynkin index one and yields the expected matter representation from F-theory models.

\begin{table}[htb]
\begin{center}
$
\arraycolsep=5pt\def\arraystretch{1.7}
\begin{array}{|c|l|l|c|}
\hline
\mathfrak{m} & \text{Chains of maximal subalgebras containing $\mathfrak{g}=\text{A}_1$} & \text{Branching Rule} & \text{Index} \\
\hline
\hline
\text{G}_2  & \begin{array}{l}
\begin{tikzcd}[column sep=1cm, ampersand replacement=\&] 
\text{G}_2 \arrow[leftarrow]{r} {(R)}   \&
\text{A}_2 \arrow[leftarrow]{r} {(R)} \& 
\text{A}_1\oplus \text{U}_1 \arrow[leftarrow]{r} {} \& 
\text{A}_1
\end{tikzcd} \\
\begin{tikzcd}[column sep=1cm, ampersand replacement=\&] 
\text{G}_2 \arrow[leftarrow]{r} {(R)}   \&
\text{A}_1\oplus\widetilde{\text{A}}_1 \arrow[leftarrow]{r} {(S)} \& 
\text{A}_1
\end{tikzcd}
\end{array}
& \mathbf{14}=\mathbf{3}\oplus \mathbf{2}^{\bigoplus 4}\oplus\mathbf{1}^{\bigoplus 3} & 1\\
\cline{2-4}
& \begin{array}{l}
\begin{tikzcd}[column sep=1cm, ampersand replacement=\&] 
\text{G}_2 \arrow[leftarrow]{r} {(R)}   \&
\text{A}_2 \arrow[leftarrow]{r} {(S)} \& 
\text{A}_1
\end{tikzcd}\\
\begin{tikzcd}[column sep=1cm, ampersand replacement=\&] 
\text{G}_2 \arrow[leftarrow]{r} {(R)}   \&
\text{A}_1\oplus\widetilde{\text{A}}_1 \arrow[leftarrow]{r} {(S)} \& 
(\text{A}_1\oplus\widetilde{\text{A}}_1)_{diag}
\end{tikzcd}
\end{array}
&  \mathbf{14}=\mathbf{3}^{\bigoplus 3}\oplus \mathbf{5} & 4\\
\cline{2-4}
 &\ \begin{tikzcd}[column sep=1cm, ampersand replacement=\&] 
\text{G}_2 \arrow[leftarrow]{r} {(R)}   \&
\text{A}_1\oplus\widetilde{\text{A}}_1 \arrow[leftarrow]{r} {(S)} \& 
\widetilde{\text{A}}_1
\end{tikzcd}
&  \mathbf{14}=\mathbf{3}\oplus \mathbf{4}^{\bigoplus 2}\oplus\mathbf{1}^{\bigoplus 3} & 3\\
\cline{2-4}
 &\ \begin{tikzcd}[column sep=1cm, ampersand replacement=\&] 
\text{G}_2 \arrow[leftarrow]{r} {(S)}   \&
\text{A}_1
\end{tikzcd}
&  \mathbf{14}=\mathbf{3}\oplus\mathbf{11} & 28\\
\hline
\end{array}
$
\end{center}
\caption{All six chains of maximal subalgebras yielding $\mathfrak{g}=\text{A}_1$ from its possible gauge enhancement $\text{G}_2$. We find four distinct branching rules and only a single branching rule corresponds to index 1.}
\label{tb:su2tog2enhancements}
\end{table}

\subsection*{$\text{A}_1\oplus\text{A}_1$: SO($4$) and Spin($4$) matter contents}
\label{sec:A1A1}

We use chains of maximal subalgebras containing $\text{A}_1\oplus\text{A}_1$ to illustrate an alternative way to obtain the matter content of SO($4$) and Spin($4$)-models studied in \cite{SO4}. This matches with the matter content found by using the method of geography of subalgebras in Section \ref{sec:collisionA1A1}.

The SO($4$) and Spin($4$)-models share the same semisimple gauge algebra $\text{A}_1\oplus\text{A}_1$, which requires a collision of singularities to study its geometry using the Kodaira fibers of type I$_2^{\text{ns}}$, I$_2^{\text{s}}$, and III. The geometric model of these two models are studied in \cite{SO4} where SO(4)-model has the Mordell--Weil group to be $\mathbb{Z}_2$ whereas the Spin(4)-model has a trivial Mordell-Weil group. The geometrically allowed enhancements of these models are derived in \cite{SO4} to be
\begin{align*}
\text{C}_2 ,\quad \text{A}_3 ,\quad \text{incomplete}\ \text{B}_3,\quad \text{incomplete}\ \text{B}_4,
\end{align*}
as depicted in Figure \ref{fig:su2su2enhancementG}. The enhanced algebras are depending on the choices of the Kodaira fibers for the two copies of $\text{A}_1$. When we model with the collision of two fibers of type I$_2^{\text{ns}}$, we get the gauge enhancement $\text{C}_2$. When we have a collision of a fiber I$_2^{\text{ns}}$ and I$_2^{\text{s}}$ or a collision of two fibers of type I$_2^{\text{s}}$, the enhanced gauge algebra is given by $\text{A}_3$. When we utilize the collision of a fiber of type III and a fiber of either type I$_2^{\text{ns}}$ or I$_2^{\text{s}}$, the enhanced fiber is an incomplete fiber of Kodaira type I$_0^*$. Similarly, when there is a collision of two fibers of type III, the enhanced fiber is an incomplete fiber of type I$_0^*$. The Kodaira fiber of type I$_0^*$ can be $\text{G}_2$, $\text{B}_3$, or $\text{D}_4$. However, we can easily conclude that this cannot be a gauge algebra $\text{G}_2$. However, for completeness, we investigate all possible the branching rules of $\text{A}_1\oplus\text{A}_1$ to $\text{G}_2$, $\text{B}_3$, and $\text{D}_4$. Lastly, we can get an enhancement to an incomplete $\text{B}_4$ when we consider a collision of a fiber of type IV$^{\text{ns}}$ and a fiber of type I$_2^{\text{ns}}$. Other than the last case which yield many possible branchings, we consider all the rest scenarios in Table \ref{tb:su2su2BR}.

For the three geometrically allowed enhancements considered,
\begin{align*}
\begin{cases}
&\text{A}_1^1\oplus\text{A}_1^1\to\text{C}_2 ,\quad \text{A}_1^1\oplus\text{A}_1^3\to\text{G}_2 ,\\ 
&\text{A}_1^1\oplus\text{A}_1^1\to\text{A}_3 ,\quad \text{A}_1^2\oplus\text{A}_1^2\to\text{A}_3 , \\
&\text{A}_1^1\oplus\text{A}_1^1\to\text{B}_3,\quad \text{A}_1^2\oplus\text{A}_1^2\to\text{B}_3,\quad \text{A}_1^1\oplus\text{A}_1^2\to\text{B}_3,\quad \text{A}_1^1\oplus\text{A}_1^3\to\text{B}_3, 
\end{cases}
\end{align*}
there are only non-maximal embeddings of $\text{A}_1+\text{A}_1$, each yielding one, two, and four distinct branching rules, respectively, which are listed in Table \ref{tb:su2su2BR}.
We also compute the Dynkin index of each branching rules. 

We  note that the unique embedding of  $\text{A}_1\oplus\text{A}_1$ in G$_2$ has Dynkin index $(1,3)$, whereas all the other enhancements of $\text{A}_1\oplus\text{A}_1$ considered include at least one branching rule of Dynkin index $(1,1)$.  The lack of the existence of a branching rule having Dynkin index one for $\text{G}_2$ is expected from the perspective of the geometry: $\text{G}_2$ is related to the fiber of type I$_0^*$, which never appears geometrically at the collision of two fibers with dual graphs $\text{A}_1$.  This hints that the proposed selection rule \ref{SR} may be used in reverse to rule out the geometrically impossible gauge enhancements or incorrect branching rules for the matter representations when they are not unique.

\renewcommand*{\arraystretch}{1.6}
\begin{longtable}{|c|l|c|}
\hline
$\mathfrak{m}$ &\ Chains of maximal subalgebras containing $\mathfrak{g}=\text{A}_1\oplus\text{A}_1$ & Index\\
\hline
\endhead
\hline
\caption{The adjoint branching rule for the subalgebras $\text{A}_1\oplus\text{A}_1$ in G$_2$, C$_2$ ,A$_3$, B$_3$, and D$_4$. } 
\label{tb:su2su2BR}
\endfirstfoot
\hline
\caption*{Table \ref{tb:su2su2BR} (continued): The adjoint branching rule for the subalgebras $\text{A}_1\oplus\text{A}_1\to$ G$_2$, C$_2$ ,A$_3$, B$_3$, and D$_4$.} 
\endfoot
$\text{G}_2$ & $\mathbf{14}=\mathbf{(3,1)}\oplus\mathbf{(1,3)}\oplus\mathbf{(2,4)}$ & $(1,3)$ \\
\cline{2-2}
 & $\begin{tikzcd}[column sep=1cm, ampersand replacement=\&] 
\text{G}_2 \arrow[leftarrow]{r} {(R)} \&
\text{A}_1\oplus\text{A}_1
\end{tikzcd}$ & \\
\hline
$\text{C}_2$ & $\mathbf{10}=\mathbf{(3,1)}\oplus\mathbf{(1,3)}\oplus\mathbf{(2,2)}$ & $(1,1)$ \\
\cline{2-2}
 & $\begin{tikzcd}[column sep=1cm, ampersand replacement=\&] 
\text{C}_2 \arrow[leftarrow]{r} {(R)} \&
\text{A}_1\oplus\text{A}_1
\end{tikzcd}$ & \\
\hline
$\text{A}_3$ & $(1)\ \mathbf{15}=\mathbf{(3,1)}\oplus\mathbf{(1,3)}\oplus \mathbf{(2,2)}^{\bigoplus 2}\oplus\mathbf{(1,1)}$ & $(1,1)$ \\
\cline{2-2}
 & $\begin{tikzcd}[column sep=1cm, ampersand replacement=\&] 
\text{A}_3 \arrow[leftarrow]{r} {(R)} \&
\text{A}_1\oplus\text{A}_1\oplus\text{U}_1 \arrow[leftarrow]{r} {} \&
\text{A}_1\oplus\text{A}_1
\end{tikzcd}$ & \\
 & $\begin{tikzcd}[column sep=1cm, ampersand replacement=\&] 
\text{A}_3 \arrow[leftarrow]{r} {(S)} \&
\text{C}_2 \arrow[leftarrow]{r} {(R)} \&
\text{A}_1\oplus\text{A}_1
\end{tikzcd}$ & \\
\cline{2-3}
 & $(2)\ \mathbf{15}=\mathbf{(3,1)}\oplus\mathbf{(1,3)}\oplus\mathbf{(3,3)}$ & $(2,2)$ \\
\cline{2-2}
 & $\begin{tikzcd}[column sep=1cm, ampersand replacement=\&] 
\text{A}_3 \arrow[leftarrow]{r} {(S)} \&
\text{A}_1\oplus\text{A}_1
\end{tikzcd}$ &\\
\hline
$\text{B}_3$ & $(1)\ \mathbf{21}=\mathbf{(3,1)}\oplus\mathbf{(1,3)}\oplus\mathbf{(2,2)}^{\bigoplus 3}\oplus\mathbf{(1,1)}^{\bigoplus 3}$ & $(1,1)$ \\
\cline{2-2}
 & $\begin{tikzcd}[column sep=1cm, ampersand replacement=\&] 
\text{B}_3 \arrow[leftarrow]{r} {(R)} \&
\text{A}_3 \arrow[leftarrow]{r} {(R)} \&
\text{A}_1\oplus\text{A}_1\oplus\text{U}_1 \arrow[leftarrow]{r} {} \&
\text{A}_1\oplus\text{A}_1
\end{tikzcd}$ & \\
 & $\begin{tikzcd}[column sep=1cm, ampersand replacement=\&] 
\text{B}_3 \arrow[leftarrow]{r} {(R)} \&
\text{A}_3 \arrow[leftarrow]{r} {(S)} \&
\text{C}_2 \arrow[leftarrow]{r} {(R)} \&
\text{A}_1\oplus\text{A}_1
\end{tikzcd}$ & \\
 & $\begin{tikzcd}[column sep=1cm, ampersand replacement=\&] 
\text{B}_3 \arrow[leftarrow]{r} {(R)} \&
\text{A}_1\oplus\text{A}_1\oplus\widetilde{\text{A}}_1 \arrow[leftarrow]{r} {} \&
\text{A}_1\oplus\text{A}_1
\end{tikzcd}$ & \\
 & $\begin{tikzcd}[column sep=1cm, ampersand replacement=\&] 
\text{B}_3 \arrow[leftarrow]{r} {(R)} \&
\text{C}_2\oplus\text{U}_1 \arrow[leftarrow]{r} {(R)} \&
\text{A}_1\oplus\text{A}_1\oplus\text{U}_1 \arrow[leftarrow]{r} {} \&
\text{A}_1\oplus\text{A}_1
\end{tikzcd}$ & \\
\cline{2-3}
 & $(2)\ \mathbf{21}=\mathbf{(3,1)}^{\bigoplus 2}\oplus\mathbf{(1,3)}^{\bigoplus 2}\oplus\mathbf{(3,3)}$ & $(2,2)$ \\
\cline{2-2}
 & $\begin{tikzcd}[column sep=1cm, ampersand replacement=\&] 
\text{B}_3 \arrow[leftarrow]{r} {(R)} \&
\text{A}_3 \arrow[leftarrow]{r} {(S)} \&
\text{A}_1\oplus\widetilde{\text{A}}_1
\end{tikzcd}$ & \\
\cline{2-3}
 & $(3)\ \mathbf{21}=\mathbf{(3,1)}\oplus\mathbf{(1,3)}\oplus\mathbf{(2,3)}^{\bigoplus 2}\oplus\mathbf{(1,1)}^{\bigoplus 3} $ & $(1,2)$ \\
\cline{2-2}
 & $\begin{tikzcd}[column sep=1cm, ampersand replacement=\&] 
\text{B}_3 \arrow[leftarrow]{r} {(R)} \&
\text{A}_1\oplus\text{A}_1\oplus\widetilde{\text{A}}_1 \arrow[leftarrow]{r} {} \&
\text{A}_1\oplus\text{A}_1
\end{tikzcd}$ & \\
\cline{2-3}
 & $(4)\ \mathbf{21}=\mathbf{(3,1)}\oplus\mathbf{(1,3)}^{\bigoplus 2}\oplus\mathbf{(2,2)}\oplus\mathbf{(2,4)}$ & $(1,3)$ \\
\cline{2-2}
 & $\begin{tikzcd}[column sep=1cm, ampersand replacement=\&] 
\text{B}_3 \arrow[leftarrow]{r} {(S)} \&
\text{G}_2 \arrow[leftarrow]{r} {(R)} \&
\text{A}_1\oplus\text{A}_1
\end{tikzcd}$ &\\
\hline
$\text{D}_4$ & $(1)\ \mathbf{28}=\mathbf{(3,1)}\oplus\mathbf{(1,3)}\oplus \mathbf{(2,2)}^{\bigoplus 4} \oplus\mathbf{(1,1)}^{\bigoplus 6}$ & $(1,1)$ \\
\cline{2-2}
 & $\begin{tikzcd}[column sep=1cm, ampersand replacement=\&] 
\text{D}_4 \arrow[leftarrow]{r} {(R)} \&
\text{A}_1\oplus\text{A}_1\oplus\text{A}_1\oplus\text{A}_1 \arrow[leftarrow]{r} {} \&
\text{A}_1\oplus\text{A}_1
\end{tikzcd}$ &\\
 & $\begin{tikzcd}[column sep=1cm, ampersand replacement=\&] 
\text{D}_4 \arrow[leftarrow]{r} {(R)} \&
\text{A}_3\oplus\text{U}_1 \arrow[leftarrow]{r} {(R)} \&
\text{A}_1\oplus\text{A}_1\oplus\text{U}_1\oplus\text{U}_1 \arrow[leftarrow]{r} {} \&
\text{A}_1\oplus\text{A}_1
\end{tikzcd}$ & \\
 & $\begin{tikzcd}[column sep=1cm, ampersand replacement=\&] 
\text{D}_4 \arrow[leftarrow]{r} {(R)} \&
\text{A}_3\oplus\text{U}_1 \arrow[leftarrow]{r} {(S)} \&
\text{C}_2\oplus\text{U}_1 \arrow[leftarrow]{r} {(R)} \&
\text{A}_1\oplus\text{A}_1\oplus\text{U}_1 \arrow[leftarrow]{r} {} \&
\text{A}_1\oplus\text{A}_1
\end{tikzcd}$ & \\
& $\begin{tikzcd}[column sep=1cm, ampersand replacement=\&] 
\text{D}_4 \arrow[leftarrow]{r} {(S)} \&
\text{B}_3 \arrow[leftarrow]{r} {(R)} \&
\text{A}_3 \arrow[leftarrow]{r} {(R)} \&
\text{A}_1\oplus\text{A}_1\oplus\text{U}_1 \arrow[leftarrow]{r} {} \&
\text{A}_1\oplus\text{A}_1
\end{tikzcd}$ & \\
 & $\begin{tikzcd}[column sep=1cm, ampersand replacement=\&] 
\text{D}_4 \arrow[leftarrow]{r} {(S)} \&
\text{B}_3 \arrow[leftarrow]{r} {(R)} \&
\text{A}_3 \arrow[leftarrow]{r} {(S)} \&
\text{C}_2 \arrow[leftarrow]{r} {(R)} \&
\text{A}_1\oplus\text{A}_1
\end{tikzcd}$ & \\
 & $\begin{tikzcd}[column sep=1cm, ampersand replacement=\&] 
\text{D}_4 \arrow[leftarrow]{r} {(S)} \&
\text{B}_3 \arrow[leftarrow]{r} {(R)} \&
\text{A}_1\oplus\text{A}_1\oplus\widetilde{\text{A}}_1 \arrow[leftarrow]{r} {} \&
\text{A}_1\oplus\text{A}_1
\end{tikzcd}$ &\\
 & $\begin{tikzcd}[column sep=1cm, ampersand replacement=\&] 
\text{D}_4 \arrow[leftarrow]{r} {(S)} \&
\text{B}_3 \arrow[leftarrow]{r} {(R)} \&
\text{C}_2\oplus\text{U}_1 \arrow[leftarrow]{r} {(R)} \&
\text{A}_1\oplus\text{A}_1\oplus\text{U}_1 \arrow[leftarrow]{r} {} \&
\text{A}_1\oplus\text{A}_1
\end{tikzcd}$ & \\
 & $\begin{tikzcd}[column sep=1cm, ampersand replacement=\&] 
\text{D}_4 \arrow[leftarrow]{r} {(S)} \&
\text{B}_3 \arrow[leftarrow]{r} {(R)} \&
\text{A}_1\oplus\text{A}_1\oplus\widetilde{\text{A}}_1 \arrow[leftarrow]{r} {} \&
\text{A}_1\oplus\text{A}_1
\end{tikzcd}$ & \\
& $\begin{tikzcd}[column sep=1cm, ampersand replacement=\&] 
\text{D}_4 \arrow[leftarrow]{r} {(S)} \&
\text{A}_1\oplus\text{C}_2 \arrow[leftarrow]{r} {(R)} \&
\widetilde{\text{A}}_1\oplus\text{A}_1\oplus\text{A}_1 \arrow[leftarrow]{r} {} \&
\widetilde{\text{A}}_1\oplus\text{A}_1
\end{tikzcd}$ & \\
\cline{2-3}
 & $(2)\ \mathbf{28}=\mathbf{(3,1)}^{\bigoplus 3}\oplus\mathbf{(1,3)}^{\bigoplus 3}\oplus\mathbf{(3,3)}\oplus\mathbf{(1,1)}$ & $(2,2)$ \\
\cline{2-2}
 & $\begin{tikzcd}[column sep=1cm, ampersand replacement=\&] 
\text{D}_4 \arrow[leftarrow]{r} {(R)} \&
\text{A}_3\oplus\text{U}_1 \arrow[leftarrow]{r} {(S)} \&
\text{A}_1\oplus\text{A}_1\oplus\text{U}_1 \arrow[leftarrow]{r} {} \&
\text{A}_1\oplus\text{A}_1
\end{tikzcd}$ & \\
 & $\begin{tikzcd}[column sep=1cm, ampersand replacement=\&] 
\text{D}_4 \arrow[leftarrow]{r} {(S)} \&
\text{B}_3 \arrow[leftarrow]{r} {(R)} \&
\text{A}_3 \arrow[leftarrow]{r} {(S)} \&
\text{A}_1\oplus\widetilde{\text{A}}_1
\end{tikzcd}$ & \\
 & $\begin{tikzcd}[column sep=1cm, ampersand replacement=\&] 
\text{D}_4 \arrow[leftarrow]{r} {(S)} \&
\text{A}_1\oplus\text{C}_2 \arrow[leftarrow]{r} {(R)} \&
\text{A}_1\oplus\text{A}_1\oplus\text{U}_1 \arrow[leftarrow]{r} {} \&
\text{A}_1\oplus\text{A}_1
\end{tikzcd}$ & \\
\cline{2-3}
 & $(3)\ \mathbf{28}=\mathbf{(3,1)}\oplus\mathbf{(1,3)}^{\bigoplus 2}\oplus\mathbf{(2,3)}^{\bigoplus 2}\oplus\mathbf{(2,1)}^{\bigoplus 2}\oplus\mathbf{(1,1)}^{\bigoplus 3}$ & $(1,2)$ \\
\cline{2-2}
 & $\begin{tikzcd}[column sep=1cm, ampersand replacement=\&] 
\text{D}_4 \arrow[leftarrow]{r} {(S)} \&
\text{B}_3 \arrow[leftarrow]{r} {(R)} \&
\text{A}_1\oplus\text{A}_1\oplus\widetilde{\text{A}}_1 \arrow[leftarrow]{r} {} \&
\text{A}_1\oplus\widetilde{\text{A}}_1
\end{tikzcd}$ & \\
& $\begin{tikzcd}[column sep=1cm, ampersand replacement=\&] 
\text{D}_4 \arrow[leftarrow]{r} {(S)} \&
\text{A}_1\oplus\text{C}_2 \arrow[leftarrow]{r} {(R)} \&
\widetilde{\text{A}}_1\oplus\text{A}_1\oplus\text{A}_1 \arrow[leftarrow]{r} {} \&
\text{A}_1\oplus\text{A}_1
\end{tikzcd}$ & \\
\cline{2-3}
 & $(4)\ \mathbf{28}=\mathbf{(3,1)}\oplus\mathbf{(1,3)}^{\bigoplus 3}\oplus\mathbf{(2,2)}^{\bigoplus 2}\oplus\mathbf{(2,4)}$ & $(1,3)$ \\
\cline{2-2}
 & $\begin{tikzcd}[column sep=1cm, ampersand replacement=\&] 
\text{D}_4 \arrow[leftarrow]{r} {(S)} \&
\text{B}_3 \arrow[leftarrow]{r} {(S)} \&
\text{G}_2 \arrow[leftarrow]{r} {(R)} \&
\text{A}_1\oplus\text{A}_1
\end{tikzcd}$ & \\
\cline{2-3}
 & $(5)\ \mathbf{28}=\mathbf{(3,1)}\oplus\mathbf{(1,3)}\oplus\mathbf{(3,5)}\oplus\mathbf{(1,7)}$ & $(2,10)$ \\
\cline{2-2}
 & $\begin{tikzcd}[column sep=1cm, ampersand replacement=\&] 
\text{D}_4 \arrow[leftarrow]{r} {(S)} \&
\text{A}_1\oplus\text{C}_2 \arrow[leftarrow]{r} {(S)} \&
\text{A}_1\oplus\text{A}_1
\end{tikzcd}$ & \\
\end{longtable}

\begin{longtable}{|c|l|c|c|}
\hline
$\mathfrak{g}$ & Branching Rules & Choice of $\text{A}_1$ & Index \\
\hline
\endhead
\hline
\caption{
The adjoint branching rule for the subalgebras $\text{A}_1$ in G$_2$, C$_2$ ,A$_3$, B$_3$, and D$_4$.
}
\label{tb:A1viaA1A1toG}
\endfirstfoot
\hline
\caption*{Table \ref{tb:A1viaA1A1toG} (continued): 
The adjoint branching rule for the subalgebras $\text{A}_1$ in G$_2$, C$_2$ ,A$_3$, B$_3$, and D$_4$. 
}
\endfoot
\hline
$\text{G}_2$ & $\begin{tikzcd}[column sep=0.7cm, ampersand replacement=\&] 
\text{G}_2 \arrow[leftarrow]{r} {(R)} \&
\text{A}_1\oplus\widetilde{\text{A}}_1\arrow[leftarrow]{r}\&
\text{A}_1
\end{tikzcd}$ & & \\
\cline{2-2}
& $(1)\ \mathbf{14}=\mathbf{3\oplus 2^{\bigoplus 4}\oplus 1^{\bigoplus 3}}$ & $\text{A}_1$ & $1$ \\
& $(2)\ \mathbf{14}=\mathbf{3\oplus 4^{\bigoplus 2}\oplus 1^{\bigoplus 3}}$ & $\widetilde{\text{A}}_1$ & $3$ \\
& $(3)\ \mathbf{14}=\mathbf{3^{\bigoplus 3}\oplus 5}$ & $\text{A}_{1,\text{diag}}$ & $4$ \\
\hline
$\text{C}_2$ & $\begin{tikzcd}[column sep=0.7cm, ampersand replacement=\&] 
\text{C}_2 \arrow[leftarrow]{r} {(R)} \&
\text{A}_1\oplus\widetilde{\text{A}}_1\arrow[leftarrow]{r}\&
\text{A}_1
\end{tikzcd}$ & & \\
\cline{2-2}
& $(1)\ \mathbf{10}=\mathbf{3\oplus 2^{\bigoplus 2}\oplus 1^{\bigoplus 3}}$ & $\text{A}_1,\, \widetilde{\text{A}}_1$ & $1$ \\
& $(2)\ \mathbf{10}=\mathbf{3^{\bigoplus 3}\oplus 1}$ & $\text{A}_{1,\text{diag}}$ & $2$ \\
\hline
$\text{A}_3$ & $\begin{tikzcd}[column sep=0.7cm, ampersand replacement=\&] 
\text{A}_3 \arrow[leftarrow]{r} {(R)} \&
\text{A}_1\oplus\text{A}_1\oplus\text{U}_1 \arrow[leftarrow]{r} {} \&
\text{A}_1\oplus\widetilde{\text{A}}_1\arrow[leftarrow]{r}\&
\text{A}_1
\end{tikzcd}$ & & \\
& $\begin{tikzcd}[column sep=0.7cm, ampersand replacement=\&] 
\text{A}_3 \arrow[leftarrow]{r} {(S)} \&
\text{C}_2 \arrow[leftarrow]{r} {(R)} \&
\text{A}_1\oplus\widetilde{\text{A}}_1\arrow[leftarrow]{r}\&
\text{A}_1
\end{tikzcd}$ & &\\
\cline{2-2}
& $(1)\ \mathbf{15}=\mathbf{3\oplus 1^{\bigoplus 4}\oplus 2^{\bigoplus 4}}$ & $\text{A}_1,\, \widetilde{\text{A}}_1$ & $1$ \\
& $(2)\ \mathbf{15}=\mathbf{3^{\bigoplus 4}\oplus 1^{\bigoplus 3}}$ & $\text{A}_{1,\text{diag}}$ & $2$ \\
\cline{2-4}
 & $\begin{tikzcd}[column sep=0.7cm, ampersand replacement=\&] 
\text{A}_3 \arrow[leftarrow]{r} {(S)} \&
\text{A}_1\oplus\widetilde{\text{A}}_1\arrow[leftarrow]{r}\&
\text{A}_1
\end{tikzcd}$ & &\\
\cline{2-2}
 & $(2)\ \mathbf{15}=\mathbf{3^{\bigoplus 4}\oplus 1^{\bigoplus 3}}$ & $\text{A}_1,\, \widetilde{\text{A}}_1$ & $2$ \\
 & $(3)\ \mathbf{15}=\mathbf{3^{\bigoplus 3}\oplus 5\oplus 1}$ & $\text{A}_{1,\text{diag}}$ & $4$ \\
\hline
$\text{B}_3$ & $\begin{tikzcd}[column sep=0.5cm, ampersand replacement=\&] 
\text{B}_3 \arrow[leftarrow]{r} {(R)} \&
\text{A}_3 \arrow[leftarrow]{r} {(R)} \&
\text{A}_1\oplus\text{A}_1\oplus\text{U}_1 \arrow[leftarrow]{r} {} \&
\text{A}_1\oplus\widetilde{\text{A}}_1\arrow[leftarrow]{r}\&
\text{A}_1
\end{tikzcd}$ &&\\
 & $\begin{tikzcd}[column sep=0.5cm, ampersand replacement=\&] 
\text{B}_3 \arrow[leftarrow]{r} {(R)} \&
\text{A}_3 \arrow[leftarrow]{r} {(S)} \&
\text{C}_2 \arrow[leftarrow]{r} {(R)} \&
\text{A}_1\oplus\widetilde{\text{A}}_1\arrow[leftarrow]{r}\&
\text{A}_1
\end{tikzcd}$ &&\\
 & $\begin{tikzcd}[column sep=0.5cm, ampersand replacement=\&] 
\text{B}_3 \arrow[leftarrow]{r} {(R)} \&
\text{A}_1\oplus\text{A}_1\oplus\widetilde{\text{A}}_1 \arrow[leftarrow]{r} {} \&
\text{A}_1\oplus\widetilde{\text{A}}_1\arrow[leftarrow]{r}\&
\text{A}_1
\end{tikzcd}$ &&\\
 & $\begin{tikzcd}[column sep=0.5cm, ampersand replacement=\&] 
\text{B}_3 \arrow[leftarrow]{r} {(R)} \&
\text{C}_2\oplus\text{U}_1 \arrow[leftarrow]{r} {(R)} \&
\text{A}_1\oplus\text{A}_1\oplus\text{U}_1 \arrow[leftarrow]{r} {} \&
\text{A}_1\oplus\widetilde{\text{A}}_1\arrow[leftarrow]{r}\&
\text{A}_1
\end{tikzcd}$ && \\
\cline{2-2}
& $(1)\ \mathbf{21}=\mathbf{3 \oplus 2^{\bigoplus 6} \oplus 1^{\bigoplus 6}}$ & $\text{A}_1,\, \widetilde{\text{A}}_1$ & $1$ \\
& $(2)\ \mathbf{21}=\mathbf{3^{\bigoplus 5}\oplus 1^{\bigoplus 6}}$ & $\text{A}_{1,\text{diag}}$ & $2$ \\
\cline{2-4}
 & $\begin{tikzcd}[column sep=0.7cm, ampersand replacement=\&] 
\text{B}_3 \arrow[leftarrow]{r} {(R)} \&
\text{A}_3 \arrow[leftarrow]{r} {(S)} \&
\text{A}_1\oplus\widetilde{\text{A}}_1
\end{tikzcd}$ &&\\
\cline{2-2}
 & $(2)\ \mathbf{21}=\mathbf{3^{\bigoplus 5}\oplus 1^{\bigoplus 6}}$ & $\text{A}_1,\, \widetilde{\text{A}}_1$ & $2$ \\
 & $(4)\ \mathbf{21}=\mathbf{3^{\bigoplus 5}\oplus 5\oplus 1}$ & $\text{A}_{1,\text{diag}}$ & $4$ \\
\hline
$\text{B}_3 $ & $\begin{tikzcd}[column sep=0.7cm, ampersand replacement=\&] 
\text{B}_3 \arrow[leftarrow]{r} {(R)} \&
\text{A}_1\oplus\text{A}_1\oplus\widetilde{\text{A}}_1 \arrow[leftarrow]{r} {} \&
\text{A}_1\oplus\widetilde{\text{A}}_1\arrow[leftarrow]{r}\&
\text{A}_1
\end{tikzcd}$&&\\
\cline{2-2}
 & $(1)\ \mathbf{21}=\mathbf{3\oplus 2^{\bigoplus 6}\oplus 1^{\bigoplus 6}}$ & $\text{A}_1$ & $1$ \\
 & $(2)\ \mathbf{21}=\mathbf{3^{\bigoplus 5}\oplus 1^{\bigoplus 6}}$ & $\widetilde{\text{A}}_1$ & $2$ \\
 & $(3)\ \mathbf{21}=\mathbf{3^{\bigoplus 2}\oplus 4^{\bigoplus 2}\oplus 2^{\bigoplus 2}\oplus 1^{\bigoplus 3}}$ & $\text{A}_{1,\text{diag}}$ & $3$ \\
\cline{2-3}
& $\begin{tikzcd}[column sep=0.7cm, ampersand replacement=\&] 
\text{B}_3 \arrow[leftarrow]{r} {(S)} \&
\text{G}_2 \arrow[leftarrow]{r} {(R)} \&
\text{A}_1\oplus\widetilde{\text{A}}_1\arrow[leftarrow]{r}\&
\text{A}_1
\end{tikzcd}$&&\\
\cline{2-2}
 & $(1)\ \mathbf{21}=\mathbf{3\oplus 2^{\bigoplus 6}\oplus 1^{\bigoplus 6}}$ & $\text{A}_1$ & $1$ \\
 & $(3)\ \mathbf{21}=\mathbf{3^{\bigoplus 2}\oplus 4^{\bigoplus 2}\oplus 2^{\bigoplus 2}\oplus 1^{\bigoplus 3}}$ & $\widetilde{\text{A}}_1$ & $3$ \\
 & $(4)\ \mathbf{21}=\mathbf{3^{\bigoplus 5}\oplus 5\oplus 1}$ & $\text{A}_{1,\text{diag}}$ & $4$ \\
\hline
$\text{D}_4$ & $\begin{tikzcd}[column sep=.5cm, ampersand replacement=\&] 
\text{D}_4 \arrow[leftarrow]{r} {(R)} \&
\text{A}_1\oplus\text{A}_1\oplus\text{A}_1\oplus\text{A}_1 \arrow[leftarrow]{r} {} \&
\text{A}_1\oplus\widetilde{\text{A}}_1\arrow[leftarrow]{r}\&
\text{A}_1
\end{tikzcd}$&&\\
 & $\begin{tikzcd}[column sep=.5cm, ampersand replacement=\&] 
\text{D}_4 \arrow[leftarrow]{r} {(R)} \&
\text{A}_3\oplus\text{U}_1 \arrow[leftarrow]{r} {(R)} \&
\text{A}_1\oplus\text{A}_1\oplus\text{U}_1\oplus\text{U}_1 \arrow[leftarrow]{r} {} \& 
\text{A}_1\oplus\widetilde{\text{A}}_1\arrow[leftarrow]{r}\&
\text{A}_1
\end{tikzcd}$&&\\
 & $\begin{tikzcd}[column sep=.5cm, ampersand replacement=\&] 
\text{D}_4 \arrow[leftarrow]{r} {(R)} \&
\text{A}_3\oplus\text{U}_1 \arrow[leftarrow]{r} {(S)} \&
\text{C}_2\oplus\text{U}_1 \arrow[leftarrow]{r} {(R)} \&
\text{A}_1\oplus\text{A}_1\oplus\text{U}_1 \arrow[leftarrow]{r} {} \& 
\text{A}_1\oplus\widetilde{\text{A}}_1\arrow[leftarrow]{r}\&
\text{A}_1
\end{tikzcd}$&&\\
& $\begin{tikzcd}[column sep=.5cm, ampersand replacement=\&] 
\text{D}_4 \arrow[leftarrow]{r} {(S)} \&
\text{B}_3 \arrow[leftarrow]{r} {(R)} \&
\text{A}_3 \arrow[leftarrow]{r} {(R)} \&
\text{A}_1\oplus\text{A}_1\oplus\text{U}_1 \arrow[leftarrow]{r} {} \& 
\text{A}_1\oplus\widetilde{\text{A}}_1\arrow[leftarrow]{r}\&
\text{A}_1
\end{tikzcd}$&&\\
 & $\begin{tikzcd}[column sep=.5cm, ampersand replacement=\&] 
\text{D}_4 \arrow[leftarrow]{r} {(S)} \&
\text{B}_3 \arrow[leftarrow]{r} {(R)} \&
\text{A}_3 \arrow[leftarrow]{r} {(S)} \&
\text{C}_2 \arrow[leftarrow]{r} {(R)} \& 
\text{A}_1\oplus\widetilde{\text{A}}_1\arrow[leftarrow]{r}\&
\text{A}_1
\end{tikzcd}$&&\\
 & $\begin{tikzcd}[column sep=.5cm, ampersand replacement=\&] 
\text{D}_4 \arrow[leftarrow]{r} {(S)} \&
\text{B}_3 \arrow[leftarrow]{r} {(R)} \&
\text{A}_1\oplus\text{A}_1\oplus\text{A}_1^\prime \arrow[leftarrow]{r} {} \&
\text{A}_1\oplus\widetilde{\text{A}}_1\arrow[leftarrow]{r}\&
\text{A}_1
\end{tikzcd}$&&\\
 & $\begin{tikzcd}[column sep=.5cm, ampersand replacement=\&] 
\text{D}_4 \arrow[leftarrow]{r} {(S)} \&
\text{B}_3 \arrow[leftarrow]{r} {(R)} \&
\text{C}_2\oplus\text{U}_1 \arrow[leftarrow]{r} {(R)} \&
\text{A}_1\oplus\text{A}_1\oplus\text{U}_1 \arrow[leftarrow]{r} {} \&
\text{A}_1\oplus\widetilde{\text{A}}_1\arrow[leftarrow]{r}\&
\text{A}_1
\end{tikzcd}$&&\\
& $\begin{tikzcd}[column sep=.5cm, ampersand replacement=\&] 
\text{D}_4 \arrow[leftarrow]{r} {(S)} \&
\text{A}_1\oplus\text{C}_2 \arrow[leftarrow]{r} {(R)} \&
\text{A}_1^\prime\oplus\text{A}_1\oplus\text{A}_1 \arrow[leftarrow]{r} {} \&
\text{A}_1^\prime\oplus\widetilde{\text{A}}_1\arrow[leftarrow]{r}\&
\text{A}_1
\end{tikzcd}$&&\\
\cline{2-2}
& $(1)\ \mathbf{28}=\mathbf{3\oplus 1^{\bigoplus 9}\oplus 2^{\bigoplus 8}}$ & $\text{A}_1,\, \widetilde{\text{A}}_1$ & $1$ \\
& $(2)\ \mathbf{28}=\mathbf{3^{\bigoplus 6}\oplus 1^{\bigoplus 10}}$ & $\text{A}_{1,\text{diag}}$ & $2$ \\
\cline{2-4}
 & $\begin{tikzcd}[column sep=.5cm, ampersand replacement=\&] 
\text{D}_4 \arrow[leftarrow]{r} {(R)} \&
\text{A}_3\oplus\text{U}_1 \arrow[leftarrow]{r} {(S)} \&
\text{A}_1\oplus\text{A}_1\oplus\text{U}_1 \arrow[leftarrow]{r} {} \&
\text{A}_1\oplus\widetilde{\text{A}}_1\arrow[leftarrow]{r}\&
\text{A}_1
\end{tikzcd}$&&\\
 & $\begin{tikzcd}[column sep=.5cm, ampersand replacement=\&] 
\text{D}_4 \arrow[leftarrow]{r} {(S)} \&
\text{B}_3 \arrow[leftarrow]{r} {(R)} \&
\text{A}_3 \arrow[leftarrow]{r} {(S)} \&
\text{A}_1\oplus\widetilde{\text{A}}_1\arrow[leftarrow]{r}\&
\text{A}_1
\end{tikzcd}$&&\\
 & $\begin{tikzcd}[column sep=.5cm, ampersand replacement=\&] 
\text{D}_4 \arrow[leftarrow]{r} {(S)} \&
\text{A}_1\oplus\text{C}_2 \arrow[leftarrow]{r} {(R)} \&
\text{A}_1\oplus\text{A}_1\oplus\text{U}_1 \arrow[leftarrow]{r} {} \&
\text{A}_1\oplus\widetilde{\text{A}}_1\arrow[leftarrow]{r}\&
\text{A}_1
\end{tikzcd}$&&\\
\cline{2-2}
& $(2)\ \mathbf{28}=\mathbf{3^{\bigoplus 6}\oplus 1^{\bigoplus 10}}$ & $\text{A}_1,\, \widetilde{\text{A}}_1$ & $2$ \\
& $(4)\ \mathbf{28}=\mathbf{3^{\bigoplus 7}\oplus 5\oplus 1^{\bigoplus 2}}$ & $\text{A}_{1,\text{diag}}$ & $4$ \\
\hline
$\text{D}_4$ & $\begin{tikzcd}[column sep=1cm, ampersand replacement=\&] 
\text{D}_4 \arrow[leftarrow]{r} {(S)} \&
\text{B}_3 \arrow[leftarrow]{r} {(R)} \&
\text{A}_1\oplus\text{A}_1\oplus\text{A}_1^\prime \arrow[leftarrow]{r} {} \&
\text{A}_1\oplus\text{A}_1^\prime
\end{tikzcd}$ &&\\
& $\begin{tikzcd}[column sep=1cm, ampersand replacement=\&] 
\text{D}_4 \arrow[leftarrow]{r} {(S)} \&
\text{A}_1\oplus\text{C}_2 \arrow[leftarrow]{r} {(R)} \&
\text{A}_1^\prime\oplus\text{A}_1\oplus\text{A}_1 \arrow[leftarrow]{r} {} \&
\text{A}_1\oplus\text{A}_1
\end{tikzcd}$&&\\
\cline{2-2}
& $(1)\ \mathbf{28}=\mathbf{3\oplus 1^{\bigoplus 9}\oplus 2^{\bigoplus 8}}$ & $\text{A}_1$ & $1$ \\
& $(2)\ \mathbf{28}=\mathbf{3^{\bigoplus 6}\oplus 1^{\bigoplus 10}}$ & $\widetilde{\text{A}}_1$ & $2$ \\
& $(3)\ \mathbf{28}=\mathbf{3^{\bigoplus 3}\oplus 4^{\bigoplus 2}\oplus 2^{\bigoplus 4}\oplus 1^{\bigoplus 3}}$ & $\text{A}_{1,\text{diag}}$ & $3$ \\
\cline{2-4}
 & $\begin{tikzcd}[column sep=1cm, ampersand replacement=\&] 
\text{D}_4 \arrow[leftarrow]{r} {(S)} \&
\text{B}_3 \arrow[leftarrow]{r} {(S)} \&
\text{G}_2 \arrow[leftarrow]{r} {(R)} \&
\text{A}_1\oplus\text{A}_1
\end{tikzcd}$&&\\
\cline{2-2}
 & $(1)\ \mathbf{28}=\mathbf{3\oplus 1^{\bigoplus 9}\oplus 2^{\bigoplus 8}}$ & $\text{A}_1$ & $1$ \\
 & $(3)\ \mathbf{28}=\mathbf{3^{\bigoplus 3}\oplus 2^{\bigoplus 4}\oplus 4^{\bigoplus 2}\oplus 1^{\bigoplus 3}}$ & $\widetilde{\text{A}}_1$ & $3$ \\
 & $(4)\ \mathbf{28}=\mathbf{3^{\bigoplus 7}\oplus 5\oplus 1^{\bigoplus 2}}$ & $\text{A}_{1,\text{diag}}$ & $4$ \\
\cline{2-4}
 & $\begin{tikzcd}[column sep=1cm, ampersand replacement=\&] 
\text{D}_4 \arrow[leftarrow]{r} {(S)} \&
\text{A}_1\oplus\text{C}_2 \arrow[leftarrow]{r} {(S)} \&
\text{A}_1\oplus\text{A}_1
\end{tikzcd}$&&\\
\cline{2-2}
& $(2)\ \mathbf{28}=\mathbf{3^{\bigoplus 6}\oplus 1^{\bigoplus 10}}$ & $\text{A}_1$ & $2$ \\
& $(5)\ \mathbf{28}=\mathbf{3\oplus 5^{\bigoplus 3}\oplus 7\oplus 1^{\bigoplus 3}}$ & $\widetilde{\text{A}}_1$ & $10$ \\
& $(6)\ \mathbf{28}=\mathbf{3^{\bigoplus 3}\oplus 5\oplus 7^{\bigoplus 2}}$ & $\text{A}_{1,\text{diag}}$ & $12$ \\
\hline
\end{longtable}

\section*{Acknowledgements} 
M.E. is supported in part by the National Science Foundation (NSF) grant DMS-1701635 ``Elliptic Fibrations and String Theory''. M.J.K. is supported by a Sherman Fairchild Postdoctoral Fellowship and the National Research Foundation of Korea (NRF) grants NRF-2020R1C1C1007591 and NRF-2020R1A4A3079707. This material is based upon work supported by the U.S. Department of Energy, Office of Science, Office of High Energy Physics, under Award Number DE-SC0011632.
\appendix

%\clearpage 


\begin{thebibliography}{10}


\bibitem{Anderson:2017zfm} 
  L.~B.~Anderson,M.~Esole,  L.~Fredrickson,  and L.~P.~Schaposnik,
  Singular Geometry and Higgs Bundles in String Theory, 
  SIGMA {\bf 14}, 037 (2018)

\bibitem{Aspinwall:1996nk} 
  P.~S.~Aspinwall and M.~Gross,
 The SO(32) heterotic string on a K3 surface,
  Phys.\ Lett.\ B {\bf 387}, 735 (1996)
%
\bibitem{Aspinwall:2000kf} 
  P.~S.~Aspinwall, S.~H.~Katz and D.~R.~Morrison,
    Lie groups, Calabi-Yau threefolds, and F theory,
  Adv.\ Theor.\ Math.\ Phys.\  {\bf 4}, 95 (2000)
  [hep-th/0002012].
%  

\bibitem{Bhardwaj:2018yhy}
L.~Bhardwaj and P.~Jefferson,
``Classifying 5d SCFTs via 6d SCFTs: Rank one,''
JHEP \textbf{07} (2019), 178


\bibitem{Bershadsky:1996nh}
M.~Bershadsky, K.~A. Intriligator, S.~Kachru, D.~R. Morrison, V.~Sadov, and
  C.~Vafa,
\newblock {Geometric singularities and enhanced gauge symmetries}.
\newblock {\em Nucl. Phys.}, B481:215--252, 1996.

\bibitem{Bershadsky:1996nu} 
  M.~Bershadsky and A.~Johansen,
  Colliding singularities in F theory and phase transitions,
  Nucl.\ Phys.\ B {\bf 489}, 122 (1997)
%



\bibitem{Borel}
A.~Borel, and J.~de~ Siebenthal,  ``Les sous-groupes ferm\'es de rang maximum des groupes de Lie clos.'' Commentarii mathematici Helvetici 23 (1949): 200-221. 






\bibitem{Bourbaki.GLA79} N.~Bourbaki, {\it Groups and Lie Algebras. Chap. 7-9.}, 
Translated from the 1975 and 1982 French originals by A.~Pressley. Elements of Mathematics (Berlin).
Springer-Verlag, Berlin, 2005.   



  
  \bibitem{DelZotto:2014hpa} 
  M.~Del Zotto, J.~J.~Heckman, A.~Tomasiello and C.~Vafa,
  6d Conformal Matter,
  JHEP {\bf 1502}, 054 (2015)
  doi:10.1007/JHEP02(2015)054
  [arXiv:1407.6359 [hep-th]].

\bibitem{Diaconescu:1998cn} 
  D.~E.~Diaconescu and R.~Entin,
  ``Calabi-Yau spaces and five-dimensional field theories with exceptional gauge symmetry,''
  Nucl.\ Phys.\ B {\bf 538}, 451 (1999).
% 


\bibitem{deGraaf} 
W.~A.~de~Graaf, 
Constructing semisimple subalgebras of semisimple Lie algebras, 
Journal of Algebra, V325, Issue 1 (2011), pp. 416-430, ISSN 0021-8693.



\bibitem{Dynkin.MSG}
E.B. Dynkin
Maximal subgroups of the classical groups
Tr. Mosk. Mat. Obs., 1 (1952), pp. 39--166
English translation in:
Amer. Math. Soc. Transl., 6 (1957), pp. 245--378

\bibitem{Dynkin.SubA}
E.~B.~Dynkin,   
semisimple subalgebras of semisimple Lie algebras, 
Mat. Sb. (N.S.), 30 (72) (1952), pp. 349--462; English translation in:
Amer. Math. Soc. Transl., 6 (1957), pp. 111--244.
%

\bibitem{Esole:2017csj} 
  M.~Esole,
  Introduction to Elliptic Fibrations,
  Math.\ Phys.\ Stud.\  {\bf 9783319654270}, 247 (2017).

\bibitem{GEO} 
M.~Esole, Matter representations and the geography of semisimple Lie subalgebras, to appear. 

\bibitem{EFY}
M.~Esole, J.~Fullwood, and S.-T. Yau.
\newblock {$D_5$ elliptic fibrations: non-Kodaira fibers and new orientifold
  limits of F-theory}.
\newblock Commun.\ Num.\ Theor.\ Phys.\  {\bf 09}, no. 3, 583 (2015).
%
\bibitem{EJJN1}
M.~Esole, S.~G. Jackson, R.~Jagadeesan, and A.~G. No{\"e}l.
\newblock {Incidence Geometry in a Weyl Chamber I: GL$_n$}, 
\newblock arXiv:1508.03038 [math.RT], 
\newblock Advances in Applied Mathematics 119: 102048, 2020. 

\bibitem{EJJN2} 
  M.~Esole, S.~G.~Jackson, R.~Jagadeesan and A.~G.~No{\"e}l,
  Incidence Geometry in a Weyl Chamber II: $SL_n$, 
  arXiv:1601.05070 [math.RT].
\newblock Advances in Applied Mathematics 119: 102049, 2020. 

\bibitem{G2} 
  M.~Esole, R.~Jagadeesan and M.~J.~Kang,
  The Geometry of G$_2$, Spin(7), and Spin(8)-models,
  arXiv:1709.04913 [hep-th].

\bibitem{SU2SU3}
M.~Esole, R.~Jagadeesan, and M.~J.~Kang, 
48 Crepant paths to SU($2$)$\times$ SU($3$), 
arXiv:1905.05174 [hep-th].

\bibitem{USP4}
M.~Esole and P.~Jefferson,
``USp(4)-models,''
[arXiv:1910.09536 [hep-th]].


\bibitem{SO356}
M.~Esole and P.~Jefferson,
``The Geometry of SO(3), SO(5), and SO(6) models,
[arXiv:1905.12620 [hep-th]].

\bibitem{F4} 
  M.~Esole, P.~Jefferson and M.~J.~Kang,
  The Geometry of F$_4$-Models,
  arXiv:1704.08251 [hep-th].
  
\bibitem{Euler} 
  M.~Esole, P.~Jefferson and M.~J.~Kang,
  Euler Characteristics of Crepant Resolutions of Weierstrass Models,
  arXiv:1703.00905 [math.AG].

\bibitem{Chernchar}
  M.~Esole and M.~J.~Kang,
  To appear.

\bibitem{CharMW}
  M.~Esole and M.~J.~Kang,
  Characteristic numbers of elliptic fibrations with non-trivial Mordell--Weil groups,
  arXiv:1808.07054 [hep-th].

\bibitem{Char}
  M.~Esole and M.~J.~Kang,
Characteristic numbers of crepant resolutions of Weierstrass models,
  arXiv:1807.08755 [hep-th].

\bibitem{SU2G2} 
  M.~Esole and M.~J.~Kang,
  The Geometry of the SU(2)$\times$ G$_2$-model,
  JHEP {\bf 1902}, 091 (2019)

\bibitem{SO4} 
  M.~Esole and M.~J.~Kang,
  Flopping and Slicing: SO(4) and Spin(4)-models,
Adv. Theor. Math. Phys. \textbf{23} (2019) no.4, 1003-1066
doi:10.4310/ATMP.2019.v23.n4.a2
[arXiv:1802.04802 [hep-th]].


\bibitem{EKY2} 
  M.~Esole, M.~J.~Kang and S.~T.~Yau,
  Mordell-Weil Torsion, Anomalies, and Phase Transitions,
  arXiv:1712.02337 [hep-th].

\bibitem{EKY} 
  M.~Esole, M.~J.~Kang and S.~T.~Yau,
  A New Model for Elliptic Fibrations with a rank one Mordell--Weil Group: I. Singular Fibers and Semi-Stable Degenerations,
  arXiv:1410.0003 [hep-th]



\bibitem{Esole:2020alo}
M.~Esole and S.~Pasterski,
``Flops and Fibral Geometry of E$_7$-models,''
[arXiv:2004.06104 [hep-th]].

\bibitem{Esole:2019rzq}
M.~Esole and S.~Pasterski,
``The suspended pinch point and SU($2$)$\times$U($1$) gauge theories,''
[arXiv:1906.07157 [hep-th]].

\bibitem{Esole:2018vnm}
M.~Esole and S.~Pasterski,
``D$_4$-flops of the E$_7$-model,''
[arXiv:1901.00093 [hep-th]].


\bibitem{ES}
M.~Esole and S.~H.~Shao,
``M-theory on Elliptic Calabi-Yau Threefolds and 6d Anomalies,''
[arXiv:1504.01387 [hep-th]].

\bibitem{ESY1}
M.~Esole, S.-H. Shao, and S.-T. Yau.
\newblock {Singularities and Gauge Theory Phases}.
\newblock {\em Adv. Theor. Math. Phys.}, 19:1183--1247, 2015.

\bibitem{ESY2} 
  M.~Esole, S.~H.~Shao and S.~T.~Yau,
  Singularities and Gauge Theory Phases II,
  Adv.\ Theor.\ Math.\ Phys.\  {20}, 683 (2016)

\bibitem{EY} 
  M.~Esole and S.~T.~Yau,
  Small resolutions of SU(5)-models in F-theory,
  Adv.\ Theor.\ Math.\ Phys.\   {17}, no. 6, 1195 (2013)
  


\bibitem{GM1}
A.~Grassi and D.~R. Morrison.
\newblock Group representations and the {E}uler characteristic of elliptically
  fibered {C}alabi-{Y}au threefolds.
\newblock {\em J. Algebraic Geom.}, 12(2):321--356, 2003.
%


%
\bibitem{Hayashi:2014kca}
H.~Hayashi, C.~Lawrie, D.~R. Morrison, and S.~Sch\"afer-Nameki.
\newblock {Box Graphs and Singular Fibers}.
\newblock {\em JHEP}, 1405:048, 2014.
%


\bibitem{Grimm:2011tb}
T.~W.~Grimm, M.~Kerstan, E.~Palti and T.~Weigand,
``Massive Abelian Gauge Symmetries and Fluxes in F-theory,''
JHEP \textbf{12} (2011), 004


\bibitem{Gu:2020fem}
J.~Gu, B.~Haghighat, A.~Klemm, K.~Sun and X.~Wang,
``Elliptic Blowup Equations for 6d SCFTs. IV: Matters,''
[arXiv:2006.03030 [hep-th]].


\bibitem{Heckman:2018jxk} 
  J.~J.~Heckman and T.~Rudelius,
 Top Down Approach to 6D SCFTs,
  J.\ Phys.\ A {\bf 52}, no. 9, 093001 (2019)



\bibitem{IMS}
K.~A. Intriligator, D.~R. Morrison, and N.~Seiberg.
\newblock {Five-dimensional supersymmetric gauge theories and degenerations of
  Calabi-Yau spaces}.
\newblock {\em Nucl.Phys.}, B497:56--100, 1997.
%
\bibitem{Intriligator:1995id}
  K.~A.~Intriligator and N.~Seiberg,
  ``Duality, monopoles, dyons, confinement and oblique confinement in supersymmetric SO(N(c)) gauge theories,''
Nucl. Phys. B \textbf{444} (1995), 125-160

\bibitem{Kan:2020lbe}
N.~Kan, S.~Mizoguchi and T.~Tani,
``Half-hypermultiplets and incomplete/complete resolutions in F-theory,''
JHEP \textbf{08} (2020), 063
doi:10.1007/JHEP08(2020)063

\bibitem{Kuramochi:2020jzz}
R.~Kuramochi, S.~Mizoguchi and T.~Tani,
``Magic square and half-hypermultiplets in F-theory,''
[arXiv:2008.09272 [hep-th]].

\bibitem{Katz:1996xe} 
  S.~H.~Katz and C.~Vafa,
   Matter from geometry,
  Nucl.\ Phys.\ B {\bf 497}, 146 (1997)

\bibitem{Klevers:2017aku} 
  D.~Klevers, D.~R.~Morrison, N.~Raghuram and W.~Taylor,
  ``Exotic matter on singular divisors in F-theory,'
  JHEP {\bf 1711}, 124 (2017)




\bibitem{LS97}
Y.~Laszlo and C.~Sorger, The line bundles on the moduli of parabolic G-bundles over curves and their sections. Ann. scient. \'{E}cole Norm. Sup. 4e series, Vol 30 n  4 (1997), p. 499--525

\bibitem{LorenteGruber}
Lorente, M., Gruber, B. (1972). Classification of semisimple subalgebras of simple Lie algebras. Journal of Mathematical Physics, 13(10), 1639-1663.


\bibitem{Marsano:2011hv}
J.~Marsano and S.~Schafer-Nameki,
``Yukawas, G-flux, and Spectral Covers from Resolved Calabi-Yau's,''
JHEP \textbf{11} (2011), 098
doi:10.1007/JHEP11(2011)098

\bibitem{McKayPatera}
 W. G. McKay and J. Patera. Tables of dimensions, indices, and branching rules for
representations of simple Lie algebras. Marcel Dekker Inc., New York, 1981.

\bibitem{Minchenko}
A.~N.~Minchenko. semisimple subalgebras of exceptional Lie algebras. Tr. Mosk.
Mat. Obs., 67:256--293, 2006. English translation in: Trans. Moscow Math. Soc. 2006,
225--259.
  
\bibitem{Miranda.smooth}
R.~Miranda.
\newblock Smooth models for elliptic threefolds.
\newblock In {\em The birational geometry of degenerations ({C}ambridge,
  {M}ass., 1981)}, volume~29 of {\em Progr. Math.}, pages 85--133.
  Birkh{\"a}user Boston, Mass., 1983.

\bibitem{Morrison:2011mb}
D.~R.~Morrison and W.~Taylor,
``Matter and singularities,''
JHEP \textbf{01} (2012), 022

\bibitem{Morrison:2012np} 
  D.~R.~Morrison and W.~Taylor,
  Classifying bases for 6D F-theory models,
  Central Eur.\ J.\ Phys.\  {\bf 10}, 1072 (2012)
 
 \bibitem{Morrison:1996na}
D.~R.~Morrison and C.~Vafa,
``Compactifications of F theory on Calabi-Yau threefolds. 1,''
Nucl. Phys. B \textbf{473} (1996), 74-92
doi:10.1016/0550-3213(96)00242-8
[arXiv:hep-th/9602114 [hep-th]].


\bibitem{Sadov:1996zm} 
  V.~Sadov,
  Generalized Green--Schwarz mechanism in F theory,
  Phys.\ Lett.\ B {\bf 388}, 45 (1996).
%

\bibitem{Ohmori:2015pia} 
  K.~Ohmori, H.~Shimizu, Y.~Tachikawa and K.~Yonekura,
  6d $\mathcal{N}=\left(1,\;0\right) $ theories on S$^{1}$ /T$^{2}$ and class S theories: part II,
  JHEP {\bf 1512}, 131 (2015)

\bibitem{Polchinski:V1}
J.~Polchinski,
``String theory. Vol. 1: An introduction to the bosonic string,''
doi:10.1017/CBO9780511816079

\bibitem{Sen:1997kz}
A.~Sen,
``A Note on enhanced gauge symmetries in M and string theory,''
JHEP \textbf{09} (1997), 001
doi:10.1088/1126-6708/1997/09/001
arXiv:hep-th/9707123 [hep-th].

\bibitem{Slansky:1981yr} 
  R.~Slansky,
  ``Group Theory for Unified Model Building,''
  Phys.\ Rept.\  {\bf 79}, 1 (1981).

\bibitem{Weigand:2018rez}
T.~Weigand,
``F-theory,''
PoS \textbf{TASI2017} (2018), 016
%
\bibitem{Witten:1995ex}
E.~Witten,
``String theory dynamics in various dimensions,''
Nucl. Phys. B \textbf{443} (1995), 85-126
doi:10.1016/0550-3213(95)00158-O
arXiv:hep-th/9503124 [hep-th].

\bibitem{Yamatsu:2015npn}
  N.~Yamatsu,
  Finite-Dimensional Lie Algebras and Their Representations for Unified Model Building,
  arXiv:1511.08771 [hep-ph].
  
  
  \bibitem{Vafa:1996xn}
C.~Vafa,
``Evidence for F theory,''
Nucl. Phys. B \textbf{469} (1996), 403-418
doi:10.1016/0550-3213(96)00172-1
  
  
\end{thebibliography}
\end{document}